\documentclass{article}

\usepackage{epubtk}

\usepackage{graphicx}
\usepackage{epsf}
\usepackage{longtable}

\usepackage{bbold}

\def\II{\hbox{I\hskip-0.6pt I}}
\def\RR{R\hskip-0.6pt R}
\def\NSNS{N\hskip-0.6pt S\hskip.6pt-\hskip-.6pt N\hskip-0.6pt S}

\font\mybb=msbm10 at 10pt
\def\bb#1{\hbox{\mybb#1}}
\def\bZ {\bb{Z}}
\def\bR {\bb{R}}
\def\bE {\bb{E}}
\def\bC {\bb{C}}
\newcommand{\Cl}{\mathrm{C}\ell}
\newcommand{\fspin}{\mathrm{spin}}

\newcommand{\be}{\begin{equation}}
\newcommand{\ee}{\end{equation}}
\newcommand{\bea}{\begin{eqnarray}}
\newcommand{\eea}{\end{eqnarray}}
\newcommand{\ba}{\begin{array}}
\newcommand{\ea}{\end{array}}
\newcommand{\nn}{\nonumber}

  \newcommand{\gam}{\gamma} 
\newcommand{\G}{\Gamma} 
\newcommand{\ep}{\epsilon}

          \newcommand{\w}{\omega}
          
           \newcommand{\T}{\Theta}

\newcommand{\SL}{\mathrm{SL}}
\newcommand{\U}{\mathrm{U}}
\newcommand{\SU}{\mathrm{SU}}
\newcommand{\SO}{\mathrm{SO}}
\newcommand{\ISO}{\mathrm{ISO}}
\newcommand{\Sp}{\mathrm{Sp}}
\newcommand{\OSp}{\mathrm{OSp}}

\def\CC{{\cal C}} \def\CS{{\cal S}}
 \def\CL{{\cal L}}
\def\CE{{\cal E}} 
 \def\CZ{{\cal Z}}
 
\def\CH{{\cal H}}

\def\5{\bar }  \def\6{\partial } \def\7{\tilde }
\def\8{\hat }  \def\l6{\partial^\leftrightarrow}
\def\hm{\hat{\mu}} \def\hn{\hat{\nu}}

\def\CF{{\cal F}} \def\CG{{\cal G}}

\newcommand{\Gn}{\Gamma_\natural}

\newcommand{\slPi}{/ {\hskip-0.27cm{\Pi}}}

\begin{document}

\title{Brane Effective Actions, Kappa-Symmetry and Applications}

\author{%
\epubtkAuthorData{Joan Sim\'on}{%
School of Mathematics and \\
Maxwell Institute for Mathematical Sciences}{%
j.simon@ed.ac.uk}{%
}}

\date{}
\maketitle

\begin{abstract}
This is a review on brane effective actions, their symmetries and some of its applications. Its first part uncovers the
Green-Schwarz formulation of single M- and D-brane effective actions focusing on kinematical aspects : the identification of their degrees of freedom, the importance of world volume diffeomorphisms and kappa symmetry, to achieve manifest spacetime covariance and supersymmetry, and the explicit construction of such actions in arbitrary on-shell supergravity backgrounds. 

Its second part deals with applications. First, the use of kappa symmetry to determine supersymmetric world volume solitons. This includes their explicit construction in flat and curved backgrounds, their interpretation as BPS states carrying (topological) charges in the supersymmetry algebra and the connection between supersymmetry and hamiltonian BPS bounds. When available, I emphasise the use of these solitons as constituents in microscopic models of black holes. Second, the use of probe approximations to infer about non-trivial dynamics of strongly coupled gauge theories using the AdS/CFT correspondence. This includes expectation values of Wilson loop operators, spectrum information and the general use of D-brane probes to approximate the dynamics of systems with small number of degrees of freedom interacting with larger systems allowing a dual gravitational description.

Its final part briefly discusses effective actions for N D-branes and M2-branes. This includes both SYM theories, their higher order corrections and partial results in covariantising these couplings to curved backgrounds, and the more recent supersymmetric Chern-Simons matter theories describing M2-branes using field theory, brane constructions and 3-algebra considerations.
\end{abstract}

\epubtkKeywords{Kappa-symmetry, supersymmetry, solitons, AdS/CFT, non-abelian gauge theories}

\newpage
\tableofcontents


\section{Introduction}
\label{sec:intro}

Branes have played a fundamental role in the main string theory developments of the last twenty years:
\begin{itemize}
\item[1.] The unification of the different perturbative string theories using {\it duality} symmetries \cite{Hull:1994ys,Witten:1995ex} relied strongly on the existence of non-perturbative supersymmetric states carrying Ramond-Ramond (RR) charge for their first tests.
\item[2.] The discovery of {\it D-branes} as being such non-perturbative states, but still allowing a perturbative description in terms of open strings \cite{Polchinski:1995mt}.
\item[3.] The existence of {\it decoupling} limits in string theory providing {\it non-perturbative} formulations in different backgrounds. This gave rise to {\it Matrix theory} \cite{Banks:1996vh} and the {\it AdS/CFT} correspondence \cite{Maldacena:1997re}. The former provides a non-perturbative formulation of string theory in Minkowski spacetime and the latter in AdS$\times$M spacetimes.
\end{itemize}
At a conceptual level, these developments can be phrased as follows:
\begin{itemize}
\item[1.] Dualities guarantee that fundamental strings are no more fundamental than other dynamical extended objects in the theory, called branes. 
\item[2.] D-branes, a subset of the latter, are non-perturbative states\footnote{Non-perturbative in the sense that their mass goes like $1/g_s$, where $g_s$ is the string coupling constant.} defined as dynamical hypersurfaces where open strings can end. Their weakly coupled dynamics is controlled by the microscopic conformal field theory description of {\it open} strings satisfying Dirichlet boundary conditions. Their spectrum contains massless {\it gauge} fields. Thus, D-branes provide a window into non-perturbative string theory that, at low energies, is governed by supersymmetric {\it gauge theories} in different dimensions. 
\item[3.] On the other hand, any source of energy interacts with gravity. Thus, if the number of branes is large enough, one expects a {\it closed} string description of the {\it same} system. The crucial realisations in \cite{Banks:1996vh} and  
\cite{Maldacena:1997re} are the existence of kinematical and dynamical regimes in which the {\it full} string theory is governed by either of these descriptions : the open or the closed string ones.
\end{itemize}

The purpose of this review is to describe the kinematical properties characterising the supersymmetric gauge theories emerging as brane effective field theories in string and M-theory, and some of their important applications. In particular, I will focus on D-branes, M2-branes and M5-branes. For a schematical representation of the review's content, see figure \ref{fig1}.

These effective theories depend on the number of branes in the system and the geometry they probe.
When a {\it single} brane is involved in the dynamics, these theories are {\it abelian} and there exists a spacetime covariant and manifestly supersymmetric formulation, extending the Green-Schwarz worldsheet one for the superstring. The main concepts I want to stress in this part are
\begin{itemize}
\item[a)] the identification of their dynamical degrees of freedom, providing a geometrical interpretation when available,
\item[b)] the discussion of the world volume gauge symmetries required to achieve {\it spacetime covariance} and {\it supersymmetry}. These will include world volume diffeomorphisms and {\it kappa symmetry},
\item[c)] the description of the couplings governing the interactions in these effective actions, their global symmetries and their interpretation in spacetime,
\item[d)] the connection between spacetime and world volume supersymmetry through gauge fixing,
\item[e)] the description of the regime of validity of these effective actions.
\end{itemize}
For {\it multiple} coincident branes, these theories are supersymmetric {\it non-abelian} gauge field theories. The second main difference from the abelian set-up is the current absence of a spacetime covariant and supersymmetric formulation, i.e. there is {\it no} known world volume diffeomorphic and kappa invariant formulation for them. As a consequence, we do not know how to couple these degrees of freedom to arbitrary (supersymmetric) curved backgrounds, as in the abelian case, and we must study these on an individual background case. 

The covariant abelian brane actions provide a generalisation of the standard charged particle effective actions describing geodesic motion to branes propagating on arbitrary {\it on-shell} supergravity backgrounds. Thus, they offer powerful tools to study the dynamics of string/M-theory in regimes that will be precisely described. In the second part of this review, I describe some of their important {\it applications}. These will be split into two categories : supersymmetric world volume solitons and dynamical aspects of the brane probe approximation. {\it Solitons} will allow me to
\begin{itemize}
\item[a)] stress the technical importance of kappa symmetry in determining these configurations, linking hamiltonian methods with supersymmetry algebra considerations,
\item[b)] prove the existence of string theory BPS states carrying different (topological) charges,
\item[c)] briefly mention microscopic constituent models for certain black holes.
\end{itemize}
Regarding the {\it dynamical} applications, the intention is to provide some dynamical interpretation to specific probe calculations appealing to the AdS/CFT correspondence \cite{Aharony:1999ti} in two main situations
\begin{itemize}
\item[a)] classical on-shell probe action calculations providing a window to strongly coupled dynamics, spectrum and thermodynamics of non-abelian gauge theories by working with appropriate backgrounds with suitable boundary conditions,
\item[b)] probes approximating the dynamics of small systems interacting among themselves and with larger systems, when the latter can be reliably replaced by supergravity backgrounds.
\end{itemize}

\epubtkImage{intro.png}{%
\begin{figure}[h]
  \centerline{\includegraphics[width=150mm]{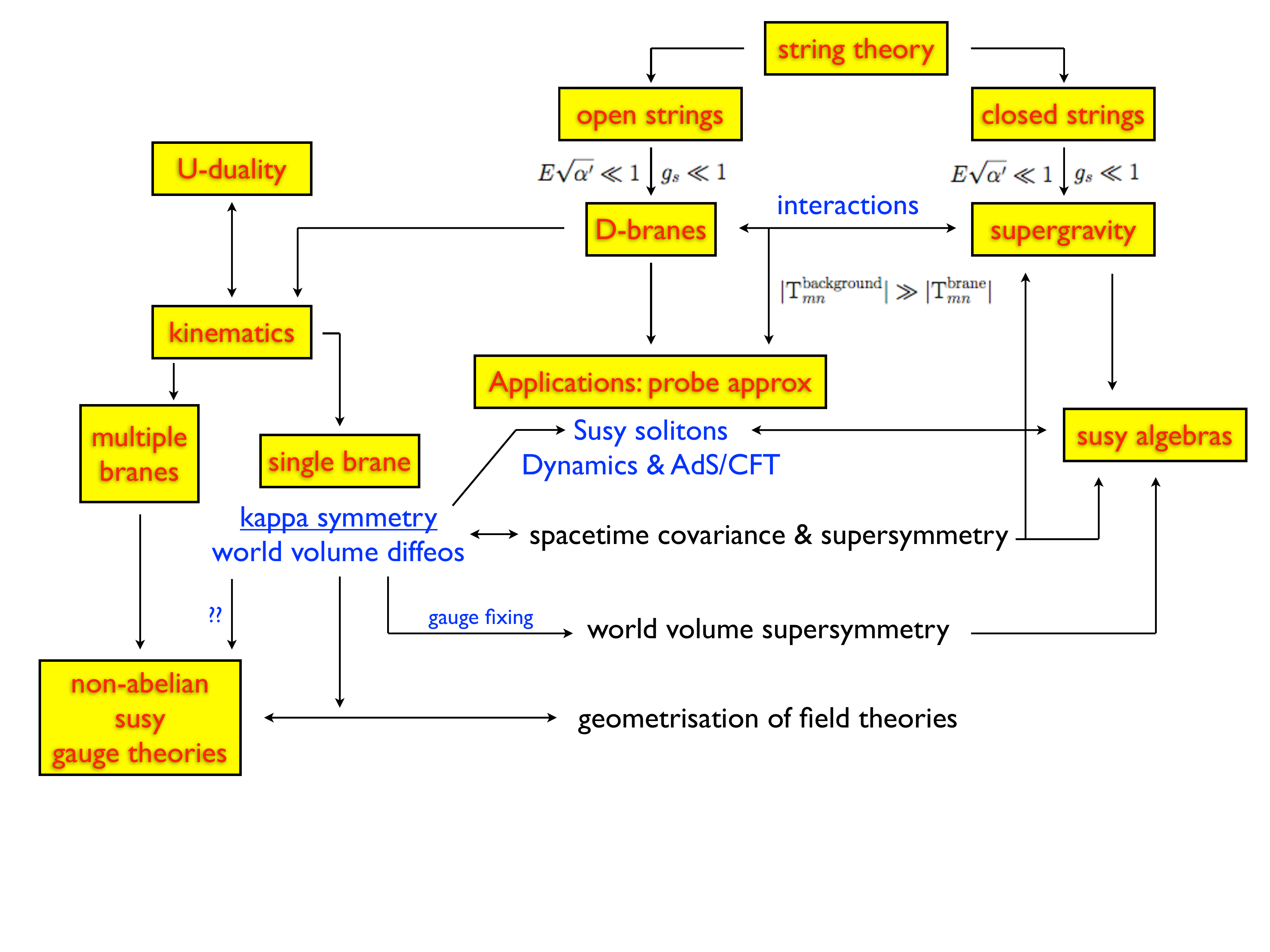}}
   \caption{Layout of the main relations uncovered in this review.}
  \label{fig1}
\end{figure}}

\paragraph{Content of the review :} I start with a very brief review of the Green-Schwarz formulation of the superstring in section \ref{sec:motiv}.This is an attempt at presenting the main features of this formulation since they are universal
in brane effective actions. This is supposed to be a reminder for those readers having a standard textbook knowledge on string theory, or simply as a brief motivation for newcomers, but it is not intended to be self-contained. It also helps to set up the notation for the rest of this review.

Section \ref{sec:bbrane} is fully devoted to the kinematic construction of brane effective actions. After describing the general string theory set-up where these considerations apply, it continues in section \ref{sec:dof} with the identification of the relevant dynamical {\it degrees of freedom}. This is done using open string considerations, constraints from world volume supersymmetry in p+1 dimensions and the analysis of Goldstone mode in supergravity. A second goal in section \ref{sec:dof} is to convey the idea that spacetime covariance and manifest supersymmetry will require these effective actions to be both diffeomorphic and {\it kappa symmetry} invariant, where at this stage the latter symmetry is just conjectured, based on our previous world sheet considerations and counting of {\it on-shell} degrees of freedom. As a warm-up exercise, in section \ref{sec:bosonic}, the {\it bosonic truncations} of these effective actions are constructed, focusing on diffeomorphism invariance, spacetime covariance, physical considerations and a set of non-trivial string theory duality checks that are carried in section \ref{sec:ccheck}. Then, I proceed to discuss the explicit construction of supersymmetric brane effective actions propagating in a fixed Minkowski spacetime in section \ref{sec:bsuperpoincare}. This has the virtue of being explicit and allows to provide a bridge towards the more technical and abstract,  but also more geometrical, superspace formalism, which provides the appropriate venue to covariantise the results in this particular background to couple the brane degrees of freedom to arbitrary curved backgrounds in section \ref{sec:susycurve}. The main result of the latter is that kappa symmetry invariance is achieved whenever the background is an on-shell supergravity background. After introducing the effective actions, I discuss both their global bosonic and fermionic symmetries in section \ref{sec:symmetries}, emphasising the difference between spacetime and world volume (super)symmetry algebras, before and after gauge fixing world volume diffeomorphisms and kappa symmetry. Last, but not least, I include a discussion on the {\it regime of validity} of these effective theories in section \ref{sec:validity}.

Section \ref{sec:solitons} develops the general formalism to study supersymmetric bosonic world volume solitons. It is proved in section \ref{sec:kcond} that any such configuration must satisfy the kappa symmetry preserving condition (\ref{spc}). Reviewing the hamiltonian formulation of these brane effective actions in \ref{sec:hamiltonian}, allows me to  establish a link between supersymmetry, kappa symmetry, supersymmetry algebra bounds and their field theory realisations in terms of hamiltonian BPS bounds in the space of bosonic configurations of these theories. The section finishes connecting these physical concepts to the mathematical notion of calibrations, and their generalisation, in section
\ref{sec:calib}.

In section \ref{sec:soliton-appl}, I apply the previous formalism in many different examples, starting with vacuum infinite branes, and ranging from BIon configurations, branes within branes, giant gravitons, baryon vertex configurations and supertubes. As an outcome of these results, I emphasise the importance of some of these in constituent models of black holes.

In section \ref{sec:adscft}, more dynamical applications of brane effective actions are considered. Here, the reader will be briefly exposed to the reinterpretation of certain on-shell classical brane action calculations in specific curved backgrounds and with appropriate boundary conditions, as holographic duals of strongly coupled gauge theory observables, the existence and properties of the spectrum of these theories, both in the vacuum or in a thermal state, and including their non-relativistic limits. This is intended to be an illustration of the power of the probe approximation technique, rather than a self-contained review of these applications, which lies beyond the scope of these notes. I provide relevant references to excellent reviews covering the material highlighted here in a more exhaustive and pedagogical way.

In section \ref{sec:nonabelian}, I summarise the main kinematical facts regarding the non-abelian description of N D-branes and M2-branes. Regarding D-branes, this includes an introduction to SuperYang-Mills theories in p+1 dimensions, a summary of statements regarding higher order corrections in these effective actions and the more relevant results and difficulties regarding the attempts to covariantise these couplings to arbitrary curved backgrounds.
Regarding M2-branes, I briefly review the more recent supersymmetric Chern-Simons matter theories describing their low energy dynamics, using field theory, 3-algebra and brane construction considerations. The latter allows to provide an explicit example of the geometrisation of supersymmetric field theories provided by brane physics.

The review closes with a brief discussion on some of the topics not uncovered in this review in section \ref{sec:open}. This  includes brief descriptions and references to the superembedding approach to brane effective actions, the description of NS5-branes and KK-monopoles, non-relatistivistic kappa symmetry invariant brane actions, blackfolds or the prospects to achieve a formulation for multiple M5-branes.

In appendices, I provide some brief but self-contained introduction to the superspace formulation of the relevant supergravity theories discussed in this review, describing the explicit constraints required to match the on-shell standard component formulation of these theories. I also include some useful tools to discuss the supersymmetry of AdS spaces and spheres, by embedding them as surfaces in higher dimensional flat spaces. I establish a one--to--one map between the geometrical Killing spinors in AdS and spheres and the covariantly constant Killing spinors in their embedding flat spaces..


\section{The Green-Schwarz superstring: a brief motivation}
\label{sec:motiv}

The purpose of this section is to briefly review the Green-Schwarz (GS) formulation of the superstring. This is not done in a self-contained way, but rather as a very swift presentation of the features that will turn out to be universal in the formulation of brane effective actions.

There exist two distinct formulations for the (super)string:
\begin{itemize}
\item[1.] The {\it worldsheet} supersymmetry formulation, the so called Ramond-Neveu-Schwarz (RNS) formulation\footnote{The discovery of the RNS model of interacting bosons and fermions in d=10 critical dimensions is due to joining the results of the original papers \cite{Ramond:1971gb,Neveu:1971rx}. This was developed further in \cite{Neveu:1971iw,Gervais:1971ji}.}, where supersymmetry in 1+1 dimensions is {\it manifest} \cite{Ramond:1971gb,Neveu:1971rx}.
\item[2.] The GS formulation, where {\it spacetime} supersymmetry is {\it manifest} \cite{Green:1980zg,Green:1981xx,Green:1981yb}.
\end{itemize}

The RNS formulation describes a 1+1 dimensional supersymmetric field theory with degrees of freedom transforming under certain representations of some internal symmetry group. After quantisation, its spectrum turns out to be arranged into supersymmetry multiplets of the internal manifold which is identified with spacetime itself. This formulation has two main disadvantages : the symmetry in the spectrum is {\it not} manifest and its extension to curved spacetime backgrounds is not obvious due to the lack of spacetime covariance.

The GS formulation is based on spacetime supersymmetry as its guiding symmetry principle. It allows a covariant extension to curved backgrounds through the existence of an extra fermionic gauge symmetry, {\it kappa symmetry}, that is 
universally linked to spacetime covariance and supersymmetry, as I will review below and in sections \ref{sec:bbrane} and \ref{sec:solitons}. Unfortunately, its quantisation is much more challenging. The first volume of the Green, Schwarz \& Witten book \cite{Green:1987sp} provides an excellent presentation of both these formulations. Below, I just review its bosonic truncation, construct its supersymmetric extension in Minkowski spacetime, and conclude with an extension to curved backgrounds. 

\paragraph{Bosonic string :} The bosonic GS string action is an extension of the covariant particle action describing geodesic propagation in a fixed curved spacetime with metric $g_{mn}$
\begin{equation}
  S_{{\rm particle}} = -m\int d\tau\,\sqrt{-\dot{X}^m\dot{X}^ng_{mn}(X)}.
\end{equation}
The latter is a one dimensional diffeomorphic invariant action equaling the physical length of the particle trajectory times its mass $m$. Its degrees of freedom $X^m(\tau)$ are the set of maps describing the embedding of the trajectory with affine parameter $\tau$ into spacetime, i.e. the local coordinates $x^m$ of the spacetime manifold become dynamical fields $X^m(\tau)$ on the world line. Diffeomorphisms correspond to the physical freedom in reparameterising the trajectory.

The bosonic string action equals its tension $T_f$ times its area
\begin{equation}
  S_{{\rm string}} = -T_f \int d^2\sigma\,\sqrt{-{\rm det}\, \CG}.
\label{eq:NGstring}
\end{equation}
This is the Nambu-Goto (NG) action \cite{Nambu:1970,Goto:1971ce} : a 1+1 dimensional field theory with coordinates $\sigma^\mu$ $\mu=0,1$ describing the propagation of a lorentzian worldsheet, through the set of embeddings $X^m(\sigma)$ $m=0,1\dots d-1$, in a fixed d-dimensional Lorentzian spacetime with metric $g_{mn}(X)$. Notice this is achieved by computing the determinant of the pullback $\CG_{\mu\nu}$ of the spacetime metric into the worldsheet
\begin{equation}
  \CG_{\mu\nu} = \partial_\mu X^m\partial_\nu X^n\,g_{mn}(X).
\end{equation}
Thus, it is a nonlinear interacting theory in 1+1 dimensions. Furthermore, it is spacetime covariant, invariant under two dimensional diffeomorphisms and its degrees of freedom $\{X^m\}$ are {\it scalars} in two dimensions, but transform as a {\it vector} in d-dimensions. 

Just as point particles can be charged under gauge fields, strings can be charged under 2-forms. The coupling to this extra field is minimal, as corresponds to an electrically charged object, and is described by a Wess-Zumino (WZ) term
\begin{equation}
  S = Q_f \int {\cal B}_{(2)},
\label{eq:WZstring}
\end{equation}
where the charge density $Q_f$ was introduced and ${\cal B}$ stands for the pull back of the d-dimensional bulk 2-form $B_{(2)}$, i.e. 
\begin{equation}
  {\cal B}_{(2)} = \frac{1}{2}\partial_\mu X^m\partial_\nu X^n\,B_{mn}(X)\,d\sigma^\mu\wedge d\sigma^\nu.
\end{equation}
Thus, the total bosonic action is:
\begin{equation}
  S_{{\rm string}} = -T_f \int d^2\sigma\,\sqrt{-{\rm det}\,\CG} + Q_f \int {\cal B}_{(2)}.
\label{eq:bos-string}
\end{equation}
Notice the extra coupling preserves worldsheet diffeomorphism invariance and spacetime covariance. In the string theory context, this effective action describes the propagation of a bosonic string in a closed string background made of a condensate of massless modes (gravitons and Neveu-Schwarz Neveu-Schwarz (NS-NS) 2-form $B_2(X)$). In that case,
\begin{equation}
  T_f = Q_f = \frac{1}{2\pi\alpha^\prime} = \frac{1}{2\pi\ell_s^2},
\end{equation}
where $\ell_s$ stands for the length of the fundamental string.

For completeness, let me stress that at the classical level, the dynamics of the background fields (couplings) is {\it not specified}. Quantum mechanically, the consistency of the interacting theory defined in (\ref{eq:bos-string}) requires the vanishing of the beta functions of the general nonlinear sigma models obtained by expanding the action around a classical configuration when dealing with the quantum path integral. The vanishing of these beta functions requires the background to solve a set of equations that are equivalent to Einstein's equations coupled to an antisymmetric tensor\footnote{The calculations of beta functions in general nonlinear sigma models were done in \cite{AlvarezGaume:1981hn,Friedan:1980jm}. For a general discussion of string theory in curved backgrounds see \cite{Callan:1985ia} or the discussions in books \cite{Green:1987sp,Polchinski:1998rq}.}. This is illustrated in figure \ref{fig5b}.

\paragraph{Supersymmetric extension: } The addition of extra internal degrees of freedom to overcome the existence of a tachyon and the absence of fermions in the bosonic string spectrum, leads to supersymmetry. Thus, besides the spacetime vector $\{X^m\}$, a set of 1+1 scalars fields $\theta^\alpha$ transforming as a spinor under the bulk (internal) Lorentz symmetry $\SO(1,d-1)$ is included.

Instead of providing the answer directly, it is instructive to go over the explicit construction, following \cite{Green:1987sp}.
Motivated by the structure appearing in supersymmetric field theories, one looks for an action invariant under the supersymmetry transformations
\begin{equation}
  \delta\theta^A = \epsilon^A\,, \quad \quad
  \delta X^m = \bar\epsilon^A\Gamma^m\theta^A\,,
\label{eq:susystring}
\end{equation}
where $\epsilon^A$ is a constant spacetime spinor, $\bar\epsilon^A = \epsilon^{At}C$ with $C$ the charge conjugation matrix and the label $A$ counts the amount of independent supersymmetries $A=1,2,\dots {\cal N}$. It is important to stress that both the dimension $d$ of the spacetime and the spinor representation are arbitrary at this stage.

In analogy with the covariant superparticle \cite{Brink:1981nb}, consider the action 
\begin{equation}
  S_1 = -\frac{T_f}{2}\int d^2\sigma\,\sqrt{h}\,h^{\mu\nu}\Pi_\mu\cdot\Pi_\nu.
\label{eq:dbistring}
\end{equation}
This uses the so called Polyakov form the action\footnote{Polyakov used the formulation of classical string theory in terms of an auxiliary world sheet metric \cite{Brink:1976sc,Deser:1976rb} to develop the modern approach to the path integral formulation of string theory in \cite{Polyakov:1981rd,Polyakov:1981re}.} involving an auxiliary 2-dimensional metric $h_{\mu\nu}$. $\Pi_\mu$ stands for the components of the supersymmetric invariant 1-forms
\begin{equation}
  \Pi^m = dX^m + \bar\theta^A\Gamma^md\theta^A,
\end{equation}
whereas $\Pi_\mu\cdot\Pi_\nu\equiv \Pi^m_\mu\Pi^n_\nu\eta_{mn}$.

Even though the constructed action is supersymmetric and 2d diffeomorphic invariant, the number of
{\it on-shell} bosonic and fermionic degrees of freedom does not generically match. To reproduce the supersymmetry in the spectrum derived from the quantisation of the NSR formulation, one must achieve such matching.
 
The standard, by now, resolution to this situation is the addition of an extra term to the action while still preserving supersymmetry. This extra term can be viewed as an extension of the bosonic WZ coupling (\ref{eq:WZstring}), a point I shall return to when geometrically reinterpreting the action so obtained \cite{Henneaux:1984mh}. Following \cite{Green:1987sp}, it turns out the extra term is
\begin{equation}
  S_2 = T_f \int d^2\sigma\left(-\epsilon^{\mu\nu}\partial_\mu X^m\left(\bar\theta^1\Gamma_m\partial_\nu\theta^1 - \bar\theta^2\Gamma_m\partial_\nu\theta^2\right) + \epsilon^{\mu\nu}\bar\theta^1\Gamma^m\partial_\mu\theta^1\bar{\theta}^2\Gamma_m\partial_\nu\theta^2\right).
\label{eq:wzstring}
\end{equation}
Invariance under {\it global} supersymmetry requires, up to total derivatives, the identity 
\begin{equation}
  \delta_\epsilon S_2 = 0 \quad \Longleftrightarrow \quad 2\bar\epsilon\Gamma_m\psi_{[1}\bar\psi_2\Gamma^m\psi_{3]}=0\,,
\end{equation}
for $\left(\psi_1,\,\psi_2,\,\psi_3\right)=(\theta,\,\theta^\prime=\partial\theta/\partial\sigma^1,\,\dot{\theta}=\partial\theta/\partial\sigma^0)$ . This condition restricts the number of spacetime dimensions $d$ and the spinor representation to be
\begin{itemize}
\item d=3 and $\theta$ is Majorana;
\item d=4 and $\theta$ is Majorana or Weyl;
\item d=6 and $\theta$ is Weyl;
\item d=10 and $\theta$ is Majorana-Weyl.
\end{itemize}

Let us focus on the last case which is well known to match the superspace formulation of ${\cal N}=2$ type IIA/B\footnote{Recently, it was pointed out in \cite{Mezincescu:2011nh} that there may exist quantum mechanically consistent superstrings in $d=3$. It remains to be seen whether this is the case.} Despite having matched the spacetime dimension and the spinor representation by the requirement of spacetime supersymmetry under the addition of the extra action term (\ref{eq:wzstring}),  the number of {\it on-shell} bosonic and fermionic degrees of freedom remains unequal. Indeed, Majorana-Weyl fermions in d=10 have 16 real components, which get reduced to 8 on-shell components by Dirac's equation. The extra ${\cal N}=2$ gives rise to a total of 16 on-shell fermionic degrees of freedom, differing from the 8 bosonic ones coming from the ten dimensional vector representation after gauge fixing worldsheet reparameterisations.

The missing ingredient in the above discussion is the existence of an additional fermionic gauge symmetry, {\it kappa symmetry}, responsible for the removal of half of the fermionic degrees of freedom\footnote{The existence of kappa symmetry as a fermionic gauge symmetry was first pointed out in superparticle actions in \cite{deAzcarraga:1982dw,deAzcarraga:1982xk,Siegel:1983hh,deAzcarraga:1987dh}. Though the term kappa symmetry was not used in these references, since it was later coined by Paul K. Townsend, the importance of WZ terms for its existence is already stated in these original works.}. This feature fixes the fermionic nature of the local parameter $\kappa(\sigma)$ and requires $\theta$ to transform by some projector operator
\begin{equation}
  \delta_\kappa\theta = (\mathbb{1}+\Gamma_\kappa)\,\kappa, \quad \quad {\rm with} \quad \quad \Gamma_\kappa^2 = \mathbb{1}.
\label{eq:stringkappa}
\end{equation}
Here $\Gamma_\kappa$ is a Clifford valued matrix depending non-trivially on $\{X^m,\,\theta\}$. The existence of such transformation is proved in \cite{Green:1987sp}.

The purpose of going over this explicit construction is to reinterpret the final action in terms of a more geometrical structure that will be playing an important role in the next section. In more modern language, one interprets $S_1+S_2$ as the action describing a superstring propagating in SuperPoincar\'e \cite{Green:1983wt}. The latter is an example of a supermanifold with local coordinates
$Z^M=\{X^m,\,\theta^\alpha\}$. It uses the analogue of the superfield formalism in global supersymmetric field theories but in supergravity, i.e. with local supersymmetry. The superstring couples to two of these superfields, the supervielbein $E_M^A(Z)$ and the NS-NS 2-form superfield $B_{AC}$, where the index M stands for curved superspace indices, i.e. $M=\{m,\,\alpha\}$, and the index A for tangent flat superspace indices, i.e. $A=\{a,\,\underline{\alpha}\}$\footnote{For a proper definition of these superfields, see appendix \ref{sec:iiab}.}.

In the case of SuperPoincar\'e, the components $E_M^{A}$ are explicitly given by 
\begin{equation}
E_m^{a}=\delta_m^{a}\,, \quad \quad E_\alpha^{\underline{\alpha}}=\delta_\alpha^{\underline{\alpha}}\,, \quad \quad E_m^{\underline{\alpha}}=0\,, \quad \quad E_\alpha^{a}=
\left({\bar \theta}\Gamma^{a}\right)_{\underline{\alpha}}
\delta_\alpha^{\underline{\alpha}}.
\end{equation}
These objects allow us to reinterpret the action $S_1+S_2$ in terms of the pullbacks of these bulk objects into the worldsheet extending the bosonic construction
\begin{eqnarray}
  \CG_{\mu\nu} &=& \Pi_\mu\cdot\Pi_\nu = \partial_\mu Z^M E_M^a(Z)\partial_\nu Z^N E_N^b(Z)\eta_{ab}\,, \nn \\
  {\cal B}_{\mu\nu} &=& \partial_\mu Z^M E_M^A(Z)\partial_\nu Z^N E_N^C(Z)\,B_{AC}(Z).
\label{eq:superspacestring}
\end{eqnarray}
Notice this allows to write both (\ref{eq:dbistring}) and (\ref{eq:wzstring}) in terms of the couplings defined in (\ref{eq:superspacestring}). This geometric reinterpretation is reassuring. If we work in standard supergravity components, Minkowski is an on-shell solution with metric $g_{mn}=\eta_{mn}$, constant dilaton and vanishing gauge potentials, dilatino and gravitino. If we work in superspace, SuperPoincar\'e is a solution to the superspace constraints having non-trivial fermionic components. The ones appearing in the NS-NS 2-form gauge potential are the ones responsible for the WZ term, as it should for an object, the superstring, that is minimally coupled to this bulk massless field.

It is also remarkable to point out that contrary to the bosonic string, where there was no a priori reason why the string tension $T_f$ should be equal to the charge density $Q_f$, its supersymmetric and kappa invariant extension fixes the relation $T_f=Q_f$. This will turn out to be a general feature in supersymmetric effective actions describing the dynamics of supersymmetric states in string theory.

\epubtkImage{RNS-GS.png}{%
\begin{figure}[h]
  \centerline{\includegraphics[width=150mm]{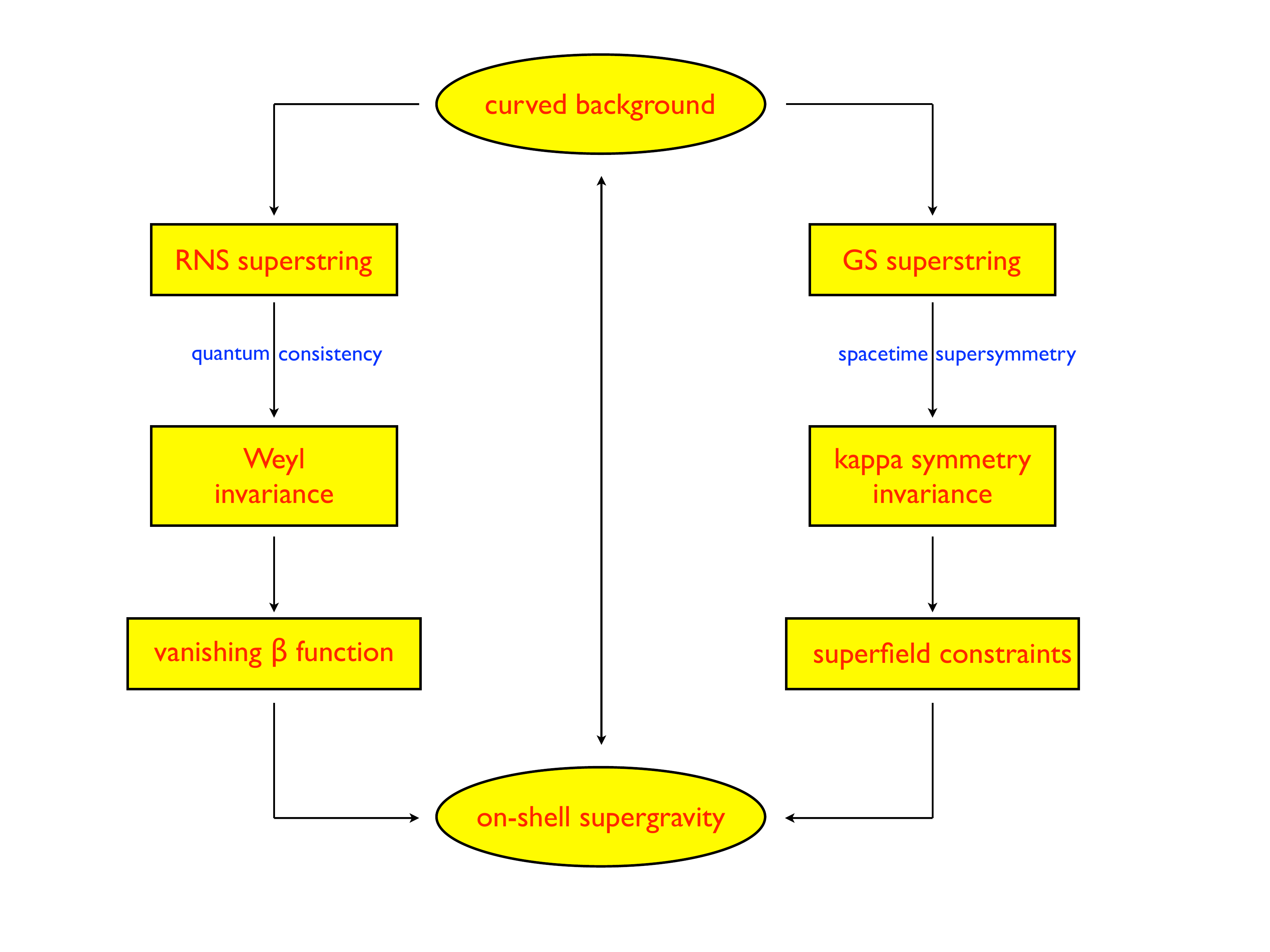}}
   \caption{Different superstring formulations require curved backgrounds to be on-shell.}
  \label{fig5b}
\end{figure}}

\paragraph{Curved background extension :} One of the spins of the superspace reinterpretation in (\ref{eq:superspacestring}) is that it allows its {\it formal} extension to {\it any} ${\cal N}=2$ type IIA/B curved background \cite{Grisaru:1985fv}
\begin{equation}
  S = -\frac{1}{2\pi\alpha^\prime}\int d^2\sigma \sqrt{-{\rm det}\, {\cal G}_{\mu\nu}} + \frac{1}{2\pi\alpha^\prime}\int {\cal B}_{(2)}.
\label{eq:gsusystring}
\end{equation}
The dependence on the background is encoded both in the superfields $E_M^{A}$ and $B_{AC}$.

The counting of degrees of freedom is {\it not} different from the one done for SuperPoincar\'e. Thus, the GS superstring (\ref{eq:gsusystring}) still requires to be kappa symmetry invariant to have an {\it on-shell} matching of bosonic and fermionic degrees of freedom. It was shown in \cite{Bergshoeff:1985su} that the effective action (\ref{eq:gsusystring}) is kappa invariant {\it only} when the ${\cal N}=2$ d=10 type IIA/B background is {\it on-shell}\footnote{See appendix \ref{sec:iiab} for a better discussion on what this means.}. In other words, superstrings can only propagate in properly on-shell backgrounds in the same theory. 

It is important to stress that in the GS formulation, {\it kappa symmetry} invariance requires the background fields to be on-shell, whereas in the RNS formulation, it is quantum {\it Weyl invariance} that ensures this self-consistency condition, as illustrated in figure \ref{fig5b}.

The purpose of the coming section is to explain how these ideas and necessary symmetry structures to achieve a manifestly spacetime covariant and supersymmetric invariant formulation extend to different half-BPS branes in string theory. More precisely, to M2-branes, M5-branes and D-branes.


\section{Brane effective actions}
\label{sec:bbrane}

This review is concerned with the dynamics of {\it low energy} string theory, or M-theory, in the presence of brane degrees of freedom in a regime in which the {\it full} string (M-) theory effective action\footnote{See \cite{Weinberg:1996kr,Pich:1998xt} for reviews and textbooks on what an effective field theory is and what the principles behind them are.} reduces to
\begin{equation}
  S \approx S_{{\rm SUGRA}} + S_{{\rm brane}}.
\label{eq:totalaction}
\end{equation}
The first term in the effective action describes the gravitational sector. It corresponds to ${\cal N}=2$ $d=10$ type IIA/IIB supergravity or ${\cal N}=1$ $d=11$ supergravity, for the systems discussed in this review. The second term 
describes both the brane excitations and their interactions with gravity.

More specifically, I will be concerned with the kinematical properties characterising $S_{{\rm brane}}$ when the latter describes a single brane, though in section \ref{sec:nonabelian}, the extension to many branes will also be briefly discussed. From the perspective of the full string theory, it is important to establish the regime in which the full dynamics is governed by $S_{{\rm brane}}$. This requires to freeze the gravitational sector to its classical on-shell description and to neglect its backreaction into spacetime. Thus, one requires
\begin{equation}
  |T_{mn}^{{\rm background}}|\gg |T_{mn}^{{\rm brane}}|,
\label{eq:noback}
\end{equation}
where $T_{mn}$ stands for the energy-momentum tensor. This is a generalisation of the argument used in particle physics by which one decouples gravity, treating Newton's constant as effectively zero.

Condition (\ref{eq:noback}) is definitely necessary, but {\it not} sufficient, to guarantee the reliability of $S_{{\rm brane}}$.
I will postpone a more thorough discussion of this important point till section \ref{sec:validity}, once the explicit details on the effective actions and the assumptions made for their derivations have been spelled out in the coming subsections. 

Below, I focus on the identification of the degrees of freedom and symmetries to describe brane physics. The distinction between world volume and spacetime symmetries and the preservation of spacetime covariance and supersymmetry will lead us, once again, to the necessity and existence of {\it kappa} symmetry. 

\subsection{Degrees of freedom \& world volume supersymmetry}
\label{sec:dof}

In this section I focus on the identification of the physical degrees of freedom describing a single brane, the constraints derived from world volume symmetries to describe their interactions and the necessity to introduce extra world volume gauge symmetries to achieve {\it spacetime supersymmetry and covariance}. I will first discuss these for Dp-branes, which allow a perturbative quantum open string description, and continue with M2 and M5-branes, applying the lessons learnt from strings and D-branes.

\paragraph{Dp-branes: } {\it Dp-branes} are p+1 dimensional hypersurfaces $\Sigma_{{\rm p+1}}$ where open strings can end. One of the greatest developments in string theory came from the realisation that these objects are dynamical, carry RR charge and allow a perturbative worldsheet description in terms of  open strings satisfying Dirichlet boundary conditions in p+1 dimensions \cite{Polchinski:1995mt}.The quantisation of open strings with such boundary conditions propagating in ten dimensional $\bR^{1,9}$ Minkowski spacetime gives rise to a perturbative spectrum containing a set of massless states that fit into an abelian {\it vector supermultiplet} of the SuperPoincar\'e group in p+1 dimensions \cite{Polchinski:1998rq,Polchinski:1998rr}. Thus, any physical process involving open strings at low enough energy, $E\sqrt{\alpha^\prime}\ll 1$, and at weak coupling, $g_s\ll 1$, should be captured by an effective supersymmetric {\it abelian} gauge theory in p+1 dimensions.

Such vector supermultiplets are described in terms of $\U(1)$ gauge theories to achieve a manifestly $\ISO(1,p)$ invariance, as is customary in gauge theories. In other words, the formulation includes additional polarisations, which are non-physical and can be gauged away. Notice the full $\ISO(1,9)$ of the vacuum is broken by the presence of the Dp-brane itself. This is manifestly reflected in the spectrum. Any attempt to achieve a {\it spacetime} supersymmetric covariant action invariant under the full $\ISO(1,9)$ will require the introduction of both extra degrees of freedom and gauge symmetries. This is the final goal of the GS formulation of these effective actions.

To argue this, analyse the field content of these vector supermultiplets. These include a set of 9-p scalar fields $X^I$ and
a gauge field $V_1$ in p+1 dimensions, describing p-1 physical polarisations. Thus, the total number of massless bosonic degrees of freedom is
\begin{center}
\begin{tabular}{cc}
Dp-brane : & 10-(p+1)+(p-1)=8\,.
\end{tabular}
\end{center}
Notice the number of world volume scalars $X^I$ matches the number of transverse translations broken by the Dp-brane and transform as a vector under the transverse Lorentz subgroup $\SO(9-p)$, which becomes an internal symmetry group. Geometrically, these modes $X^I(\sigma)$ describe the transverse excitations of the brane. This phenomena is rather universal in brane physics and constitutes the essence in the geometrisation of field theories provided by branes in string theory.

Since Dp-branes propagate in ten dimensions, any covariant formalism {\it must} involve a set of ten scalar fields $X^m(\sigma)$, transforming like a vector under the full Lorentz group $\SO(1,9)$. This is the same situation we encountered for the superstring. As such, it should be clear the extra bosonic gauge symmetries required to remove these extra scalar fields are p+1 dimensional diffeomorphisms describing the freedom in embedding $\Sigma_{{\rm p+1}}$ in $\bR^{1,9}$. Physically, the Dirichlet boundary conditions used in the open string description did fix these diffeomorphisms, since they encode the brane location in $\bR^{1,9}$.

What about the fermionic sector ? The discussion here is entirely analogous to the superstring one. This is because spacetime supersymmetry forces us to work with two copies of Majorana-Weyl spinors in ten dimensions. Thus, matching the eight {\it on-shell} bosonic degrees of freedom requires the effective action to be invariant under a new fermionic gauge symmetry. I will refer to this as {\it kappa} symmetry, since it will share all the characteristics of the latter for the superstring.

\paragraph{M-branes :} {\it M-branes} do not have a perturbative quantum formulation. Thus, one must appeal to alternative arguments to identify the relevant degrees of freedom governing their effective actions at low energies. In this subsection, I will appeal to the constraints derived from the existence of supermultiplets in p+1 dimensions satisfying the geometrical property that their number of scalar fields matches the number of transverse dimensions to the M-brane, extending the notion already discussed for the superstring and Dp-branes. Later, I shall review more stringy arguments to check the conclusions obtained below, such as consistency with string/M theory dualities.

Let us start by the more geometrical case of an {\it M2-brane}. This is a 2+1 surface propagating in d=1+10 dimensions. One expects the massless fields to include 8 scalar fields in the bosonic sector describing the M2-brane transverse excitations. Interestingly, this is precisely the bosonic content of an {\it scalar supermultiplet} in d=1+2 dimensions. Since the GS formulation also fits into an scalar supermultiplet in d=1+1 dimensions for a long string, it is natural to expect this is the right supermultiplet for an M2-brane. To achieve spacetime covariance, one must increase the number of scalar fields to eleven$X^m(\sigma)$, transforming as a vector under $\SO(1,10)$ by considering a d=1+2 dimensional diffeomorphic invariant action. If this holds, how do fermions work out ?

First, target space covariance requires the background to allow a superspace formulation in d=1+10 dimensions\footnote{I will introduce this notion more thoroughly in section \ref{sec:bsuperpoincare} and appendix \ref{sec:appcons}.}. Such formulation involves a single copy of d=11 Majorana fermions, which gives rise to a pair of d=10 Majorana-Weyl fermions, matching the superspace formulation for the superstring described in section \ref{sec:motiv}. d=11 Majorana spinors have $2^{[11/2]}=32$ real components, which are further reduced to 16 due to the Dirac equation. Thus, a further gauge symmetry is required to remove {\it half} of these fermionic degrees of freedom, matching the eight bosonic on-shell ones.
Once again, {\it kappa} symmetry will be required to achieve this goal.

What about the M5-brane ? The fermionic discussion is equivalent to the M2-brane one. The bosonic one {\it must} contain a new ingredient. Indeed, geometrically, there are only five scalars describing the transverse M5-brane excitations. These do not match the eight on-shell fermionic degrees of freedom. This is reassuring because there is {\it no} scalar supermultiplet in d=6 dimensions with such number of scalars. Interestingly, there exists a {\it tensor} supermultiplet in d=6 dimensions whose field content involves 5 scalars and a two form gauge potential $V_2$ with {\it self-dual} field strength. The latter involves 6-2 choose 2 physical polarisations, with self-duality reducing these to 3 on-shell degrees of freedom. To keep covariance and describe the right number of polarisations, the d=1+5 theory must be invariant under $\U(1)$ gauge transformations for the 2-form gauge potential. I will later discuss how to keep covariance while satisfying the self-duality constraint.

\paragraph{Brane scan :} World volume supersymmetry generically constrains the low energy dynamics of supersymmetric branes. Even though our arguments were concerned with M2, M5 and D-branes, they clearly are of a more general applicability. This gave rise to the brane scan programme \cite{Achucarro:1987nc,Duff:1992hu,Duff:1994an,Duff:1996zn}. The main idea was to classify the set of supersymmetric branes in different dimensions by matching the number of their transverse dimensions with the number of scalar fields appearing in the list of existent supermultiplets. For an exhaustive classification of all unitary representations of supersymmetry with maximum spin 2, see \cite{Strathdee:1986jr}. Given the importance of scalar, vector and tensor supermultiplets, I list below the allowed multiplets of these kinds in different dimensions indicating the number of scalar fields in each of them \cite{Bergshoeff:1996wk}. 

Let me start with scalar supermultiplets containing
$X$ scalars in $d=p+1$ dimensions, the results being summarised in table
\ref{tab:smultiplet}. Notice we recover the field content of the M2-brane in $d=3$ and $X=8$ and of the superstring in
$d=2$ and $X=8$. 

\begin{longtable}{|c|c|c|c|c|}
\hline
p+1 & X & X & X & X \\ 
\hline
1&1&2&4&8\\
2&1&2&4&8\\
3&1&2&4&8\\
4&&2&4&\\
5&&&4&\\
6&&&4&\\ 
\hline
\caption{Scalar multiplets with X scalars in p+1 worldvolume dimensions.}
\label{tab:smultiplet}
\end{longtable}

Concerning vector supermultiplets with $X$ scalars in $d=p+1$ dimensions,
the results are summarised in table \ref{tab:vmultiplet}. Note that the last column 
describes the field content of all Dp-branes,
starting from the D0-brane $(p=0)$ and finishing with the D9 brane $(p=9)$
filling in all spacetime. Thus, the field content of all Dp-branes matches
with the one corresponding to the different vector supermultiplets in
$d=p+1$ dimensions. This point agrees with the open string conformal
field theory description of D branes. 

\begin{longtable}{|c|c|c|c|c|}
\hline
p+1 & X & X & X & X \\ 
\hline
1&2&3&5&9\\
2&1&2&4&8\\
3&0&1&3&7\\
4&&0&2&6\\
5&&&1&5\\
6&&&0&4\\ 
7&&&&3\\
8&&&&2\\
9&&&&1\\
10&&&&0\\
\hline
\caption{Vector multiplets with X scalar degrees of freedom in 
p+1 worldvolume dimensions.}
\label{tab:vmultiplet}
\end{longtable}

Finally, there 
is just one interesting tensor multiplet with $X=5$ scalars in six dimensions,
corresponding to the forementioned M5 brane, among the six dimensional tensor 
supermultiplets listed in table \ref{tab:tmultiplet}.

\vspace{0.5cm}
\begin{longtable}{|c|c|c|}
\hline
p+1 & X & X \\ 
\hline
6&1&5\\
\hline
\caption{Tensor multiplets with X scalar degrees of freedom in 
p+1 world volume dimensions.}
\label{tab:tmultiplet}
\end{longtable}

\paragraph{Summary :} all half-BPS Dp-branes, M2-branes and M5-branes are described at low energies by effective actions written in terms of supermultiplets in the corresponding world volume dimension. The number of {\it on-shell} bosonic degrees of freedom is 8. Thus, the fermionic content in these multiplets satisfies
\begin{equation}
  8 = \frac{1}{4}M\,{\cal N}\,,
\end{equation}
where $M$ is the number of real components for a minimal spinor representation in D spacetime dimensions and ${\cal N}$ the number of spacetime supersymmetry copies. 

These considerations identified an ${\cal N}=8$ supersymmetric field theory in $d=3$ dimensions (M2 brane), ${\cal N}=(2,0)$ supersymmetric gauge field theory in $d=6$ (M5 brane) and an ${\cal N}=4$ supersymmetric gauge field theory in $d=4$ (D3 brane), as the low energy effective field theories describing their dynamics\footnote{Here, ${\cal N}$ stands for the number of world volume supersymmetries.}. The addition of interactions must be consistent with such $d$ dimensional supersymmetries. 

By construction, an effective action written in terms of these on-shell degrees of freedom can {\it neither} be spacetime covariant {\it nor} $\ISO(1,D-1)$ invariant (in the particular case when branes propagate in Minkowski, as I have assumed all along so far). Effective actions satisfying these two symmetry requirements involve the addition of both extra, non-physical, bosonic and fermionic degrees of freedom. To preserve their non-physical nature, these supersymmetric brane effective actions {\it must} be invariant under additional gauge symmetries
\begin{itemize}
\item world volume diffeomorphisms, to gauge away the extra scalars,
\item kappa symmetry, to gauge away the extra fermions. 
\end{itemize}

\subsubsection{Supergravity Goldstone modes}
\label{sec:goldstone}

Branes carry energy, consequently, they gravitate. Thus, one expects to find gravitational configurations (solitons) carrying the same charges as branes solving the classical equations of motion capturing the effective dynamics of the gravitational sector of the theory. The latter is the effective description provided by type IIA/B supergravity theories, describing the low energy and weak coupling regime of closed strings, and ${\cal N}=1$ d=11 supergravity. The purpose of this section is to argue the existence of the same world volume degrees of freedom and symmetries from the analysis of massless fluctuations of these solitons, applying collective coordinate techniques that are a well-known notion for solitons in standard, non-gravitational, gauge theories. 

In field theory, given a soliton solving its classical equations of motion, there exists a notion of effective action for its small excitations. At low energies, the latter will be controlled by massless excitations, whose number matches the number of broken symmetries by the background soliton \cite{Goldstone:1962es} \footnote{The first examples of this phenomena were reported by Nambu \cite{Nambu:1960xd} and Goldstone \cite{Goldstone:1961eq}.}. These symmetries are {\it global}, whereas all brane solitons are on-shell configurations in supergravity, whose relevant symmetries are {\it local}. To get some intuition for the mechanism operating in our case, it is convenient to review the study of the moduli space of monopoles or instantons in abelian  gauge theories. The collective coordinates describing their small excitations include not only the location of the monopole/instanton, which would match the notion of transverse excitation in our discussion given the pointlike nature of these gauge theory solitons, but also a fourth degree of freedom associated with the breaking of the gauge group \cite{Rajaraman:1982is,Harvey:1996ur}. The reason the latter is particularly relevant to us is because whereas the first set of massless modes are indeed related to the breaking of Poincar\'e invariance, a global symmetry in these gauge theories, the latter has its origin on a {\it large} $\U(1)$ gauge transformation. 

This last observation points out that the notion of collective coordinate can generically be associated to {\it large} gauge transformations, and not simply to global symmetries. It is precisely in this sense how it can be applied to gravity theories and their soliton solutions. In the string theory context, the first work where these ideas were applied was \cite{Callan:1991ky} in the particular set-up of 5-brane solitons in heterotic and type II strings. It was later extended to M2-branes and M5-branes in \cite{Kaplan:1995cp}. In this section, I follow the general discussion in \cite{Adawi:1998ta} for the M2, M5 and D3-branes. These brane configurations are the ones interpolating between Minkowski, at infinity, and Anti-de Sitter (AdS) times a sphere, near their horizons. Precisely for these cases, it was shown in \cite{Gibbons:1993sv} that the world volume theory on these branes is a supersingleton field theory on the corresponding AdS space.

Before discussing the general strategy, let me introduce the on-shell bosonic configurations to be analysed below.
All of them are described by a non-trivial metric and a gauge field carrying the appropriate brane charge.
The multiple M2-brane solution, first found in \cite{Duff:1990xz}, is
\begin{eqnarray}
  ds^2&=&U^{-\frac{2}{3}}\eta_{\mu\nu}dx^\mu dx^\nu+ U^{\frac{1}{3}}\delta_{pq}dy^pdy^q\,, \nonumber \\
	A_3 &=& \pm\frac{1}{3!}\,U^{-1}\varepsilon_{\mu\nu\rho}dx^\mu\wedge dx^\nu\wedge dx^\rho\,.
\label{eq:m2sol}
\end{eqnarray}
Here, and in the following examples, $x^\mu$ describe the longitudinal brane directions, i.e. $\mu = 0,1,2$ for the M2-brane, whereas the transverse cartesian coordinates are denoted by $y^p$, $p=3,\dots 10$. The solution is invariant under $\ISO(1,2)\times \SO(8)$ and is characterised by a single harmonic function $U$ in $\bR^8$
\begin{equation}
  U = 1 + \left(\frac{R}{r}\right)^6, \qquad \quad r^2 = \delta_{pq}y^py^q.
\end{equation}
The structure for the M5-brane, first found in \cite{Gueven:1992hh}, is analogous but differs in the dimensionality of the tangential and transverse subspaces to the brane and in the nature of its charge, electric for the M2-brane and magnetic for the M5-brane below
\begin{eqnarray}
ds^2 &=& U^{-\frac{1}{3}}\eta_{\mu\nu}dx^\mu dx^\nu+U^{\frac{2}{3}}\delta_{mn}dy^mdy^n, \nonumber \\
R_4 = dA_3 &=& \pm\frac{1}{4!}\,\delta^{mn}\partial_m U \varepsilon_{npqsu}dy^p\wedge dy^q\wedge dy^s\wedge dy^s.
\label{eq:m5sol}
\end{eqnarray}
In this case, $\mu=0,1\dots, 5$ and $p=6,\dots ,10$. The isometry group is $\ISO(1,5)\times\SO(5)$ and again it is characterised by a single harmonic function $U$ in  $\bR^5$
\begin{equation}
  U = 1 + \left(\frac{R}{r}\right)^3\,, \qquad \quad r^2 = \delta_{pq}y^py^q\,.
\end{equation}
The D3-brane, first found in \cite{Duff:1991pea}, similarly has a non-trivial metric and self-dual five form RR field strength
\begin{eqnarray}
  ds^2 &=& U^{-\frac{1}{2}}\eta_{\mu\nu}dx^\mu dx^{\nu}+U^{\frac{1}{2}}\delta_{mn}dy^mdy^n, \nonumber \\
  F_5 &=& \pm\frac{1}{5!}(\delta^{mn}\partial_m U \varepsilon_{npqstu}dy^p\wedge dy^q\wedge dy^s\wedge dy^t \wedge dy^u
  \nonumber \\
   & & \quad\thinspace+5\partial_m U^{-1}\varepsilon_{\mu\nu\rho\sigma}dy^m\wedge dx^{\mu}\wedge dx^{\nu}\wedge dx^{\rho}   \wedge dx^{\sigma}),
\end{eqnarray}
with isometry group $\ISO(1,3)\times \SO(6)$. It is characterised by a single harmonic function $U$ in $\bR^6$
\begin{equation}
  U = 1 + \left(\frac{R}{r}\right)^4, \qquad \quad r^2 = \delta_{pq}y^py^q.
\end{equation}
All these brane configurations are half-BPS supersymmetric. The subset of sixteen supercharges being preserved in each case is correlated with the choice of sign in the gauge potentials fixing their charges. I shall reproduce this correlation in the effective brane action in section \ref{sec:susycurve}. 

Let me first sketch the argument behind the generation of massless modes in supergravity theories, where all relevant symmetries are gauge, before discussing the specific details below. Consider a background solution with field content  $\varphi_i^{(0)}$,
where $i$ labels the field, including its tensor character, having an isometry group $G^\prime$. Assume the configuration has some fixed asymptotics with isometry group $G$, so that  $G^\prime\subset G$. The relevant {\it large} gauge transformations $\xi_i(y^p)$ in our discussion are those that act non-trivially at infinity, matching a broken global transformation asymptotically $\epsilon_i$, but differing otherwise in the bulk of the background geometry
\begin{equation}
  \lim_{r\to\infty} \xi_i(y) = \epsilon_i\,.
 \label{eq:largeg}
\end{equation}
In this way, one manages to associate a gauge transformation with a global one, only asymptotically. The idea is then to perturb the configuration $\varphi_i^{(0)}$ by such pure gauge, $\delta_{\xi_i} \varphi_i$ and finally introduce some world volume dependence on the parameter $\epsilon_i$, i.e. $\epsilon_i(x^\mu)$. At that point, the transformation $\delta_{\xi_i} \varphi_i$ is {\it no} longer pure gauge. Plugging the transformation in the initial action and expanding, one can compute the first order correction to the equations of motion fixing some of the ambiguities in the transformation by requiring the perturbed equation to correspond to a massless normalisable mode.

In the following, I explain the origin of the different bosonic and fermionic massless modes in the world volume supermultiplets discussed in the previous section by analysing large gauge diffeomorphisms, supersymmetry and abelian tensor gauge transformations.

\paragraph{Scalar modes : } These are the most intuitive geometrically. They correspond to the breaking of translations along the transverse directions to the brane. The relevant gauge symmetry is clearly a diffeomorphism. Due to the required asymptotic behaviour, it is natural to consider $\epsilon^p = U^s\,\bar\phi^p$, where $\bar\phi^p$ is some constant parameter. Notice the dependence on the harmonic function guarantees the appropriate behaviour at infinity, for any $s$. Dynamical fields transform under diffeomorphisms through Lie derivatives. For instance, the metric would give rise to the {\it pure} gauge transformation
\begin{equation}
  h_{mn} = \CL_\epsilon g_{mn}^{(0)}\,.
\end{equation}
If we allow $\bar\phi^p$ to arbitrarily depend on the world volume coordinates $x^\mu$, i.e. $\bar\phi^p \to \phi^p(x^\mu)$, the perturbation $h_{mn}$ will no longer be pure gauge. If one computes the first order correction to Einstein's equations in supergravity, including the perturbative analysis of the energy momentum tensor, one discovers the lowest order equation of motion satisfied by $\phi^p$ is
\begin{equation}
  \partial^\mu\partial_\mu \phi^p = 0\,,
\end{equation}
for $s=-1$. This corresponds to a massless mode and guarantees its normalisability when integrating the action in the directions transverse to the brane. Later, we will see that the lowest order contribution (in number of derivatives) to the gauge fixed world volume action of M2, M5 and D3-branes in flat space is indeed described by the Klein-Gordon equation.

\paragraph{Fermionic modes : } These must correspond to the breaking of supersymmetry.  Consider the supersymmetry transformation of the eleven dimensional gravitino $\Psi_m$
\begin{equation}
  \delta \Psi_m = \tilde{D}_m \zeta\,,
\end{equation}
where $\tilde{D}$ is some non-trivial connection involving the standard spin connection and some contribution from the gauge field strength. The search for massless fermionic modes leads us to consider the transformation $\zeta = U^s\,\bar\lambda$ for some constant spinor $\bar\lambda$. First, one needs to ensure that such transformation matches, asymptotically, with the supercharges preserved by the brane. Consider the M5-brane, as an example. The preserved supersymmetries are those satisfying $\delta\Psi_m=0$. This forces $s=-\frac{7}{12}$ and fixes the six dimensional chirality of $\bar\lambda$ to be positive, i.e. $\bar\lambda_+$. Allowing the latter to become an arbitrary function of the world volume coordinates $\lambda_+(x^\mu)$, $\delta \Psi_m$ becomes non-pure gauge. Plugging the latter into the original Rarita-Schwinger equation, the linearised equation for the perturbation reduces to
\begin{equation}
  \Gamma^\mu\partial_\mu \lambda_+ = 0\,.
\end{equation}
The latter is indeed the massless Dirac equation for a chiral six dimensional fermion. A similar analysis holds for the M2 and D3-branes. The resulting perturbations are summarised in table \ref{tab:sugragoldstone}

\paragraph{Vector modes : } The spectrum of open strings with Dirichlet boundary conditions includes a vector field. Since the origin of such massless degrees of freedom must be the breaking of some abelian supergravity gauge symmetry, it must be the case that the degree form of the gauge parameter must coincide with the one-form nature of the gauge field. Since this must hold for {\it any} D-brane, the natural candidate is the abelian gauge symmetry associated with the \NSNS\ 2-form
\begin{equation}
  \delta B_2 = d\Lambda_1\,.
\end{equation}
Proceeding as before, one considers a transformation with $\Lambda_1 = U^k\,\bar{V}_1$ for some number $k$ and constant one-form $\bar{V}_1$. When $\bar{V}_1$ is allowed to depend on the world volume coordinates, the perturbation
\begin{equation}
  \delta B_2 = dU^k\wedge V_1(x^\mu)\,,
\end{equation}
becomes physical. Plugging this into the \NSNS\ 2-form equation of motion, one derives $dF=0$ where $F=dV_1$ for both of the four dimensional duality components, for either $k=\pm 1$. Clearly, only $k=-1$ is allowed by the normalisability requirement. 

\paragraph{Tensor modes :} The presence of five transverse scalars to the M5-brane and the requirement of world volume supersymmetry in six dimensions allowed us to identify the presence of a two form potential with self-dual field strength. This must have its supergravity origin in the breaking of the abelian gauge transformation
\begin{equation}
  \delta A_3 = d\Lambda_2\,,
\end{equation}
where indeed the gauge parameter is a two-form. Consider then $\Lambda_2 = U^k\,\bar{V}_2$ for some number $k$ and constant two form $\bar{V}_2$. When $\bar{V}_2$ is allowed to depend on the world volume coordinates, the perturbation
\begin{equation}
  \delta A_3 = dU^k\wedge V_2(x^\mu)\,,
\end{equation}
becomes physical. Plugging this into the $A_3$ equation of motion, we learn that each world volume duality component $\star_x F_3 = \pm F_3$ with $F_3=dV_2$ satisfies the bulk equation of motion if $dF_3=0$ for a specific choice of $k$. More precisely, self-dual components require $k=1$, whereas antiself-dual ones require $k=-1$. Normalisability would fix $k=-1$. Thus, this is the origin of the extra three bosonic degrees of freedom forming the tensor supermultiplet in six dimensions.

\vspace{0.5cm}
The matching between supergravity Goldstone modes and the physical content of world volume supersymmetry multiplets is illustrated in figure \ref{fig7}. Below, a table presents the summary of supergravity Goldstone modes 
\begin{longtable}[c]{p{4 cm}p{2 cm}p{2 cm}p{2 cm}p{2 cm}}
\hline
\hline
 {\bf Symmetry} & & {\bf M2} & {\bf M5} & {\bf D3} \endhead
\hline
\hline
Reparametrisations: & $\varepsilon^m = $ & $U^{-1}
{\bar \phi}^m$ & $U^{-1}{\bar \phi}^m$ & $U^{-1}{\bar \phi}^m$\\
\hline
Local supersymmetry : & $\zeta=$ & $U^{-{2/3}}
{\bar \lambda}_-$ & $U^{-{7/12}}{\bar \lambda}_+ $ & $U^{-{5/8}}{\bar \lambda}_+$ \\
\hline
Tensor gauge symmetry: & $\Lambda =$ & & $U^{-1}
{\bar V}_{(2)}$ & $U^{-1}{\bar V}_{(1)}$ \\
 & & & $({\star}{\bar H}={\bar H})$ & $(i{\star}{\bar F}={\bar F})$\\
\hline
\hline
\\
\caption{\it Summary of supergravity Goldstone modes.}
\label{tab:sugragoldstone}
\end{longtable}
where $\pm$ indeces stand for the chirality of the fermionic zero modes.
In particular, for the M2 brane it describes negative eight dimensional
chirality of the eleven dimensional spinor $\lambda$, while for the
M5 and D3 branes, it describes positive six dimensional and four dimensional
one.

Thus, using purely effective field theory techniques, one is able to derive the spectrum of massless excitations of brane supergravity solutions. This method only provides the lowest order contributions to their equations of motion. The approach followed in this review is to use other perturbative and non-perturbative symmetry considerations in string theory to determine some of the higher order corrections to these effective actions. Our current conclusion, from different perspectives, is that the physical content of these theories must be describable in terms of the massless fields in this section.

\subsection{Bosonic actions}
\label{sec:bosonic}

After the identification of the relevant degrees of freedom and gauge symmetries governing brane effective actions, I focus on the construction of their bosonic truncations, postponing their supersymmetric extensions to sections \ref{sec:bsuperpoincare} and \ref{sec:susycurve}. The main goal below will be to couple brane degrees of freedom to arbitrary curved backgrounds in a world volume diffeomorphic invariant way.

I shall proceed in order of increasing complexity, starting with the M2-brane effective action which is purely geometric, continuing with D-branes and their one form gauge potentials and finishing with M5-branes including their self-dual three form field strength\footnote{For earlier reviews on D-brane effective actions and on M-brane interactions, see \cite{Johnson:2000ch} and \cite{Berman:2007bv}, respectively.}.

\paragraph{Bosonic M2-brane :} In the absence of world volume gauge field excitations, all brane effective actions must satisfy two physical requirements
\begin{itemize}
\item[1.] Geometrically, branes are p+1 hypersurfaces $\Sigma_{p+1}$ propagating in a fixed background with metric $g_{mn}$. Thus, their effective actions should account for their world volumes.
\item[2.] Physically, all branes are electrically charged under some appropriate spacetime p+1 gauge form $C_{p+1}$. Thus, their effective actions should contain a minimal coupling accounting for the brane charges.
\end{itemize}
Both requirements extend the existent effective action describing either a charged particle $(p=0)$ or a string $(p=1)$. Thus, the universal description of the purely scalar field $X^m$ brane degrees of freedom must be of the form
\begin{equation}
  S_p = -T_p\,\int_{\Sigma_{p+1}} d^{p+1}\sigma\,\sqrt{-{\rm det}\,\CG} + Q_p\,\int_{\Sigma_{p+1}}{\cal C}_{p+1}\,,
\label{eq:mstructure}
\end{equation}
where $T_p$ and $Q_p$ stand for the brane tension and charge density\footnote{Since I am not considering supersymmetric branes at this point, $|Q_p|=T_p$ is not a necessary condition.}. The first term computes the brane world volume from the induced metric $\CG_{\mu\nu}$
\begin{equation}
  \CG_{\mu\nu}=\partial_\mu X^m\partial_\nu X^n g_{mn}(X),
\end{equation}
whereas the second WZ term ${\cal C}_{p+1}$ describes the pullback of the target space p+1 gauge field $C_{p+1}(X)$ under which the brane is charged
\begin{equation}
  {\cal C}_{(p+1)} = \frac{1}{(p+1)!}\epsilon^{\mu_1\dots\mu_{p+1}}\partial_{\mu_1}X^{m_1}\dots \partial_{\mu_{p+1}}X^{m_{p+1}}\,C_{m_1\dots m_{p+1}}(X).
\end{equation}
At this stage, one assumes all branes propagate in a background with lorentzian metric $g_{mn}(X)$ coupled to other matter fields, such as $C_{p+1}(X)$, whose dynamics are neglected in this approximation. In string theory, these background fields correspond to the bosonic truncation of the supergravity multiplet and their dynamics at low energy is governed by supergravity theories. More precisely, M2 and M5-branes propagate in d=11 supergravity backgrounds, i.e. $m,n=0,1,\dots 10$, and they are electrically charged under the gauge potential $A_3(X)$ and its 6-form dual potential $A_6$, respectively (see appendix \ref{sec:appcons} for conventions). D-branes propagate in d=10 type IIA/B backgrounds and the set $\{C_{p+1}(X)\}$ correspond to the set of RR gauge potentials in these theories, see (\ref{eq:fsum}).

The relevance of the minimal charge coupling can be understood by considering the full effective action involving both brane and gravitational degrees of freedom (\ref{eq:totalaction}). Restricting ourselves to the kinetic term for the target space gauge field, i.e. $R=dC_{p+1}$, the combined action can be written as
\begin{equation}
\int_{{\cal M}_D}\left(\,\frac{1}{2} R\wedge\star R+Q_p\hat n\wedge C_{p+1}\,\right)
\label{Haction}\,.
\end{equation}
Here ${\cal M}_D$ stands for the D-dimensional spacetime, whereas $\hat n$ is a $(D-p-1)$-form whose components are those of an epsilon tensor normal to the brane having a $\delta$-function 
support on the world volume\footnote{This is the correct way to compute energy the momentum tensor due to the coupling of branes to gravity. The energy carried by such a brane must be localised on its p+1 dimensional world volume.}. Thus, the bulk equation of motion for the gauge potential $C_{p+1}$ acquires a source term whenever a brane exists. Since the brane charge is computed as the integral of $\star R$ over any topological $(D-p-2)$-sphere surrounding it, one obtains
\begin{equation}
\int_{\Sigma_{D-p-2}}\star R=\int_{B_{D-p-1}}d\star R=\int_{B_{D-p-1}}Q_p\hat n=Q_p,
\end{equation}
where the equation of motion was used in the last step. Thus, minimal WZ couplings do capture the brane physical charge.

Since M2-branes do not involve any gauge field degree of freedom, the above discussion uncovers {\it all} its bosonic degrees of freedom. Thus, one expects its bosonic effective action to be
\begin{equation}
S_{M2}=-T_{{\rm M2}}\int d^3\sigma\,\sqrt{-{\rm det}\,\CG} + Q_{{\rm M2}}\int {\cal A}_{3}\,,
\label{bosonicm2}
\end{equation} 
in analogy with the bosonic worldsheet string action. If (\ref{bosonicm2}) is viewed as the bosonic truncation of a supersymmetric M2-brane, then $|Q_{{\rm M2}}|=T_{{\rm M2}}$. Besides its manifest spacetime covariance and its invariance under world volume diffeomorphisms infinitesimally generated by
\begin{equation}
  \delta_\xi X^m = {\cal L}_\xi X^m = \xi^\mu\partial_\mu X^m,
\end{equation}
this action is also quasi-invariant (invariant up to total derivatives) under the target space gauge transformation  $\delta_\Lambda A_3=d\Lambda_2$ leaving ${\cal N}=1$ d=11 supergravity invariant, as reviewed in equation (\ref{aonea}) of appendix \ref{sec:d11app}. This is reassuring given that the full string theory effective action (\ref{eq:totalaction}) describing both gravity and brane degrees of freedom involves both actions.

\paragraph{Bosonic D-branes :} Due to the perturbative description in terms of open strings \cite{Polchinski:1995mt}, D-brane effective actions can, in principle, be determined by explicit calculation of appropriate open string disk amplitudes.
Let me first discuss the dependence on gauge fields in these actions. Early bosonic open string calculations in background gauge fields \cite{Abouelsaood:1986gd}, allowed to determine the effective action for the gauge field, with purely Dirichlet boundary conditions \cite{Fradkin:1985qd} or with mixed boundary conditions \cite{Leigh:1989jq}, giving rise to a non-linear generalisation of Maxwell's electromagnetism originally proposed by Born \& Infeld in \cite{Born:1934gh}:
\begin{equation}
-\int_{\Sigma_{p+1}}d^{p+1}\sigma \,\sqrt{-\det(\eta_{\mu\nu}+ 2\pi\alpha^\prime F_{\mu\nu})}.
\end{equation}
I will refer to this non-linear action as the Dirac-Born-Infeld (DBI) action. Notice this is an exceptional situation in string theory in which an infinite sum of different $\alpha^\prime$ contributions is analytically computable. This effective action ignores {\it any} contribution from the derivatives of the field strength $F$, i.e. $\partial_\mu F_{\nu\rho}$ terms or higher derivative operators. Importantly, it was shown in  \cite{Abouelsaood:1986gd} that the first such corrections, for the bosonic open string, are compatible with the DBI structure.

Having identified the non-linear gauge field dependence, one is in a position to include the dependence on the embedding scalar fields $X^m(\sigma)$ and the coupling with non-trivial background closed string fields. Since in the absence of world volume gauge field excitations, D-brane actions should reduce to (\ref{eq:mstructure}), it is natural to infer the right answer should involve
\begin{equation}
\sqrt{-\det(\CG_{\mu\nu}+ 2\pi\alpha^\prime F_{\mu\nu})},
\label{eq:bosdbi}
\end{equation}
using the general arguments of the preceding paragraphs. Notice this action does {\it not} include any contribution from acceleration and higher derivative operators involving scalar fields, i.e. $\partial_{\mu\nu}X^m$ terms and/or higher derivative terms\footnote{The importance of these assumptions will be stressed when discussing the regime of validity of brane effective actions in section \ref{sec:validity}.}. This proposal has nice properties under T-duality \cite{Bachas:1995kx,Bergshoeff:1996ui,Alvarez:1996up,Bergshoeff:1996cy}, which I will explore in detail in section \ref{sec:tcov} as a non-trivial check on
(\ref{eq:bosdbi}). In particular, it will be checked that absence of acceleration terms is compatible with T-duality. 

The DBI action is a natural extension of the NG action for branes, but it does not capture all the relevant physics, even in the absence of acceleration terms, since it misses important background couplings, responsible for the WZ terms appearing for strings and M2-branes. Let me stress the main two issues separately:
\begin{itemize}
\item[1.] The functional dependence on the gauge field $V_1$ in a general closed string background. D-branes are hypersurfaces where open strings can end. Thus, open strings {\it do} have endpoints. This means that the WZ term describing such open string is not invariant under the target space gauge transformation $\delta B_2=d\Lambda_1$
\begin{equation}
\delta\int_{\Sigma_2}b=\int_{\Sigma_2}d\Lambda=\int_{\6\Sigma_2}\Lambda,
\label{eq:noninvar}
\end{equation}
due to the presence of boundaries. These are the D-branes themselves, which see these endpoints as charge point sources. The latter has a minimal coupling of the form $\int_{\6\Sigma_2}V_{1}$, whose variation cancels (\ref{eq:noninvar}) if the gauge field transforms as $\delta V_{1}=dX^m(\sigma)\Lambda_m$ under the bulk gauge transformation. Since D-brane effective actions must be invariant under these target space gauge symmetries, this physical argument determines that {\it all} the dependence on the gauge field $V_{1}$ should be through the gauge invariant combination $\CF=2\pi\alpha^\prime dV_{1}-{\cal B}$.
\item[2.] The coupling to the dilaton. The D-brane effective action is an open string tree level action, i.e. the self-interactions of open strings and their couplings to closed string fields come from conformal field theory disk amplitudes. Thus, the brane tension should include a $g_s^{-1}$ factor coming from the expectation value of the closed string dilaton $e^{-\phi}$. Both these considerations lead us to consider the DBI action
\begin{equation}
  S_{DBI} = -T_{{\rm D_p}} \int d^{p+1}\sigma\,e^{-\phi}\sqrt{-{\rm det}\,(\CG + \CF)},
\end{equation}
where $T_{{\rm D_p}}$ stands for the D-brane tension.
\item[3.] The WZ couplings. Dp-branes are charged under the RR potential $C_{p+1}$. Thus, their effective actions should include a minimal coupling to the pullback of such form. Such coupling would not be invariant under the target space gauge transformations (\ref{TargetSpaceGaugeTransf}). To achieve this invariance in a way compatible with the bulk Bianchi identities (\ref{BianchiIdentities}),  the D-brane WZ action must be of the form
\begin{equation}
\int_{\Sigma_{p+1}} \CC\wedge e^{\CF},
\end{equation}
where $\CC$ stands for the corresponding pullbacks of the target space RR potentials $C_r$ to the world volume, according to the definition given in (\ref{eq:fsum}). Notice this involves more terms than the mere minimal coupling to the bulk RR potential $C_{p+1}$. An important physical consequence of this fact will be that turning on non-trivial gauge fluxes on the brane can induce non-tivial lower dimensional D-brane charges, extending the argument given above for the minimal coupling \cite{Douglas:1995bn}. This property will be discussed in more detail in the second part of this review. For a discussion on how to extend these couplings to massive type IIA supergravity, see \cite{Green:1996bh}. 
\end{itemize}

Putting together all previous arguments, one concludes the final form of the bosonic D-brane action is
\footnote{There actually exist further gravitational interaction 
terms necessary for the cancellation of anomalies \cite{Green:1996dd}, but we 
will always omit them in our discussions concerning D-brane effective 
actions.} :
\begin{equation}
S_{{\rm Dp}}=-T_{{\rm Dp}}\int_{\Sigma_{p+1}}d^{p+1}\sigma \,e^{-\phi}\sqrt{-\det(\CG+\CF)}+
Q_{{\rm Dp}}\int_{\Sigma_{p+1}} \CC\wedge e^{\CF}.
\label{bosonicDp}
\end{equation}
If one views this action as the bosonic truncation of a supersymmetric D-brane,  the D-brane charge density equals its tension in absolute value, i.e. $|Q_{{\rm Dp}}|= T_{{\rm Dp}}$. The latter can be determined from first principles to be \cite{Polchinski:1995mt,Bachas:1995kx}
\begin{equation}
T_{{\rm Dp}}= \frac{1}{g_s\sqrt{\alpha'}}  \frac{1}{ (2 \pi \sqrt{\alpha'})^{p}}.
\label{eq:tdbrane}
\end{equation}

\paragraph{Bosonic covariant M5-brane :}  The bosonic M5-brane degrees of freedom involve scalar fields and a world volume 2-form with self-dual field strength. The former are expected to be described by similar arguments to the ones presented above.
The situation with the latter is more problematic given the tension between Lorentz covariance and the self-duality constraint.
This problem has a fairly long history, starting with electric--magnetic duality and the Dirac monopole
problem in Maxwell theory, see \cite{Blagojevic:1985sh} and references therein, and more recently, in connection with the formulation of supergravity theories such as type IIB, with the self-duality of the field strength of the \RR\ 4-form gauge potential. 
There are several solutions in the literature based on different formalisms :
\begin{itemize}
\item[1.] One natural option is to give-up Lorentz covariance and work with non-manifestly Lorentz invariant actions. This was the approach followed in \cite{Perry:1996mk} for the M5-brane, building on previous work \cite{Floreanini:1987as,Henneaux:1988gg,Schwarz:1993vs}.
\item[2.] One can introduce an infinite number of auxiliary (non-dynamical) fields to achieve a covariant formulation. This is the approach followed in \cite{McClain:1990sx,Wotzasek:1990zr,Martin:1994wf,Devecchi:1996cp,Bengtsson:1996fm,Berkovits:1996tn,Berkovits:1996rt,Berkovits:1996em}.
 \item[3.] One can follow the covariant approach due to Pasti, Sorokin and Tonin (PST-formalism) \cite{Pasti:1995tn,Pasti:1996vs}, in which a single auxiliary field is introduced in the action with a non-trivial non polynomial dependence on it. The resulting action has extra gauge symmetries. These allow to recover the structure in \cite{Perry:1996mk} as a gauge fixed version of the PST formalism.
 \item[4.] Another option is to work with a lagrangian that does not imply  the self-duality condition but allows it, leaving the implementation of this condition to the path integral. This is the approach followed by Witten \cite{Witten:1996hc}, which was extended to include non-linear interactions in \cite{Cederwall:1997gg}. The latter work includes kappa symmetry and a proof that their formalism is equivalent to the PST one.
\end{itemize}
 
In this review, I follow the PST formalism. This assigns the following bosonic action to the M5-brane \cite{Pasti:1997gx}
\begin{eqnarray}
S_{{\rm M5}} &=& -T_{{\rm M5}}\int d^6\sigma\left(\sqrt{-{\rm det}\,(\CG_{\mu\nu}+\tilde H_{\mu\nu})}-\sqrt{-{\rm det}\,\CG}\frac{1}{4\partial_\mu a\partial^\mu a}\partial_\delta a(\sigma)\CH^{*\mu\nu\delta}\CH_{\mu\nu\rho}\partial^\rho a(\sigma)\right)\nonumber \\
& & +T_{{\rm M5}}\int \left({\cal A}_{6}+\frac{1}{2}\CH_3 \wedge {\cal A}_{3}\right),
\label{eq:m5boson}
\end{eqnarray}
As in previous effective actions, all the dependence on the scalar fields $X^m$ is through the bulk fields and their pullbacks to the 6-dimensional world volume. As in D-brane physics, {\it all} the dependence on the world volume gauge potential $V_2$ is not just simply through its field strength $dV_2$, but through the gauge invariant 3-form
\begin{equation}
   \CH_3 = dV_2 -{\cal A}_{3}\,.
\end{equation}
The physics behind this is analogous. $\CF$ describes the ability of open strings to end on D-branes, whereas $\CH_3$ describes the possibility of M2-branes to end on M5-branes \cite{Strominger:1995ac,Townsend:1996em}\footnote{For a discussion on the interpretation of an M5-brane as a 'D-brane' for an open membrane, see \cite{Becker:1996my}.}. Its world volume Hodge dual and the tensor $\tilde H_{\mu\nu}$ are then defined as
\begin{eqnarray}
  \CH^{*\mu\nu\rho}&=& \frac{1}{6\sqrt{-{\rm det}\, \CG}} \varepsilon^{\mu\nu\rho\alpha_1\alpha_2\alpha_3}\CH_{\alpha_1\alpha_2\alpha_3}\,, \\
\tilde H_{\mu\nu} &=& \frac{1}{\sqrt{|(\partial a)^2|}}\CH^{*}_{\mu\nu\rho}\partial^\rho a(\sigma)\,.
\end{eqnarray}
The latter involves an auxiliary field $a(\sigma)$ responsible for keeping covariance and implementing the self-duality constraint through the second term in the action (\ref{eq:m5boson}). Its auxiliary nature was proved in \cite{Pasti:1996vs,Pasti:1995tn}, where it was shown that its equation of motion is {\it not} independent from the generalised self-duality condition. The full action also includes a DBI-like term, involving the induced world volume metric $\CG_{\mu\nu}=\partial_\mu X^m\partial_\nu X^n g_{mn}(X)$, and a WZ term, involving the pullbacks ${\cal A}_{3}$ and ${\cal A}_{6}$ of the 3-form gauge potential and its Hodge dual in ${\cal N}=1$ d=11 supergravity \cite{Aharony:1996wp}. 

Besides being manifestly invariant under 6-dimensional world volume diffeomorphisms and ordinary abelian gauge transformations $\delta V_2= d\Lambda_1$, the action (\ref{eq:m5boson}) is also invariant under the transformation 
\begin{equation}
\delta a(\sigma)=\Lambda(\sigma)\,, \quad \quad \delta V_{\mu\nu}=\frac{\Lambda(\sigma)}{\sqrt{|(\partial
a)^2|}}\left(2\frac{\delta\CL_{{\rm DBI}}}{\delta\tilde H^{\mu\nu}}-(dV_2)_{\mu\nu\rho}\frac{\partial^\rho a}{\sqrt{|(\partial a)^2|}}\right)\,,
\label{eq:m5extragauge}
\end{equation} 
Given the {\it non-dynamical} nature of $a(\sigma)$, one can always fully remove it from the classical action by gauge fixing the symmetry (\ref{eq:m5extragauge}). It was shown in \cite{Pasti:1997gx} that for an M5-brane propagating in Minkowski, the non-manifest Lorentz invariant formulation in \cite{Perry:1996mk} emerges after gauge fixing (\ref{eq:m5extragauge}). This was achieved by working in the gauge $\partial_\mu a(\sigma)=\delta^5_\mu$ and $V_{\mu 5}=0$. Since $\partial_\mu a$ is a world volume vector, 6-dimensional Lorentz transformations do {\it not} preserve this gauge slice. One must use a compensating gauge transformation (\ref{eq:m5extragauge}), which also acts on $V_{\mu\nu}$. The overall gauge fixed action {\it is} invariant under the full 6-dimensional Lorentz group but in a non-linear non-manifestly Lorentz covariant way as discussed in \cite{Perry:1996mk}. 

As a final remark, notice the charge density $Q_{{\rm M5}}$ of the bosonic M5-brane has already been set equal to its tension $T_{{\rm M5}}=1/(2\pi)^5\ell_p^6$. 

\subsection{Consistency checks}
\label{sec:ccheck}

The purpose of this section is to check the consistency of the kinematic structures governing classical bosonic brane effective actions with {\it string dualities} \cite{Hull:1994ys,Witten:1995ex}. At the level of supergravity, these dualities are responsible for the existence of a non-trivial web of relations among their classical lagrangians. Here, I describe the realisation of some of these dualities on classical bosonic brane actions. This will allow us to check the consistency of all brane couplings. Alternatively, one can also view the discussions below as independent ways of deriving the latter.

The specific dualities I will be appealing to are the strong coupling limit of type IIA string theory, its relation to M-theory and the action of T-duality on type II string theories and D-branes. Figure \ref{fig6} summarises the set of relations between the brane tensions discussed in this review under these symmetries.

\epubtkImage{tensions.png}{%
\begin{figure}[h]
  \centerline{\includegraphics[width=150mm]{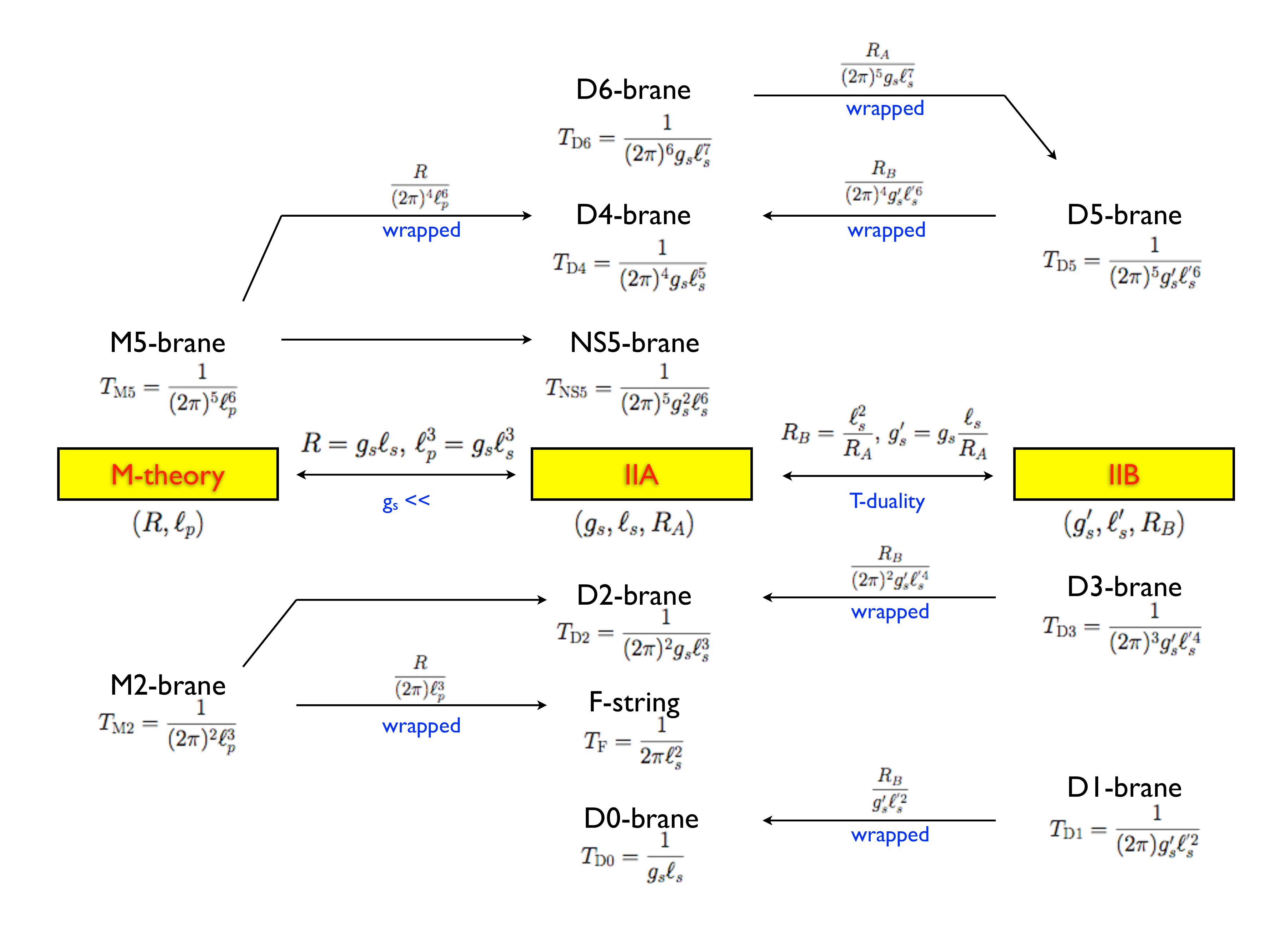}}
   \caption{Set of half-BPS branes discussed in this review, their tensions and some of their connections under T-duality and the strongly coupled limit of type IIA.}
  \label{fig6}
\end{figure}} 

\paragraph{M-theory as the strong coupling limit of type IIA:} From the spectrum of 1/2-BPS states in string theory and M-theory, an M2/M5-brane in $\bR^{1,9}\times S^1$ has a weakly coupled description in type IIA 
\begin{itemize}
\item either as a long string or a D4-brane, if the M2/M5-brane wraps the M-theory circle, respectively
\item or as a D2-brane/NS5 brane, if the M-theory circle is transverse to the M2/M5-brane world volume.
\end{itemize}
The question to ask is : how do these statements manifest in the classical effective action ? The answer is by now well known. They involve a {\it double or a direct dimensional reduction}, respectively. The idea is simple. The bosonic effective action describes the coupling of a given brane with a fixed supergravity background. If the latter involves a circle and one is interested in a description of the physics nonsensitive to this dimension, one is entitled to replace the d-dimensional supergravity description by a d-1 one using a Kaluza--Klein (KK) reduction (see \cite{Duff:1986hr} for a review on KK compactifications). In the case at hand, this involves using the relation between d=11 bosonic supergravity fields and the type IIA bosonic ones summarised below \cite{Obers:1998fb}
\begin{eqnarray}
  ds^2_{11} &=& e^{-\frac{2}{3}\phi}\,ds^2_{10} + e^{\frac{4}{3}\phi}\left(dy+C_{1}\right)^2\,, \nonumber \\
  A_{3} &=& C_{3} + dy\wedge B_{2}\,,
\label{eq:KK}
\end{eqnarray}
where the left hand side 11-dimensional fields are rewritten in terms of type IIA fields. The above reduction involves a low energy limit in which one only keeps the zero mode in a Fourier expansion of all background fields on the bulk S${}^1$. In terms of the parameters of the theory, the relation between the M-theory circle $R$ and the eleven dimensional Planck scale $\ell_p$ with the type IIA string coupling $g_s$ and string length $\ell_s$ is
\begin{equation}
  R = g_s\ell_s, \quad \quad \ell_p^3 = g_s\ell_s^3.
\label{eq:mIIA}
\end{equation}

The same principle should hold for the brane degrees of freedom $\{\Phi^A\}$. The distinction between a double and a direct dimensional reductions comes from the physical choice on whether the brane wraps the internal circle or not :
\begin{itemize}
\item If it does, one partially fixes the world volume diffeomorphisms by identifying the bulk circle direction $y$ with one of the world volume directions $\sigma^p$, i.e. $Y(\sigma)=\sigma^p$, and keeps the zero mode in a Fourier expansion of all the remaining brane fields, i.e. $\Phi^A=\Phi^A(\sigma^\prime)$ where $\{\sigma\}=\{\sigma^\prime,\,\sigma^p\}$. This procedure is denoted as a {\it double} dimensional reduction \cite{Duff:1987bx}, since both the bulk and the world volume get their dimensions reduced by one.
\item If it does not, there is no need to break the world volume diffeomorphisms and one simply truncates the fields to their bulk zero modes. This procedure is denoted as a {\it direct} reduction since the brane dimension remains unchanged while the bulk one gets reduced.
\end{itemize}

\paragraph{T-duality on closed and open strings:} From the quantisation of open strings satisfying Dirichlet boundary conditions, all D-brane dynamics are described by a massless vector supermultiplet, whose number of scalar fields depends on the number of transverse dimensions to the D-brane. Since D-brane states are mapped among themselves under T-duality \cite{Dai:1989ua,Polchinski:1996na}, one expects the existence of a transformation mapping their classical effective actions under this duality. The question is how such transformation acts on the action. This involves two parts : the transformation of the background and the one of the brane degrees of freedom.

Let me focus on the bulk transformation. T-duality is a perturbative string theory duality \cite{Giveon:1994fu}. It says that type IIA string theory on a circle of radius $R$ and string coupling $g_s$ is equivalent to type IIB on a dual circle of radius $R^\prime$ and string coupling $g_s^\prime$ related as \cite{Buscher:1987sk,Buscher:1987qj,Ginsparg:1986wr}
\begin{equation}
R^\prime = \frac{\alpha^\prime}{R}\,, \quad \quad
g^\prime_s = g_s\frac{\sqrt{\alpha^\prime}}{R}\,,
\label{eq:wstduality}
\end{equation}
when momentum and winding modes are exchanged in both theories. This leaves the free theory spectrum invariant \cite{Kikkawa:1984cp}, but it has been shown to be an exact perturbative symmetry when including interactions \cite{Nair:1986zn,Giveon:1994fu}. Despite its stringy nature, there exists a clean field theoretical realisation of this symmetry. The main point is that any field theory on a circle of radius $R$ has a discrete momentum spectrum. Thus, in the limit $R\to 0$, all non-vanishing momentum modes decouple, and one only keeps the original vanishing momentum sector. Notice this is effectively implementing a KK compactification on this circle. This is in contrast with the stringy realisation where in the same limit, the spectrum of winding modes opens up a dual circle of radius $R^\prime$.

Since Type IIA and Type IIB supergravities are field theories, the above field theoretical realisation applies. Thus, the $R\to 0$ compactification limit should give rise to two separate ${\cal N}=2$ d=9 supergravity theories. But it is known \cite{Meessen:1998qm} that there is a unique such supergravity theory. In other words, given the type IIA/B field content $\{\varphi_{A/B}\}$ and their KK reduction to d=9 dimensions, i.e. $\varphi_A= \varphi_A (\varphi_9)$ and $\varphi_B = \varphi_B (\varphi_9)$, the {\it uniqueness} of ${\cal N}=2$ d=9 supergravity guarantees the existence of a non-trivial map between type IIA and type IIB fields in the subset of backgrounds allowing an S${}^1$ compactification
\begin{equation}
\varphi_A = \varphi_A (\varphi_B)\,.
\label{embed3}
\end{equation}

\epubtkImage{sugra-tduality.png}{%
\begin{figure}[h]
  \centerline{\includegraphics[width=150mm]{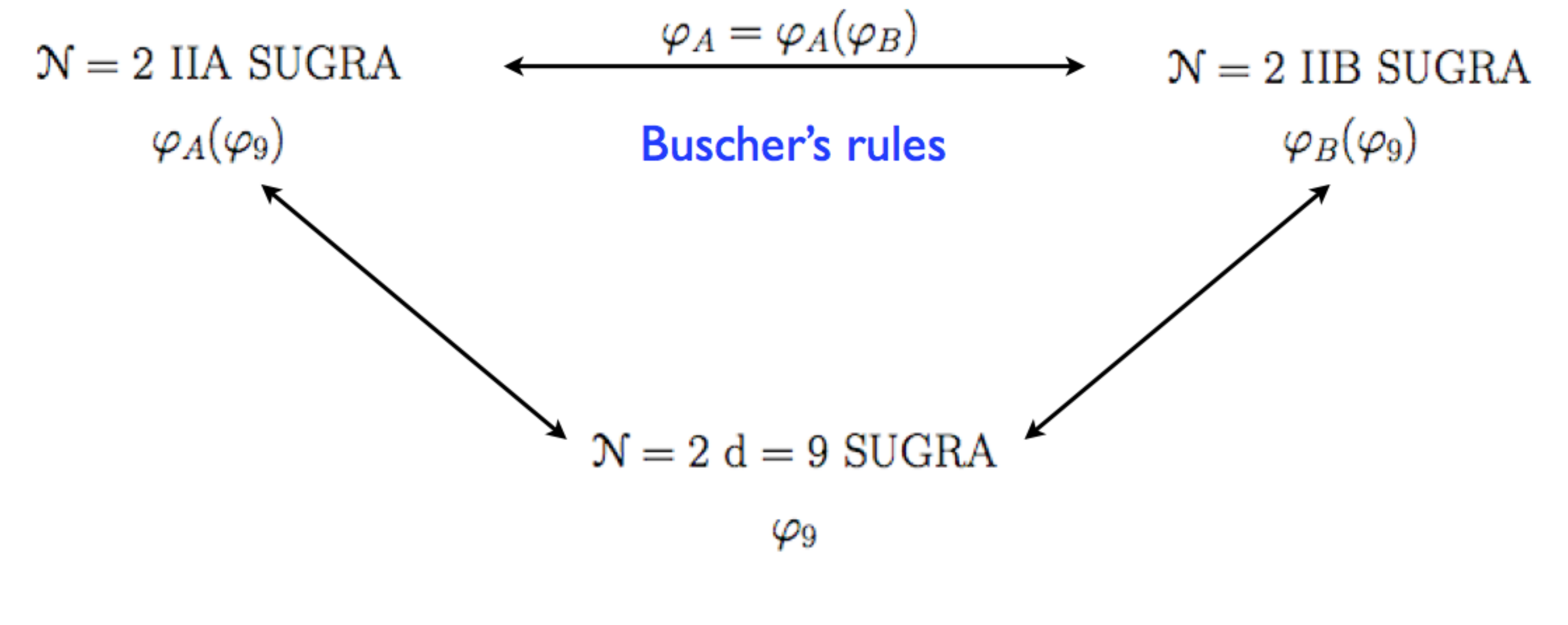}}
   \caption{Schematic diagram describing the derivation of Buscher's T-duality rules using type IIA/IIB supergravities.}
  \label{fig5}
\end{figure}}

This process is illustrated in the diagram \ref{fig5}. These are the T-duality rules. When expressed in terms of explicit field components, they become \cite{Bergshoeff:1995as,Meessen:1998qm}
\begin{eqnarray}
g_{zz} & = & \frac{1}{g'_{z'z'}} \nonumber \\
\phi & = & \phi' - \frac{1}{2}{\rm log}\vert g'_{z'z'}\vert \nonumber \\
B_{nz} & = & -\frac{g'_{nz'}}{g'_{z'z'}} \nonumber \\
g_{nz} & = & -\frac{B'_{nz'}}{g'_{z'z'}} \nonumber \\
g_{mn} & = & g'_{mn} - \frac{g'_{mz'}g'_{nz'}-B'_{mz'}B'_{nz'}}{g'_{z'z'}} 
\nonumber \\
B_{mn} & = & B'_{mn} - \frac{B'_{mz'}g'_{nz'}-B'_{nz'}g'_{mz'}}{g'_{z'z'}} \nn \\
C^{(p+1)}_{m_1 \ldots m_p z} & = & C'^{(p)}_{m_1 \ldots m_p} -
p\frac{C'^{(p)}_{[m_1 \ldots m_{p-1}z'}g'_{m_p] z'}}{g'_{z'z'}} \nn \\
C^{(p)}_{m_1 \ldots m_p} & = & C'^{(p+1)}_{m_1 \ldots m_p z'}
-pC'^{(p-1)}_{[m_1 \ldots m_{p-1}}B'_{m_p ]z'} \nonumber \\
& & -p(p-1)\frac{C'^{(p-1)}_{[m_1 \ldots m_{p-2}z'}B'_{m_{p-1}z'}g'_{m_p ]z'}}{g'_{z'z'}}\,.
\label{eq:tdualityrules}
\end{eqnarray}
These correspond to the bosonic truncations of the superfields introduced in appendix \ref{sec:iiab}.
Prime and unprimed fields correspond to both T-dual theories. The same notation applies to the tensor components where $\{z,\,z'\}$ describe both T-dual circles. Notice the dilaton and the $g_{zz}$ transformations do capture the worldsheet relations (\ref{eq:wstduality}). 

Let me move to the brane transformation. A D(p+1)-brane wrapping the original circle is mapped under T-duality to a Dp-brane where the dual circle is transverse to the brane \cite{Polchinski:1996na}. It must be the case that one of the gauge field components in the original brane maps into a transverse scalar field describing the dual circle. At the level of the effective action, implementing the $R\to 0$ limit must involve, first, a partial gauge fixing of the world volume diffeomorphisms, to explicitly make the physical choice that the brane wraps the original circle, and second, keeping the zero modes of all the remaining dynamical degrees of freedom. This is precisely the procedure described as a double dimensional reduction. The two differences in this D-brane discussion will be : the presence of a gauge field and the fact that the KK reduced supergravity fields $\{\varphi_9\}$ will be rewritten in terms of the T-dual ten dimensional fields using the T-duality rules (\ref{eq:tdualityrules}). 

In the following, it will be proved that the classical effective actions described in the previous section are interconnected in a way consistent with our T-duality and strongly coupled considerations. Our logic is as follows. The M2-brane is linked to our starting worldsheet action through double dimensional reduction. The former is then used to derive the D2-brane effective by direct dimensional. T-duality covariance extends this result to any non-massive D-brane. Finally, to check the consistency of the PST covariant action for the M5-brane, its double dimensional reduction will be shown to match the D4-brane effective action. This will complete the set of classical checks on the bosonic brane actions discussed so far.

It is worth mentioning that the self-duality of the D3-brane effective action under S-duality could have also been included as a further test. For discussions on this point, see \cite{Tseytlin:1996it,Green:1996qg}.

\subsubsection{M2-branes and its classical reductions}

In the following, I discuss the double and direct dimensional reductions of the bosonic M2-brane effective action (\ref{bosonicm2}) to match the bosonic worldsheet string action (\ref{eq:bos-string}) and the D2-brane effective action, i.e. the $p=2$ version of (\ref{bosonicDp}). This analysis will also allow us to match/derive the tensions of the different branes.

\paragraph{Connection to the string worldsheet :}

Consider the propagation of an M2-brane in an eleven dimensional backgrounds of the form (\ref{eq:KK}).
Decompose the set of scalar fields as $\{X^M\}=\{X^m,\,Y\}$, identify one of the world volume directions $(\rho)$ with the KK circle, i.e. partially gauge fix the world volume diffeomorphisms by imposing
$Y=\rho$, and keep the zero modes in the Fourier expansion of all remaining scalar fields $\{X^m\}$ along the world volume circle, i.e. $\partial_\rho X^m= 0$. Under these conditions, which mathematically characterise a double dimensional reduction, the Wess-Zumino coupling becomes
\begin{equation}
\int_{\Sigma_2\times S_1} {\cal A}_{3} = \int_{\Sigma_2\times S_1} d^3\sigma\,
\frac{1}{2}\epsilon^{\hm\hn\rho}\6_{\hm}X^m\6_{\hn}X^nA_{mny} = 
\left(\int_{S_1} d\rho\right)\int_{\Sigma_2}{\cal B}_{2},
\end{equation}
where I already used the KK reduction ansatz (\ref{eq:KK}). Here ${\cal B}_2$ stands for the pull-back of the NS-NS two form into the surface $\Sigma_2$ parameterised by $\{\sigma^{\hm}\}$. The DBI action is reduced using the identity satisfied by the induced world volume metric
\begin{equation}
\CG_{\mu\nu} = \left(\matrix{ e^{-2\phi/3}(\CG_{\hm\hn}+e^{2\phi}
\CC_{\hm} \CC_{\hn}) & e^{4\phi/3}\CC_{\hm} \cr
e^{4\phi/3}\CC_{\hn} & e^{4\phi/3}}\right)\quad \Longrightarrow \quad \mbox{det}\,\CG_{\mu\nu}=\mbox{det}\,\CG_{\hm\hn}\,.
\end{equation}
Since the integral over $\rho$ equals the length of the M-theory circle,
\begin{equation}
 \int_{S_1} d\rho=2\pi R = 2\pi g_s l_s \quad \Longrightarrow \quad T_f = T_{M2}\int_{S_1} d\rho= \frac{1}{2\pi\alpha'},
\end{equation}
where I used (\ref{eq:mIIA}), $T_{M2}=1/(2\pi)^2l_p^3$ and absorbed the overall circle length, expressed in terms of type IIA data, in a new energy density scale, matching the fundamental string tension $T_f$ defined in section \ref{sec:motiv}. The same argument applies to the charge density leading to $Q_f = Q_{M2}\,2\pi R$.

Altogether, the double reduced action reproduces the bosonic effective action (\ref{eq:bos-string}) describing the string propagation in a type IIA background. Thus, our classical bosonic M2-brane action is consistent with the relation between half-BPS  M2-brane and fundamental strings in the spectrum of M-theory and type IIA.

\paragraph{Connection to the D2-brane :} The direct dimensional reduction of the bosonic M2 brane describes a three dimensional diffeomorphism invariant theory propagating in ten dimensions, with eleven scalars as its field content. The latter disagrees with the bosonic field content of a D2-brane which includes a vector field. Fortunately, an scalar field is Hodge dual, in three dimensions, to a one form. Thus, one expects that by direct dimensional reduction of the bosonic M2-brane action and after {\it world volume dualisation} of the scalar field $Y$ along the M-theory circle, one should reproduce the classical D2-brane action \cite{Schmidhuber:1996fy,Townsend:1995af,Bergshoeff:1996tu,Townsend:1996xj}.

To describe the direct dimensional reduction, consider the lagrangian \cite{Townsend:1996xj}
\begin{equation}
S= \frac{T_{{\rm M2}}}{2}\int d^3\sigma\left( v^{-1}\det \CG^{(11)}_{\mu\nu} - v +
\frac{1}{3}\epsilon^{\mu\nu\rho}{\cal A}_{\mu\nu\rho}\right).
\label{eq:bosmema}
\end{equation}
This is classically equivalent to (\ref{bosonicm2}) after integrating out the auxiliary scalar density $v$ by solving its algebraic equation of motion. Notice I already focused on the relevant case for later supersymmetric considerations, i.e. $Q_{{\rm M2}}=T_{{\rm M2}}$. The induced world volume fields are
\begin{eqnarray}
\label{eq:tentoel}
\CG_{\mu\nu}^{(11)} &=& e^{-\frac{2}{3}\phi} \CG_{\mu\nu} 
+ e^{\frac{4}{3}\phi}Z_\mu Z_\nu \\
{\cal A}_{\mu\nu\rho} &=& \CC_{\mu\nu\rho} + 3{\cal B}_{[\mu\nu}Z_{\rho]} 
-3{\cal B}_{[\mu\nu}\CC_{\rho]} \,,
\end{eqnarray}
where 
\begin{equation}
Z \equiv dY + \CC_{1}\ .
\label{eq:ydef}
\end{equation}
Using the properties of $3\times3$ matrices, 
\begin{equation}
\det \CG_{\mu\nu}^{(11)} = e^{-2\phi} \det [\CG_{\mu\nu} + e^{2\phi}
Z_\mu Z_\nu]= (\det \CG_{\mu\nu})\big[ e^{-2\phi} + |Z|^2\big] \,,
\label{eq:matid}
\end{equation}
where $|Z|^2 = Z_\mu Z_\nu \CG^{\mu\nu}$, the action (\ref{eq:bosmema}) can be written as
\begin{eqnarray}
S &=& \frac{T_{{\rm M2}}}{2}\int\! d^3\sigma\, \left(v^{-1}e^{-2\phi}\det 
\CG_{\mu\nu} -v
 + \frac{1}{3}\epsilon^{\mu\nu\rho}[\CC_{\mu\nu\rho}- 3{\cal B}_{\mu\nu} 
\CC_\rho ] \right. \nonumber \\
& & \qquad \left. +\ v^{-1}(\det \CG_{\mu\nu}) |Z|^2 + \epsilon^{\mu\nu\rho}
{\cal B}_{\mu\nu}Z_\rho\right).
\label{eq:bosmemc} 
\end{eqnarray}

The next step is to describe the world volume dualisation and the origin of the $\U(1)$ gauge symmetry on the D2 brane effective action \cite{Townsend:1996xj}. By definition, the identity 
\begin{equation}
d(Z -\CC_1)\equiv 0
\label{eq:abosm}
\end{equation}
holds. Adding the latter to the action through an exact two-form $F=dV$ Lagrange multiplier
\begin{equation}
-\frac{1}{2\pi}\int\!  F\wedge (Z-\CC_1),
\label{eq:lagrangema}
\end{equation}
allows to treat $Z$ as an independent field. For a more thorough discussion on this point and the nature of the $\U(1)$ gauge symmetry, see \cite{Townsend:1996xj}.  Adding (\ref{eq:lagrangema}) to (\ref{eq:bosmemc}), one obtains
\begin{eqnarray}
\label{eq:bosmemd}
S &= \frac{T_{{\rm M2}}}{2}\int\! d^3\sigma\, \left( v^{-1}e^{-2\phi}\det 
\CG_{\mu\nu} -v + \frac{1}{3}\epsilon^{\mu\nu\rho}\big[\CC_{\mu\nu\rho}+ 3{\cal F}_{\mu\nu}
\CC_\rho\big] \right. \nonumber \\
& \left. +\ v^{-1}(\det \CG_{\mu\nu}) |Z^2| - \epsilon^{\mu\nu\rho}{\cal F}_{\mu\nu} Z_\rho \right).
\end{eqnarray}
Notice I already introduced the same gauge invariant quantity introduced in D-brane lagrangians 
\begin{equation}
{\cal F}_{\mu\nu} = F_{\mu\nu} - {\cal B}_{\mu\nu}.
\label{eq:moddeff}
\end{equation}
Since $Y$ is now an independent field, it can be eliminated 
by solving its algebraic equation of motion
\begin{equation}
Z^\mu= \frac{v}{2\det \CG} \epsilon^{\mu\nu\rho} {\cal F}_{\mu\nu}.
\label{eq:yeq}
\end{equation}
Inserting this back into the action and integrating out the auxiliary field $\tilde{v}=-\det(\CG_{\mu\nu})/v$ by solving its equation of motion, yields
\begin{equation}
S =  -T_{{\rm D2}}\int d^3\sigma\, e^{-\phi}\sqrt{- \det (\CG_{\mu\nu}+ 
{\cal F}_{\mu\nu})}\ 
+\  T_{{\rm D2}}\int_w\! (\CC_{3} + {\cal F}\wedge \CC_{1}).
\label{eq:bosmemf}
\end{equation}
This matches the proposed D2-brane effective action, since $T_{{\rm M2}}=T_{{\rm D2}}$ as a consequence of (\ref{eq:mIIA}) and
(\ref{eq:tdbrane}).

\subsubsection{T-duality covariance}
\label{sec:tcov}

In this section, I extend the D2-brane's functional form to any Dp-brane using T-duality covariance. My goal is to show that the bulk T-duality rules (\ref{eq:tdualityrules}) guarantee the covariance of the D-brane effective action functional form \cite{Simon:1998az} and to review the origin in the interchange between scalar fields and gauge fields on the brane\footnote{Relevant work on the subject includes \cite{Bachas:1995kx,Bergshoeff:1996ui,Alvarez:1996up,Bergshoeff:1996cy}.}.

The second question can be addressed by an analysis of the D-brane action bosonic symmetries. These act infinitesimally as
\begin{eqnarray}
s\,X^M & = & \xi^{\nu}\6_{\nu}X^M + \Delta X^M,
\label{trafo1} \\
s\,V_\mu & = & \xi^{\nu}\6_{\nu}V_\mu + V_\nu\6_\mu \xi^\nu + \6_\mu c +
\Delta V_\mu.
\label{trafo2}
\end{eqnarray} 
They involve world volume diffeomorphisms $\xi^\nu$, a $\U(1)$ gauge transformation $c$ and global transformations $\Delta\phi^i$. Since the background will undergo a T-duality transformation, by assumption, this set of global transformations {\it must} include translations along the circle, i.e. $\Delta Z  = \epsilon$, $\Delta X^m = \Delta V_\mu = 0$, where the original $X^M$ scalar fields were split into $\{X^m,Z\}$.

I argued that the realisation of T-duality on the brane action requires to study its double dimensional reduction. The latter involves a partial gauge fixing $Z=\sigma^p \equiv \rho$, identifying one world volume direction with the starting S${}^1$ bulk circle and a zero-mode Fourier truncation in the remaining degrees of freedom, $\6_\rho X^m = \6_\rho V_\mu = 0$. Extending this functional truncation to the p-dimensional diffeomorphisms $\xi^{\8\mu}$, where I split the world volume indices according to $\{\mu\}=\{\8\mu,\rho\}$ and the space of global transformations, i.e. $\6_z \Delta x^M = \6_z \Delta V_\mu = 0$, the consistency conditions requiring the infinitesimal transformations to preserve the subspace of field configurations defined by the truncation and the partial gauge fixing, i.e. $\6_z s\phi^i \mid_{g.f. + trunc}=0$, determines
\begin{equation}
c(\sigma^{\8\mu},\rho) = \7c (\sigma^{\8\mu}) + a + \frac{\epsilon^{\prime}}{2\pi\alpha^{\prime}}\rho
\label{eq:partu1}
\end{equation}
where $a,\epsilon^{\prime}$ are constants, the latter having
mass dimension minus one. The set of transformations the double dimensional reduction are
\begin{eqnarray}
\7s X^m & = & \xi^{\8\nu}\6_{\8\nu}X^m + \7\Delta X^m
\label{trafo11} \\
\7s V_{\8\mu} & = & \xi^{\8\nu}\6_{\8\nu}V_{\8\mu} + V_{\8\nu}\6_{\8\mu}
\xi^{\8\nu} + \6_{\8\mu}\7c + \7\Delta V_{\8\mu}
\label{trafo21} \\
\7s V_\rho & = & \xi^{\8\nu}\6_{\8\nu}V_\rho + \7\Delta V_\rho 
\label{trafo22}
\end{eqnarray}
where $\7\Delta V_{\8\mu} = \Delta V_{\8\mu} - 
V_\rho\6_{\8\mu} \Delta Z$, $\7\Delta V_\rho = \Delta V_\rho + 
\epsilon^{\prime}/2\pi\alpha^{\prime}$ and $\7\Delta x^m$ satisfies
$\6_z \7\Delta x^m = 0$.

Let me comment on (\ref{trafo22}). $V_\rho$ was a gauge field component in the original action. But in its gauge fixed functionally truncated version, it transforms like a world volume {\it scalar}. Furthermore, the constant piece $\epsilon^\prime$ in the original $\U(1)$ transformation (\ref{eq:partu1}), describes a {\it global} translation along the scalar direction. The interpretation of both observations is that under double dimensional reduction
\begin{equation}
(2\pi\alpha')\,V_\rho \equiv Z^\prime
\label{eq:tdualrule}
\end{equation}
$Z^\prime$ becomes the {\it T-dual} target space direction along the T-dual circle and $\epsilon^\prime$ describes the corresponding translation isometry. This discussion reproduces the well known massless open string spectrum when exchanging a Dirichlet boundary condition with a Neumann boundary condition.

Having clarified the origin of symmetries in the T-dual picture, let me analyse the functional dependence of the effective action. First, consider the couplings to the NS sector in the DBI action. Rewrite the induced metric $\CG$ and the gauge invariant $\CF$ in terms of the T-dual background $(g^\prime,\,B^\prime)$ and degrees of freedom $(\{X^{M'}\}=\{X^{m'},\,Z^\prime\})$, which will be denoted by primed quantities. This can be achieved by adding and subtracting the relevant pullback quantities. The following identities hold
\begin{eqnarray}
\CG_{\hm \rho} & = & \6_{\hm} X^m g_{mz} \\
\CG_{\rho\rho} & = & g_{zz} \\
\CG_{\hm\hn} & = & \CG'_{\hm\hn} + \6_{\hm}X^m \6_{\hn}X^n (g_{mn} -
g'_{mn}) - \6_{\hm}X^m \6_{\hn}Z^\prime g'_{z'm} \nonumber \\
& & - \6_{\hm}Z^\prime \6_{\hn}X^{M'}g'_{z'M'} \\
\CF_{\hm \rho} & = & \6_{\hm}Z^\prime - \6_{\hm}X^mB_{mz} \\
\CF_{\hm \hn} & = & \CF'_{\hm \hn} -\6_{\hm}X^m\6_{\hn}X^n(B_{mn}-
B'_{mn}) + \6_{\hm}Z^\prime\6_{\hn}X^n B'_{z'n} \nonumber \\
& & + \6_{\hm}X^m\6_{\hn}Z^\prime B'_{mz'}
\end{eqnarray}
It is a consequence of our previous symmetry discussion that $X^{m'}=X^m$ and $V_{\hat{\mu}}=V'_{\hat{\mu}}$, i.e. there is {\it no} change in the description of the dynamical degrees of freedom not involved in the circle directions. The determinant appearing in the DBI action can now be computed to be
\begin{eqnarray}
& & {\rm det}\,(\CG_{\mu\nu}+\CF_{\mu \nu}) = 
g_{zz}\,{\rm det}\,\left(\CG'_{\hm\hn}+\CF'_{\hm \hn} \right.\nonumber \\
& & \left. +\6_{\hm}X^m\6_{\hn}X^n\,[(g_{mn}-g'_{mn}) - (B_{mn}-B'_{mn})
-(g_{mz}-B_{mz})(g_{nz}+B_{nz})/g_{zz}] 
\right. \nonumber \\
& & \left.-\6_{\hm}X^m\6_{\hn}Z^\prime\,[(g'_{mz'}-B'_{mz'}) - (g_{mz}-B_{mz})
/g_{zz}] \right.\nonumber \\
& & \left.-\6_{\hm}Z^\prime\6_{\hn}X^n\,[(g'_{z'n}+B'_{nz'}) + (g_{nz}+B_{nz})
/g_{zz}] \right.\nonumber \\
& & \left. -\6_{\hm}Z^\prime\6_{\hn}Z^\prime\,(g'_{z'z'}-\frac{1}{g_{zz}})\right)
\label{sufficient}
\end{eqnarray}
Notice that whenever the bulk T-duality rules (\ref{eq:tdualityrules}) are satisfied, the functional form of the effective action remains covariant, i.e. of the form
\begin{equation}
-T'_{{\rm D(p-1)}}\int d^p \sigma\, e^{-\phi'}\,\sqrt{-{\rm det}\,(\CG'_{\hm\hn}+
\CF'_{\hm \hn})}\, .
\end{equation}
This is because all terms in the determinant vanish except for those in the first line. Finally, $e^{-\phi}\,\sqrt{g_{zz}}$ equals the T-dual dilaton coupling $e^{-\phi^\prime}$ and the original Dp-brane tension $T_{{\rm Dp}}$ becomes the D(p-1)-brane tension in the T-dual theory due to the worldsheet defining properties (\ref{eq:wstduality}) after the integration over the world volume circle
\begin{equation}
  T_{{\rm Dp}} \int d\rho = \frac{1}{(2\pi)^p\,g_sl_s^{p+1}}\,2\pi\,R = \frac{1}{(2\pi)^{p-1}\,g^\prime_sl_s^p} = T'_{{\rm D(p-1)}}\,.
\end{equation}

Just as covariance of the DBI action is determined by the NS-NS sector, one expects the RR sector to do the same for the WZ action. Here I follow similar techniques to the ones developed in \cite{Green:1996bh,Simon:1998az}. First, decompose the WZ lagrangian density as
\begin{equation}
{\cal L}_{WZ} = {\cal L}^+_{WZ} + {\cal L}^-_{WZ}
\equiv d\rho\wedge i_{\6{_\rho}} {\cal L}_{WZ} +
i_{\6{_\rho}}(d\rho\wedge {\cal L}_{WZ})\,.
\label{WRpaar}
\end{equation}
Due to the functional truncation assumed in the double dimensional reduction, the second term vanishes. 
The D-brane WZ action then becomes
\begin{equation}
T_p \int_{\Sigma_{p+1}} {\cal L}_{WZ} = 
T_p \int_{\Sigma_{p+1}} d\rho\wedge\,e^{\CF^-}\wedge\left( i_{\6_{\rho}} C + 
i_{\6_{\rho}} \CF \wedge \, C^{-}\right) 
\label{WZ}
\end{equation}
where $\CF^-  \equiv i_{\6_{\rho}}(d\rho\wedge \CF)$ and the following conventions are used
\begin{eqnarray}
i_{\6_{\rho}}\Omega_{(n)} & = & \frac{1}{(n-1)!} 
\Omega_{\rho\mu_2 \ldots \mu_n} d\sigma^{\mu_2}\wedge \ldots d\sigma^{\mu_n} \nonumber \\
i_{\6_{\rho}} (\Omega_{(m)}\wedge \Omega_{(n)}) & = & i_{\6_{\rho}}
\Omega_{(m)}\wedge \Omega_{(n)} +  (-1)^m \Omega_{(m)}\wedge i_{\6_{\rho}}\Omega_{(n)}\,.
\label{convention}
\end{eqnarray}
Using the T-duality transformation properties of the gauge invariant quantity $\CF$, derived from our DBI analysis,
\begin{eqnarray}
& \CF^- \longrightarrow \CF' - \left(i_{\6_{z'}} B' 
\wedge i_{\6_{z'}} g'\right)/g'_{z'z'} & 
\label{FT1} \\
& i_{\6_{\rho}} \CF \longrightarrow -\, i_{\6_{z'}} g'/g'_{z'z'} &
\label{FT2}
\end{eqnarray}
it was shown in \cite{Simon:1998az} that the functional form of the WZ term is preserved, i.e.  $T'_{{\rm D(p-1)}}\int_{\6 \Sigma} e^{\CF'}\wedge C'$, whenever the condition
\begin{equation}
(-1)^pC'_{p} =  i_{\6_{\rho}}C_{p+1} - \frac{i_{\6_{z'}} 
B'\wedge i_{\6_{z'}} g'}{g'_{z'z'}}\wedge i_{\6_{\rho}}
C_{p-1} - \frac{i_{\6_{z'}} g'}{g'_{z'z'}}\wedge C^{-}_{p-1}\, .
\label{RR}
\end{equation}
holds (the factor $(-1)^p$ is due to our conventions (\ref{convention}) and the choice of orientation $\epsilon^{\7\mu_1 \ldots \7\mu_p} \equiv \epsilon^{\mu_1 \ldots \mu_p \rho}$ and $\epsilon^{01 \ldots p}=1$). 

Due to our gauge fixing condition, $Z=\rho$, the $\pm$ components of the pullbacked 
world volume forms appearing in (\ref{RR}) can be lifted to $\pm$ components
of the spacetime forms. The condition (\ref{RR}) is then solved by
\begin{eqnarray}
i_{\6_{z}}C_{p+1} & = & (-1)^p \left(C'_{(p)} - \frac{i_{\6_{z'}} g'}{g'_{z'z'}}\wedge i_{\6_{z'}} C'_{p}\right) 
\label{TRR1} \\
C^{-}_{p-1} & = & (-1)^{(p-1)}\left(i_{\6_{z'}} C'_{p} -
i_{\6_{z'}} B' \wedge \left( C'_{p-2} - \frac{i_{\6_{z'}} g'}{g'_{z'z'}}\wedge i_{\6_{z'}} 
C'_{p-2} \right)\right)\,.
\label{TRR2}
\end{eqnarray}
These are entirely equivalent to the T-duality rules (\ref{eq:tdualityrules}) but written in an intrinsic way.

The expert reader may have noticed that the RR T-duality rules do not coincide with the ones 
appearing in part of the literature \cite{Eyras:1998hn}. The reason behind this is the freedom to redefine the fields in our theory. In particular, there exist different choices for the RR potentials, depending on their transformation properties under S-duality. For example, the 4-form $C_{4}$ appearing in our WZ couplings is not S self-dual, but transforms as 
\begin{equation}
C_{4} \to C_{4} - C_{2}\wedge B_{2}\,.
\end{equation}
Using a superindex S to denote an S-dual self-dual 4-form, the latter must be
\begin{equation}
 C^S_{4} = C_{4} -\frac{1}{2}C_{2}\wedge B_{2}\,.
\end{equation}
Similarly, $C_{6}$ does not transform as a doublet under S-duality, whereas
\begin{equation}
C^S_{6} = C_{6} -\frac{1}{4} C_{2}\wedge B_{2}\wedge B_{2}\,,
\end{equation}
does. It is straightforward to check that equations (\ref{TRR1}) and (\ref{TRR2}) are equivalent to the ones appearing in \cite{Eyras:1998hn} using the above redefinitions. Furthermore, one finds
\begin{eqnarray}
C^S_{m_1 \ldots m_6} & = & C'_{m_1 \ldots m_6 z'} -6
C'_{[m_1 \ldots m_5}\frac{g'_{m_6 ]z'}}{g'_{z'z'}} \nonumber \\
& & -45\left( C'_{[m_1} - C'_{z'}\frac{g'_{[m_1 z'}}{g'_{z'z'}}\right)
B'_{m_2m_3}B'_{m_4 m_5}B'_{m_6]z'} \nonumber \\
& & -45C'_{[m_1m_2z'}B'_{m_3m_4}\left(B'_{m_5m_6} -4B'_{m_5z'}
\frac{g'_{m_6 ]z'}}{g'_{z'z'}}\right) \nonumber \\
& & -30C'_{[m_1\ldots m_4 z'}B'_{m_5z'}\frac{g'_{m_6 ]z'}}{g'_{z'z'}}
\end{eqnarray}
which was not computed in \cite{Eyras:1998hn}. 

In section \ref{sec:mdbrane}, I will explore the consequences that can be extracted from the requirement of T-duality covariance for the covariant description of the effective dynamics of N overlapping parallel D-branes in curved backgrounds, following \cite{Myers:1999ps}.

\subsubsection{M5-brane reduction}

The double dimensional reduction of the M5-brane effective action, both in its covariant \cite{Pasti:1997gx,Aganagic:1997zq} and non-covariant formulations \cite{Perry:1996mk,Perry:1996mk,Bergshoeff:1996ev} was checked to agree with the D4-brane effective action. It is important to stress that the outcome of this reduction may not be in the standard D4-brane action form given in (\ref{bosonicDp}), but in the so called dual formulation. The two are related through the world volume dualisation procedure described in \cite{Tseytlin:1996it,Aganagic:1997zk}.

\subsection{Supersymmetric brane effective actions in Minkowski}
\label{sec:bsuperpoincare}

In the study of global supersymmetric field theories, one learns the {\it superfield} formalism is the most manifest way of writing {\it interacting} manifestly supersymmetric lagrangians \cite{Wess:1992cp}. One extends the manifold $\bR^{1,3}$ to a supermanifold through the addition of Grassmann fermionic coordinates $\theta$. Physical fields $\phi(x)$ become components of superfields $\Phi (x,\theta)$, the natural objects in this mathematical structure defined as finite polynomials in a Taylor-like $\theta$ expansion
\begin{equation}
  \Phi(x,\theta) = \phi(x) + \theta^\alpha \phi_\alpha (x) + \dots \nonumber
\end{equation}
that includes auxiliary (non-dynamical) components allowing to close the supersymmetry algebra off-shell. Generic superfields do not transform irreducibly under the SuperPoincar\'e group. Imposing constraints on them, i.e. $f_i(\Phi)=0$,  gives rise to the different irreducible supersymmetric representations. For a standard reference on these concepts,
see \cite{Wess:1992cp}.

These considerations also apply to the p+1 dimensional supermultiplets describing the {\it physical} brane degrees of freedom propagating in $\bR^{1,9}$, since these correspond to supersymmetric field theories in $\bR^{1,p}$.
The main difference in the GS formulation of brane effective actions is that it is {\it spacetime} itself that must be formulated in a manifestly supersymmetric way. By the same argument used in global supersymmetric theories, one would require to work in a ten or eleven dimensional superspace, with standard bosonic coordinates $x^m$ and the addition of fermionic ones $\theta$, whose representations will depend on the dimension of the bosonic submanifold. There are two crucial points to appreciate for our purposes 
\begin{itemize}
\item[1.] the superspace coordinates $\{x^m,\,\theta\}$ will become the brane {\it dynamical} degrees of freedom $\{X^m(\sigma),\,\theta(\sigma)\}$, besides any additional gauge fields living on the brane;
\item[2.] the couplings of the latter to the fixed background where the brane propagates must {\it also} be described in a manifestly spacetime supersymmetric way. The formulation achieving precisely that is the superspace formulation of supergravity theories \cite{Wess:1992cp}. 
\end{itemize}
Both these points were already encountered in our review of the GS formulation for the superstring. The same features will hold for all brane effective actions discussed below. After all, both strings and branes are different objects in the same theory. Consequently, any manifestly spacetime supersymmetric and covariant formulation should refer to the same superspace. It is worth emphasising the world volume manifold $\Sigma_{p+1}$ with local coordinates $\sigma^\mu$ remains bosonic in this formulation. This is {\it not} what occurs in standard superspace formulations of supersymmetric field theories. There exists a classically equivalent formulation to the GS one, the so called {\it superembedding} formulation that extends both the spacetime and the world volume to supermanifolds. Though I will briefly mention this alternative and powerful formulation in section \ref{sec:open}, I refer readers to the review \cite{Sorokin:1999jx}. 

\epubtkImage{symmetry.png}{%
\begin{figure}[h]
  \centerline{\includegraphics[width=150mm]{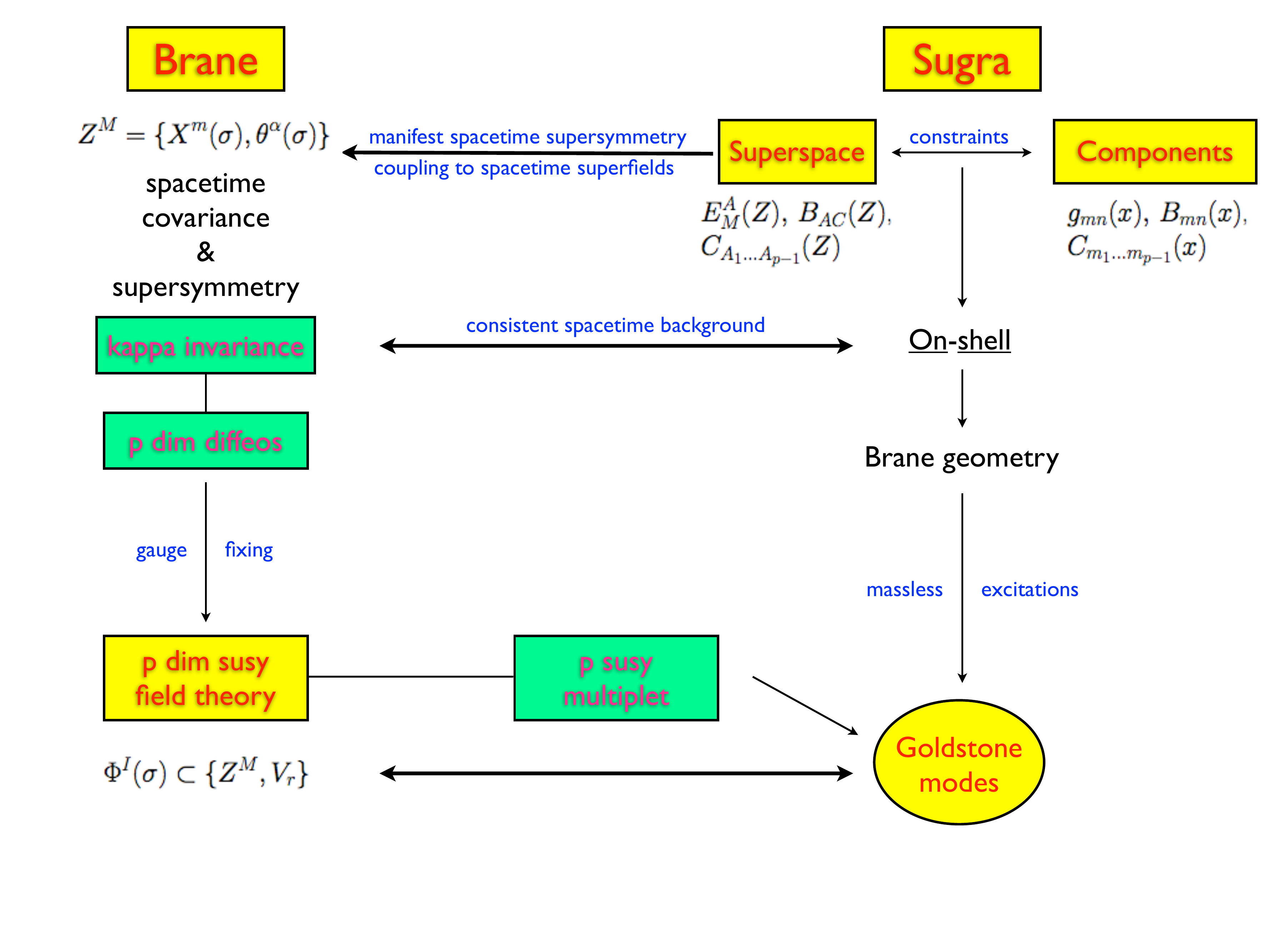}}
   \caption{Kappa symmetry and world volume diffeomorphisms allow to couple the brane degrees of freedom to the superfield components of supergravity in a manifestly covariant and supersymmetric way. Invariance under kappa symmetry forces the background to be on-shell. The gauge fixing of these symmetries connects the GS formulation with world volume supersymmetry, whose on-shell degrees of freedom match the Goldstone modes of the brane supergravity configurations.}
  \label{fig7}
\end{figure}}

As in global supersymmetric theories, supergravity superspace formulations involve an increase in the number of degrees of freedom describing the spacetime dynamics (to preserve supersymmetry covariance). Its equivalence with the more standard component formalism is achieved through the satisfaction of a set of non-trivial constraints imposed on the supergravity superfields. These guarantee the on-shell nature of the physical superfield components. I refer the reader to a brief but self-contained appendix \ref{sec:appcons} where this superspace formulation is reviewed for ${\cal N}=2$ type IIA/B d=10 and ${\cal N}=1$ $d=11$ supergravities, including the set of constraints that render them on-shell. These will play a very important role in the self-consistency of the supersymmetric effective actions I am about to construct. 

Instead of discussing the supersymmetric coupling to an arbitrary curved background at once, my plan is to review the explicit construction of supersymmetric D-brane and M2-brane actions propagating in Minkowski spacetime, or its superspace extension, SuperPoincar\'e\footnote{For a discussion of the supersymmetric and kappa invariant M5-brane covariant action propagating in SuperPoincar\'e, see \cite{Claus:1997cq}.}. The logic will be analogous to the one presented for the superstring. First, I will construct these supersymmetric and kappa invariant actions {\it without} using the superspace formulation, i.e. using a more explicit component approach. Afterwards, I will rewrite those action in superspace variables, pointing in the right direction to achieve a covariant extension of these results to curved backgrounds in section \ref{sec:susycurve}.

\subsubsection{D-branes in flat superspace}

In this section, I am aiming to describe the propagation of D-branes in a fixed Minkowski target space preserving {\it all} spacetime supersymmetry and being world volume kappa symmetry invariant. Just as for bosonic open strings, the gauge field dependence was proved to be of the DBI form by explicit open superstring calculations  \cite{Tseytlin:1986ti,Metsaev:1987qp,Bergshoeff:1987at}\footnote{Following the same philosophy as for their bosonic truncations, this functional dependence can be derived from the double dimensional reduction of the supersymmetric M2-brane action to be discussed in subsection \ref{sec:m2flatsuper} \cite{Townsend:1995af,Schmidhuber:1996fy}. This also provides a derivation of the WZ couplings to be constructed in this subsection. Of course, this consideration would only apply for the D2-brane, but T-duality would allow to extend this conclusion for any Dp-brane \cite{Hassan:1999bv,Kamimura:2000bi}}.

Here I follow the strategy in \cite{Aganagic:1996pe}. First, I will construct a supersymmetric invariant DBI action building on the superspace results reported in section \ref{sec:motiv}. Second, I will determine the WZ couplings by requiring both supersymmetry and kappa symmetry invariance. Finally, as in our brief review of the GS superstring formulation, I will reinterpret the final action in terms of superspace quantities and their pullback to p+1 world volume hypersurfaces. This step will identify the correct structure to be generalised to arbitrary curved backgrounds.

Let me first set my conventions. The field content includes a set of p+1 dimensional world volume scalar fields $\{Z^M(\sigma)\}=\{X^m(\sigma),\,\theta^\alpha(\sigma)\}$ describing the embedding of the brane into the bulk supermanifold. Fermions depend on the theory under consideration
\begin{itemize}
\item ${\cal N}=2$ d=10 type IIA superspace involves two fermions of different chiralities $\theta_{\pm}$, i.e. $\Gamma_\sharp \theta_\pm= \pm \theta_\pm$, where $\Gamma_\sharp=\Gamma_0\Gamma_1\dots\Gamma_9$. I describe them jointly by a unique fermion $\theta$, satisfying $\theta=\theta_+ + \theta_-$.
\item ${\cal N}=2$ d=10 type IIB superspace contains two fermions of the same chirality (positive by assumption), $\theta^i$ $i=1,2$. The index $i$ is an internal $\SU(2)$ index keeping track of the doublet structure on which Pauli matrices $\tau_a$ act.
\end{itemize}
In either case, one defines ${\bar \theta}= \theta^t\,C$, in terms of an antisymmetric charge conjugation matrix $C$ satisfying
\begin{equation}
\Gamma_m^t = -C\Gamma_m C^{-1}\,, \quad \quad C^t = -C\,,
\end{equation}
with $\Gamma_m$ satisfying the standard Clifford algebra $\{\Gamma_m,\,\Gamma_n\}= 2\eta_{mn}$ with mostly plus eigenvalues. I am not introducing an special notation above to refer to the tangent space, given the flat nature of the bulk.
This is not accurate but will ease the notation below. I will address this point when reinterpreting our results in terms of a purely superspace formulation.

Let me start the discussion with the DBI piece of the action. This involves couplings to the NS-NS bulk sector, a sector that is also probed by the superstring. Thus, both the supervielbein $(E^m,\,E^\alpha)$ and the NS-NS 2-form $B_{2}$ were already identified to be
\begin{eqnarray}
E^m = \Pi^m &=& dX^m + d{\bar \theta}\Gamma^m d\theta \quad, \quad
E^\alpha = d\theta^\alpha \\
B_{2} &=& -{\bar \theta}\Gamma_\sharp\Gamma_m d\theta\,(dX^m +\frac{1}{2}{\bar \theta}
\G^m d\theta)\,,
\end{eqnarray}
in type IIA, whereas in type IIB one replaces $\Gamma_\sharp$ by $\tau_3$. The DBI action 
\begin{equation}
S_{{\rm DBI}} = -T_{{\rm Dp}}\int d^{p+1}\sigma\,\sqrt{-{\rm det}\,(\CG + \CF)}
\label{superdbi}
\end{equation}
will therefore be invariant under the spacetime supersymmetry transformations
\begin{equation}
\delta_\epsilon \theta = \epsilon \,,\quad \delta_\epsilon X^m = 
{\bar \epsilon}\Gamma^m\theta
\label{eq:flatsusy}
\end{equation}
if both, the induced world volume metric $\CG$ and the gauge invariant 2-form $\CF$, are. These are defined by
\begin{eqnarray}
\CG_{\mu\nu} &=& \Pi^m_\mu \Pi^ n_\nu \eta_{mn},\quad
\Pi^m_\mu = \6_\mu X^m - {\bar \theta}\Gamma^m\6_\mu \theta \\
\CF_{\mu\nu} &=& 2\pi\alpha^\prime F_{\mu\nu}-{\cal B}_{\mu\nu},
\end{eqnarray}
where ${\cal B}$ stands for the pull back of the superspace 2-form $B_{2}$ into the worldvolume, i.e. ${\cal B}_{\mu\nu}=\partial_\mu Z^M\partial_\nu Z^N\,B_{MN}$. Since $B_2$ is quasi-invariant under (\ref{eq:flatsusy}), one chooses
\begin{equation}
\delta_\epsilon V = \bar\epsilon \Gamma_\sharp \Gamma_m \theta d X^m + 
\frac{1}{6} (\bar\epsilon \Gamma_\sharp \Gamma_m \theta \bar\theta \Gamma^m 
d\theta + \bar\epsilon \Gamma_m \theta\bar\theta \Gamma_\sharp \Gamma^m  
d\theta),
\label{Vsusy}
\end{equation}
so that $\delta_\epsilon\CF=0$, guaranteeing the invariance of the action (\ref{superdbi}) since the set of 1-forms $\Pi^m$ are supersymmetric invariant.

Let me consider the WZ piece of the action
\begin{equation}
S_{{\rm WZ}} = \int \Omega_{p+1}\,.
\end{equation}
Since invariance under supersymmetry allows total derivatives, the lagrangian can be characterised in terms of a (p+2)- form
\begin{equation}
I_{p+2}=d\Omega_{p+1},
\end{equation}
satisfying
\begin{equation} 
\delta_\epsilon I_{p+2}=0 \quad \Longrightarrow \quad \delta_\epsilon \Omega_{p+1}
=d\Lambda_{p}.
\end{equation}
Thus, mathematically, $I_{(p+2)}$ must be constructed out of supersymmetry invariants $\{\Pi^m,\, d\theta,\,\CF\}$.

The above defines a cohomological problem whose solution is {\it not} guaranteed to be kappa invariant. Since our goal is to construct an action invariant under both symmetries, let me first formulate the requirements due to the second invariance. The strategy followed in \cite{Aganagic:1996pe} consists in two steps :
\begin{itemize}
\item First, parameterise the kappa transformation of the bosonic fields $\{X^m,\,V_{1}\}$ in terms of an arbitrary  $\delta_\kappa \theta$. Experience from supersymmetry and kappa invariance for the superparticle and superstring suggest
\begin{eqnarray}
\delta_\kappa X^m &=& -\delta_\kappa {\bar \theta}\Gamma^m \theta \nonumber \\
\delta_\kappa V_1 &=& - \delta_\kappa\bar\theta \Gamma_\sharp \Gamma_m \theta 
\Pi^m + \frac{1}{2}\delta_\kappa \bar\theta \Gamma_\sharp \Gamma_{m} \theta 
\bar \theta \Gamma^m d\theta
- \frac{1}{2} \delta_\kappa \bar\theta \Gamma^m \theta \bar\theta \Gamma_\sharp
\Gamma_m d\theta\,. \label{Vkappa}
\end{eqnarray}
Notice $\delta_\kappa V$ is chosen to remove the exact form coming from the kappa symmetry variation of $B_{2}$, i.e. $\delta_\kappa B_{2}= -2\delta_\kappa{\bar \theta}\Gamma_\sharp\Gamma_m d\theta \Pi^m + d\delta_\kappa V_1$.
\item Second, kappa symmetry {\it must} be able to remove half of the fermionic degrees of freedom. Thus, as in the superstring discussion, one expects $\delta_\kappa\theta$ to involve some non-trivial projector. This fact can be used to conveniently parameterise the kappa invariance of the total lagrangian. The idea in \cite{Aganagic:1996pe} was to parameterise the DBI kappa transformation as
\begin{equation}
\delta_\kappa \CL_{{\rm DBI}} = 2\delta_\kappa {\bar \theta} \gamma^{(p)}
T^\nu_{(p)}\6_\nu \theta \,, \quad \quad {\rm with} \quad \quad (\gamma^{(p)})^2=\mathbb{1}\,,
\end{equation}
requiring the WZ kappa transformation to be
\begin{equation}
\delta_\kappa \CL_{{\rm WZ}} = 2\delta_\kappa {\bar \theta}T^\nu_{(p)}\6_\nu 
\theta\,.
\end{equation} 
In this way, the kappa symmetry variation of the full lagrangian equals
\begin{equation}
\delta_\kappa (\CL_{{\rm DBI}}+\CL_{{\rm WZ}})=
2\delta_\kappa {\bar \theta}(\mathbb{1}+\gamma^{(p)})T^\nu_{(p)}\6_\nu \theta\,.
\end{equation}
This is guaranteed to vanish choosing $\delta_\kappa {\bar \theta}={\bar \kappa}(\mathbb{1}-\gamma^{(p)})$, given the projector nature of $\frac{1}{2}(\mathbb{1}\pm \gamma^{(p)})$.
\end{itemize}

The question is whether $T^\nu_{(p)}$, $\gamma^{(p)}$ and $I_{(p+2)}$ exist satisfying all the above requirements.
The explicit construction of these objects was given in \cite{Aganagic:1996pe}. Here, I simply summarise their results. The WZ action was found to be
\begin{equation}
d\CL_{{\rm WZ}}=-T_{\rm Dp} {\cal R}e^{\CF}\,,
\end{equation}
where ${\cal R}$ is the pullback of the field strength of the RR gauge potential $C$, as defined in (\ref{eq:fsum}). Using $\,\slPi=\Pi^m\Gamma_m$, this can be written as  \cite{Hatsuda:1998by}
\begin{eqnarray}
{\cal R}&=&{\bar E}{\cal C}_A(\slPi)E,~~~~~~~~~~~
{\cal C}_A(\slPi)~=~\sum_{l=0}~(\Gamma_\sharp)^{l+1}~\frac{\slPi^{2l}}{(2l)!}
\end{eqnarray}
in type IIA, whereas in type IIB \cite{Kamimura:1997ju}
\begin{eqnarray}
{\cal R} &=&-{\bar E} ~{\cal S}_B(\slPi)~\tau_1~E,~~~~~~~~~~~
{\cal S}_B(\slPi)~=~\sum_{l=0}
(\tau_{3})^{l}~\frac{\slPi^{2l+1}}{(2l+1)!}\,.
\end{eqnarray}
Two observations are in order :
\begin{itemize}
\item[1.] $d{\cal L}_{{\rm WZ}}$ is indeed manifestly supersymmetric, since it only depends on supersymmetric invariant quantities, but  ${\cal L}_{{\rm WZ}}$ is quasi-invariant. Thus, when computing the algebra closed by the set of conserved charges, one can expect the appearance of non-trivial charges in the right hand side of the supersymmetry algebra. This is a universal feature of brane effective actions that will conveniently interpreted in section \ref{sec:symmetries}.
\item[2.] This analysis has determined the explicit form of {\it all} the RR potentials $C_p$ as superfields in superspace. This was achieved by world volume symmetry considerations, but it is reassuring to check the expressions found above do satisfy the superspace constraints reported in appendix \ref{sec:iiab}. I will geometrically reinterpret the derived action as one describing a Dp-brane propagating in a fixed SuperPoincar\'e target space shortly.
\end{itemize}

Let me summarise the global and gauge symmetry structure of the full action. The set of gauge symmetries involves world volume diffeomorphisms $(\xi^\mu)$, an abelian $\U(1)$ gauge symmetry $(c)$ and kappa symmetry $(\kappa)$. Their infinitesimal transformations are 
\begin{eqnarray}
s X^m& = & \CL_\xi X^m  + \delta_\kappa X^m 
= \xi^\mu\6_\mu X^m - \delta_{\kappa}\5\theta \Gamma^m\theta\,, \\
s\theta^\alpha & = & \CL_\xi \theta^\alpha + \delta_{\kappa}\theta^\alpha =  \xi^\mu\6_\mu \theta^\alpha + 
\delta_{\kappa}\theta^\alpha\,, \\
sV_\mu & = & \CL_\xi V_\mu + \6_\mu c + \delta_\kappa V_\mu = \xi^\nu\6_\nu V_\mu + V_\nu \6_\mu \xi^\nu + \6_\mu c + \delta_\kappa V_\mu\,,
\end{eqnarray}
where $\delta_\kappa V_\mu$ is given in (\ref{Vkappa}) and $\delta_\kappa \theta$ was determined in
\cite{Aganagic:1996pe}
\begin{equation}
\delta_{\kappa}\bar\theta = {\bar \kappa}(\mathbb{1} -\gam^{(p)}),~~~~~~~
\gam^{(p)}=\frac{\rho^{(p)}}{\sqrt{-{\rm det}(\CG+\CF)}}\,.
\label{eq:kapfer}
\end{equation}
In type IIA, the matrix $\rho^{(p)}$ stands for the p+1 world volume form coefficient of 
$\CS_A(\slPi)e^\CF$, where 
\begin{equation}
\rho^{(p)} =[\CS_A(\slPi)e^\CF]_{p+1},~~~~~~~~
 {\cal S}_A(\slPi)~=~\sum_{l=0}(\Gamma_\sharp)^{l+1}\frac{\slPi^{2l+1}}{(2l+1)!}
\end{equation}
while in type IIB, it is given by  
\begin{equation}
\rho^{(p-1)}=-[\CC_B(\slPi)e^{\CF}\tau_1]_{p},~~~~~~~~
{\cal C}_B(\slPi)~=~\sum_{l=0}(\tau_3)^{l+1}\frac{\slPi^{2l}}{(2l)!}\,.
\end{equation}
It was proved in \cite{Aganagic:1996pe} that $\rho^2 = -{\rm det}(\CG+\CF)\mathbb{1}$. This proves $\gamma_{(p)}^2$ equals the identity, as required in our construction.

The set of global symmetries includes supersymmetry $(\epsilon)$, bosonic
translations $({\rm a}^m)$ and Lorentz transformations $(\omega^{mn})$. The field infinitesimal transformations are
\begin{eqnarray}
\Delta X^m &=& \delta_\epsilon X^m + \delta_{\rm a} X^m +
\delta_\omega X^m =  \5\epsilon\Gamma^m \theta +
{\rm a}^m+\w^m\,_n X^n, \\
\Delta \theta^\alpha &=& \delta_\epsilon \theta^\alpha + \delta_\omega
\theta^\alpha = \epsilon^\alpha + \frac{1}{4}\omega^{mn}\left(
\Gamma_{mn}\theta\right)^\alpha, \\
\Delta V_\mu &=& \delta_\epsilon V_{\mu},
\end{eqnarray}
with $\delta_\epsilon V_{\mu}$ given in (\ref{Vsusy}) and  $\omega^m\,_n\equiv \omega^{mp}\eta_{pn}$.

\paragraph{Geometrical reinterpretation of the effective action :} the supersymmetric action was constructed out of  the supersymmetric invariant forms $\{\Pi^m,\, d\theta,\,\CF\}$. These can be reinterpreted as the pullback of ten dimensional superspace tensors to the p+1 brane world volume. To see this, it is convenient to introduce the explicit supervielbein components $E_M^A(Z)$, defined in appendix \ref{sec:iiab}, where the index M stands for curved superspace indices, i.e. $M=\{m,\,\alpha\}$, and the index A for tangent flat superspace indices, i.e. $A=\{a,\,\underline{\alpha}\}$. In this language, the SuperPoincar\'e supervielbein components equal
\begin{equation}
E_m^{a}=\delta_m^{a}\,, \quad \quad E_\alpha^{\underline{\alpha}}=\delta_\alpha^{\underline{\alpha}}\,, \quad \quad E_m^{\underline{\alpha}}=0\,, \quad \quad E_\alpha^{a}=
\left({\bar \theta}\Gamma^{a}\right)_{\underline{\alpha}}
\delta_\alpha^{\underline{\alpha}} \,.
\label{eq:fsuperv}
\end{equation}
This notation makes manifest that {\it all} Clifford matrices $\Gamma^a$ act in the tangent space, as they should. The components (\ref{eq:fsuperv}) allow us to rewrite all couplings in the effective action as pullbacks
\begin{eqnarray}
\CG_{\mu\nu}(Z) &=& \partial_\mu Z^M E_M^a(Z)\partial_\nu Z^N E_N^b(Z)\eta_{ab}, \nonumber \\
{\cal B}_{\mu\nu}(Z) &=& \partial_\mu Z^M E_M^A(Z)\partial_\nu Z^N E_N^C(Z)\,B_{AC}(Z), \\
{\cal C}_{\mu_1\dots \mu_{p+1}}(Z) &=& \partial_{\mu_1}Z^{M_1}E_{M_1}^{A_1}(Z)\dots \partial_{\mu_{p+1}}Z^{M_{p+1}}E_{M_{p+1}}^{A_{p+1}}(Z)C_{A_1\dots A_{p+1}}(Z), \nonumber
\end{eqnarray}
of the background superfields $E_M^A$, $B_{AC}$ and $C_{A_1\dots A_{p+1}}$ to the brane world volume. Furthermore, the kappa symmetry transformations (\ref{Vkappa}) and (\ref{eq:kapfer}) also allow a natural superspace description as
\begin{equation}
\delta_\kappa Z^M\,E_M^{a} = 0, \quad \quad \delta_\kappa Z^M\,E_M^{\underline{\alpha}} 
= (\mathbb{1} +\Gamma_\kappa)\kappa
\end{equation}
where the kappa symmetry matrix  $\Gamma_\kappa$ is nicely repackaged
\begin{eqnarray}
\left(\Gamma_\kappa\right)_{(p+1)} &=& 
\frac{1}{\sqrt{-{\rm det}\,(\CG + \CF)}}\sum_{l=0}^k
\gamma_{(2l+1)}\,\Gamma_\sharp^{l+1}\wedge e^{\CF} \quad \mbox{type}\, 
\mbox{IIA}\,\, \mbox{p=2k} 
\label{kappasuperdpa} \\
\left(\Gamma_{\kappa}\right)_{(p+1)} &=& 
\frac{1}{\sqrt{-{\rm det}\,(\CG + \CF)}}\sum_{l=0}^{k+1}
\gamma_{(2l)}\tau_3^l \wedge e^{\CF}\,i\tau_2 \quad \mbox{type}\, 
\mbox{IIB}\,\, \mbox{p=2k+1}\,,
\end{eqnarray}
in terms of the induced Clifford algebra matrices $\gamma_\mu$ and the gauge invariant tensor ${\cal F}$
\begin{eqnarray}
\gamma_{(1)}&\equiv & d\sigma^\mu\gamma_\mu = d\sigma^\mu \6_\mu Z^M 
E_M^{a}(Z)\Gamma_{a}, \\
\CF &=& 2\pi\alpha^\prime F - {\cal B}_2,
\end{eqnarray}
whereas $\gamma_{(l)}$ stands for the wedge product of the 1-forms $\gamma_{(1)}$.

\paragraph{Summary :} We have constructed an effective action describing the propagation of Dp-branes in ten dimensional Minkowski spacetime being invariant under p+1 dimensional diffeomorphisms,  ten dimensional supersymmetry and kappa symmetry. The final result resembles the bosonic action (\ref{bosonicDp}) in that it is written in terms of pullbacks of the components of the different superfields $E_M^A(Z)$, $B_{AC}(Z)$ and $C_{A_1\dots A_{p+1}}(Z)$ encoding the non-trivial information about the {\it non-dynamical} background where the brane propagates in a manifestly supersymmetric way. These superfields are {\it on-shell} supergravity configurations, since they satisfy the set of constraints listed in appendix \ref{sec:iiab}. It is this set of features that will allow us to generalise these couplings to arbitrary {\it on-shell} superspace backgrounds in section
\ref{sec:susycurve}, while preserving the same kinematic properties.

\subsubsection{M2-brane in flat superspace}
\label{sec:m2flatsuper}

Let me consider an M2-brane as an example of an M-brane propagating in d=11 SuperPoincar\'e.
Given the lessons from the superstring and D-brane discussions, my presentation here will be much more economical. 

First, let me describe d=11 SuperPoincar\'e as a solution of eleven dimensional supergravity using the superspace formulation introduced in appendix \ref{sec:d11app}. In the following, all fermions will be eleven dimensional Majorana fermions $\theta$ as corresponds to ${\cal N}=1$ $d=11$ superspace. Denoting the full set of superspace coordinates as $\{Z^M\}=\{X^m,\theta^\alpha\}$ with $m=0,\dots,10$ and $\alpha=1,\dots ,32$, the superspace description of ${\cal N}=1$ d=11 SuperPoincar\'e is \cite{DeAzcarraga:1989gm,Claus:1997cq}
\begin{eqnarray}
E^{a}&=& dX^{a} + d{\bar \theta}\Gamma^{a}\theta \,, \quad E^{\underline{\alpha}}=d\theta^{\underline{\alpha}} \nonumber \\
R_{4} &=& \frac{1}{2}E^{a}\wedge E^{b}\wedge
d\theta^{\underline{\alpha}}\wedge d\theta^{\underline{\beta}}\left(
\Gamma_{ab}\right)_{\underline{\alpha\beta}} \nonumber \\
R_{7} &=& \frac{1}{5!}E^{a_1}\wedge E^{a_2}\wedge
E^{a_3}\wedge E^{a_4}\wedge E^{a_5}\wedge d\theta^{\underline{\alpha}}\wedge 
d\theta^{\underline{\beta}}\left(\Gamma_{a_1a_2a_3a_4a_5}\right)_{\underline{\alpha\beta}}
\end{eqnarray}
It includes the supervielbein $E^{A} = \{E^{a},\,E^{\underline{\alpha}}\}$ and the gauge invariant field strengths $R_{4}=dA_{3}$
and its Hodge dual $R_{7}=dA_{6}+\frac{1}{2}A_{3}\wedge R_{4}$, defined as (\ref{onea}) in appendix \ref{sec:d11app}.

The full effective action can be written as \cite{Bergshoeff:1987qx}
\begin{eqnarray}
S &=& -T_{{\rm M2}}\int d^3\sigma\,\sqrt{-{\rm det}\,\CG_{\mu\nu}} + T_{{\rm M2}}\int {\cal A}_{3}
\label{superm2} \\
\CG_{\mu\nu} &=& E_\mu^{a}(X,\theta)E_\nu^{b}(X,\theta)\eta_{ab}, \quad E_\mu^{A}\equiv \partial_\mu Z^M\,E_M^{A}(X,\theta)\nonumber \\
{\cal A}_{3} &=& \frac{1}{3!}\varepsilon^{\mu\nu\rho}E_\mu^{B}E_\nu^{C}E_\rho^{D} A_{BCD}(X,\theta). \nonumber
\end{eqnarray}
Notice it depends on the supervielbeins $E_M^{A}(X,\theta)$
and the three form potential $C_{ABC}(x,\theta)$ superfields only
through their pullbacks to the world volume. 

Its symmetry structure is analogous to the one described for D-branes. Indeed, the action (\ref{superm2}) is gauge invariant under world volume diffeomorphisms $(\xi^\mu)$ and kappa symmetry $(\kappa)$ with infinitesimal transformations given by
\begin{eqnarray}
s X^m& = & \CL_\xi X^m  + \delta_\kappa X^m 
= \xi^\mu\6_\mu X^m + \delta_{\kappa}\5\theta \Gamma^m\theta\,, \\
s\theta^\alpha & = & \CL_\xi \theta^\alpha + \delta_{\kappa}\theta^\alpha =  \xi^\mu\6_\mu \theta^\alpha + 
(1+\G_\kappa)\kappa \,, \\
\Gamma_\kappa &=& \frac{1}{3!\,\sqrt{-{\rm det}\,\CG}}\varepsilon^{\mu\nu\rho}
E_\mu^{a}E_\nu^{b} E_\rho^{c}\, \Gamma_{abc}\,.
\label{eq:kmatrix-m2}
\end{eqnarray}
The kappa matrix (\ref{eq:kmatrix-m2}) satisfies $\G_\kappa^2=\mathbb{1}$. Thus, $\delta_\kappa\theta$ is a projector that will allow to gauge away half of the fermionic degrees of freedom.

The action (\ref{superm2}) is also invariant under global SuperPoincar\'e transformations
\begin{eqnarray}
\delta \theta &=& \epsilon + \frac{1}{4}\omega_{mn}\Gamma^{mn}\theta \\
\delta X^m &=& {\bar \epsilon}\Gamma^m \theta + a^m + \omega^m\,_n X^n.
\end{eqnarray}
Supersymmetry quasi-invariance can be easily argued for since $R_4$ is manifestly invariant. Thus, its gauge potential pullback variation will be a total derivative
\begin{equation}
\delta_\epsilon A_{3} = d[{\bar \epsilon}\Delta_2]
\label{susyc3}
\end{equation} 
for some spinor valued two form $\Delta_2$. 

It is worth mentioning that just as the bosonic membrane action reproduces the string world sheet action under double dimensional reduction, the same statement is true for their supersymmetric and kappa invariant formulations \cite{Duff:1987bx,Townsend:1995kk}. 

\subsection{Supersymmetric brane effective actions in curved backgrounds}
\label{sec:susycurve}

In this section, I extend the supersymmetric and kappa invariant D-brane and M2-brane actions in SuperPoincar\'e to D-branes, M2-branes and M5-branes in arbitrary curved backgrounds. The main goal, besides introducing the formalism itself, is to highlight that the existence of kappa symmetry invariance forces the supergravity background to be {\it on-shell}.

In all effective actions under consideration, the set of degrees of freedom includes scalars $Z^M=\{X^m,\theta^\alpha\}$ and it may include some gauge field $V_p$, whose dependence is always through the gauge invariant combination $dV_p-{\cal B}_{p+1}$\footnote{For $p=1$, $B_2$ is the NS-NS 2-form, whereas for $p=2$, $B_3=A_3$ is the d=11 3-form gauge potential.}. The set of kappa symmetry transformations will universally be given by
\begin{eqnarray}
  \delta_\kappa Z^M\,E_M^{a}(X,\theta) &=& 0\,, \nonumber \\
  \delta_\kappa Z^M\,E_M^{\underline{\alpha}}(X,\theta) &=& (\mathbb{1}+\Gamma_\kappa)\kappa\,, \nonumber \\   
  \delta_\kappa  V_p &=& Z^\star i_\kappa B_{p+1}. \label{eq:genkappatrafo}
\end{eqnarray}
The last transformation is a generalisation of the one encountered in SuperPoincar\'e. Indeed, the kappa variation of the pullback of any $T_n$ n-form satisfies
\begin{equation}
  \delta_\kappa {\cal T}_n \equiv \delta_\kappa Z^\star T_n = Z^\star \CL_\kappa T_n =  
  Z^\star\{d,\,i_\kappa\}T_n,
\end{equation}
where $Z^\star$ stands for the pullback of $T_n$ to the world volume. The choice in (\ref{eq:genkappatrafo}) guarantees the kappa transformation of $dV_p$ removes the total derivative in $\delta_\kappa {\cal B}_{p+1}$.

The structure of the transformations (\ref{eq:genkappatrafo}) is universal, but the details of the kappa symmetry matrix $\Gamma_\kappa$ depend on the specific theory, as described below. A second universal feature, associated with the projection nature of kappa symmetry transformations, i.e. $\Gamma_\kappa^2=\mathbb{1}$, is the correlation between the brane charge density and the sign of $\Gamma_\kappa$ in (\ref{eq:genkappatrafo}). More specifically, any brane effective action will have the structure
\begin{equation}
  S_{{\rm brane}} = -T_{{\rm brane}}\int d^{p+1}\sigma \left(\CL_{DBI}-\epsilon_1\CL_{WZ}\right).
\end{equation}
Notice this is equivalent to requiring $T_{{\rm brane}} = |Q_{{\rm brane}}|$, a property that is just reflecting the half-BPS nature of these branes. It can be shown that
\begin{equation}
\delta_\kappa S_{{\rm brane}} \propto (1+\epsilon_1 \Gamma_\kappa)\delta_\kappa \theta \quad
\Longrightarrow \quad \delta_\kappa \theta=(1-\epsilon_1 \Gamma_\kappa)\kappa.
\label{eq:charskap}
\end{equation}
The choice of $\epsilon_1$ is correlated to the distinction between a brane and an anti-brane. Both are supersymmetric, but preserve complementary supercharges. This ambiguity explains why some of the literature has apparently different conventions, besides the possibility of working with different Clifford algebra realisations\footnote{The derivation of the property (\ref{eq:charskap}) is made more manifest in formalisms in which the tension is generated dynamically by the addition of an auxiliary volume density \cite{Bergshoeff:1992gq,Lindstrom:1997uj,Bergshoeff:1998ha}.}.

\subsubsection{M2-branes}

The effective action describing a single M2-brane in an arbitrary eleven dimensional background
is formally the same as in (\ref{superm2})
\begin{equation}
  S_{M2} = -T_{{\rm M2}}\int d^3\sigma\,\sqrt{-{\rm det}\,\CG_{\mu\nu}} + T_{{\rm M2}}\int {\cal A}_{3},
\label{eq:m2gcurve}
\end{equation}
with the same definitions for the induced metric $\CG$ and the pull back 3-form ${\cal A}_3$. The information regarding {\it different} eleven dimensional backgrounds is encoded in the different couplings described by the supervielbein $E_M^A(X,\,\theta)$ and 3-form $A_{ABD}(X,\,\theta)$ superfields.

The action (\ref{eq:m2gcurve}) is manifestly 3d diffeomorphism invariant. It was shown to be kappa invariant under the transformations (\ref{eq:genkappatrafo}), without any gauge field, whenever the background superfields satisfy the constraints reviewed in appendix \ref{sec:d11app}, i.e. whenever they are {\it on-shell}, for a kappa symmetry matrix given by \cite{Bergshoeff:1987cm}
\begin{equation}
\Gamma_\kappa = \frac{1}{3!\sqrt{-{\rm det}\,\CG}}\varepsilon^{\mu\nu\rho}
E_\mu^{a}(X,\theta)\,E_\nu^{b}(X,\theta)\,E_\rho^{c}(X,\theta)\,
\Gamma_{abc},
\label{eq:gkappam2}
\end{equation}
where $E_\mu^{a}(X,\theta)=\partial_mu X^m E_m^a(X,\theta)$ is the pullback of the curved supervielbein to the world volume.

\subsubsection{D-branes}

Proceeding in an analogous way for Dp-branes, their effective action in an arbitrary type IIA/B background is
\begin{eqnarray}
S_{{\rm Dp}} &=& -T_{{\rm Dp}}\int d^{p+1}\sigma\, e^{-\phi}\,\sqrt{-{\rm det}\,(\CG + \CF)}
+ T_{{\rm Dp}}\int {\cal C}\wedge e^{\CF} \label{kappadpbrane} \\
\CF_{\mu\nu} &=& 2\pi\alpha^\prime F_{\mu\nu} - E_\mu^{A}\,E_\nu^{C}\,B_{AC}, \quad
{\cal C}_{r} =\frac{1}{r!}\epsilon^{\mu_1\dots\mu_{r}}E_{\mu_1}^{A_1}\dots E_{\mu_{r}}^{A_{r}}C_{A_1\dots A_{r}}.
\end{eqnarray}
It is understood that $\CG_{\mu\nu}(X,\theta)=E_\mu^aE_\nu^b\eta_{ab}$ and ${\cal C}$ is defined using the same notation as in (\ref{eq:fsum}), i.e. as a formal sum of forms, so that the WZ term picks {\it all} contributions coming from the wedge product of this sum and the Taylor expansion of $e^{{\cal F}}$ that saturate the $p+1$ world volume dimension. Notice {\it all} information on the background spacetime is encoded in the superfields $E_M^{A}(X,\theta)$, $\phi(X,\theta)$, $B_{AC}(X,\theta)$ and the set of RR potentials $\{C_{A_1\dots A_r}(X,\theta)\}$. 

The action (\ref{kappadpbrane}) is p+1 dimensional diffeomorphic invariant and it was shown to be kappa invariant under the transformations (\ref{eq:genkappatrafo}) for $V_1$ in \cite{Cederwall:1996ri,Bergshoeff:1996tu} when the kappa symmetry matrix equals
\begin{eqnarray}
\left(\Gamma_\kappa\right)_{(p+1)} &=& 
\frac{1}{\sqrt{-{\rm det}\,(\CG + \CF)}}\sum_{l=0}
\gamma_{(2l+1)}\,\Gamma_{11}^{l+1}\wedge e^{\CF} \quad \mbox{type}\, 
\mbox{IIA}\,\,\, \mbox{p=2k} 
\label{eq:kappasuperdpa} \\
\left(\Gamma_{\kappa}\right)_{(p+1)} &=& 
\frac{1}{\sqrt{-{\rm det}\,(\CG + \CF)}}\sum_{l=0}
\gamma_{(2l)}\tau_3^l \wedge e^{\CF}\,i\tau_2 \quad \mbox{type}\, 
\mbox{IIB}\,\,\, \mbox{p=2k+1}\,,
\end{eqnarray}
and the background is {\it on-shell}, i.e. satisfies the constraints reviewed in appendix \ref{sec:iiab}. In the expressions above $\gamma_{(1)}$ stands for the pullback of the bulk tangent space Clifford matrices
\begin{equation}
\gamma_{(1)}= d\sigma^\mu\gamma_\mu = d\sigma^\mu E_\mu^{a}(X,\theta)\Gamma_{a},
\end{equation}
and $\gamma_{(r)}$ stands for the wedge product of r of these 1-forms. In \cite{Bergshoeff:1998ha}, readers can find an extension of the results reviewed here when the background includes a mass parameter, i.e. it belongs to massive IIA \cite{Romans:1985tz}.

\subsubsection{M5-branes}

The six dimensional diffeomorphic and kappa symmetry invariant M5-brane \cite{Bandos:1997ui} is a formal extension of the bosonic one
\begin{eqnarray}
S_{{\rm M5}} & = & T_{{\rm M5}}\int\! d^6\xi\, (\CL_0 + \CL_{{\rm WZ}})\,, \nonumber \\
\CL_0 &=& -\sqrt{-{\rm \det} (\CG_{\mu\nu}+{\tilde H}_{\mu\nu})}
+ \frac{\sqrt{-{\rm det}\, \CG}}{4(\partial a \cdot \partial a)}
({\partial}_\mu a) (\CH^*)^{\mu\nu\rho} \CH_{\nu\rho\iota}({\partial}^\iota a) \label{23a}\\
\CL_{{\rm WZ}} &=& {\cal A}_6 + \frac{1}{2} \CH_3 \wedge {\cal A}_{3} \,,
\end{eqnarray}
where all pullbacks refer to superspace. This is kappa invariant under the transformations (\ref{eq:genkappatrafo}) for $V_2$, including the extra transformation law
\begin{equation}
  \delta_\kappa a = 0,
\end{equation}
for the auxiliary scalar field introduced in the PST formalism. These transformations are determined by the kappa symmetry
matrix
\begin{equation}
\Gamma_\kappa =  \frac{v_\mu  \gamma^\mu}{\sqrt{-{\rm \det} (\CG_{\mu\nu}+{\tilde H}_{\mu\nu})}} 
\bigg[\gamma_\nu t^\nu + \frac{\sqrt{-{\rm det}\, \CG}}{2}\,\gamma^{\nu\rho} \tilde H_{\nu\rho}
- \frac{1}{5!}\,{\varepsilon}^{\mu_1\dots \mu_5 \nu} v_\nu \gamma_{\mu_1\dots \mu_5}\bigg]\,.
\label{eq:kmatrixm5}
\end{equation}
where $\gamma_\mu=E_\mu{}^a\Gamma_a$ and the vector fields $t^\mu$ and $v_\mu$ are defined by
\begin{equation}
t^\mu = \frac{1}{8}\, {\varepsilon}^{\mu\nu_1\nu_2\rho_1\rho_2\iota}
{\tilde H}_{\nu_1\nu_2}{\tilde H}_{\rho_1\rho_2} v_\iota \quad {\rm with} \quad v_\mu \equiv \frac{\partial_\mu a}{\sqrt{-\partial a\cdot \partial a}}.
\end{equation}

\paragraph{Further comments on kappa symmetry :} $\kappa$--symmetry is a fermionic local symmetry for which {\it no} gauge field is necessary. Besides its defining projective nature when acting on fermions, i.e. $\delta_\kappa \theta = \left(\mathbb{1} + \Gamma_\kappa\right)\kappa$ with $\Gamma_\kappa^2=\mathbb{1}$, there are two other distinctive features it satisfies \cite{Sezgin:1993xg}
\begin{itemize}
\item[1.] the algebra of $\kappa$--transformations only closes on-shell,
\item[2.] $\kappa$--symmetry is an infinitely reducible symmetry.
\end{itemize}
The latter statement uses the terminology of Batalin and Vilkovisky \cite{Batalin:1984jr} and it is a direct consequence of its projective nature, since the existence of the infinite chain of transformations
\begin{equation}
\kappa \rightarrow (1-\Gamma_\kappa)\kappa_1\,, \kappa_1 \rightarrow 
(1+\Gamma_\kappa)\kappa_2 \dots
\end{equation}
gives rise to an {\it infinite} tower of ghosts when attempting to follow the Batalin-Vilkovisky  quantization procedure, which is also suitable to handle the first remark above. Thus, covariant quantisation of kappa invariant actions is a subtle problem.
For detailed discussions on problems arising from the regularisation of infinite sums and dealing with Stueckelberg type residual gauge symmetries, readers are referred to \cite{Kallosh:1989cj,Kallosh:1989yv,Green:1989ab,Gates:1989hg,Bergshoeff:1989rf}. 

It was later realised, using the Hamiltonian formulation, that kappa symmetry does allow covariant quantisation provided the ground state of the theory is massive \cite{Kallosh:1997nr}. The latter is clearly consistent with the brane interpretation of these actions, by which these vacua capture the half-BPS nature of the (massive) branes themselves\footnote{I will prove this explicitly in section \ref{sec:vacuum}.}. 

For further interesting kinematical and geometrical aspects of kappa symmetry, see the reviews \cite{Sezgin:1993xg,deAzcarraga:2004df,deAzcarraga:2001fi} and references therein.

\subsection{Symmetries : spacetime vs world volume}
\label{sec:symmetries}

The main purpose of this section is to discuss the global symmetries of brane effective actions, the algebra they close and to emphasise the interpretation of some of the conserved charges appearing in these algebras before and after gauge fixing of the world volume diffeomorphisms and kappa symmetry
\begin{itemize}
\item {\it before} gauge fixing, the p+1 field theory will be invariant under the full superisometry of the background where the brane propagates. This is a natural extension of the SuperPoincar\'e invariance when branes propagate in Minkowski. As such, the algebra closed by the brane conserved charges will be a subalgebra of the maximal spacetime superalgebra one can associate to the given background.
\item {\it after} gauge fixing, only the subset of symmetries preserved by the brane embedding will remain linearly realised. This subset determines the world volume (supersymmetry) algebra. In the particular case of brane propagation in Minkowski, this algebra corresponds to a subalgebra of the maximal  SuperPoincar\'e algebra in p+1 dimensions.
\end{itemize}

To prove that background symmetries give rise to brane global symmetries, one must first properly define the notion of superisometry of a supergravity background. This involves a Killing superfield $\xi(Z)$ satisfying the properties
\begin{eqnarray}
  {\cal L}_\xi (E^a \otimes_s E^b)\eta_{ab} &=& 0\,, \label{eq:metricisometry} \\
  {\cal L}_\xi R_4 &=& {\cal L}_\xi R_7 = 0\,, \quad \quad \mbox{M-theory} \label{eq:misometry} \\
  {\cal L}_\xi H_3 &=& {\cal L}_\xi \phi = {\cal L}_\xi R_k = 0\,, \quad \quad \mbox{Type IIA/B} \label{eq:abisometry}
\end{eqnarray}
${\cal L}_\xi$ denotes the Lie derivative with respect to $\xi$, $\eta$ is either the d=11 or d=10 Minkowski metric on the tangent space, depending on which superspace we are working on and $\{R_k, H_3\}$ are the different M-theory or type IIA/B field strengths satisfying the generalised Bianchi identities defined in appendix \ref{sec:appcons}. Notice these are the superfield versions of the standard bosonic Killing isometry equations. Invariance of the field strengths allows the corresponding gauge potentials to have non-trivial transformations
\begin{eqnarray}
  {\cal L}_\xi A_3 &=& d\Delta_2\,, \quad \quad {\cal L}_\xi A_6 = d\Delta_5 - \frac{1}{2}\Delta_2\wedge R_4\,, \quad \quad \mbox{M-theory} \label{eq:mgpisometry} \\
  {\cal L}_\xi B_2 &=& d\lambda_1\,, \quad \quad {\cal L}_\xi C_{p+1} = d\omega_p - d\omega_{p-2}\wedge H_3\,, \quad \quad \mbox{Type IIA/B} 
\label{eq:abgpisometry}
\end{eqnarray}
for some set of superfield forms $\{\Delta_2,\,\Delta_5,\,\omega_i\}$.

The invariance of brane effective actions under the global transformations
\begin{equation}
  \delta_\xi Z^M = \xi^M(Z)\,,
\label{eq:scalartrafo}
\end{equation}
was proved in \cite{Bergshoeff:1998ha}. The proof can be established by analysing the DBI and WZ terms of the action separately. If the brane has gauge field degrees of freedom, one can always choose its infinitesimal transformation
\begin{eqnarray}
  \delta V_2 &=&  {\rm Z}^\star(\Delta_2)\,, \quad \quad \mbox{M-theory}  \\
  \delta V_1 &=& {\rm Z}^\star(\lambda_1)\,, \quad \quad \mbox{Type IIA/B} 
\end{eqnarray}
where ${\rm Z}^\star$ stands for pullback to the world volume, i.e. ${\rm Z}^\star(\lambda_1) = dZ^M (\lambda_1)_M$. This guarantees the invariance of the gauge invariant forms, i.e.  ${\cal L}_\xi \CF = {\cal L}_\xi \CH_3 = 0$. Furthermore, the transformation of the induced metric
\begin{equation}
  {\cal L}_\xi \CG_{\mu\nu} = \partial_\mu Z^M \partial_\nu Z^N\,{\cal L}_\xi (E_M^aE_N^b\eta_{ab})\,,
\end{equation}
vanishes because of (\ref{eq:metricisometry}). This establishes the invariance of the DBI action. On the other hand, the WZ action is quasi-invariant by construction due to (\ref{eq:mgpisometry}) and (\ref{eq:abgpisometry}). Indeed,
\begin{eqnarray}
  \delta {\cal L}_{{\rm WZ}} &=& {\rm Z}^\star\left(d\Delta_2\right)\,, \quad \quad \mbox{M2-brane} \nonumber \\
  \delta {\cal L}_{{\rm WZ}} &=& {\rm Z}^\star\left(d(\Delta_5 + \frac{1}{2}\CH_3\wedge \Delta_2)\right)\,, \quad \quad \mbox{M5-brane} \nonumber  \\
  \delta {\cal L}_{{\rm WZ}} &=& {\cal L}_\xi {\cal C}\wedge e^\CF = {\rm Z}^\star\left(d\omega\right)\,. \quad \quad \mbox{D-branes}
\label{eq:WZqinvariance}
\end{eqnarray}

\paragraph{Summary :} Brane effective actions include the supergravity superisometries $\xi(Z)$ as a subset of their global symmetries. It is important to stress that kappa symmetry invariance is {\it necessary} to define a supersymmetric field theory on the brane, but {\it not sufficient}. Indeed, {\it any} on-shell supergravity background having {\it no} Killing spinors, i.e. some superisometry in which fermions are shifted as $\delta\theta = \epsilon$, breaks supersymmetry, and consequently, will never support a supersymmetric action on the brane.

The derivation discussed above does not exclude the existence of further infinitesimal transformations leaving the effective action invariant. The question of determining the full set of continuous global symmetries of a given classical field theory is a well posed mathematical problem in terms of cohomological methods \cite{Barnich:1994db,Barnich:2000zw}. Applying these to the bosonic D-string \cite{Brandt:1997cy} gave rise to the discovery of the existence of an infinite number of global symmetries \cite{Brandt:1997pk,Brandt:1998ri}. These were also proved to exist for the kappa invariant D-string action \cite{Brandt:1998wx}.

\subsubsection{Supersymmetry algebras} 
\label{sec:susyalgebra}

Since spacetime superisometries generate world volume global symmetries,  Noether's theorem \cite{Noether:1918zz} 
guarantees a field theory realisation of the spacetime (super)symmetry algebra using Poisson brackets. It is by now well known that such (super)algebras contain more bosonic charges than the ones geometrically realised as (super)isometries. There are several ways of reaching this conclusion:
\begin{itemize}
\item[1.] Group theoretically, the anticommutator of two supercharges $\{Q_\alpha,\,Q_\beta\}$ defines a symmetric matrix belonging to the adjoint representation of the some symplectic algebra $\Sp(N,\bR)$, whose order $N$ depends on the spinor representation $Q_\alpha$ belongs to. One can decompose this representation into irreducible representations of the bosonic spacetime isometry group.This can explicitly be done by using the completeness of the basis of antisymmetrised Clifford algebra gamma matrices as follows 
\begin{equation}
  \{Q_\alpha,\,Q_\beta\} = \sum_p (\Gamma^{m_1\dots m_p}C^{-1})_{\alpha\beta}Z_{m_1\dots m_p}\,,
\label{eq:susyt}
\end{equation}
where the allowed values of $p$ depend on symmetry considerations. The right hand side defines a set of bosonic charges $\{Z_{m_1\dots m_p}\}$ that typically goes beyond the spacetime bosonic isometries.
\item[2.] Physically, BPS branes in a given spacetime background have masses equal to their charges by virtue of the amount of supersymmetry they preserve. This would not be consistent with the supersymmetry algebra if the latter would not include extra charges, the set $\{Z_{m_1\dots m_p}\}$ introduced above, besides the customary spacetime isometries among which the mass (time translations) always belongs to. Thus, some of the extra charges {\it must} correspond to such brane charges. The fact these charges have non-trivial tensor structure means they are typically {\it not} invariant under the spacetime isometry group. This is consistent with the fact that the presence of branes breaks the spacetime isometry group, as I already explicitly discussed in SuperPoincar\'e.
\item[3.] All brane effective actions reviewed above are quasi-invariant under spacetime superisometries, since the WZ term transformation equals a total derivative (\ref{eq:WZqinvariance}). Technically, it is a well known theorem that such total derivatives can induce extra charges in the commutation of conserved charges through Poisson brackets. This is the actual field theory origin of the group theoretically allowed set of charges $\{Z_{m_1\dots m_p}\}$.
\end{itemize}

Let me review how these structures emerge in both supergravity and brane effective actions. Consider the most general superPoincar\'e algebra in eleven dimensions. This is spanned by a Majorana spinor supercharge $Q_{\alpha}~ (\alpha=1,\dots,32)$ satisfying the anti-commutation relations\footnote{All our charge conjugation matrices are antisymmetric and unitary, i.e. $C^T=-C$ and $C^\dag C=\mathbb{1}$. Furthermore, all Clifford matrices satisfy the symmetry relation $\Gamma_m^T = -C\Gamma_mC^{-1}$.} \cite{vanHolten:1982mx,Townsend:1995gp,Townsend:1997wg}
\begin{equation}
\{ Q_{\alpha}, Q_{\beta} \} =(\Gamma^mC^{-1})_{\alpha \beta} P_m +
\frac{1}{2}(\Gamma^{mn}C^{-1})_{\alpha \beta}Z_{mn}+ \frac{1}{5!}
(\Gamma ^{m_1\dots m_5}C^{-1})_{\alpha \beta}Y_{m_1\dots m_5}\,.
\label{eq:msusyalgebra}
\end{equation}
That this superalgebra is maximal can be argued using the fact that its left hand side defines a symmetric tensor with 528 independent components. Equivalently, it can be interpreted as belonging to the adjoint representation of the Lie algebra of $\Sp(32,\bR)$. The latter decomposes under its subgroup $\SO(1,10)$, the spacetime Lorentz isometry group, as
\begin{equation}
  {\bf 528} \rightarrow {\bf 11} \oplus {\bf 55} \oplus {\bf 462}\,.
\end{equation}
The irreducible representations appearing in the direct sum do precisely correspond to the bosonic tensor charges appearing in the right hand side : the 11-momentum $P_m$, a 2-form charge $Z_{mn}$, which is 55 dimensional, and a 5-form charge $Y_{m_1\dots m_5}$, which is 462 dimensional.

The above is merely based on group theory considerations that may or may not be realised in a given physical theory. In eleven dimensional supergravity, the extra bosonic charges are realised in terms of electric $Z_{{\rm e}}$ and magnetic $Z_{{\rm m}}$ charges, the Page charges \cite{Page:1984qv}, that one can construct out of the 3-form potential $A_3$ equation of motion, as reviewed in \cite{Stelle:1998xg,Stelle:1996tz}
\begin{eqnarray}
  Z_{{\rm e}} &=& \frac{1}{4\Omega_7}\int_{\partial{\cal M}_8} (\star R_4 + \frac{1}{2}A_3\wedge R_4)\,, \label{eq:mec} \\
  Z_{{\rm m}} &=& \frac{1}{\Omega_4}\int_{\partial{\cal M}_5} R_4\,. \label{eq:mmc}
\end{eqnarray}
The first integral is over the boundary at infinity of an arbitrary infinite 8-dimensional spacelike manifold ${\cal M}_8$, with volume $\Omega_7$. Given the conserved nature of this charge, it does not depend on the time slice chosen to compute it. But there still are many ways of embedding ${\cal M}_8$ in the corresponding 
ten dimensional spacelike hypersurface ${\cal M}_{10}$. Thus, $Z_{{\rm e}}$ represents a set of charges parameterised by the volume element 2-form describing how ${\cal M}_8$ is embedded in ${\cal M}_{10}$. This precisely matches the 2-form $Z_{mn}$ in (\ref{eq:msusyalgebra}). There is an analogous discussion for $Z_{{\rm m}}$ which corresponds to the 5-form charge $Y_{m_1\dots m_5}$. As an example, consider the M2 and M5-brane configurations in (\ref{eq:m2sol}) and (\ref{eq:m5sol}). If one labels the M2-brane tangential directions as 1 and 2, there exists a non-trivial charge $Z_{12}$ computed from (\ref{eq:mec}) by plugging in (\ref{eq:m2sol}) and evaluating the integral over the transverse 7-sphere at infinity. The reader is encouraged to read the lecture notes by K.S. Stelle \cite{Stelle:1998xg} where these issues are discussed very explicitly in a rather general framework including all standard half-BPS branes. For a more geometric construction of these maximal superalgebras in AdS$\times$S backgrounds, see \cite{FigueroaO'Farrill:2008ka} and references therein.

The above is a very brief reminder regarding spacetime superalgebras in supergravity. For a more thorough presentation of these issues, the reader is encouraged to read the lectures notes by P.K. Townsend \cite{Townsend:1997wg}, where similar considerations are discussed for both type II and heterotic supergravity theories. Given the importance given to the action of dualities on effective actions, the reader may wonder how these same dualities act on superalgebras. It was shown in
\cite{Bergshoeff:2000qu} that these actions correspond to picking different complex structures of an underlying $\OSp(1|32)$ superalgebra.

Consider the perspective offered by the M5-brane effective action propagating in d=11 superPoincar\'e. The latter is invariant both under supersymmetry and bulk translations. Thus, through Noether's theorem, there exist field theory realisations of these charges. Quasi-invariance of the WZ term will be responsible for the generation of extra terms in the calculation of the Poisson bracket of these charges \cite{DeAzcarraga:1989gm}. This was confirmed for the case at hand in \cite{Sorokin:1997ps}, where the M5-brane superalgebra was explicitly computed. The supercharges $Q_\alpha$ are
\begin{equation}
  Q_\alpha = i\int\! d^5\sigma\,\big[(\pi+\bar\theta\Gamma^{m}P_{m})_\alpha
+i({\cal P}^{i_1i_2} + \frac{1}{4}
{\CH}^{*0i_1i_2})(\Delta^2_{i_1i_2})_\alpha
-i\varepsilon^{i_1\dots i_5}
(\Delta^5_{i_1 \dots i_5})_\alpha\big]\,,
\end{equation}
where $\pi$, $P_m$ and ${\cal P}^{ij}$ are the variables canonically conjugate to $\theta$, $X^m$ and $V_{ij}$. As in any hamiltonian formalism, world volume indices were split according to $\sigma^\mu = \{t,\,\sigma^i\}$ $i=1,\dots 5$. Notice the pullbacks of the forms  $\Delta_2$ and $\Delta_5$ appearing in $\delta{\CL}_{{\rm WZ}}$ for the M5-brane in (\ref{eq:WZqinvariance}) do make an explicit appearance in this calculation. The anti-commutator of the M5 brane world volume supercharges equals (\ref{eq:msusyalgebra}) with
\begin{eqnarray}
P_{m} &=& \int d^5\sigma\, \frac{\delta \CL}{\delta(\partial_t X^m)}\,, \\
Z^{mn} &=& -\int_{{\cal M}_5} dX^m\wedge dX^n \wedge dV_2\,, \\
Y^{m_1\dots m_5} &=& \int_{{\cal M}_5} dX^{m_1}\wedge \cdots \wedge dX^{m_5}\,, \label{eq:5-charge}
\end{eqnarray}
where all integrals are computed on the 5-dimensional spacelike hypersurface ${\cal M}_5$ spanned by the M5-brane. Notice the algebra of supercharges depends on the brane dimensionality. Indeed, a single M2-brane has a two dimensional spacelike surface that can {\it not} support the pullback of an spacetime 5-form as a single M5-brane can (see (\ref{eq:5-charge})). This conclusion could be modified if the degrees of freedom living on the brane would be {\it non-abelian}.

Even though my discussion above only applies to the M5-brane in the superPoincar\'e background, my conclusions are general given the quasi-invariance of their brane WZ action, a point first emphasised in \cite{DeAzcarraga:1989gm}. The reader is encouraged to read \cite{DeAzcarraga:1989gm,DeAzcarraga:1991tm} for similar analysis carried for super p-branes, \cite{Hammer:1997ts} for D-branes in superPoincar\'e and general mathematical theorems based on the structure of brane effective actions and \cite{Sato:1998yu,Sato:1998ax}, for superalgebra calculations in some particular curved backgrounds.

\subsubsection{World volume supersymmetry algebras} 
\label{sec:worldsusyalgebra}

Once the physical location of the brane is given, the spacetime superisometry group $\mathrm{G}$ is typically broken to
\begin{equation}
  \mathrm{G} \to \mathrm{G}_0 \times \mathrm{G}_1
\end{equation}
The first factor $\mathrm{G}_0$ corresponds to the world volume symmetry group in (p+1)-dimensions, i.e. the analogue of the Lorentz group in a supersymmetric field theory in (p+1)-dimensions, whereas the second factor $\mathrm{G}_1$ is interpreted as an internal symmetry group acting on the dynamical fields building (p+1)-dimensional supermultiplets. The purpose of this subsection is to relate the superalgebras before and after this symmetry breaking process \cite{Kallosh:1997ky}\footnote{For earlier work, see \cite{Achucarro:1988qb}, who extended the original Volkov-Akulov approach in \cite{Volkov:1973ix}.}.

The link between both superalgebras is achieved through the gauge fixing of world volume diffeomorphisms and kappa symmetry, the gauge symmetries responsible for the covariance of the original brane action in the GS formalism. Focusing on the scalar content in these theories $\{X^m,\,\theta\}$, these transform as
\begin{eqnarray}
  sX^m &=& k^m(X) + \CL_\xi X^m + \delta_\kappa X^m + \delta_\epsilon X^m\,, \\
  s\theta &=& \epsilon + \delta_k \theta + \CL_\xi \theta + (\mathbb{1}+\Gamma_\kappa)\kappa + \delta_k\theta\,.
\end{eqnarray}
The general Killing superfield was decomposed into a supersymmetry translation denoted by $\epsilon$ and a bosonic Killing vector fields $k^M(X)$. World volume diffeomorphisms were denoted as $\xi$. At this stage, the reader should already notice the inhomogeneity of the supersymmetry transformation acting on fermions (the same is true for bosons if the background spacetime has a constant translation as an isometry, as it happens in Minkowski). 

Locally, one can always impose the {\it static} gauge :  $X^\mu=\sigma^\mu$, where one decomposes the scalars fields $X^m$ into world volume directions $X^\mu$ and transverse directions $X^I\equiv \Phi^I$. For infinite branes, this choice is valid globally and does describe a vacuum configuration. To diagnose which symmetries act, and how, on the physical degrees of freedom $\Phi^i$, one must make sure to work in the subset of symmetry transformations preserving the gauge slice $X^\mu=\sigma^\mu$. This forces to act with a compensating world volume diffeomorphism
\begin{equation}
  sX^\mu|_{X^\mu=\sigma^\mu} = 0 \quad \Rightarrow \quad \xi^\mu = -k^\mu  - \delta_\kappa X^\mu - \delta_\epsilon X^\mu\,.
\end{equation}
The latter acts on the physical fields giving rise to the following set of transformations preserving the gauge fixed action
\begin{eqnarray}
  s\Phi^I|_{X^\mu=\sigma^\mu} &=& k^I - k^\mu\partial_\mu \Phi^I + \dots \,, \\
  s\theta|_{X^\mu=\sigma^\mu} &=& -k^\mu\partial_\mu\theta + \CL_k\theta + \dots \,.
\end{eqnarray}
There are two important comments to be made at this point
\begin{itemize}
\item[1.] The physical fields $\Phi^I$ transform as proper {\it world volume} scalars \cite{Achucarro:1987nc}. Indeed, $\Phi^I(\sigma)=(\Phi^\prime)^I(\sigma^\prime)$ induces the infinitesimal transformation $k^\mu\partial_\mu \Phi^I$ for any $k^\mu(\sigma)$ preserving the p+1 dimensional world volume. Below, the same property will be checked for fermions.
\item[2.] If the spacetime background allows for any {\it constant} $k^I$ isometry, it would correspond to an inhomogeneous symmetry transformation for the physical field $\Phi^I$. In field theory, the latter would be interpreted as an spontaneous broken symmetry and the corresponding $\Phi^I$ would be its associated massless Goldstone field. This is precisely matching our previous discussions regarding the identifications of the appropriate brane degrees of freedom.
\end{itemize}

There is a similar discussion regarding the gauge fixing of kappa symmetry and the emergence of a subset of linearly realised supersymmetries on the (p+1)-dimensional world volume field theory. Given the projector nature of the kappa symmetry transformations, it is natural to assume ${\cal P}\theta = 0$ as a gauge fixing condition, where ${\cal P}$ stands for some projector. Preservation of this gauge slice, determines the kappa symmetry parameter $\kappa$ as a function of the background Killing spinors $\epsilon$
\begin{equation}
  s\theta|_{{\cal P}\theta = 0}=0 \quad \Longrightarrow \quad \kappa= \kappa(\epsilon)\,.
\end{equation}
When analysing the supersymmetry transformations for the remaining dynamical fermions, only certain linear combinations of the original supersymmetries $\epsilon$ will be linearly realised. The difficulty in identifying the appropriate subset depends on the choice of ${\cal P}$.

\paragraph{Branes in superPoincar\'e :} The above discussion can be made explicit in this case. Consider a p+1 dimensional brane propagating in d dimensional superPoincar\'e. For completeness, let me remind the reader of the full set of transformations leaving the brane actions invariant
\begin{eqnarray}
  sX^m &=& a^m + a^m{}_nX^n + \CL_\xi X^m + \bar\epsilon\Gamma^m\theta + \delta_\kappa X^m\,, \\
  s\theta &=& \frac{1}{4}a_{mn}\Gamma^{mn}\theta + \CL_\xi\theta + \epsilon + \delta_\kappa \theta\,,
\end{eqnarray}
where I ignored possible world volume gauge fields. Decomposing the set of bosonic scalar fields $X^m$ $m=0,1,\dots d-1$ into world volume directions $X^\mu$ $\mu=0,1,\dots p$ and transverse directions $X^I\equiv \Phi^I$ $I=p+1,\dots d-1$, one can now explicitly solve for the preservation of the static gauge slice $X^\mu=\sigma^\mu$, which does globally describe the vacuum choice of a p-brane extending in the first p spacelike directions and time. This requires some compensating world volume diffeomorphism
\begin{equation}
  \xi^\mu = -a^\mu -a^\mu{}_\nu\sigma^\nu - a^\mu{}_I \Phi^I - \bar\epsilon\Gamma^\mu\theta -\delta_\kappa X^\mu\,,
\end{equation}
inducing the following transformations for the remaining degrees of freedom
\begin{eqnarray}
  s\Phi^I & = & -a^\mu\partial_\mu \Phi^I -a^\mu{}_\nu\sigma^\nu\6_\mu \Phi^I - a^\mu{}_J \Phi^J\6_\mu \Phi^I + a^J +
a^I{}_J \Phi^J + a^I{}_\mu\sigma^\mu + \mbox{fermions}\,, \\
  s\theta &=& - a^\mu{}_\nu\sigma^\nu\partial_\mu \theta + \frac{1}{4}a_{\mu\nu}\Gamma^{\mu\nu}\theta + \frac{1}{4}a_{IJ}\Gamma^{IJ}\theta 
\label{gfbosonm2}
\end{eqnarray}
The subset of linearly realised symmetries is $\ISO(1,p)\times \SO(D-(p+1))$. The world volume "Poincar\'e" group is indeed $\ISO(1,p)$, under which $\Phi^I$ are scalars whereas $\theta$ are fermions, including the standard spin connection transformation giving them their spinorial nature. $\SO(D-(p+1))$, the transverse rotational group to the brane is reinterpreted as an internal symmetry, under which $\Phi^I$ transforms as a vector. The parameters $a^\mu{}_I$ describing the coset $\SO(1,D-1)/(\SO(1,p)\times \SO(D-p-1))$ are generically non-linearly realised, whereas the transverse translations $a^I$ act inhomogeneously on the dynamical fields $\Phi^I$, identifying the latter as Goldstone massless fields, as corresponds to the spontaneous symmetry breaking of these symmetries due to the presence of the brane in the chosen directions.

There is a similar discussion for the 32 spacetime supersymmetries $(\epsilon)$. {\it Before} gauge fixing {\it all} fermions $\theta$ transform inhomogeneously under supersymmetry. After gauge fixing ${\cal P}\theta=0$, the compensating kappa symmetry transformation $\kappa(\epsilon)$ required to preserve the gauge slice in configuration space will induce an extra supersymmetry transformation for the dynamical fermions, i.e. $(1-{\cal P})\theta$. On general grounds, there must exist sixteen linear combinations of supersymmetries being linearly realised, whereas the sixteen remaining will be spontaneously broken by the brane. There are many choices for ${\cal P}\theta=0$. In \cite{Aganagic:1996nn}, where they analysed this aspect for D-branes in superPoincar\'e, they set one of the members of the ${\cal N}=2$ fermion pair to zero, leading to fairly simple expressions for the gauge fixed lagrangian. Another natural choice corresponds to picking the projector describing the preserved supersymmetries  by the brane from the spacetime perspective. For instance, the supergravity solution describing M2-branes has 16 Killing spinors satisfying
\begin{equation}
  \Gamma_{012}\epsilon = \pm \epsilon\,,
\end{equation}
where the $\pm$ is correlated with the $R_4$ flux carried by the solution. If one fixes kappa symmetry according to
\begin{equation}
  {\cal P}\theta = (1+\Gamma_\star)\theta = 0\,, \qquad \qquad {\rm with} \qquad \qquad \Gamma_\star = \Gamma_3\dots \Gamma_9\Gamma_\sharp\,,
\end{equation}
where $\Gamma_\sharp$ stands for the eleven dimensional Clifford algebra matrix, then the physical fermionic degrees of freedom are not only 3-dimensional spinors, but they are chiral spinors from the internal symmetry $\SO(8)$ perspective. They actually transform in the $({\bf {\underline 2}},\,{\bf {\underline 8}^s})$ \cite{Bergshoeff:1987qx}. Similar considerations would apply for any other brane considered in this review. 

Having established the relation between spacetime and world volume symmetries, it is natural to close our discussion by revisiting the superalgebra closed by the linearly realised world volume (super)symmetries once both diffeomorphisms and kappa symmetry have been fixed. Since spacetime superalgebras included extra bosonic charges due to the quasi-invariance of the brane WZ action, the same will be true for their gauge fixed actions. Thus, these (p+1)-dimensional world volume superalgebras will include as many extra bosonic charges as allowed by group theory and by the dimensionality of the brane world spaces \cite{Bergshoeff:1997bh}. Consider the M2-brane discussed above. Supercharges transform in the $({\bf {\underline 2}},\,{\bf {\underline 8}^s})$ representation of the $\SO(1,2)\times \SO(8)$ bosonic isometry group. Thus, the most general supersymmetry algebra compatible with these generators, ${\cal N}=8$ $d=3$, is \cite{Bergshoeff:1997bh}
\begin{equation}
  \{Q^I_\alpha,\,Q^J_\beta\} = \delta^{IJ}P_{(\alpha\beta)} + Z^{(IJ)}_{(\alpha\beta)} + \varepsilon_{\alpha\beta}Z^{[IJ]} \qquad {\rm with} \qquad (\delta_{IJ}Z^{(IJ)}=0)\,.
\end{equation}
$P_{(\alpha\beta)}$ stands for a 3-dimensional one-form, the momentum on the brane; $Z^{(IJ)}_{(\alpha\beta)}$ transforms in the ${\bf 35^+}$ under the R-symmetry group $\SO(8)$, or equivalently, as a self-dual 4-form in the transverse space to the brane; $
Z^{[IJ]}$ is a world volume scalar while transforms in the ${\bf 28}$ of $\SO(8)$, i.e. as a 2-form in the transverse space. The same superalgebra is realised on the non-abelian effective action describing N coincident M2-branes \cite{Passerini:2008qt} to be reviewed in section \ref{sec:mm2}. Similar structures exist for other infinite branes. For example, the M5-brane gives rise to the d=6 $(2,0)$ superalgebra \cite{Bergshoeff:1997bh}
\begin{equation}
  \{Q^I_\alpha,\, Q^J_\beta\} = \Omega^{IJ}P_{[\alpha\beta]} + Z^{(IJ)}_{(\alpha\beta)} + Y^{[IJ]}_{[\alpha\beta]} \qquad {\rm with} \qquad (\Omega_{IJ}Y^{[IJ]}=0)\,.
\end{equation}
Here $\alpha,\,\beta=1,\dots ,4$ is an index of $\SU^*(4)\simeq \mathrm{Spin}(1,5)$, the natural Lorentz group for spinors in d=6 dimensions, $I,J=1,\dots 4$ is an index of $\Sp(2)\simeq \mathrm{Spin}(5)$, which is the double cover of the geometrical isometry group $\SO(5)$ acting on the transverse space to the M5-brane and $\Omega^{IJ}$ is an $\Sp(2)$ invariant antisymmetric tensor. Thus, using appropriate isomorphisms, these superalgebras allow a geometrical reinterpretation in terms of brane world volumes and transverse isometry groups becoming R-symmetry groups. The last decomposition is again maximal since $P_{[\alpha\beta]}$ stands for 1-form in d=6 (momentum), $Z^{(IJ)}_{(\alpha\beta)}$ transforms as a self-dual 3-form in d=6 and a 2-form in the transverse space and $Y^{[IJ]}_{[\alpha\beta]}$ as a 1-form both in d=6 and in the transverse space. For an example of a non-trivial world volume superalgebra in a curved background, see \cite{Craps:1999nc}.

I would like to close this discussion with a remark that is usually not stressed in the literature. By construction, any diffeomorphism and kappa symmetry gauge fixed brane effective action describes an interacting supersymmetric field theory in p+1 dimensions\footnote{I have assumed both the background and the brane preserve some supersymmetry.}. As such, if there are available superspace techniques in these dimensions involving the relevant brane supermultiplet, the gauge fixed action can {\it always} be rewritten in that language. The matching between both formulations generically involves non-trivial field redefinitions. To be more precise,  consider the example of ${\cal N}=1$ $d=4$ supersymmetric abelian gauge theories coupled to matter fields. Their kinetic terms are fully characterised by a K\"ahler potential. If one considers a D3-brane in a background breaking the appropriate amount of supersymmetry, the expansion of the gauge fixed D3-brane action must match the standard textbook description. The reader can find an example of the kind of non-trivial bosonic field redefinitions that is required in \cite{Johnson:2001ze,OferJoan}. The matching of fermionic components is expected to be harder.

\subsection{Regime of validity}
\label{sec:validity}

After thoroughly discussing the kinematic structure of the effective action describing the propagation of single branes in arbitrary {\it on-shell} backgrounds, I would like to reexamine the regime of validity under which the dynamics of the full string (M-) theory reduces to $S_{{\rm brane}}$.

As already stressed at the beginning of section \ref{sec:bbrane}, working at low energies, allows us to consider
the action
\begin{equation}
  S \approx S_{{\rm SUGRA}} + S_{{\rm brane}}.
\end{equation}
In string theory, {\it low energies} means energies $E$ satisfying $E\sqrt{\alpha^\prime}\ll 1$. This guarantees that {\it no} on-shell states will carry energies above that scale allowing to write an effective action in terms of the fields describing massless excitations and their derivatives. The argument is valid for both the open and the closed string sectors. Furthermore, to ensure the validity of this {\it perturbative} description, one must ensure the {\it weak} coupling regime is satisfied, i.e. $g_s\ll 1$, to suppress higher loop world sheet contributions.

Dynamically, all brane effective actions reviewed previously, describe the propagation of a brane in a {\it fixed} on-shell spacetime background solving the classical supergravity equations of motion. Thus, to justify neglecting the dynamics of the gravitational sector, focusing on the brane dynamics, one must guarantee condition (\ref{eq:noback})
\begin{equation}
  |T_{mn}^{{\rm background}}|\gg |T_{mn}^{{\rm brane}}|,
\end{equation}
but also to work in a regime where the effective Newton's constant tends to zero. Given the low energy and weak coupling approximations, the standard lore condition for the absence of quantum gravity effects, i.e. $E\ell_p^{(10)}\ll 1$, is naturally satisfied since $E\ell_p^{(10)}\sim \left(E\sqrt{\alpha^\prime}\right)g_s^{1/4}\ll1$. The analogous condition for eleven dimensional supergravity is $E\ell_p\ll 1$.

The purpose of this section is to spell out more precisely the conditions that make the above requirements not sufficient.
As in any effective field theory action, one must check the validity of the assumptions made in their derivation. In our discussions, this includes 
\begin{itemize}
\item[1.] conditions on the derivatives of brane degrees of freedom, both geometrical $X^m$ and world volume gauge fields, such as the value of the electric field;
\item[2.] the reliability of the supergravity background;
\item[3.] the absence of extra massless degrees of freedom emerging in string theory under certain circumstances.
\end{itemize}
I will break the discussion below into background and brane considerations.

\paragraph{Validity of the background description:} Whenever the supergravity approximation is not reliable, the brane description will also break down. Assuming no extra massless degrees of freedom arise, any on-shell ${\cal N}=2$ type IIA/IIB supergravity configuration satisfying the conditions described above, must also avoid
\begin{equation}
   e^\phi\sim 1, \quad \quad {\cal R}\,(\ell_p^{(10)})^2 \simeq 1.
\end{equation}
Since the string coupling constant $g_s$ is defined as the expectation value of $e^\phi$, the first condition determines the regions of spacetime where string interactions become strongly coupled. The second condition, or any dimensionless scalar quantity constructed out of the Riemann tensor, determines the regions of spacetime where curvature effects can not be neglected. Whenever there are points in our classical geometry where any of the two conditions are satisfied, the assumptions leading to the classical equations of motion being solved by the background under consideration are {\it violated}. Thus, our approximation is not self-consistent in these regions.

Similar considerations apply to eleven dimensional supergravity. In this case, the first natural condition comes from the absence of strong curvature effects, which would typically occur whenever
\begin{equation}
  {\cal R}\ell_p^2\simeq 1
\end{equation}
where once more the scalar curvature can be replaced by other curvature invariants constructed out of the eleven dimensional Riemann tensor in appropriate units of the eleven dimensional Planck scale $\ell_p$.

Since the strong coupling limit of type IIA string theory is M-theory, which at low energies is approximated by ${\cal N}=1$ $d=11$ supergravity, it is clear that there should exist further conditions. This connection involves a compactification on a circle, and it is natural to examine whether our approximations hold as soon as its size $R$ is comparable to $\ell_p$. Using the relations (\ref{eq:mIIA}, one learns
\begin{equation}
  R\sim \ell_p \quad \Longleftrightarrow \quad g_s\sim 1
\end{equation}
Thus, as soon as the M-theory circle explores subPlanckian eleven dimensional scales, which would not allow a reliable eleven dimensional classical description, the type IIA string coupling becomes weakly coupled, opening a possible window of reliable classical geometrical description in terms of the KK reduced configuration (\ref{eq:KK}).

The above discussion also applies to type IIA and IIB geometries. As soon as the scale of some compact submanifold, such as a circle, explores substringy scales, the original metric description stops being reliable. Instead, its T-dual description  (\ref{eq:tdualityrules}) does, using (\ref{eq:wstduality}). 

Finally, the strong coupling limit of type IIB may also allow a geometrical description given the $\SL(2,\bR)$ invariance of its supergravity effective action, which includes the S-duality transformation
\begin{equation}
  e^\phi \to e^{-\phi}.
\end{equation}
The latter maps an strongly coupled region to a weakly coupled one, but it also rescales the string metric. Thus, one must check whether the curvature requirements ${\cal R}\,(\ell_p^{(10)})^2 \ll 1$ hold or not.

It is important to close this discussion reminding the reader that any classical supergravity description assumes the only relevant massless degrees of freedom are those included in the supergravity multiplet. The latter is {\it not} always true in string theory. For example, string winding modes become massless when the circle radius the string wraps goes to zero size. This is precisely the situation alluded to above, where the T-dual description, in which such modes become momentum modes, provides a T-dual reliable description in terms of supergravity multiplet fluctuations. The emergence of extra massless modes in certain classical singularities in string theory is far more general, and it can be responsible for the resolution of the singularity. The existence of extra massless modes is a quantum mechanical question that requires going beyond the supergravity approximation. What certainly remains universal is the geometrical breaking down associated with the divergence of scalar curvature invariants due to a singularity, independently of whether the latter is associated with extra massless modes or not.

\paragraph{Validity of the brane description:} Besides the generic low energy and weak coupling requirements applying to D-brane effective actions (\ref{kappadpbrane}), the microscopic derivation of the DBI action assumed the world volume field strength $F_{\mu\nu}$ was {\it constant}. Thus, kappa symmetric invariant D-brane effective actions ignore corrections in derivatives of this field strength, i.e. terms like $\partial_\rho F_{\mu\nu}$ or higher in number of derivatives. Interestingly, these corrections map to acceleration and higher order derivative corrections in the scalar fields $X^m$ under T-duality, see (\ref{eq:tdualrule}). Thus, 
there exists the further requirement that {\it all} dynamical fields in brane effective actions are {\it slowly varying}. In Minkowski, this would correspond to conditions like
\begin{equation}
  \sqrt{\alpha^\prime}\partial^2 X \ll \partial X,
\end{equation} 
or similar tensor objects constructed with the derivative operator in appropriate string units. In a general curved background, these conditions must be properly covariantised, although locally, the above always applies.

Notice these conditions are analogous to the ones we would encounter in the propagation of a point particle in a fixed background.
Any corrections to geodesic motion would be parameterised by an expansion in derivatives of the scalar fields parameterising the particle position, this time in units of the mass particle.

Brane effective actions carrying electric fields $E$ can manifestly become ill defined for values above a certain {\it critical} electric field $E_{\rm crit}$ for which the DBI determinant vanishes. It was first noticed for the bosonic string in \cite{Burgess:1986dw,Nesterenko:1989pz} that such critical electric field is the value for which the rate of Schwinger charged-string pair production \cite{Schwinger:1951nm} diverges. The divergence captures a divergent density of string states in the presence of such critical electric field. These calculations were extended to the superstring in \cite{Bachas:1992bh}. The conclusion is the same, though in this latter case the divergence applies to {\it any} pair of charge-conjugate states. Thus, there exists a correlation between this pathological behaviour of the DBI action and the existence of string instabilities\footnote{There exists some similar phenomena on the M5-brane dynamics with the self-dual 3-form field strength. See \cite{Bergshoeff:2000ai} for a discussion on the emergence of noncommutative gauge theories when the self-dual 3-form field strength is close to its critical value.}. Heuristically, one interprets the regime with $E>E_{\rm crit}$ as one where the string tension can no longer hold the string together\footnote{There are several claims in the literature advocating that extra massless degrees of freedom emerge in brane effective actions when the latter probe black holes very close to their horizons. See \cite{Kabat:1998vc,Kabat:1999yq,Iizuka:2001cw,Horowitz:2009wm} for interesting work in this direction.}.


\section{World volume solitons : generalities}
\label{sec:solitons}

Brane effective actions capture the relevant dynamics of M-theory or string theory in some appropriate regimes of validity. Thus, they contain reliable information about its spectrum and its dynamics in those regimes. In this section, I will develop the tools to study the world volume realisation of supersymmetric states carrying the extra bosonic (topological) charges appearing in the maximal supersymmetry algebras introduced in sections \ref{sec:susyalgebra} and \ref{sec:worldsusyalgebra}.

One such realisation is in terms of classical bosonic on-shell configurations. As it often occurs with supersymmetric configurations, instead of focusing on the integration of the equations of motion, I will focus on the conditions ensuring preservation of supersymmetry and on their physical interpretation. In particular,
\begin{itemize}
\item I will argue the existence of a necessary condition that any bosonic supersymmetric configuration must satisfy involving the kappa symmetry matrix $\Gamma_\kappa$ and the background Killing spinors $\epsilon$.
\item I will review the hamiltonian formulation for brane effective actions to compute the energy of these configurations. The latter will minimise the energy for a given set of charges carried by the state. The existence of energy bounds can be inferred from merely algebraic considerations and I will discuss their field theory realisations as BPS bounds\footnote{BPS stands for Bogomolny, Prasad and Sommerfield and their work on stable solitonic configurations \cite{Bogomolny:1975de,Prasad:1975kr}.}. Furthermore, the relation between their saturation and the solution to the necessary kappa symmetry condition will also be explained.
\item I will discuss the relation between these physical considerations and the mathematical notion of calibration, which is a purely geometric formulation of the problem of finding volume minimising surfaces. Since the latter corresponds to
a subset of bosonic brane supersymmetric configurations, this connection will allow us to review the notion of generalised calibration which, in physical terms, includes world volume gauge field excitations.
\end{itemize}
The framework and set of relations uncovered in this section are summarised in figure \ref{fig4}.

\epubtkImage{kappa.png}{%
\begin{figure}[h]
  \centerline{\includegraphics[width=150mm]{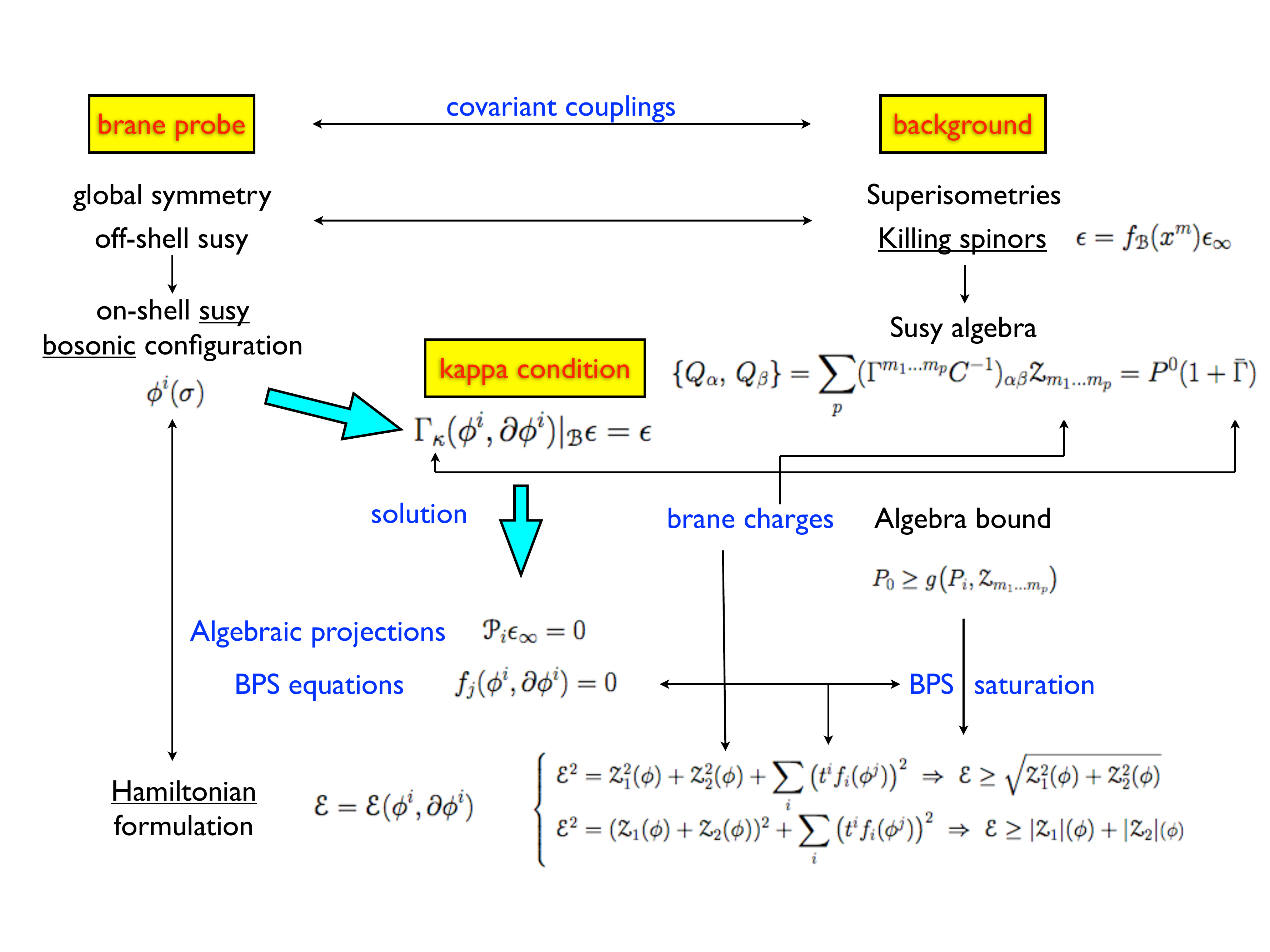}}
   \caption{Set of relations involving kappa symmetry, spacetime supersymmetry algebras, their bounds and their realisation as field theory BPS bounds in terms of brane solitons using the Hamiltonian formulation of brane effective actions.}
  \label{fig4}
\end{figure}}

\subsection{Supersymmetric bosonic configurations \& kappa symmetry} 
\label{sec:kcond}

To know whether any given {\it on-shell} bosonic brane configuration is supersymmetric, and if so, how many supersymmetries does preserve, one must develop some tools analogous to the ones for bosonic supergravity configurations. I will review these first.

Consider any supergravity theory having bosonic $({\cal B})$ and fermionic $({\cal F})$ degrees of freedom. It is consistent with the equations of motion to set ${\cal F}=0$. The question of whether the configuration ${\cal B}$ preserves supersymmetry reduces to study whether there exists any supersymmetry transformation $\epsilon$ preserving the bosonic nature of the on-shell configuration, i.e. $\delta {\cal F}|_{{\cal F}=0}=0$, without transforming ${\cal B}$, i.e. $\delta {\cal B}|_{{\cal F}=0}=0$. Since the structure of the local supersymmetry transformations in supergravity is
\begin{equation}
  \delta {\cal B} \propto {\cal F}\,, \quad \quad \delta {\cal F} = {\cal P}({\cal B})\,\epsilon\,,
\end{equation}
these conditions reduce to ${\cal P}({\cal B})\,\epsilon=0$. In general, the Clifford valued operator ${\cal P}({\cal B})$ is not higher than first order in derivatives, but it can also be purely algebraic. Solutions to this equation involve
\begin{itemize}
\item[1.] Differential constraints on the subset of bosonic configurations ${\cal B}$. Given the first order nature of the operator ${\cal P}({\cal B})$, these are simpler than the second order equations of motion and help to reduce the complexity of the latter.
\item[2.] Differential and algebraic constraints on $\epsilon$. These reduce the infinite dimensional character of the original arbitrary supersymmetry transformation parameter $\epsilon$ to a {\it finite} dimensional subset, i.e. $\epsilon = f_{{\cal B}}(x^m)\epsilon_\infty$, where the function $f_{{\cal B}}(x^m)$ is uniquely specified by the bosonic background ${\cal B}$ and the {\it constant} spinor $\epsilon_\infty$ typically satisfies a set of conditions ${\cal P}_i\epsilon_\infty=0$, where ${\cal P}_i$ are projectors satisfying ${\cal P}_i^2={\cal P}_i$ and ${\rm tr}{\cal P}_i=0$. These $\epsilon$ are the {\it Killing spinors} of the bosonic background ${\cal B}$. They can depend on the spacetime point, but they are {\it no} longer arbitrary. Thus, they are understood as {\it global} parameters.
\end{itemize}
This argument is general and any condition derived from it is necessary. Thus, one is instructed to analyse the condition ${\cal P}({\cal B})\,\epsilon=0$ before solving the equations of motion. As a particular example, and to make contact with the discussions in section \ref{sec:goldstone}, consider ${\cal N}=1$ d=11 supergravity. The only fermionic degrees of freedom are the gravitino
components $\Psi_a = E^M{}_a\Psi_M$. Their supersymmetry transformation is \cite{Stelle:1996tz}
\begin{equation}
  \delta \Psi_a = \left(\partial_a + \frac{1}{4}\omega_a{}^{bc}\Gamma_{bc}\right)\epsilon -\frac{1}{288}\left(\Gamma_a{}^{bcde}-8\delta_a{}^b\Gamma^{cde}\right)R_{bcde}\epsilon\,.
\end{equation}
Solving the supersymmetry preserving condition $\delta\Psi_a=0$ in the M2-brane and M5-brane backgrounds determines the Killing spinors of these solutions to be \cite{Stelle:1996tz,Stelle:1998xg}
\begin{eqnarray}
  {\rm M2-brane}  \quad \epsilon &=& U^{-1/6}\epsilon_\infty \quad \quad {\rm with} \quad \quad \Gamma_{012}\epsilon_\infty=\pm\epsilon_\infty\,, \\
   {\rm M5-brane}  \quad \epsilon &=& U^{-1/12}\epsilon_\infty \quad \quad {\rm with} \quad \quad \Gamma_{012345}\epsilon_\infty=\pm\epsilon_\infty\,.
\end{eqnarray}
A similar answer is found for all D-branes in ${\cal N}=2$ d=10 type IIA/B supergravities.

The same question for brane effective actions is treated in a conceptually analogous way. The subspace of bosonic configurations ${\cal B}$ defined by $\theta=0$ is compatible with the brane equations of motion. Preservation of supersymmetry requires $s\theta|_{{\cal B}} = 0$. The total transformation $s\theta$ is given by
\begin{equation}
s\theta=\delta_\kappa \theta + \epsilon + \Delta\theta + \xi^\mu\partial_\mu\theta\,,
\end{equation}
where $\delta_\kappa\theta$ and $\xi^\mu\partial_\mu\theta$ stand for the kappa symmetry and world volume diffeomorphism infinitesimal transformations and $\Delta\theta$ for any global symmetries different from supersymmetry, which is generated by the Killing spinors $\epsilon$. When restricting to the subspace 
${\cal B}$ of bosonic configurations,
\begin{eqnarray}
\delta_\kappa \theta |_{{\cal B}} &=& \left(\mathbb{1} + \Gamma_\kappa |_{{\cal B}}
\right)\kappa \\
\Delta \theta |_{{\cal B}} &=& 0\,,
\end{eqnarray}
one is left with
\begin{equation}
s\theta |_{{\cal B}} = \left(1 + \Gamma_\kappa |_{{\cal B}}
\right)\kappa + \epsilon\,.
\label{fvar}
\end{equation}
\noindent
This is because $\Delta\theta$ describes {\it linearly} realised symmetries. Thus, kappa symmetry and supersymmetry transformations do generically not leave the subspace ${\cal B}$ invariant. 

We are interested in deriving a general condition for {\it any} bosonic configuration to preserve supersymmetry. Since not
all fermionic fields $\theta$ are physical, working on the subspace $\theta=0$ is not precise enough for our purposes. We must work in the subspace of field configurations being both physical and bosonic \cite{Bergshoeff:1997kr}.This forces us to work at the intersection of $\theta=0$ and some kappa symmetry gauge fixing condition. Because of this, I find it convenient to break the general argument in two steps
\begin{enumerate}
\item {\sl Invariance under kappa symmetry}. Consider the kappa symmetry gauge fixing condition ${\cal P}\theta=0$, where ${\cal P}$ stands for any field independent projector. This allows us to decompose the original fermions according to
\begin{equation}
\theta = {\cal P}\theta + (\mathbb{1}-{\cal P})\theta\,.
\end{equation}
To preserve the kappa gauge slice in the subspace ${\cal B}$ requires 
\begin{equation}
s{\cal P}\theta|_{{\cal B}}={\cal P}(\mathbb{1}+\Gamma_\kappa|_{{\cal B}})\kappa
+ {\cal P}\epsilon = 0\,.
\label{derive1}
\end{equation}
This determines the necessary compensating kappa symmetry transformation $\kappa(\epsilon)$ as a function of the background Killing spinors.

\item {\sl Invariance under supersymmetry}. Once the set of dynamical fermions $(\mathbb{1}-{\cal P})\theta$ is properly defined, we ask for the set of global supersymmetry transformations preserving them 
\begin{equation}
s(\mathbb{1}-{\cal P})\theta|_{{\cal B}}=0\,.
\end{equation}
This is equivalent to
\begin{equation}
(\mathbb{1}+\Gamma_\kappa|_{{\cal B}})\kappa(\epsilon) + \epsilon = 0
\end{equation}
once equation (\ref{derive1}) is taken into account. Projecting this equation into the $(\mathbb{1}-\Gamma_\kappa|_{{\cal B}})$ subspace, gives condition
\begin{equation}
\Gamma_\kappa|_{{\cal B}} \epsilon = \epsilon\,.
\label{spc}
\end{equation}
No further information can be gained by projecting to the orthogonal subspace $(\mathbb{1}+\Gamma_\kappa|_{{\cal B}})$.
\end{enumerate}

I will refer to equation (\ref{spc}) as the {\it kappa symmetry preserving condition}. It was first derived in \cite{Bergshoeff:1997kr}. 
This is the universal necessary condition that  {\it any} bosonic {\it on-shell} brane configuration $\{\phi^i\}$ must satisfy to preserve some supersymmetry. 

\begin{longtable}{|p{3.5cm}|p{8.0 cm}|}
  \hline
  Brane & Bosonic kappa symmetry matrix \\
  \hline
  \hline
  {\it M2-brane} & $\Gamma_{\kappa}|_{{\cal B}} = \frac{1}{3!\,\sqrt{-{\rm det}\,\CG}}\epsilon^{\mu\nu\rho}
\gamma_{\mu\nu\rho}$ \\
  {\it M5-brane} & $\Gamma_\kappa|_{{\cal B}} =\frac{v_\mu\gamma^\mu}{\sqrt{-{\rm det}\,(\CG + \7H)}}\left[\frac{\sqrt{-{\rm det}\,\CG}}{2}\gamma^{\mu_1\mu_2}\7H_{\mu_1\mu_2} + \gamma_\nu t^\nu \right. $  \\
 & $\left.  - \frac{1}{5!}\,\epsilon^{\mu_1\ldots\mu_5\nu}v_\nu\,\gamma_{\mu_1\ldots\mu_5}\right]$ \\
  {\it  IIA Dp-branes} & $\Gamma_\kappa|_{{\cal B}} = \frac{1}{\sqrt{-{\rm det}\,(\CG+\CF)}}\sum_{l=0}\gamma_{2l+1}
\Gamma_{\sharp}^{l+1}\wedge e^{\CF}$  \\
  {\it  IIB Dp-branes} &  $\Gamma_\kappa|_{{\cal B}} = \frac{1}{\sqrt{-{\rm det}\,(\CG+\CF)}}\sum_{l=0}\gamma_{2l}
\tau_3^l\,i\tau_2 \wedge e^{\CF}$ \\
  \hline
  \caption{\it Set of kappa symmetry matrices $\Gamma_\kappa$ evaluated in the bosonic subspace of configurations ${\cal B}$.}
\label{tab:bosonickappa}
\end{longtable}

In table \ref{tab:bosonickappa}, I evaluate all kappa symmetry matrices $\Gamma_\kappa$ in the subspace of bosonic configurations ${\cal B}$ for future reference. This matrix encodes information 
\begin{itemize}
\item[1.] on the background, both explicitly through the induced world volume Clifford valued matrices $\gamma_\mu = E_\mu{}^a\Gamma_a = \partial_\mu X^m E_m{}^a\Gamma_a$ and the pullback of spacetime fields, such as $\CG$, ${\cal F}$ or $\tilde{H}$, but also implicitly through the background Killing spinors $\epsilon$ solving the supergravity constraints ${\cal P}(\epsilon)=0$, which also depend on the remaining background gauge potentials.
\item[2.] on the brane configuration $\{\phi^i\}$, including scalar fields $X^m(\sigma)$ and gauge fields, either $V_1$ or $V_2$, depending on the brane under consideration. 
\end{itemize}

Just as in supergravity, any solution to equation (\ref{spc}) involves two sets of conditions, one on the space of configurations $\{\phi^i\}$ and one on the amount of supersymmetries. More precisely,
\begin{enumerate}
\item
a set of constraints among dynamical fields and their derivatives, $f_j(\phi^i,\partial\phi^i)=0$, 
\item
a set of supersymmetry projection conditions, ${\cal P}^\prime_i\epsilon_\infty=0$, with ${\cal P}^\prime_i$ being projectors, reducing the dimensionality of the vector space spanned by the original $\epsilon_\infty$.
\end{enumerate}
The first set will turn out to be  {\it BPS equations}, whereas the second will determine the amount of supersymmetry preserved by the combined background \& probe system.

\subsection{Hamiltonian formalism}
\label{sec:hamiltonian}

In this subsection, I review the hamiltonian formalism for brane effective actions. This will allow us not only to compute the energy of a given supersymmetric on-shell configuration solving (\ref{spc}), but also to interpret the constraints $f_j(\phi^i)=0$ as {\it BPS bounds} \cite{Bogomolny:1975de,Prasad:1975kr}. This will lead us to interpret these configurations as brane-like excitations supported on the original brane world volume.

The existence of energy bounds in supersymmetric theories can already be derived from purely superalgebra considerations. For example, consider the M-algebra (\ref{eq:msusyalgebra}). Due to the positivity of its left hand side, one derives the energy bound
\begin{equation}
\label{eq:susybound}
 P_0 \ge f\big( P_i, Z_{ij}, Y_{i_1\dots i_5};
Z_{0i}, Y_{0i_1\dots i_4} \big),
\end{equation}
where the charge conjugation matrix was chosen to be $C=\Gamma^0$ and the spacetime indices were split as $m=\{0,\,i\}$. For simplicity, let us set the time components $Y_{0i_1\dots i_4}$ and $Z_{0i}$ to zero. The superalgebra reduces to
\begin{equation}
 \label{eq:bgamma} 
 \{ Q, Q\} = P^0(1 +\bar\Gamma)\,, \quad \quad {\rm with} \quad \quad  \bar\Gamma =
(P^0)^{-1}\big[\Gamma^{0i}P_i + \frac{1}{2} \Gamma^{0ij} Z_{ij} +
\frac{1}{5!}\Gamma^{0i_1\dots i_5} Y_{i_1\dots i_5}\big]\,.
\end{equation} 
The bound (\ref{eq:susybound}) is now equivalent to the statement that no eigenvalue of
$\bar\Gamma^2$ can exceed unity. Any bosonic charge (or distribution of them) for which the corresponding $\bar\Gamma$ satisfies
\begin{equation}
\bar\Gamma^2=\mathbb{1}\,,
\end{equation}
defines a projector $\frac{1}{2}(\mathbb{1}+\bar\Gamma)$. The eigenspace of $\bar\Gamma$ with
eigenvalue 1 coincides with the one spanned by the Killing spinors $\epsilon_\infty$ determining the supersymmetries preserved by supergravity configurations corresponding to individual brane states. In other words, there is a one--to--one map between half BPS branes, the charges they carry and the precise supersymmetries they preserve. This allowed to interpret all the charges appearing in $\bar\Gamma$ is terms of brane excitations : the 10-momentum $P_i$ describes d=11 massless superparticles \cite{Bergshoeff:1996tu}, the 2-form charges $Z_{ij}$ M2-branes \cite{Bergshoeff:1987cm,Bergshoeff:1987qx}, whereas the 5-form charges $Y_{i_1\dots i_5}$, M5-branes \cite{Sorokin:1997ps}. This correspondence extends to the time components $\{Y_{0i_1\dots i_4},\,Z_{0i}\}$. These describe branes appearing in Kaluza-Klein vacua \cite{Hull:1997kt,Townsend:1997wg}. Specifically, $Y_{0i_1\dots i_4}$ is carried by type IIA D6-branes (the M-theory KK monopole), while $Z_{0i}$ can be related to type IIA D8-branes.

That these algebraic energy bounds should allow a field theoretical realisation is a direct consequence of the brane effective action global symmetries and Noether's theorem \cite{Noether:1918zz}. If the system is invariant under time translations, energy will be preserved, and it can be computed using the Hamiltonian formalism, for example. Depending on the amount and nature of the charges turned on by the configuration, the general functional dependence of the bound (\ref{eq:susybound}) changes. This is because each charge appears in $\bar\Gamma$ multiplied by different antisymmetric products of Clifford matrices. Depending on whether these commute or anticommute, the bound satisfied by the energy $P_0$ changes, see for example a discussion on this point in \cite{Molins:2000xe}. Thus, one expects to be able to decompose the hamiltonian density for these configurations as sums of the other charges and positive definite extra terms such that when they vanish, the bound is saturated. More precisely,
\begin{itemize}
\item[1.] For {\it non-threshold} bound states, or equivalently, when the associated Clifford matrices anticommute, one expects the energy density to satisfy
\begin{equation}
\CE^2 = \CZ_1^2+ \CZ_2^2 + \sum_i \left(t^i f_i(\phi^j)\right)^2 \,.
\label{nonthreshold}
\end{equation}
\item[2.] For bound states {\it at threshold}, or equivalently, when the associated Clifford matrices commute, one expects
\begin{equation}
\CE^2 = (\CZ_1+ \CZ_2)^2 + \sum_i \left(t^i f_i(\phi^j)\right)^2 \,.
\label{threshold}
\end{equation}
\end{itemize}
In both cases, the set $\{t^i\}$ involves non-trivial dependence on the dynamical fields and their derivatives.
Due to the positivity of the terms in the right hand side, one can derive lower bounds on the energy, or {\it BPS bounds},
\begin{eqnarray}
\CE &\geq& \sqrt{\CZ_1^2+ \CZ_2^2} \label{nonthresholdbound} \\
\CE &\geq& |\CZ_1| + |\CZ_2| \label{thresholdbound}
\end{eqnarray}
being saturated precisely when $f_i(\phi^j)=0$ are satisfied, justifying
their interpretation as {\it BPS equations} \cite{Bogomolny:1975de,Prasad:1975kr}. Thus, saturation of the bound matches the energy ${\cal E}$ with some {\it charges} that may usually have some topological origin \cite{DeAzcarraga:1989gm}.

In the current presentation, I assumed the existence of two non-trivial charges, ${\cal Z}_1$ and $\CZ_2$. The argument can be extended to any number of them. This will change the explicit saturating function in (\ref{eq:susybound}) (see \cite{Molins:2000xe}), but not the conceptual difference between the two cases outlined above. It is important to stress that just as in supergravity solving the gravitino/dilatino equations, i.e. $\delta {\cal F}=0$, does {\it not} guarantee the resulting configuration to be on-shell, the same is true in brane effective actions. In other words, not all configurations solving (\ref{spc}) and saturating a BPS bound are guaranteed to be on-shell. For example, in the presence of non-trivial gauge fields, one must still impose the Gauss' law independently. 

After these general arguments, I review the relevant phase space reformulation of the effective brane lagrangian dynamics discussed in section \ref{sec:bbrane}.

\subsubsection{D-brane hamiltonian}

As in any hamiltonian formulation\footnote{For a complete and detailed discussion of the supersymmetric and kappa invariant D-brane hamiltonian formalism, see \cite{Bergshoeff:1998ha}, which extends the bosonic results in \cite{Lindstrom:1997uj,Kallosh:1997nr} and the type IIB superMinkowski ones in \cite{Kamimura:1997ju}. Here, I follow \cite{Bergshoeff:1998ha}, even though the analysis restricts to the bosonic sector.}, the first step consists in breaking covariance to allow a proper treatment of time evolution. Let me split the world volume coordinates as $\sigma^\mu=\{t,\,\sigma^i\}$ for $i=1,\dots ,p$ and rewrite the bosonic D-brane lagrangian by singling out all time derivatives using standard conjugate momenta variables
\begin{equation}
\CL= \dot X^m P_m + \dot V_{i} E^{i}+ \dot\psi T_{{\rm Dp}} - H\,.
\label{eq:hamp}
\end{equation}
Here $P_m$ and $E^i$ are the conjugate momentum to $X^m$  and $V_i$, respectively,
while $H$ is the Hamiltonian density. $\psi$ is the Hodge dual of a p-form potential introduced in \cite{Bergshoeff:1998ha} to generate the tension $T_{{\rm Dp}}$ dynamically \cite{Bergshoeff:1992gq,Lindstrom:1997uj}. This is convenient to study the tensionless limit in these actions as a generalisation of the massless particle action limit. It was shown in \cite{Bergshoeff:1998ha} that $H$ can be written as a sum of constraints
\begin{equation}
H = \psi^{i} {\cal T}_{i} + V_t {\cal K} + s^{i} {\cal H}_{i} + \lambda {\cal H} \,,
\label{hamt}
\end{equation}
where
\begin{eqnarray}
{\cal T}_{i} &=& -\partial_{i} T_{{\rm Dp}} \nonumber\\
{\cal K} &=& - \partial_{i} {\tilde E}^{i} + (-1)^{p+1} T_{{\rm Dp}} {\cal S} \qquad {\rm with} \qquad {\cal S} = *({\cal R}e^\CF)_{p} \nonumber \\
{\cal H}_{i} &=& {\tilde P}_aE_i^a + {\tilde E}^{j} {\cal F}_{ij} \qquad {\rm with} \qquad E_i^a = E_m^a\partial_i X^m \nonumber \\ 
{\cal H} &=& \frac{1}{2}\big[{\tilde P}^2 + {\tilde E}^{i} {\tilde E}^{j} \CG_{ij} + T_{{\rm Dp}}^2 e^{-2\phi}
{\rm det}\,(\CG_{ij}+{\cal F}_{ij}) \big]\,.
\label{hamu}
\end{eqnarray}
The first constraint is responsible for the constant tension of the brane. It generates abelian gauge transformations for the p-form potential generating the tension dynamically. The second generates gauge field transformations and it implements the Gauss' law constraint ${\cal K}=0$. Notice its dependence on ${\cal R}$, the pullback of the RR field strengths $R=dC - C\wedge H_3$, coming from the WZ couplings and acting as {\it sources} in Gauss' law. Finally, ${\cal H}_{a}$ and ${\cal H}$ generate world space diffeomorphisms and time translations, respectively. 

The modified conjugate momenta ${\cal P}_a$ and $\tilde E^i$ determining all these constraints are defined in terms of the original conjugate momenta as
\begin{eqnarray}
\tilde P_a &=& E_a{}^m \big(P_m + E^{i}Z^\star\left(i_m B\right)_{i} + T_{{\rm Dp}}{\cal C}_m\big),\qquad {\rm with} \qquad 
{\cal C}_m = * \left(Z^\star(i_m C) \wedge e^\CF\right)_{p},  \nonumber \\
\tilde E^{i} &=& E^{i} + T{\cal C}^{i}\,, \qquad {\rm with} \qquad {\cal C}^i= [*({\cal C}e^\CF)_{p-1}]^i.
\label{hamr}
\end{eqnarray}
$Z^\star\left(i_m B\right)_{i}$ stands for the pullback to the world volume of the contraction of $B_2$ along the vector field $\partial/\partial X^m$. Equivalently, $Z^\star\left(i_m B\right)_{i}=\partial_i X^n B_{mn}$. $Z^\star(i_m C)$ is defined analogously. Notice $\star$ stands for the Hodge dual in the p-dimensional D-brane world space.

In practice, given the equivalence between the lagrangian formulation and the one above, one solves the equations of motion on the subspace of configurations solving (\ref{spc}) in phase space variables and finally computes the energy density of the configuration $P_0={\cal E}$ by solving the Hamiltonian constraint i.e. ${\cal H}=0$, which is a quadratic expression in the conjugate momenta, as expected for a relativistic dynamical system.

\subsubsection{M2-brane hamiltonian}

The hamiltonian formulation for the M2-brane can be viewed as a particular case of the analysis provided above, but in the absence of gauge fields.  It was originally studied in \cite{Bergshoeff:1987in}. One can check the full bosonic M2-brane lagrangian is equivalent to
\begin{equation}
{\cal L} = \dot{X}^mP_m - s^i{\tilde P}_aE_i^a-\frac{1}{2}\lambda\left[{\tilde P}^2 + T^2_{{\rm M2}}\,{\rm det}\, \CG_{ij}\right],
\label{eq:m2phasespace}
\end{equation}
where the modified conjugate momentum ${\tilde P}_a$ is related to the standard conjugate momentum $P_m$ by
\begin{equation}
  {\tilde P}_a = E_a^m\left(P_m + T_{{\rm M2}}\,{\cal C}_m\right) \qquad {\rm with} \qquad {\cal C}_m = 
  *\left(Z^\star(i_m  C_3)\right),
\end{equation}
where $*$ describes the Hodge dual computed in the 2-dimensional world space spanned by $i,j=1,2$. Notice no dynamically generated tension was considered in the formulation above.

As before, one usually solves the equations of motion $\delta{\cal L}/\delta s^a=\delta{\cal L}/\delta v=0$ in the subspace of phase space configurations solving (\ref{spc}), and computes its energy by solving the hamiltonian constraint, i.e. $\delta{\cal L}/\delta \lambda = 0$.

\subsubsection{M5-brane hamiltonian}

It turns out the hamiltonian formulation for the M5-brane dynamics is more natural than its lagrangian one since it is easier to deal with the self-duality condition in phase space \cite{Bergshoeff:1998vx}. One follows the same strategy and notation as above, splitting the world volume coordinates as $\sigma^\mu = \{t,\,\sigma^i\}$ with $i=1,\dots 5$. Since the hamiltonian formulation is expected to break $\SO(1,5)$ into $\SO(5)$, one works in the gauge $a=\sigma^0=t$. It is convenient to work with the world space metric $\CG_{ij}$ and its inverse $\CG^{ij}_5$\footnote{This notation is introduced to emphasise $\CG^{ij}_5$ does not correspond to the world space components of $\CG^{\mu\nu}$, but to the inverse matrix of the restriction of $\CG_{\mu\nu}$ to the world space subspace.}. Then, the following identities hold
\begin{eqnarray}
\tilde H^{ij} &=& \frac{1}{6\,\sqrt{{\rm det}\,\CG_5}}\, \varepsilon^{ijk_1k_2k_3} {\cal H}_{k_1k_2k_3}\,, \nonumber \\
{\rm det}\, (\CG_{\mu\nu} + \tilde H_{\mu\nu}) &=& 
(\CG_{00} - \CG_{0i}\CG_5^{ij} \CG_{0j}) {\rm det} {}^5\!(\CG+ \tilde H)\,
\end{eqnarray}
where ${\rm det}\,\CG_5$ is the determinant of the world space components $\CG_{ij}$, ${\rm det} {}^5\!(\CG+ \tilde H) = {\rm det}\, (\CG_{ij} + \tilde H_{ij} )$ and $\tilde H_{ij} = \CG_{ik}\CG_{jl}\tilde H^{kl}$. 

It was shown in \cite{Bergshoeff:1998vx} that the full bosonic M5-brane lagrangian in phase space equals
\begin{equation}
\label{eq:m5phase}
\CL= \dot X^m P_m + \frac{1}{2}\Pi^{ij} \dot V_{ij}  -\lambda {\cal H} -s^{i} {\cal H}_{i}  + \sigma_{ij} {\cal K}^{ij}\,, 
\end{equation}
where $P_m$ and $\Pi^{ij}$ are the conjugate momenta to $X^m$ and the 2-form $V_{ij}$
\begin{eqnarray}
  P_m &=& E_m{}^a \tilde P_a + T_{{\rm M5}} \hat{\cal C}_m\,, \nonumber \\
  \hat{\cal C}_m &=& *\big[Z^\star\left(i_m C_{6}\right) -\frac{1}{2} Z^\star\left(i_m C_{3}\right)\wedge ({\cal C}_{3} + 2{\cal H}_3)\big]\,, \nonumber \\
\Pi^{ij} &=& \frac{1}{4}T_{{\rm M5}}\varepsilon^{ijk_1k_2k_3} \partial_{k_1} V_{k_2k_3}\,.
\end{eqnarray}
Notice the last equation is equivalent to $\Pi = \frac{1}{2}T_{{\rm M5}} \star (dV)$, from which we conclude $d\star\Pi=0$, using the Bianchi identify for $dV_2$. The last three functionals appearing in (\ref{eq:m5phase})
\begin{eqnarray}
{\cal H} &=& \frac{1}{2} \big[{\cal P}^2 + T_{{\rm M5}}^2 \det{}^5\!(\CG+ \tilde H)\big] \nonumber \\
{\cal H}_{i} &=& \partial_{i} X^m P_m + T_{{\rm M5}}(V_{i} -\hat{\cal C}_{i}) \nonumber \\
{\cal K}^{ij} &=& \Pi^{ij}-\frac{1}{4}T_{{\rm M5}}\varepsilon^{ijk_1k_2k_3}\partial_{k_1} V_{k_2k_3}
\end{eqnarray}
correspond to constraints generating time translations, world space diffeomorphisms and the self-duality condition. The following definitions were used in the expressions above
\begin{eqnarray}
  V_{i} &=& \frac{1}{24} \varepsilon^{i_1i_2i_3i_4i_5}H_{i_3i_4i_5} H_{i_1i_2i}\,, \nonumber \\
  {\cal P}_a &=& E_a{}^mP_m + 
T_{{\rm M5}}(V^{i}\partial_{i} X^m E_m{}^b\eta_{ba} - \hat{\cal C}_a)\,, \nonumber \\
 \hat{\cal C}_{a} &=& E^m_a\hat{\cal C}_m\,.
\end{eqnarray}
As for D-branes and M2-branes, in practice one solves the equations of motion in the subspace of phase space configurations solving (\ref{spc}) and eventually computes the energy of the system by solving the quadratic constraint coming from the hamiltonian constraint ${\cal H}=0$.

\subsection{Calibrations}
\label{sec:calib}

In the absence of WZ couplings and brane gauge field excitations, the energy of a brane configuration equals its volume. The problem of identifying minimal energy configurations is equivalent to that of minimising the volumes of p-dimensional submanifolds embedded in an n-dimensional ambient space. The latter is a purely geometrical question that can in principle be  mathematically formulated independently of supersymmetry, kappa symmetry or brane theory. This is what the notion of {\it calibration} achieves. In this subsection, I review the close relation between this mathematical topic and a subset of supersymmetric brane configurations \cite{Gibbons:1998hm,Gauntlett:1998vk,Acharya:1998en}. I start with static brane solitons in $\bR^n$, for which the connection is more manifest, leaving their generalisations to the appropriate literature quoted below.

Consider the space of oriented p dimensional subspaces of $\bR^n$, i.e. the grassmannian ${\rm G}(p,\bR^n)$.
For any $\xi\in {\rm G}(p,\bR^n)$, one can always find an orthonormal basis $\{e_1,\dots ,e_n\}$ in $\bR^n$ such that $\{e_1,\dots ,e_p\}$ is a basis in $\xi$ so that its co-volume is
\begin{equation}
  \vec{\xi} = e_1\wedge \dots \wedge e_p\,.
\end{equation}
A p-form $\varphi$ on an open subset $U$ of $\bR^n$ is a calibration of degree $p$ if
\begin{itemize}
\item[(i)] $d\varphi=0$
\item[(ii)] for every point $x\in U$, the 
form $\varphi_x$ satisfies $\varphi_x(\vec{\xi})\leq 1$ for {\it all} 
$\xi\in  {\rm G}(p,\bR^n)$ and such that the {\it contact set}
\begin{equation}
{\rm G}(\varphi)=\{ \xi\in {\rm G}(p, \bR^n): \varphi(\vec{\xi})=1\}
\label{ntwo}
\end{equation}
is not empty.
\end{itemize}

One of the applications of calibrations is to provide a bound for the volume
of p-dimensional submanifolds of $\bR^n$. Indeed, the fundamental 
theorem of calibrations \cite{Harvey:1982xk} states 
\paragraph{Theorem: } Let $\varphi$ be a calibration of degree 
p on $\bR^n$. The p-dimensional submanifold $N$ for which
\begin{equation}
\varphi(\vec{N})=1\,,
\label{nthree}
\end{equation}
is volume minimising. One refers to such minimal submanifolds as
calibrated submanifolds, or as calibrations for short, of degree $p$.

The proof of this statement is fairly elementary. Choose an open subset $U$ of 
$N$ with boundary $\partial U$ and assume the existence of a second open subset $V$ in another subspace $W$ of $\bR^n$ with the same boundary, i.e. $\partial U=\partial V$. By Stokes' theorem,
\begin{equation}
{\rm vol} (U)=\int_U \varphi=\int_V \varphi=
\int \varphi(\vec{V}) \mu_V\leq \int_V
\mu_V ={\rm vol} (V)\,,
\label{nfour}
\end{equation}
where $\mu_V = \alpha_1\wedge \dots \alpha_p$ is the volume form constructed out of the dual basis $\{\alpha_1,\dots ,\alpha_p\}$ to $\{e_1,\dots ,e_n\}$.

Two remarks can motivate why these considerations should have a relation to brane solitons and supersymmetry:
\begin{itemize}
\item[1.] For static brane configurations with no gauge field excitations and in the absence of WZ couplings, the energy of the brane soliton equals the volume of the brane submanifold embedded in $\bR^n$. Thus, bounds on the volume correspond to brane energy bounds, which are related to supersymmetry saturation, as previously reviewed. Indeed, the dynamical field $X^i(\sigma)$ do mathematically describe the map from the world volume $\bR^p$ into $\bR^n$. The above bound can then be re-expressed as
\begin{equation}
\int d^p\sigma \sqrt{\rm{det} \,\CG_{\mu\nu}} \geq \int X^*\varphi\,,
\end{equation}
where $X^*\varphi$ stands for the pullback of the p-form $\varphi$.
\item[2.] There exists an explicit spinor construction of calibrations emphasising the connection between calibrated submanifolds, supersymmetry and kappa symmetry.
\end{itemize}

Let me review this spinor construction \cite{DadokReeseHarvey1993,Harvey1990}. For $p=1,2$ mod $4$, the p-form calibration takes the form
\begin{equation}
\label{eq:pcalib}
\varphi = dX^{i_1} \wedge \dots \wedge dX^{i_p} \epsilon^T
\Gamma_{0i_1\dots i_p}\epsilon 
\end{equation}
where the set $X^i$ ($i=1,\dots,n$) stands for the transverse scalars to the brane parameterising $\bR^n$, $\epsilon$ is a constant real spinor normalised so that $\epsilon^T \epsilon =1$ and $\Gamma_{i_1\dots i_k}$ are antisymmetrised products of Clifford matrices in $\bR^n$. Notice that given a tangent p-plane $\xi$, one can write $\varphi|_\xi$ as
\begin{equation}
\label{eq:pcalibb}
\varphi|_\xi = \sqrt{{\rm det}\, \CG}\, \epsilon^T \Gamma_\xi \epsilon
\end{equation}
where the matrix $\Gamma_\xi$
\begin{equation}
\label{gamma}
\Gamma_\xi =  \frac{1}{p! \sqrt{{\rm det}\, \CG}}\,\varepsilon^{\mu_1\dots \mu_p}
\partial_{\mu_1}X^{i_1}\cdots
\partial_{\mu_p}X^{i_p} \Gamma_{0i_1\dots i_p}\, ,
\end{equation}
is evaluated at the point to which $\xi$ is tangent. Given the restriction on the values of $p$, 
\begin{equation}
\label{eq:gamcon}
\Gamma_\xi^2=\mathbb{1} \, .
\end{equation}
It follows that $\varphi|_\xi \le {\rm vol}_\xi$ for all $\xi$. Since $\varphi$ is
also closed, one concludes it is a calibration. Its contact set is the set of p-planes for 
which this inequality is saturated. Using (\ref{eq:pcalibb}), the latter is equally characterised by the set of p-planes $\xi$ for which
\begin{equation}
\label{eq:contact}
\Gamma_\xi \epsilon =\epsilon \,.
\end{equation}
Because of (\ref{eq:gamcon}) and the fact that ${\rm tr}\,\Gamma_\xi=0$, the solution space to this equation is always half the dimension of the spinor space spanned by $\epsilon$ for any {\it given} tangent p-plane $\xi$.  However, this solution space generally varies as $\xi$ varies over the contact set, so that the solution space of the set is generally smaller.

So far the discussion involved no explicit supersymmetry. Notice, however, that the matrix $\Gamma_\xi$ in (\ref{gamma}) matches the kappa symmetry matrix $\Gamma_\kappa$ for branes in the static gauge with no gauge field excitations propagating in Minkowski. This observation allows us to identify the saturation of the calibration bound with the supersymmetry preserving condition (\ref{spc}) derived from the gauge fixing analysis of kappa symmetry.

Let me close the logic followed in section \ref{sec:solitons} by pointing out a very close relation between the supersymmetry algebra and kappa symmetry that all my previous considerations suggest.  Consider a single infinite flat M5-brane propagating in d=11 Minkowski and fix the extra gauge symmetry of the PST formalism by $a(\sigma^\mu)=t$ (temporal gauge).
The kappa symmetry matrix (\ref{eq:kmatrixm5}) reduces to
\begin{equation}
\Gamma_\kappa =
\frac{1}{\sqrt{{\rm det}\,(\delta_{ij}+ \tilde H_{ij})}}
[\Gamma^0 \Gamma_i t^i + \frac{1}{2}\Gamma^0 \Gamma^{ij}{\tilde H}_{ij}
- \frac{1}{5!}\Gamma^0 \Gamma_{i_1,\dots, i_5} {\varepsilon}^{i_1\dots i_5}]\,,
\end{equation}
where all $\{i,j\}$ indices stand for world space M5 indices. Notice that the structure of this matrix is equivalent to the one appearing  in (\ref{eq:bgamma}) for $\bar\Gamma$ by identifying 
\begin{eqnarray}
Y^{i_1\dots i_5} =- \varepsilon^{i_1\dots i_5}\,, & & \tilde H_{ij}= Z_{ij} \nonumber \\
P^i =\frac{1}{8} \varepsilon^{i\, j_1 j_2 j_3 j_4} \, Z_{j_1j_2}Z_{j_3j_4}\,, & &
P^0 =\sqrt{ {\rm det}\, (\delta_{ij} + Z_{ij})}\,.
\label{29}
\end{eqnarray}
Even though, this was only argued for the M5-brane and in a very particular background, it does provide some preliminary evidence for the existence of such connection. In fact, a stronger argument can be provided by developing a phase space formulation of the kappa symmetry transformations that allows to write the supersymmetry anticommutator as \cite{Gutowski:1999tu}
\begin{equation}
\label{eq:susyguess}
\{Q,Q\} = \Gamma^{0}
\int d^p\sigma\,\left[ \Gamma^a \tilde p_a + \gamma\right]\,, \quad{\rm with} \quad \gamma = \frac{1}{p!}\,\varepsilon^{a_1\dots a_p}
\partial_{a_1}X^{i_1}\cdots
\partial_{a_p}X^{i_p} \Gamma_{i_1\dots i_p}\, .
\end{equation}
This result has not been established in full generality but it agrees with the flat space case \cite{DeAzcarraga:1989gm} and 
those non-flat cases that have been analysed \cite{Sato:1998yu,Sato:1998ax}. I refer the reader to \cite{Gutowski:1999tu}
where they connect the functional form in the right hand side of (\ref{eq:susyguess}) with the kappa symmetry transformations for fermions in its hamiltonian form.

The connection between calibrations, supersymmetry and kappa symmetry goes beyond the arguments given above. The original mathematical notion of calibration was extended in \cite{Gutowski:1999iu,Gutowski:1999tu} relaxing its first condition $d\varphi\neq 0$. Physically, this allowed to include the presence of non-trivial potential energies due to background fluxes coming from the WZ couplings. Some of the applications derived from this notion include \cite{Gauntlett:2001ur,Gauntlett:2002sc,Gauntlett:2003cy,Martelli:2003ki,Cascales:2004qp}. Later, the notion of {\it generalised calibration} was introduced in \cite{Koerber:2005qi}, where it was shown to agree with the notion of calibration defined in generalised Calabi-Yau manifolds \cite{2004math1221G} following the seminal work \cite{Hitchin:2004ut}. This general notion allows to include the effect of non-trivial magnetic field excitations on the calibrated submanifold, but still assumes the background and the calibration to be static. Some applications of these notions in the physics literature can be found in \cite{Koerber:2005qi,Martucci:2005ht,Papadopoulos:2006hk}. More recently, this formalism was generalised to include electric field excitations \cite{Martucci:2011dn}, establishing a precise correspondence between generic supersymmetric brane configurations and generalised geometry.

\paragraph{Summary :}  A necessary condition for a bosonic brane configuration to preserve supersymmetry is to solve the kappa symmetry preserving condition (\ref{spc}). In general, this is not sufficient for being an on-shell configuration, though it can be, if there are no gauge field excitations. Solutions to (\ref{spc}) typically impose a set of constraints on the field configuration, which can be interpreted as BPS equations by computing the hamiltonian of the configuration, and a set of projection conditions on the constant parts $\epsilon_\infty$ of the background Killing spinors $\epsilon$. The energy bounds saturated when the BPS equations hold are a field theory realisation of the algebraic bounds derived from the supersymmetry algebra. An attempt to summarise the essence of these relations is illustrated in figure \ref{fig4}.

\section{World volume solitons : applications}
\label{sec:soliton-appl}

There are two natural set of applications involving brane effective actions :  kinematical and dynamical.
In this section, I will discuss the application of the general formalism developed in section \ref{sec:solitons} 
to study the existence of certain string theory BPS states realised as world volume supersymmetric bosonic solitons, leaving more AdS/CFT dynamically oriented applications to the next section. 

The main goals in this section include:
\begin{itemize}
\item[1.] In a Minkowski background, the identification of the vacuum of all the p+1 dimensional supersymmetric field theories discussed before as half-BPS flat infinite branes, and the discussions of some of their excitations carrying topological charges, which are interpretable as brane intersections or branes within branes.
\item[2.] Supertubes, as examples of supersymmetric bound states realised as expanded branes without carrying charge under the gauge potential which the world volume brane minimally couples to. 
\item[3.] As examples of solitons in curved backgrounds, I will discuss the baryon vertex and giant gravitons in AdS${}_5\times$S${}^5$.
\item[4.] I will stress the relevance of supertubes and giant gravitons as constituents of small supersymmetric black holes, their connection to fuzzball ideas and the general use of probe techniques to identify black hole constituents in more general situations.
\end{itemize}

\subsection{Vacuum infinite branes}
\label{sec:vacuum}

There exist half-BPS branes in ten and eleven dimensional Minkowski spacetime. Since their effective actions were discussed in section \ref{sec:bbrane}, we can check their existence and the amount of supersymmetry they preserve, by solving the brane classical equations of motion and the kappa symmetry preserving condition (\ref{spc}).

First, one works with the bosonic truncation $\theta=0$. The background, in cartesian coordinates, involves the metric
\begin{equation}
  ds^2 = \eta_{mn} dx^mdx^n\,, \qquad \qquad m,n=0,1,\dots D-1
\end{equation}
and all remaining bosonic fields vanish, except for the dilaton, in type IIA/B, which is constant. This supergravity configuration is maximally supersymmetric, i.e. it has Killing spinors spanning a vector space which is 32-dimensional. In cartesian coordinates, these are constant spinors $\epsilon=\epsilon_\infty$. 

Half-BPS branes should correspond to vacuum configurations in these field theories describing infinite branes breaking the isometry group $\ISO(1,D-1)$ to $\ISO(1,p)\times \SO(D-p-1)$ and preserving half of the supersymmetries. Geometrically, these configurations are specified by the brane location. This is equivalent to first, splitting the scalar fields $X^m(\sigma)$ into longitudinal $X^\mu$ and transverse $X^I$ directions, setting the latter to constant values $X^I=c^I$ (the transverse brane location). Second, one identifies the world volume directions with the longitudinal directions, $X^\mu=\sigma^\mu$. The latter can also be viewed as fixing the world volume diffeomorphisms to the so called  {\it static gauge}. This information can be encoded as an array
\begin{equation}
\begin{array}{cccccccccccl}
p-brane: &1&2&.&.&p&\_&\_&\_&\_& \
\end{array}
\end{equation}

It is easy to check that the above is an {\it on-shell} configuration given the structure of the Euler-Lagrange equations and the absence of non-trivial couplings except for the induced world volume metric $\CG_{\mu\nu}$ which equals $\eta_{\mu\nu}$ in this case.

To analyse the supersymmetry preserved, one must solve (\ref{spc}). Notice that in the static gauge and in the absence of any further excitations, the induced gamma matrices equal
\begin{equation}
\gamma_\mu = \partial_\mu x^m E_m^{a}\Gamma_{a} = \Gamma_{\mu} 
\Longrightarrow
\gamma_{\mu_0\ldots \mu_p}=\Gamma_{\mu_0\ldots \mu_p},
\end{equation}
where I already used $E_m^{a}=\delta_m^{a}$. Thus, $\Gamma_\kappa$ reduces to a constant Clifford valued matrix standing for the volume of the brane, $\Gamma_{{\rm vol}}$, up to the chirality of the background spinors which is parameterised by the matrix $\tau$
\begin{equation}
  \Gamma_\kappa = \Gamma_{{\rm vol}}\tau\,.
\end{equation}
The specific matrices for the branes discussed in this review are summarised in table \ref{table:infinitebranesusy}. Since $\Gamma_\kappa^2 = \mathbb{1}$ and ${\rm Tr} \Gamma_\kappa = 0$, only half of the vector space spanned by $\epsilon_\infty$ preserves these bosonic configurations, i.e. all infinite branes preserve half of the supersymmetries. These projectors match the ones derived from bosonic supergravity backgrounds carrying the same charges as these infinite branes.

\begin{longtable}[c]{p{3 cm}p{3 cm}}
\hline
\hline
{\bf BPS state} & {\bf Projector} \endhead
\hline
\hline
M2-brane & $\Gamma_{012}\epsilon= \epsilon$ \\
\hline
M5-brane & $\Gamma_{012345}\epsilon=\epsilon$ \\
\hline
IIA ${\rm D}_{2n}$-brane & $\Gamma_{0\ldots 2n}\Gamma_\sharp^{n+1}\epsilon=\epsilon$ \\
\hline
IIB ${\rm D}_{2n-1}$-brane & $\Gamma_{0\ldots 2n-1}\tau_3^n\,i\tau_2\epsilon=\epsilon$  \\
\hline
\hline
\\
\caption{\it Half-BPS branes and the supersymmetries they preserve.}
\label{table:infinitebranesusy}
\end{longtable}

All these configurations have an energy density equaling the brane tension $T$ since the hamiltonian constraint is always solved by
\begin{equation}
  \CE^2 = {\rm T}^2\,{\rm det}\,\CG = {\rm T}^2\,.
\end{equation}
From the spacetime superalgebra perspective, these configurations saturate a bound between the energy and the p-form bosonic charge carried by the volume form defined by the brane
\begin{equation}
  {\cal E} = {\cal Z}_{\mu_1\dots \mu_p} = {\rm T}\epsilon_{\mu_1\dots\mu_p}.
\end{equation}
The saturation corresponds to the fact that {\it any} excitation above the infinite brane configuration would increase the energy. From the world volume perspective, the solution is a vacuum, and consequently, it is annihilated by {\it all} sixteen world volume supercharges. These are precisely the ones solving the kappa symmetry preserving condition (\ref{spc}).

\subsection{Intersecting M2-branes}

As a first example of an excited configuration, consider the intersection of two M2-branes in a point corresponding to the array
\begin{equation}
\begin{array}{cccccccccccl}
M2: &1&2&\_&\_&\_&\_&\_&\_&\_&\_ & \nn \\
M2: &\_&\_&3&4&\_&\_&\_&\_&\_&\_ & \,.  
\end{array}
\label{arraym2m2}
\end{equation}
In the probe approximation, the M2-brane effective action describes the first M2-brane by fixing the static gauge and the second M2-brane as an excitation above this vacuum by turning on two scalar fields $(X^3,\,X^4)$ according to the ansatz
\begin{eqnarray}
X^\mu &=& \sigma^\mu \quad , \quad X^i=c^i \nn \\
X^3(\sigma^a) &\equiv & y(\sigma^a) \quad , \quad
X^4(\sigma^a) \equiv z(\sigma^a)
\label{m2m2}
\end{eqnarray}
\noindent
where $a$ runs over the spatial world volume directions and $i$ over
the transverse directions not being excited.

\paragraph{Supersymmetry analysis :} Given the ansatz (\ref{m2m2}), the induced metric components equal ${\cal G}_{00}=-1,\,{\cal G}_{0a}=0,\,{\cal G}_{ab}=\delta_{ab} + \6_a X^r \6_b X^s \delta_{rs}$ (with $r,s=3,4$), whereas its determinant and the induced gamma matrices reduce to
\begin{eqnarray}
-\det\CG &=& 1 + |\vec \nabla y|^2 + |\vec \nabla z|^2 + (\vec \nabla y \times \vec \nabla z)^2
\label{indmetm2} \\
\gamma_0 &=& \Gamma_{0} \quad , \quad \gamma_a=
\Gamma_{a} + \6_a X^r \Gamma_{r}\, .
\label{indgamm2}
\end{eqnarray}
Altogether, the kappa symmetry preserving condition (\ref{spc}) is
\begin{equation}
\sqrt{-{\rm det}\,\CG}\,\epsilon = \left(\Gamma_{012} + \varepsilon^{ab}\6_a y \6_b z \Gamma_{034} - \varepsilon^{ab}\6_a x^r \Gamma_{0br}\right)\epsilon \,.
\label{km21}
\end{equation}

If the excitation given in (\ref{m2m2}) must describe the array in (\ref{arraym2m2}), the subspace of Killing spinors $\epsilon$ spanned by the solutions to (\ref{km21}) must be characterised by two projection conditions
\begin{equation}
\Gamma_{012}\epsilon = \Gamma_{034}\epsilon =\epsilon,
\label{ksm2}
\end{equation}
one for each M2-brane in the array (\ref{arraym2m2}). Plugging these projections into (\ref{km21})
\begin{equation}
\left(\sqrt{-{\rm det}\,\CG}- (1+\varepsilon^{ab}\6_a y \6_b z)\right)\,\epsilon = \varepsilon^{ab}\6_a X^r \Gamma_{0br}\epsilon, 
\label{km22}
\end{equation}
one obtains an identity involving two different Clifford valued contributions : the left hand side is proportional to the
identity matrix acting on the Killing spinor, while the right hand side involves some subset of antisymmetric products of gamma matrices. Since these Clifford valued matrices are independent, each term must vanish independently.
This is equivalent to two partial differential equations
\begin{equation}
\6_2 y = -\6_1 z\,, \quad \quad \6_1 y = \6_2 z\,.
\label{cauchy}
\end{equation}
Notice this is equivalent to the {\it holomorphicity} of the complex function $U(\sigma^+)=y+iz$ in terms of the complex world space coordinates $\sigma^\pm = \sigma^1\pm i\sigma^2$, since (\ref{cauchy}) are equivalent to the Cauchy-Riemann equations for $U(\sigma^+)$.

When conditions (\ref{cauchy}) are used in the remaining left hand side of (\ref{km22}), one recovers an identity.
Thus, the solution to (\ref{spc}) in this particular case involves the two supersymmetry projections (\ref{ksm2}) and the BPS equations (\ref{cauchy}) satisfied by {\it holomorphic} functions $U(\sigma^+)$.

\paragraph{Hamiltonian analysis :} Since this is the first non-trivial example of a supersymmetric soliton discussed in this review, it is pedagogically constructive to rederive equations (\ref{cauchy}) from a purely hamiltonian point of view \cite{Gauntlett:1997ss}. This will also convince the reader that holomorphicity is the only requirement to be on-shell. To ease notation below, rewrite (\ref{cauchy}) as
\begin{equation}\label{cauchy2}
\vec \nabla y = \star\vec\nabla z \,,
\end{equation}
where standard vector calculus notation for $\bR^2$ is used, i.e. $\vec\nabla = (\partial_1,\partial_2)$ and $\star \vec \nabla =
(\partial_2,-\partial_1)$.

Consider the phase space description for the M2-brane lagrangian given in (\ref{eq:m2phasespace}) in a Minkowski background. The Lagrange multiplier fields $s^a$ impose the world space diffeomorphism constraints. In the static gauge, these reduce to
\begin{equation}
P_a = P_I\cdot\partial_a X^I,
\end{equation}
where $P_I$ are the conjugate momenta to the eight world volume scalars $X^I$ describing transverse fluctuations. For static configurations carrying no momentum, i.e. $P_I=0$, the world space momenta will also vanish, i.e. $P_a=0$.

Solving the hamiltonian constraint imposed by the Lagrange multiplier $\lambda$ for the energy density $\CE = P_0$, one obtains
\cite{Gauntlett:1997ss}
\begin{equation}
(\CE/T_{{\rm M2}})^2 = 1 + |\vec \nabla y|^2 + |\vec \nabla z|^2 + 
(\vec \nabla y \times \vec \nabla z)^2 = (1 - \vec \nabla y \times \vec \nabla z)^2 + 
|\vec \nabla y - \star\vec\nabla z|^2\,.
\end{equation}
This already involved the computation of the induced world space metric determinant and its rewriting in a suggestive way to derive the bound
\begin{equation}
\CE/T_{{\rm M2}} \geq 1 + | \vec \nabla y \times \vec \nabla z |\,.
\end{equation}
The latter is saturated if and only if equation (\ref{cauchy2}) is satisfied. This proves the BPS character of the constraint derived from solving (\ref{spc}) in this particular case and justifies that any solution to (\ref{cauchy2}) is on-shell, since it extremises the energy and there are {\it no} further gauge field excitations.

Integrating over the world space of the M2 brane, allows us to derive a bound on the charges carried by these subset of configurations
\begin{equation}
E \geq E_0 + | Z |\,.
\label{eq:m2m2bound}
\end{equation}
$E_0$ stands for the energy of the infinite M2-brane vacuum, whereas $Z$ is the topological charge
\begin{equation}
Z = T_{M2}\int_{M2} dy\wedge dz = T_{M2}
\frac{i}{2}\int_{M2} dU\wedge d{\bar U}\,,
\label{m2m2topo}
\end{equation}
accounting for the second M2-brane in the system. 

The bound (\ref{eq:m2m2bound}) matches the spacetime supersymmetry algebra bound : the mass (E) of the system is larger than the sum of the masses of the two M2-branes. Field theoretically, the first M2-brane charge corresponds to the vacuum energy $(E_0)$ while the second to the topological charge $(Z)$ describing the excitation. When the system is supersymmetric, the energy saturates the bound $E=E_0+|Z|$ and preserves 1/4 of the original supersymmetry. From the world volume superalgebra perspective, the energy is always measured with respect to the vacuum. Thus, the bound corresponds to the excitation energy $E-E_0$ equalling $|Z|$. This preserves 1/2 of the world volume supersymmetry preserved by the vacuum, matching the spacetime 1/4 fraction.

For more examples of M2-brane solitons see \cite{Bergshoeff:1999jc} and for a related classification of D2-brane supersymmetric soltions see \cite{Bak:2006yb}.

\subsection{Intersecting M2 and M5-branes}

As a second example of BPS excitation, consider the $1/4$ BPS configuration $M5\perp M2(1)$ corresponding to the brane array
\begin{equation}
\begin{array}{cccccccccccl}
M5: &1&2&3&4&5&\_&\_&\_&\_&\_ & \nn \\
M2: &\_&\_&\_&\_&5&6&\_&\_&\_&\_ & \,.  
\end{array}
\label{arraym5}
\end{equation}
The idea is to describe an infinite M5-brane by the static gauge and to turn on a transverse scalar field $X^6$ to account for the M2-brane excitation. However, $X^6$ is not enough to support an M2-brane interpretation, since the latter is electrically charged under the eleven dimensional supergravity three form $A_{3}$. Thus, the sought M5-brane soliton must source the $A_{056}$ components. From the Wess-Zumino coupling
\begin{equation}
\int dV_{2}\wedge {\cal A}_{3}\,,
\end{equation} 
one learns that the magnetic $(dV)_{\hat{a}\hat{b}\hat{c}}$ components, where hatted indices stand for world space directions different from $\sigma^5$, i.e. $\hat{a}\neq 5$, must also be excited.

The full ansatz will assume delocalisation along the $\sigma^5$ direction, so that the string-like excitation in the $X^6$ direction can be viewed as a membrane:
\begin{eqnarray}
& X^\mu=\sigma^\mu \quad , \quad X^{i}=c^i & \nonumber \\
& X^6(\sigma^{\hat{a}})=y(\sigma^{\hat{a}}) & \nonumber \\
& \CH_{5\hat{a}\hat{b}}=0 \,.&
\label{m5m2ansatz}
\end{eqnarray}

\paragraph{Supersymmetry analysis :} The M5-brane kappa symmetry matrix (\ref{eq:kmatrixm5}) in the temporal gauge $a=\tau$ reduces to
\begin{equation}
\Gamma_\kappa = \frac{1}{\sqrt{-{\rm det}\,({\cal G} + \tilde{H})}}
\left[\frac{1}{5!}\epsilon^{a_1\ldots a_5}\Gamma_{0}
\gamma_{a_1\ldots a_5} - \frac{1}{2}\sqrt{-{\rm det}\,{\cal G}}\,
\Gamma_{0}\gamma_{ab}\tilde{H}^{ab} -\Gamma_{0}\gamma_a t^a \right].
\label{a0kappam5}
\end{equation}
For the subset of configurations described by the ansatz (\ref{m5m2ansatz}), it follows
\begin{eqnarray}
&t_a = 0 \quad , \quad \tilde{H}^{\hat{a}\hat{b}}=0 & \nonumber \\
& \tilde{H}^{5\hat{a}} = \frac{\Pi^{\hat{a}}}{\sqrt{-{\rm det}\,{\cal G}}} 
\quad , \quad
\Pi^{\hat{a}}=\frac{1}{3!}\epsilon^{\hat{a}\hat{a}_1\hat{a}_2\hat{a}_3}
\CH_{\hat{a}_1\hat{a}_2\hat{a}_3} \,.&
\end{eqnarray}
This reduces (\ref{a0kappam5}) to
\begin{equation}
\Gamma_\kappa = \frac{1}{\sqrt{-{\rm det}\,({\cal G} + \tilde{H})}}
\left[\Gamma_{012345} + \partial_{\hat{a}}y
\Gamma_{05y}
\Gamma_{05}\Gamma^{\hat{a}}
\Gamma_{012345}-
\Gamma_{05}\Gamma_{\hat{a}}
\Pi^{\hat{a}} - \partial_{\hat{a}}y \Pi^{\hat{a}}
\Gamma_{05y}\right].
\label{a0kappam51}
\end{equation}
To solve the kappa symmetry preserving condition (\ref{spc}), I impose two projection conditions
\begin{eqnarray}
\Gamma_{012345}\epsilon&=&\epsilon \nonumber \\
\Gamma_{05y}\epsilon &=& \epsilon \,,
\label{m5m2susy}
\end{eqnarray}
on the constant Killing spinors $\epsilon$. The 8 supercharges satisfying them match the ones preserved by $M5\perp M2(1)$. Using (\ref{m5m2susy}) into (\ref{a0kappam51}), $\Gamma_\kappa$ keeps a non-trivial dependence on $\Gamma_{05}\Gamma_{\hat{a}}$. Requiring its coefficient to vanish gives rise to the BPS condition
\begin{equation}
\Pi_{\hat{a}}=-\partial_{\hat{a}}y.
\label{m5m2bps}
\end{equation}
Overall, the kappa symmetry preserving condition (\ref{spc}) reduces to the purely algebraic condition
\begin{equation}
\sqrt{-{\rm det}\,({\cal G} + \tilde{H})}\epsilon=\left(1+\delta^{\hat{a}\hat{b}}
\partial_{\hat{a}}y\partial_{\hat{b}}y\right)\epsilon.
\label{m5m2last}
\end{equation}
To check this holds, notice the only non-vanishing components of $\tilde{H}_{\mu\nu}$ are $\tilde{H}_{5\hat{a}}$
\begin{equation}
\tilde{H}_{5\hat{a}}={\cal G}_{55}{\cal G}_{\hat{a}\hat{b}}
\frac{\Pi^{\hat{b}}}{\sqrt{-{\rm det}\,{\cal G}}}\,.
\end{equation}
This allows us to compute the determinant
\begin{equation}
-{\rm det}\,({\cal G} + \tilde{H})=\det\,({\cal G}_{\hat{a}\hat{b}} + \tilde{H}
_{5\hat{a}}\tilde{H}_{5\hat{b}}) = \det\,({\cal G}_{\hat{a}\hat{b}})\left(
1+{\cal G}^{\hat{a}\hat{b}}\tilde{H}_{5\hat{a}}\tilde{H}_{5\hat{b}}\right),
\end{equation}
which becomes a perfect square once the BPS equation (\ref{m5m2bps}) is used 
\begin{equation}
-\det\,({\cal G} + \tilde{H})=\left(1+ \delta^{\hat{a}\hat{b}}
\partial_{\hat{a}}y\partial_{\hat{b}}y\right)^2\,.
\end{equation}
This shows equation (\ref{m5m2last}) holds automatically. Thus, the solution to the kappa symmetry preserving condition (\ref{spc}) for the ansatz (\ref{m5m2ansatz}) on an M5-brane action is solved by the supersymmetry projection conditions (\ref{m5m2susy}) and the BPS equation (\ref{m5m2bps}). Since the soliton involves a non-trivial world volume gauge field, the Bianchi identity $d\CH_3=0$ must still be imposed. This determines  the harmonic character for the excited transverse scalar in the
four dimensional world space $\omega_4$
\begin{equation}
\6^{\8a}\6_{\8a}y=0.
\end{equation}

\paragraph{Hamiltonian analysis :}  The hamiltonian analysis for this system was studied in \cite{Gauntlett:1997ss} following  the M5-brane phase space formulation given in (\ref{eq:m5phase}). For static configurations, the hamiltonian constraint can be solved by the energy density ${\cal E}$ as
\begin{equation}
\frac{{\cal E}^2}{T_{\rm M5}^2} = 1 + (\partial y)^2 + \frac{1}{2} |\tilde {\cal H}|^2 + 
|\tilde {\cal H}\cdot \partial y|^2 + |V|^2 
\end{equation}
where
\begin{eqnarray}
|\tilde {\cal H}|^2 &=& \tilde {\cal H}^{ab} \tilde {\cal H}^{cd} \delta_{ac} \delta_{bd}\,, \qquad \qquad \tilde {\cal H}^{ab} = \frac{1}{6}\varepsilon^{abcde}\CH_{cde} \nonumber \\
|\tilde {\cal H}\cdot \partial y|^2 &=& 
\tilde {\cal H}^{ab}\tilde {\cal H}^{cd}\partial_b y\partial_d y\delta_{ac} 
\nonumber \\
|V|^2 &=& V_aV_b \delta^{ab}\,,
\end{eqnarray}
and world space indices were denoted by latin indices $\sigma^a$ $a=1,\dots ,5$. It was noted in \cite{Gauntlett:1997ss} that by introducing a unit length world space 5-vector $\zeta$, i.e. $\zeta^a \zeta^b\delta_{ab} =1$, the energy density could be written in the suggestive form
\begin{eqnarray}
\frac{{\cal E}^2}{T_{\rm M5}^2}  &=& \big|\zeta^a \pm \tilde {\cal H}^{ab}\partial_b y\big| ^2 
+ 2\bigg|\partial_{[a} y \zeta_{b]} \pm \frac{1}{2} \delta_{ac}\delta_{bd} \tilde
{\cal H}^{cd}\bigg|^2 \nn \\
&& +\, (\zeta^a\partial_a y)^2 + |V|^2\,.
\end{eqnarray}
The unit vector provides a covariant way of introducing a preferred direction in the 5-dimensional world space. Choosing $\zeta ^5=1$ and $\zeta^{\hat a}=0$, to match the delocalisation direction in our bosonic ansatz, one derives the inequality
\begin{equation}
\frac{{\cal E}}{T_{\rm M5}} \ge 1 \pm \frac{1}{6}\Pi^{\hat a}\partial_{\hat a}y.
\end{equation}
The latter is saturated if and only if
\begin{equation}\label{stringsola}
\partial_5 y=0 \qquad \CH_{5\hat a \hat b} =0
\end{equation}
and
\begin{equation}\label{stringsolb}
\CH_3 = \pm \star dy
\end{equation}
where $\CH_3$ is only defined on the 4-dimensional subspace $\omega_4$, orthogonal to $\zeta$, and $\star$ is its Hodge dual. 
This confirms the BPS nature of equation (\ref{m5m2bps}). Since ${\cal H}_3$ is closed, $y$ is harmonic in $\omega_4$.

To regulate the divergent energy, one imposes periodic boundary conditions in the 5-direction making the orbits of the
vector field $\zeta$ have finite length $L$.  Then, the total energy satisfies 
\begin{equation}
{\it E} \ge E_0 + L\cdot |Z|
\end{equation}
where $Z$ is the topological charge
\begin{equation}
Z= \int_{w_4} {\cal H}_3\wedge dy .
\end{equation}
The tension of the soliton, i.e. energy per unit of length, equals $T={\it E-E_0}/L$. It is bounded by $Z$. It only equals the latter for configurations satisfying (\ref{stringsolb}). Singularities in the harmonic function match
the strings found in \cite{Howe:1997ue}. To check this interpretation, consider a solution with a single isolated point
singularity at the origin. Its energy can be rewritten as the small radius limit of a surface integral over a 3-sphere surrounding the
origin. Since $y$ is constant on this integration surface, one derives the string tension equals \cite{Gauntlett:1997ss} 
\begin{equation}
T = \mu \lim_{\delta\to 0}  y(\delta) \quad {\rm where} \quad \mu = \int_{S^3} {\cal H}_3
\end{equation}
is the string charge. Even though this tension diverges, it does so consistently as being the boundary of a semi-infinite membrane.

\subsection{BIons}

Perhaps one of the most pedagogical examples of brane solitons are BIons. These were first described in \cite{Callan:1997kz,Gibbons:1997xz} and correspond to on-shell supersymmetric D-brane configurations representing a fundamental
string ending on the D-brane, i.e. the defining property of the D-brane itself. They correspond to the array of branes
\begin{equation}
\begin{array}{ccccccccccl}
Dp: &1&.&.&.&p&\_&\_&\_&\_& \nn \\
F : &\_&\_&\_&\_&\_&p+1&\_&\_&\_& \,.  
\end{array}
\end{equation} 
Working in the static gauge describes the vacuum infinite Dp-brane. The static soliton excites a transverse scalar field  $(y=y(\sigma^a))$ and the electric field $(V_0=V_0(\sigma^a))$, while setting the magnetic components of the gauge field $(V_a)$ to zero
\begin{eqnarray}
X^\mu &=& \sigma^\mu \quad , \quad X^i=c^i \nonumber \\
 X^{p+1}(\sigma^a) &=& y(\sigma^a) \quad , \quad V_0 = V_0(\sigma^a)
\label{bionansatz}
\end{eqnarray}
The gauge invariant character of the scalar ensures its physical observability as a deformation of the flat world volume geometry described by the global static gauge, whereas the electric field can be understood 
as associated to the end of the open string, which is seen as a charged particle from the world volume perspective. A second way of arguing the necessity for such electric charge is to remember that fundamental strings are electrically charged under the \NSNS\ two form. The latter appears in the effective action through the gauge invariant form $\CF$. Thus, turning on $V_0$ is equivalent to turning such charge\footnote{There are many papers studying the dynamics of BIons, including \cite{Lee:1997xh,Bak:1998xp,Kastor:1999ag} and \cite{Thorlacius:1997zd}, where the solution to the Born-Infeld action reviewed here is proved to solve the equations of motion derived from higher order corrections to the effective action.}.

\paragraph{Supersymmetry analysis :} Let me analyse the amount of supersymmetry preserved by configurations (\ref{bionansatz}) in type IIA and type IIB, separately. In both cases, the matrix $\CG_{\mu\nu}+ \CF_{\mu\nu}$ equals 
\begin{equation}
\CG_{\mu\nu}+\CF_{\mu\nu} = \left(\matrix{
-1 & F_{0b} \cr
-F_{0a}  & \delta_{ab}  + \6_a y \6_b y}\right) \quad \Longrightarrow \quad -{\rm det}\,(\CG_{\mu\nu}+\CF_{\mu\nu})={\rm det}\,(\delta_{ab}  + \6_a y \6_b y -F_{0a}F_{0b})
\end{equation}
while the induced gamma matrices are decomposed as
\begin{equation}
\gamma_0 =\Gamma_{0}, \qquad \qquad
\gamma_a = \Gamma_{a} + \6_a y \Gamma_{y},
\label{indbion}
\end{equation}
where $a$ stands for world space indices. Due to the electric ansatz for the gauge field, the kappa symmetry matrix $\Gamma_\kappa$ has only two contributions. In particular, for type IIA $(p=2k)$
\begin{equation}
\Gamma_\kappa = \frac{1}{\sqrt{-{\rm det}\,(\CG_{\mu\nu}+\CF_{\mu\nu})}}
\frac{1}{(p+1)!}\varepsilon^{\mu_0\ldots \mu_p}\left(\gamma_{\mu_0\ldots
\mu_p}\Gamma_{\sharp}^{k+1} +
\left(\matrix{ p+1\cr
2}\right)F_{\mu_0\mu_1}\gamma_{\mu_3\ldots\mu_p}\Gamma_{\sharp}^k\right)\,.
\end{equation}
Summing over world volume time, one obtains
\begin{equation}
\G_\kappa =\frac{1}{\sqrt{-{\rm det}\,(\CG_{\mu\nu}+\CF_{\mu\nu})}}
\frac{1}{p!}\varepsilon^{a_1\ldots a_p}\left(\Gamma_{0}
\gamma_{a_1\ldots a_p}\Gamma_{\sharp}^{k+1} 
+\frac{p}{2}F_{0a_1}\gamma_{a_2\ldots a_p}\Gamma_{\sharp}^k\right)\,.
\label{kappabion}
\end{equation} 
Using the duality relation
\begin{equation}
\varepsilon^{i_1\ldots i_k j_{k+1}\ldots j_{p+1}}\gamma_{j_{k+1}\ldots
j_{p+1}}= (-1)^{k(k-1)/2}(p+1-k)!\gamma^{i_1\ldots i_k}\,\sqrt{-{\rm det}\,\CG}\,
\Gamma_{0\ldots p},
\label{gamduality}
\end{equation}
one can write the first term on the right hand side of (\ref{kappabion}) as
\begin{equation}
\Gamma_{0\ldots p}\Gamma_{\sharp}^{k+1}-
\Gamma^{b}\6_b y \Gamma_{y}
\Gamma_{0\ldots p}\Gamma_{\sharp}^{k+1}.
\label{bion1}
\end{equation}
Using the same duality relation and proceeding in an analogous way, the second term equals
\begin{equation}
F_{0a}\Gamma^{a}\Gamma_{1\ldots p}\Gamma_{\sharp}^k +
F_{0a}\6_b y \Gamma_{y}\Gamma^{ab}\Gamma_{1\ldots p}\Gamma_{\sharp}^k.
\label{bion2}
\end{equation}
Inserting equations (\ref{bion1}) and (\ref{bion2}), the kappa symmetry
preserving condition can be expressed as
\begin{equation}
\sqrt{-{\rm det}\,(\CG+\CF)}\epsilon = \left[1+\Gamma^{a}\Gamma_{0}
\Gamma_{\sharp}(F_{0a}-\6_a y \Gamma_{0y}\Gamma_{\sharp})
- \Gamma^{ab}F_{0a}\6_b y \Gamma_{0y}\Gamma_{\sharp}\right]
\Gamma_{0\ldots p}\Gamma_{\sharp}^{k+1}\epsilon\,.
\label{kappabion2}
\end{equation}
Given the physical interpretation of the sought soliton, one imposes the following two supersymmetry projection conditions
\begin{eqnarray}
\Gamma_{0\ldots p}\Gamma_{\sharp}^{k+1} \epsilon &=&
\epsilon \label{susybion1} \\
\Gamma_{0y}\Gamma_{\sharp}\epsilon &=&\epsilon
\label{susybion2}
\end{eqnarray}
corresponding to having a type IIA Dp-brane along directions $1,\ldots,p$
and a fundamental string along the transverse direction $y$. Since both Clifford valued matrices commute, the dimensionality of the subspace of solutions is 8, as corresponds to preserving $\nu=1/4$ of the bulk supersymmetry.
Plugging these projections into (\ref{kappabion2}), the kappa symmetry preserving condition reduces to
\begin{equation}
\sqrt{-{\rm det}\,(\CG+\CF)}\epsilon = \left(1+ \Gamma^{a} \Gamma_{0}
\Gamma_{\sharp}(F_{0a}-\6_a y) - \Gamma^{ab}F_{0a}\6_b y\right)\epsilon\,.
\label{kappabion3}
\end{equation}
It is clear that the BPS condition
\begin{equation}
F_{0a}=\6_a y
\label{BPSbion}\,,
\end{equation}
derived from requiring the coefficient of $\Gamma^{a}\Gamma_{0}\Gamma_\sharp$ to vanish solves (\ref{kappabion3}).
Indeed, the last term in (\ref{kappabion3}) vanishes due to antisymmetry, whereas the square root
of the determinant equals one, whenever (\ref{BPSbion}) holds.

The analysis for type IIB Dp-branes $(p=2k+1)$ works analogously by appropriately dealing with the different bulk fermion chiralities, i.e. one should replace $\Gamma_{\sharp}^k$ by $\tau_3^k\,i\tau_2$. Thus, the supersymmetry projection conditions (\ref{susybion1}) and
(\ref{susybion2}) are replaced by
\begin{eqnarray}
\Gamma_{0\ldots p}\tau_3^{k+1}\,i\tau_2 \epsilon &=&
\epsilon \label{susybion1b} \\
\Gamma_{0y}\tau_3\epsilon &=&\epsilon
\label{susybion2b}
\end{eqnarray}
corresponding to having a type IIB Dp-brane along the directions $1,\ldots,p$
and a fundamental string along the transverse direction $y$.

Satisfying the BPS equation (\ref{BPSbion}) does not guarantee the on-shell nature of the configuration. Given the non-triviality of the gauge field, Gauss' law $\partial_a E^a=0$ must be imposed, where $E^a$ is the conjugate momentum to the electric field, which reduces to
\begin{equation}
E^a=\frac{\6\CL}{\6\dot{V}_a}=\delta^{ab}F_{0b},
\end{equation}
when (\ref{BPSbion}) is satisfied. Thus, the transverse scalar $y$ must be an harmonic function on
the p-dimensional D-brane world space
\begin{equation}
\6_a\6^a y =0\,.
\label{eq:bioneom}
\end{equation}

\paragraph{Hamiltonian analysis : } Using  
the phase space formulation of the D-brane lagrangian in (\ref{eq:hamp}) and (\ref{hamt}), I will reproduce the BPS bound (\ref{BPSbion}) and interpret the charges carried by BIons. Working in static gauge, the world space diffeomorphism constraints are trivially solved for {\it static} configurations, i.e. $P_i=0$, and in the absence of magnetic gauge field excitations, i.e. $F_{ab}=0$.
The Hamiltonian constraint can be solved for the energy density \cite{Gauntlett:1997ss}
\begin{equation}\label{electrica}
\frac{{\cal E}^2}{T^2_{\rm Dp}} = E^aE^b\CG_{ab} + \det \CG_{ab}\,.
\end{equation}
Since $\det \CG_{ab} = 1 + (\partial y)^2$, (\ref{electrica}) is equivalent to \cite{Gauntlett:1997ss}
\begin{equation}\label{electrica1}
\frac{{\cal E}^2}{T^2_{\rm Dp}} = (1\pm E^a\partial_a y)^2  + (E\mp\partial y)^2.
\end{equation}
There exists an energy bound
\begin{equation}
\frac{{\cal E}}{T_{\rm Dp}}\ge 1+\big| E^a\partial_a y \big|,
\end{equation}
being saturated if and only is 
\begin{equation}\label{bionBPS}
E_a = \pm \partial_a y.
\end{equation}
This is precisely the relation (\ref{BPSbion}) derived from the solution to the kappa symmetry preserving condition (\ref{spc}).
(the sign is related to the sign of the fundamental string charge). Thus, the total energy integrated over the D-brane world space $\omega$ satisfies
\begin{equation}\label{bionbound}
{\rm E} \ge {\rm E_0} + |Z_{el}|
\end{equation}
where $Z_{el}$ is the charge
\begin{equation}\label{bioncharge}
Z_{el}= \int_{\omega} E^a\partial_a y.
\end{equation}

To interpret this charge as the charge carried by a string, consider the most symmetric solution to (\ref{eq:bioneom}), for D$p$-branes with $p\ge 3$, depending on the radial coordinate in world space $r$, i.e. $r^2=\sigma^a\sigma^b\delta_{ab}$,
\begin{equation}
y(\sigma^a) = \frac{q}{\Omega_{p-1} r^{p-2}},
\end{equation}
where $\Omega_{p}$ stands for the volume of the unit $p$-sphere. This describes a charge $q$ at the origin. Gauss's law allows us to express the energy as an integral over a (hyper)sphere of radius $\delta$ surrounding
the charge. Since $y=y(\delta)$ is constant over this (hyper)sphere, one has
\begin{eqnarray}
 {\rm E} &=&  \lim_{\delta \rightarrow 0}\big| y(\delta)
\int_{r=\delta} \vec {dS}\cdot \vec E \big| \nn \\
&=& q\; \lim_{\delta \to 0} y(\delta) 
\end{eqnarray}
Thus, the energy is infinite since $y\to\infty$ as $\delta\to 0$, but this divergence has its physical origin on the infinite length of a string of {\it finite} and {\it constant} tension $q$ \cite{Callan:1997kz,Gibbons:1997xz}. See \cite{Dasgupta:1997pu} for a discussion on the D-string case, corresponding to string junctions.

\subsection{Dyons}

Dyons are on-shell supersymmetric D3-brane configurations describing a (p,q) string bound state ending on the brane. They are described by the array
\begin{equation}
\begin{array}{ccccccccccl}
D3: &1&2&3&\_&\_&\_&\_&\_&\_& \nn \\
F : &\_&\_&\_&4&\_&\_&\_&\_&\_& \nn \\
D1: &\_&\_&\_&4&\_&\_&\_&\_&\_& \,. 
\end{array}
\end{equation} 
Since the discussion is analogous to the one for BIons, I shall be brief. The ansatz is as in (\ref{bionansatz}) but including some  magnetic components for the gauge field. This is both because a (p,q) string is seen as a dyonic particle on the brane and a D-string is electrically charged under the \RR\ two form. The latter can be induced from the Wess-Zumino coupling
\begin{equation}
\int \CC_2 \wedge \CF \,.
\end{equation}
This shows that magnetic components in $\CF$ couple to electric components in $C_2$. Altogether, the dyonic ansatz is
\begin{eqnarray}
X^\mu &=& \sigma^\mu \quad , \quad X^i=c^i \nonumber \\
X^4(\sigma^a) &=& y(\sigma^a) \quad , \quad
V_0 = V_0(\sigma^a) \quad , \quad V_a = V_a(\sigma^b)\,.
\label{dyon}
\end{eqnarray}

\paragraph{Supersymmetry analysis :} In this case, the matrix elements $\CG_{\mu\nu}+\CF_{\mu\nu}$ are
\begin{eqnarray}
\CG_{\mu\nu}+\CF_{\mu\nu} &=& \left(\matrix{
-1 & F_{0b} \cr
-F_{0a}  & \delta_{ab}  + \6_a y \6_b y+ F_{ab}}\right), \nonumber \\
-{\rm det}\,(\CG_{\mu\nu}+\CF_{\mu\nu}) &=& {\rm det}\,(\delta_{ab}  + \6_a y \6_b y
-F_{0a}F_{0b}+F_{ab}),
\end{eqnarray}
while the induced gamma matrices are exactly those of (\ref{indbion}). Due to the electric and magnetic components of the gauge field, the bosonic kappa matrix has a quadratic term in $F_{\mu\nu}$
\begin{equation}
\Gamma_\kappa=\frac{1}{4!\sqrt{-{\rm det}\,(\CG+\CF)}}
\varepsilon^{\mu_0\ldots\mu_3}\left(
\gamma_{\mu_0\ldots\mu_3}\,i\tau_2 + 6F_{\mu_0\mu_1}\gamma_{\mu_3\mu_4}\,
\tau_1 + 3F_{\mu_0\mu_1}F_{\mu_2\mu_3}\,i\tau_2\right) \,.
\label{kappadyon}
\end{equation} 
To correctly capture the supersymmetries preserved by such a physical system, we impose the projection conditions
\begin{eqnarray}
\Gamma_{0123}\, i\tau_2 
\epsilon &=& \epsilon, 
\label{susydyon1} \\
\Gamma_{0y}(\cos\alpha\,\tau_3 + \sin\alpha\,
\tau_1)\epsilon &=&\epsilon,
\label{susydyon2}
\end{eqnarray}
on the constant Killing spinor $\epsilon$, describing a D3-brane and a (p,q)-string bound state, respectively. Defining $B^a=\frac{1}{2}\epsilon^{abc}F_{bc}$ as the magnetic field and inserting equations (\ref{susydyon1}) and
(\ref{susydyon2}) into the resulting kappa symmetry preserving condition, one obtains
\begin{eqnarray}
\sqrt{-{\rm det}\,(\CG+\CF)}\ep&=& \left(1+\Gamma^{a}\Gamma_{0}
\6_a y (\cos\alpha \tau_3 + \sin\alpha \tau_1) - 
\Gamma^{a}\Gamma_{0}\tau_3 F_{0a} \right. \nn \\
&& \left. + \Gamma^{ab}F_{0a}\6_b y (\cos\alpha \tau_3 + 
\sin\alpha \tau_1) - \Gamma^{a}\Gamma_{0}B_a \tau_1
\right. \nn \\
&& \left. + B^a\6_a y(\cos\alpha \tau_3 + \sin\alpha \tau_1) +
B^aF_{0a}i\tau_2\right].
\label{kappadyonw}
\end{eqnarray} 
This equation is trivially satisfied when the following BPS conditions hold
\begin{equation}
F_{0a} =\cos\alpha\, \6_a y, \qquad \qquad
B^a = \sin\alpha\, \delta^{ab}\6_b y \label{dyonbps}.
\end{equation}

\paragraph{Hamiltonian analysis : } Following \cite{Gauntlett:1997ss},  the hamiltonian constraint can be solved and rewritten as a sum of positively definite terms\footnote{For simplicity I am setting the D3-brane tension to one.}
\begin{eqnarray}
{\cal E}^2 &=& 1 + |\vec\nabla y|^2 + |\vec E|^2 + |\vec B|^2 + (\vec E\cdot
\vec\nabla y)^2 + (\vec B\cdot \vec\nabla y)^2 + |\vec E\times \vec B|^2\nn \\
&=& (1+ \sin \alpha \; \vec E \cdot \vec\nabla y + \cos\alpha\; 
\vec B \cdot \vec\nabla y)^2 + |\vec E - \sin\alpha\; \vec\nabla y|^2 +
|\vec B - \cos\alpha\; \vec\nabla y|^2 \nn \\
&&+\, |\cos\alpha\; \vec E\cdot \vec\nabla y - \sin\alpha\; \vec B \cdot
\vec\nabla y|^2 + |\vec E\times \vec B|^2
\label{endyon2}
\end{eqnarray}
where the last equality holds for any angle $\alpha$. This allows to derive the bound
\begin{equation}
{\cal E}^2 \ge (1+ 
\sin \alpha \; \vec E \cdot \vec\nabla y + \cos\alpha\; 
\vec B \cdot \vec\nabla y)^2.
\label{dyonbound}
\end{equation}
Thus the total energy satisfies
\begin{equation}
{\rm E} \ge {\rm E_0}+\sin\alpha Z_{{\rm el}} + \cos\alpha Z_{{\rm mag}}\,,
\label{dyonbpsbound}
\end{equation}
with
\begin{equation}
Z_{{\rm el}} = \int_{D3} \vec E \cdot \vec\nabla y, \qquad
Z_{{\rm mag}} = \int_{D3} \vec B \cdot \vec\nabla y .
\end{equation}
The bound (\ref{dyonbound}) is extremised when
\begin{equation}
\tan\alpha = Z_{{\rm el}}/Z_{{\rm mag}},
\end{equation}
for which the final energy bound reduces to
\begin{equation}
E \geq E_0 + \sqrt{Z_{{\rm el}}^2+ Z_{{\rm mag}}^2}.
\label{dyonbpsbound2}
\end{equation}
Here $E_0$ corresponds to the energy of the vacuum configuration (infinite D3-brane).
The bound (\ref{dyonbpsbound2}) is saturated when
\begin{equation}\label{dyonBPS}
\vec E = \sin\alpha\; \vec\nabla y, \qquad 
\vec B = \cos\alpha \vec\nabla y. 
\end{equation}
These are precisely the conditions (\ref{dyonbps}) derived from supersymmetry considerations, confirming their BPS nature.
Using the divergence free nature of both $\vec E$ and $\vec B$, $y$ must be harmonic, i.e.
\begin{equation}\label{harm}
\nabla^2 y =0.
\end{equation}
The interpretation of the isolated point singularities in this harmonic function as the endpoints of (p,q) string carrying electric and magnetic charge is analogous to the BIon discussion.

In fact, all previous results can be understood in terms of the $\SL(2,\bZ)$
symmetry of type IIB string theory. In particular, a $(1,0)$ string, or fundamental string, is mapped
into a $(p,q)$ string by an $\SO(2)$ transformation
rotating the electric and magnetic fields. The latter is a non-local transformation in terms
of the gauge field $V$, but leaves the energy density (\ref{endyon2})
invariant
\begin{equation}
\left(\matrix{ E'^a \cr B'^a}\right)=
\left(\matrix{ \cos\alpha & -\sin\alpha \cr
	       \sin\alpha & \cos\alpha}\right)
\left(\matrix{ E^a \cr B^a}\right)\,.
\end{equation}
Applying this transformation to the BIon solution, one reproduces (\ref{dyonBPS}).

\subsection{Branes within branes}

The existence of Wess-Zumino couplings of the form
\begin{equation}
  \int_{{\rm D_{p+4}}} \CC_{p+1}\wedge \CF \wedge\CF\,, \qquad\qquad \int_{{\rm D_{p+2}}} \CC_{p+1}\wedge \CF\,,
\end{equation}
suggests that on-shell non-trivial magnetic flux configurations can source the electric components of the corresponding \RR\ potentials. Thus, one may speculate with the existence of D(p+4)-Dp and D(p+2)-Dp bound states realised as on-shell solutions in the higher dimensional D-brane effective action. In this section, I will review the conditions the magnetic fluxes must satisfy to describe such supersymmetric bound states. 

The analysis below should be viewed as a further application of the techniques described previously, and {\it not} as a proper derivation for the existence of such bound states in string theory. The latter can be a rather subtle quantum mechanical question, which typically involves non-abelian phenomena \cite{Witten:1995im,Douglas:1995bn}. For general discussions on D-brane bound states, see \cite{Sethi:1997pa,Polchinski:1996na,Polchinski:1998rq}, on marginal D0-D0 bound states \cite{Sen:1995vr}, on D0-D4 bound states \cite{Sen:1995hb,Vafa:1995zh} while for D0-D6, see \cite{Taylor:1997ay}. D0-D6 bound states in the presence of B-fields, which can be supersymmetric \cite{Mihailescu:2000dn}, were considered in \cite{Witten:2000mf}. There exist more general analysis for the existence of supersymmetric D-branes with non-trivial gauge fields in backgrounds with non-trivial NS-NS 2-forms in \cite{Marino:1999af}.

\subsubsection{Dp-D(p+4) systems}

These are bound states at threshold corresponding to the brane array
\begin{equation}
\begin{array}{ccccccccccl}
D(p+4): &1&.&p&.&.&.&p+4&\_&\_& \nn \\
Dp: &1&.&p&\_&\_&\_&\_&\_&\_& \,.
\end{array}
\end{equation} 
Motivated by the Wess-Zumino coupling $\CC\wedge\CF\wedge \CF$, one considers the ansatz on the D(p+4)-brane effective action
\begin{eqnarray}
& X^\mu = \sigma^\mu \quad \mu=0,\ldots ,p+4 \quad X^i=c^i \quad 
i=p+5, \ldots , 9 & \nonumber \\
& V_a = V_a(\sigma^b) \quad a,b = p+1,\ldots , p+4 &
\label{dpdp4ansatz}
\end{eqnarray}

Let me first discuss when such configurations preserve supersymmetry. Consider type IIA $(p=2k)$, even though there is an analogous analysis for type IIB. $\Gamma_\kappa$ reduces to
\begin{eqnarray}
\Gamma_\kappa &=& \frac{1}{\sqrt{-{\rm det}\,(\eta_{\mu\nu}+F_{\mu\nu})}}\frac{1}{(2k+5)!}
\varepsilon^{\mu_1\ldots \mu_{2k+5}}\left(\Gamma_{\mu_1\ldots \mu_{2k+5}}
\Gamma_{\sharp}^{k+1}
\right.  \nonumber \\
& & \left. + \left(\ba{cccc}
2k+5 \\ 2 \ea\right)\,F_{\mu_1\mu_2}\Gamma_{\mu_3\ldots \mu_{2k+5}}\Gamma_{\sharp}^k 
\right. \nn \\
&& \left. + \frac{1}{2}\left(\ba{cccc}
2k+5 \\ 4 \ea\right)\left(\ba{c}
4 \\ 2 \ea\right)\, F_{\mu_1\mu_2}F_{\mu_3\mu_4}\Gamma_{\mu_5\ldots \mu_{2k+5}}
\Gamma_{\sharp}^{k+1}\right),
\end{eqnarray}
where I already used the static gauge and the absence of excited transverse scalars, so that $\gamma_\mu = \Gamma_\mu$. For the same reason, ${\rm det}\,(\eta_{\mu\nu}+F_{\mu\nu}) = {\rm det}\,(\delta_{ab}+F_{ab})$, involving a $4\times 4$ determinant.

Given our experience with previous systems, it is convenient to impose the supersymmetry projection conditions on the constant Killing spinors that are appropriate for the system at hand. These are
\begin{eqnarray}
\Gamma_{0\ldots p+4}\Gamma_{\sharp}^{k+1} \epsilon 
& = & \epsilon, \\
\Gamma_{0\ldots p}\Gamma_{\sharp}^{k+1} \epsilon 
& = & \epsilon.
\label{dpdp4susy}
\end{eqnarray}
Notice that commutativity of both projectors is guaranteed due to the dimensionality of both constituents, which is what selects the Dp-D(p+4) nature of the bound state in the first place. Inserting these into the kappa symmetry preserving condition,
the latter reduces to
\begin{equation}
\sqrt{{\rm det}\,(\delta_{ab}+F_{ab})} \epsilon = \left(1 + \frac{1}{4}
\7F^{ab}F_{ab} -\frac{1}{2}\Gamma^{\underline{a}\underline{b}}
\Gamma_{\sharp}F_{ab}\right) \epsilon,
\label{dpdp4kappa2}
\end{equation}
where $\7F^{ab}=\frac{1}{2}\varepsilon^{abcd}F_{cd}$. Requiring the last term in (\ref{dpdp4kappa2}) to vanish is equivalent to the self-duality condition
\begin{equation}
\7F^{ab}=F^{ab}\,.
\label{dpdp4bps}
\end{equation}
When the latter holds, equation (\ref{dpdp4kappa2}) is trivially satisfied. Equation (\ref{dpdp4bps}) is the famous instanton equation in four dimensions\footnote{This equation has a huge history in mathematical physics. For a self-contained presentation on all the mathematical developments regarding this equation, see \cite{MR1079726}. For generalisations to higher dimensions, see \cite{MR1634503,2009arXiv0902.3239D}.}. The hamiltonian analysis done in \cite{Gauntlett:1997ss} again confirms its BPS nature.

\subsubsection{Dp-D(p+2) systems}

These are non-threshold bound states corresponding to the brane array
\begin{equation}
\begin{array}{ccccccccccl}
D(p+2): &1&.&p&p+1&p+2&\_&\_&\_&\_& \nn \\
Dp: &1&.&p&\_&\_&\_&\_&\_&\_& 
\end{array}
\end{equation}
Motivated by the Wess-Zumino coupling $\CC\wedge\CF$, one considers the ansatz on the D(p+4)-brane effective action
\begin{eqnarray}
& X^\mu = \sigma^\mu \quad \mu=0,\ldots , p+2 \quad X^i= c^i \quad 
i=p+3, \ldots ,9 & \nonumber \\
& V_a = V_a(\sigma^b).&
\label{dpdp2ansatz}
\end{eqnarray}
Since there is a single non-trivial magnetic component, I will denote it by $F_{ab}\equiv F$ to ease the notation. The Dirac-Born-Infeld determinant reduces to
\begin{equation}
-{\rm det}\,(\CG_{\mu\nu}+\CF_{\mu\nu}) = 1+F^2\,,
\end{equation}
whereas the kappa symmetry preserving condition in type IIA is
\begin{equation}
\sqrt{1+F^2}\epsilon = \left(\Gamma_{0\ldots p+2}
\Gamma_{\sharp}^k + \Gamma_{0\ldots p}\Gamma_{\sharp}^{k+1} F\right)\epsilon
\label{dpdp2kappa}
\end{equation}
for $p=2k$. This is solved by the supersymmetry projection
\begin{equation}
\left(\cos\alpha\Gamma_{0\ldots p+2}
\Gamma_{\sharp}^k + \sin\alpha\Gamma_{0\ldots p}
\Gamma_{\sharp}^{k+1}\right)\epsilon=\epsilon,
\label{dpdp2susy}
\end{equation}
for any $\alpha$, for the magnetic flux satisfying
\begin{equation}
F=\tan \alpha.
\end{equation}

To interpret the solution physically, assume the world space of the D(p+2)-brane is of the form $\bR^p\times
T^2$. This will quantise the magnetic flux threading the 2-torus according to
\begin{equation}
\int_{T^2}F = 2\pi k \,\, \Longrightarrow \,\, F = \frac{(2\pi)^2 k\alpha'}{L_1 L_2}.
\end{equation}
To derive this expression, I used the 2-torus has area $L_1L_2$ and I rescaled the magnetic field according to $F\to 2\pi\alpha^\prime\,F$, since it is in the latter units that it appears in brane effective actions. Since the energy density satisfies $\CE^2 = T^2_{{\rm D(p+2)}}(1+F^2)$, flux quantisation allows us to write the latter as
\begin{equation}
{\cal E}^2 = T_{D(p+2)}^2 + T_{Dp}^2\left(\frac{k}{ L_1 L_2}\right)^2,
\end{equation}
matching the non-threshold nature of the bound state
\begin{equation}
E= \sqrt{E_{D(p+2)}^2 + E_{kDp}^2}\,,
\end{equation}
where the last term stands for the energy of $k$ Dp-branes.

\subsubsection{F-Dp systems}

These are non-threshold bound states corresponding to the brane array
\begin{equation}
\begin{array}{ccccccccccl}
Dp: &1&.&.&.&p&\_&\_&\_&\_& \nn \\
F: &\_&\_&\_&\_&p&\_&\_&\_&\_& 
\end{array}
\end{equation}
Following previous considerations, one looks for bosonic configurations with the ansatz
\begin{eqnarray}
& X^\mu =\sigma^\mu \quad , \quad X^i=c^i & \nonumber \\
& F_{0\rho}=F_{0\rho}(\sigma^a) &
\label{fdpansatz}
\end{eqnarray}
Given the absence of transverse scalar excitations, $\gamma_\mu = \Gamma_{\mu}$ and
$\sqrt{-{\rm det}\,({\cal G} +{\cal F})}=\sqrt{1-F^2}$, where $F_{0\rho}\equiv F$. The kappa symmetry preserving condition reduces to
\begin{equation}
\sqrt{1-F^2}\epsilon = \left(\Gamma_{0\ldots p}
\Gamma_{\sharp}^k - F\Gamma_{0\rho}\Gamma_{\sharp}
\Gamma_{0\ldots p}\Gamma_{\sharp}^k\right)\epsilon 
+ \left(1-F\Gamma_{0\rho}\Gamma_{\sharp}\right)
\Gamma_{0\ldots p}\Gamma_{\sharp}^k\epsilon\,.
\label{kappafdp1}
\end{equation}
This is solved by the supersymmetry projection condition
\begin{equation}
\left(\cos\alpha\Gamma_{0\ldots p}\Gamma_{\sharp}^k
+\sin\alpha \Gamma_{0\rho}\Gamma_{\sharp}\right)
\epsilon=\epsilon\,,
\label{fdpsusy}
\end{equation} 
whenever
\begin{equation}
F=-\sin\alpha\,.
\label{fdpbps}
\end{equation}

To physically interpret the solution, compute its energy density
\begin{equation}
\CE^2 = \CE_0^2 + F^2\,,
\end{equation}
where I already used that $F_{0\rho}=F=E^\rho$. These configurations
are T-dual to a system of D0-branes moving on a compact space. In this
T-dual picture, it is clear that the momentum along the compact direction
is quantised in units of $1/L$. Thus, the electric flux along the T-dual circle must also be quantised, leading to
the condition
\begin{equation}
F=\frac{1}{2\pi\alpha'}\frac{n}{L}\,,
\end{equation}
where the world volume of the D$p$-brane is assumed to be $\bb{R}^p\times S_1$.
In this way, one can rewrite the energy for the F-D$p$ system as
\begin{equation}
E=\sqrt{E^2_{Dp} + T^2_{f}\left(\frac{n}{L}\right)^2}\,,
\end{equation}
which corresponds to the energy of a non-threshold bound state made of a D$p$-brane
and $n$ fundamental strings ($T_f$).

\subsection{Supertubes}
\label{sec:supertube}

All reviewed solitonic configurations carry charge under the p+1-dimensional gauge potential they minimally couple to. In this section, I want to consider an example where this is not the case. This phenomena may occur when a collection of lower dimensional branes finds it energetically favourable to {\it expand} into higher dimensional ones. The stability of these is due to either an {\it external force}, typically provided by non-trivial fluxes in the background, or presence of angular momentum preventing the brane from collapse. A IIA superstring blown-up to a tubular D2-brane \cite{Emparan:1997rt}, a collection of D0-branes turning into a fuzzy 2-sphere \cite{Myers:1999ps} or wrapping D-branes with quantised non-trivial world volume gauge fields in AdS${}_m\times$S${}^n$ \cite{Pawelczyk:2000hy} are examples of the first kind, whereas giant gravitons \cite{McGreevy:2000cw}, to be reviewed in section \ref{eq:giants}, are examples of the second.

{\it Supertubes} are tubular D2-branes of arbitrary cross-section in a {\it Minkowski vacuum} spacetime supported against collapse by the angular momentum generated by a non-trivial Poynting vector on the D2-brane world volume due to non-trivial electric and magnetic Born-Infeld (BI) fields. They were discovered in \cite{Mateos:2001qs} and its arbitrary cross-section reported in \cite{Mateos:2001pi}, generalising some particular non-circular cross-sections discussed in \cite{Bak:2001xx,Bak:2001tt}. Their stability is definitely {\it not} due to an external force, since these states exist in Minkowski spacetime. Furthermore, supertubes can be supersymmetric, preserving $1/4$ of the vacuum supersymmetry. At first, the presence of non-trivial angular momentum may appear to be in conflict with supersymmetry, since the latter requires a time-independent energy density. This point, and its connection with the expansion of lower dimensional branes, will become clearer once I have reviewed the construction of these configurations.

Let me briefly review the arbitrary cross-section supertube from \cite{Mateos:2001pi}. Consider a D2-brane with world volume coordinates $\sigma^\mu=\{t,z,\sigma\}$ in the type IIA Minkowski vacuum
\begin{equation}
ds_{\rm10}^2 = - dT^2 + dZ^2 + d\vec{Y} \cdot d\vec{Y} \,,
\label{st-metric}
\end{equation}
where $\vec{Y}=\{Y^i\}$ are Cartesian coordinates on $\bR^8$. We are interested in describing a tubular D2-brane of arbitrary cross-section extending along the Z direction. To do so, consider the set of bosonic configurations
\begin{eqnarray}
& &T =t,  \quad Z=z, \quad \vec{Y}=\vec{y}(\sigma), \nonumber \\
& & F = E \, dt \wedge dz + B(\sigma) \, dz \wedge d\sigma.
\label{eq:tubeansatz}
\end{eqnarray}
The static gauge guarantees the tubular nature of the configuration, whereas the arbitrary embedding functions $\vec{Y}=\vec{y}(\sigma)$ describe its cross-section. Notice the Poynting vector will not vanish, due to the choice of electric and magnetic components, i.e. the world volume electromagnetic field will indeed carry angular momentum.

To study the preservation of supersymmetry, one solves (\ref{spc}). Given the ansatz (\ref{eq:tubeansatz}) and the flat background (\ref{st-metric}), this condition reduces to \cite{Mateos:2001pi}
\begin{equation}
y_i' \, \Gamma_{i}  \, \Gamma_\sharp \, 
\left( \Gamma_{TZ} \Gamma_\sharp + E \right) \, \epsilon 
+ \left( B \, \Gamma_{T}\Gamma_\sharp - 
\sqrt{(1-E^2) |\vec{y}\,'|^2 + B^2} \right) \, \epsilon =0,
\label{susy}
\end{equation}
where the prime denotes differentiation with respect to $\sigma$. For generic curves, that is, without imposing extra constraints on the embedding functions $\vec{Y}=\vec{y}(\sigma)$, supersymmetry requires both to set $|E|=1$ and to impose the projection conditions 
\begin{equation}
\Gamma_{TZ} \Gn \, \epsilon = -\mbox{sgn} (E) \, \epsilon \, , \qquad
\Gamma_{T}\Gn \, \epsilon = \mbox{sgn} (B) \, \epsilon 
\label{projections}
\end{equation}
on the constant background Killing spinors $\epsilon$. These conditions have solutions, preserving 1/4 of the vacuum supersymmetry, if $B(\sigma)$ is a constant-sign, but otherwise completely arbitrary, function of
$\sigma$. Notice the two projections \ref{projections} correspond to string charge along the $Z$-direction and to D0-brane charge, respectively.

In order to improve our understanding on the arbitrariness of the cross-section, it is instructive to compute the charges carried by supertubes and its energy momentum tensor, to confirm the absence of any pull (tension) along the different spacelike directions where the tube is embedded in ten dimensions. First, the conjugate momentum $P_i$ and the conjugate variable to the electric field, $\Pi$, are
\begin{eqnarray}
  P_i & =& \frac{\partial{\cal L}_{D2}}{\partial \dot{Y}^i} = \frac{B E y_i'}{\sqrt{(1-E^2)|\vec{y}\,'|^2 + B^2}} =   \rm{sgn}(\Pi B) \, y_i' \,, \label{pxpy-susy} \\
  \Pi (\sigma) &=& \frac{\partial{\cal L}_{D2}}{\partial E} = \frac{E |\vec{y}\,'|^2}{\sqrt{(1-E^2) |\vec{y}\,'|^2 + B^2}}= \rm{sgn}(E)\frac{|\vec{y}\,'|^2}{|B|}\,,
\end{eqnarray}
where in the last step the supersymmetry condition $|E|=1$ was imposed. Notice supertubes satisfy the identity
\begin{equation}
|\vec{P}|^2 = |\Pi B| \,.
\label{pib-susy}
\end{equation}
Second, the fundamental string $q_{{\rm F1}}$ and D0-brane $q_{{ D0}}$ charges are 
\begin{equation} 
q_{F1} =  \int d\sigma \, \Pi \,, \quad
q_{D0} =  \int d\sigma \, B \,.
\label{charges}
\end{equation}
Finally, the supertube energy-momentum tensor \cite{Mateos:2001pi}
\begin{equation}
T^{mn}(x) = \left. \frac{2}{\sqrt{-{\rm det}\,g}} \, \frac{\delta S_{{\rm D2}}}{\delta g_{mn}(x)} 
\right|_{g_{mn}=\eta_{mn}} = - \sqrt{-{\rm det}\,(\CG+F)} \, \left[ (\CG+F)^{-1} \right]^{(\mu\nu)} \partial_\mu X^m \partial_\nu X^n,
\label{st-tensor}
\end{equation}
with $X^m= \{T, Z, Y^i\}$, has only non-zero components
\begin{equation}
\mbox{${\cal T}$}^{TT} = |\Pi| + |B| \,, \quad \mbox{${\cal T}$}^{ZZ} = - |\Pi| \,, \quad 
\mbox{${\cal T}$}^{Ti} = \mbox{sgn}(\Pi B) \, y_i'.
\end{equation}
Some comments are in order :
\begin{itemize}
\item[1.] As expected, the linear momentum density \ref{pxpy-susy} carried by the tube is responsible for the off-diagonal components $\mbox{${\cal T}$}^{Ti}$. 
\item[2.] The absence of non-trivial components $\mbox{${\cal T}$}^{ij}$ confirms the absence of tension along the cross-section, providing a more technical explanation of why an arbitrary shape is stable. 
\item[3.] The tube tension $-\mbox{${\cal T}$}^{ZZ}=|\Pi|$ 
in the $Z$-direction is {\it only} due to the string density, since D0-branes behave like dust.
\item[4.] The expanded D2-brane does not contribute to the tension in any direction.
\end{itemize} 
Integrating the energy momentum tensor along the cross-section, one obtains the net energy 
of the supertube per unit length in the $Z$-direction
\begin{equation}
\mbox{${\cal E}$} = \int d\sigma \, \mbox{${\cal T}$}^{TT} = |q_{F1}| + |q_{D0}|\,,
\label{energy}
\end{equation}
matching the expected energy bound from supersymmetry considerations. 

Let me make sure the notion of supersymmetry is properly tied with the expansion mechanism. Supertubes involve a uniform electric field along the tube and some magnetic flux. Using the language and intuition from previous subsections, the former can be interpreted as "dissolved" IIA superstrings and the latter as "dissolved" D0-branes, that have expanded into a tubular D2-brane. Their charges are the ones appearing in the supersymmetry algebra allowing the energy to be minimised. Notice the expanded D2-brane couples locally to the \RR\ gauge potential $C_3$ under which the string and D0-brane constituents are neutral. This is precisely the point made at the beginning of the section : supertubes do {\it not} carry D2-brane charge\footnote{Strictly speaking if the supertube cross-section is open, they can carry D2-brane charge. The arguments given above do only apply for {\it closed} cross-sections. The reader is encouraged to read the precise original discussion in \cite{Mateos:2001pi} concerning this point and the bounds on angular momentum derived from it.}. When the number of constituents is large, one may expect an effective description in terms of the higher-dimensional D2-brane in which the original physical charges become fluxes of various types. 

The energy bound (\ref{energy}) suggests supertubes are marginal bound states of D0s and fundamental strings (Fs). This was further confirmed by studying the spectrum of BPS excitations around the circular shape supertube by quantising the linearised perturbations of the DBI action \cite{Palmer:2004gu,Bak:2004kz}. The quantization of the space of configurations with fixed angular momentum $J$ \cite{Palmer:2004gu,Bak:2004kz} allowed to compute the entropy associated with states carrying these charges
\begin{equation}
  S = 2\pi\sqrt{2(q_{D0}q_{F1}-J)}\,.
\end{equation}
This entropy reproduces the microscopic conjecture made in \cite{Lunin:2002qf} where the Bekenstein-Hawking entropy was computed using an stretched horizon. These considerations do support the idea that supertubes are typical D0-F bound states.

\paragraph{Supergravity description and fuzzball considerations : }  The fact that world volume quantization reproduces the entropy of a macroscopic configuration and the presence of arbitrary profiles, at the classical level, suggest that supersymmetric supertubes may provide a window to understand the origin of gravitational entropy in a regime of parameters where gravity is reliable. This is precisely one of the goals of the fuzzball programme \cite{Lunin:2001jy,Lunin:2002iz}\footnote{For a list of reviews on this subject, see \cite{Mathur:2005zp,Mathur:2008nj,Bena:2007kg,Skenderis:2008qn,Balasubramanian:2008da,Simon:2011zz}.}.

A first step towards this connection was provided by the supergravity realisation of supertubes given in \cite{Emparan:2001ux}. These are smooth configurations described in terms of harmonic functions whose sources allow arbitrary profiles, thus matching the arbitrary cross-section feature in the world volume description \cite{Mateos:2001pi}. 

The notion of supertube is more general than the one described above. Different encarnations of the same stabilising mechanism provide U-dual descriptions of the famous string theory D1-D5 system. To make this connection more apparent, consider supertubes with arbitrary cross-sections in $\bR^4$ and with an S${}^1$ tubular direction, allowing the remaining 
4-spacelike directions to be a 4-torus. These supertubes are U-dual to D1-D5 bound states with angular momentum $J$ \cite{Lunin:2002iz}, or to winding ondulating strings \cite{Lunin:2001fv} obtained from the original work \cite{Callan:1995hn,Dabholkar:1995nc}. It was pointed out in \cite{Lunin:2002iz} that in the D1-D5 frame, the actual supertubes correspond to KK monopoles wrapping the 4-torus, the circle also shared by D1 and D5-branes and the arbitrary profile in $\bR^4$\footnote{By arbitrary, it is meant a general curve that is not self-intersecting and whose tangent vector never vanishes.}. Smoothness of these solutions is then due to the KK monopole smoothness.

Since the U-dual D1-D5 description involves an AdS${}_3\times$S${}^3$ near horizon, supertubes were interpreted in the dual CFT : the maximal angular momentum configuration corresponding to the circular profile is global AdS${}_3$, whereas non-circular profile configurations are chiral excitations above this vacuum \cite{Lunin:2002iz}.

Interestingly, geometric quantization of the classical moduli space of these D1-D5 smooth configurations was carried in \cite{Rychkov:2005ji}, using the covariant methods originally developed in \cite{Crnkovic:1986ex,Zuckerman:1989cx}.  The Hilbert space so obtained produced a degeneracy of states that was compatible with the entropy of the extremal black hole in the limit of large charges, i.e. $S=2\pi\sqrt{2(q_{D0}q_{F1})}$. Further work on the quantization of supergravity configurations in AdS${}_3\times$S${}^3$ and its relation to chiral bosons can be found in \cite{Donos:2005vs}. The conceptual framework described above corresponds to a particular case of the one illustrated in figure \ref{fig8}.

\subsection{Baryon vertex}

As a first example of a supersymmetric soliton in a non-trivial background, I will review the baryon vertex \cite{Witten:1998xy,Gross:1998gk}. Technically, this will provide an example on how to deal with non-constant Killing spinors. Conceptually, it is a nice use of the tools explained in this review having an interesting AdS/CFT interpretation.

Let me first try to conceptually motivate the entire set-up. Consider a {\it closed} D5-brane surrounding $N$ D3-branes, i.e.  such that the D3-branes thread the D5-brane. The Hanany-Witten (HW) effect \cite{Hanany:1996ie} allows us to argue that each of these $N$ D3-branes will be connected to the D5-brane by a fundamental type IIB string. Consequently, the lowest energy configuration should {\it not} allow the D5-brane to contract to a single point, but should describe these $N$ D3-branes with $N$ strings attached to them allowing to connect the D3 and D5-branes. In the large $N$ limit, one can replace the D3-branes by their supergravity backreaction description. The latter has an AdS${}_5\times$S${}^5$ near horizon. One can think of the D5-brane as wrapping the 5-sphere and the $N$ strings emanating from it can be pictured as having their endpoints on the AdS${}_5$ boundary. This is the original configuration interpreted in \cite{Witten:1998xy,Gross:1998gk} as a baryon-vertex of the ${\cal N}=4$ $d=4$ superYang-Mills (SYM) theory.

At a technical level and based on our previous discussions regarding BIons, one can describe the baryon vertex as a single D5-brane carrying $N$ units of world volume electric charge \cite{Imamura:1998gk,Callan:1998iq} to account for the $N$ type IIB strings. If one assumes all the electric charge is concentrated at one point, then one expects the minimum energy configuration to preserve the $\SO(5)$ rotational invariance around it. Such configuration will be characterised by the radial position of the D5-brane in AdS${}_5$ as a function $r(\theta)$ of the co-latitude angle $\theta$ on S${}^5$. This is the configuration studied in \cite{Imamura:1998gk,Callan:1998iq,Craps:1999nc}. Since it is, a priori, not obvious whether the requirement of minimal energy forces the configuration to be $\SO(5)$ invariant, one can relax this condition and look for configurations where the charge is distributed through different points. One can study whether these configurations preserve supersymmetry and saturate some energy bound. This is the approach followed in \cite{Gomis:1999xs}, where the term {\it baryonic branes} was coined for all this kind of configurations, and the one I will follow below.

\paragraph{Set-up :}  One is interested in solving the equations of motion of a {\it single} D5-brane in the background of $N$ D3-branes carrying some units of electric charge to describe type IIB strings. The background is described by a constant dilaton, a non-trivial metric and self-dual 5-form field strength $R_{5}$ \cite{Duff:1991pea}
\begin{eqnarray}
\label{eq:baryonbackground}
ds^2_{{\rm 10}} & = & U^{-1/2}\,ds^2(\bb{E}^{(1,3)}) + U^{1/2}\left[ dr^2 + r^2
d\Omega_5^2\right]\\
R_{5} & = & 4R^4\left[ \omega_{5} + \star \omega_5\right]
\end{eqnarray}
where $d\Omega_5^2$ is the $\SO(6)$-invariant metric on the unit
5-sphere, $\omega_5$ is its volume 5-form and $\star \omega_5$ its Hodge dual.
The function $U$ is
\begin{equation}
U= a + \left(\frac{L_4}{r}\right)^4 \qquad \left(L_4^4= 4\pi g_s
N(\alpha')^2\right)\,.
\label{harmonic}
\end{equation}
Notice $a=1$ corresponds to the full D3-brane background solution, whereas $a=0$ to its near-horizon limit.

Consider a probe D5-brane of unit tension wrapping the 5-sphere. Let $\xi^\mu=(t,\theta^i)$ be the world volume
coordinates, so that $\theta^i$ ($i=1,\dots,5$) are coordinates for the worldspace 5-sphere. This will be achieved by the static gauge
\begin{equation}
X^0 = t\, , \quad \T^i=\theta^i \, .
\end{equation}
Since one is only interested in {\it radial} deformations of the world space carrying electric charge, one considers the ansatz
\begin{equation}
X^1 = X^2 = X^3 = 0 \,, \qquad \qquad r=r(\theta^i)\,, \qquad \quad F = \frac{1}{2}F_{0i}(\theta^i)\,dt\wedge  d\theta^i\,.
\label{eq:baryonansatz}
\end{equation}
Even though the geometry will be curved, it can give some intuition to think of this system in terms of the array
\begin{equation}
\ba{ccccccccccl}
D3: &1&2&3&\_&\_&\_&\_&\_&\_ &\quad \mbox{background}    \nonumber \\
D5: &\_&\_&\_&4&5&6&7&8&\_ &\quad \mbox{probe}     \nonumber \\
F1: &\_&\_&\_&\_&\_&\_&\_&\_&9 &\quad \mbox{soliton}
\ea
\label{triple}
\end{equation}
viewing the 9-direction as the radial one.

\paragraph{Supersymmetry analysis :} Given the electric nature of the world volume gauge field, the kappa symmetry matrix reduces to
\begin{equation}
\Gamma_\kappa=\frac{1}{6!}\frac{1}{\sqrt{-{\rm det}(\CG +F)}}
\varepsilon^{\mu_1 ...\mu_6}~\left[\gamma_{\mu_1 ...\mu_6}\tau_1 +
15 ~F_{\mu_1\mu_2}\gamma_{\mu_3 ...\mu_6}(i\tau_2)\right]\,.
\label{eq:kappabaryon}
\end{equation}
Given the ansatz (\ref{eq:baryonansatz}) and the background (\ref{eq:baryonbackground}), the induced world volume metric equals
\begin{equation}
\CG_{\mu\nu}=\pmatrix{-\,U^{-1/2} & 0\cr 
0& g_{ij}}
\end{equation}
where
\begin{equation}
g_{ij}=U^{1/2}\left(r^2\bar g_{ij} + \partial_i r\partial_j r\right)\,,
\end{equation}
and $\bar g_{ij}$ stands for the $\SO(6)$-invariant metric on the unit 5-sphere. Taking into account the non-trivial vielbeins, the induced gamma matrices equal
\begin{equation}
\gamma_0 =U^{-\,1/4}\Gamma_0 \,, \qquad \qquad
\gamma_i =U^{1/4}\,r\hat\gamma_i + U^{1/4}\,\partial_i
r\Gamma_r\,.
\end{equation}
where the matrices $\hat\gamma_i$ are defined as
\begin{equation}
\hat\gamma_i = e_i{}^{a}\Gamma_{a}\,,
\end{equation}
in terms of the f\"unfbein $e_i{}^{a}$ in the 5-sphere. Thus, $\{\hat\gamma_i,\hat\gamma_j\}= 2\bar g_{ij}$.

To solve the kappa symmetry preserving condition (\ref{spc}), one requires the background Killing spinors. These are of the form
\begin{equation}
\epsilon = U^{-\frac{1}{8}}\chi
\end{equation}
where $\chi$ is a covariantly constant spinor on $\bb{E}^{(1,3)}\times \bb{E}^6$
subject to the projection condition
\begin{equation}
\Gamma_{0123}\,i\,\tau_2\chi =\chi \,,
\label{eq:d3proj}
\end{equation}
describing the D3-branes in the background. Importantly, $\chi$ is {\it not} constant when using polar coordinates in $\bE^6$. Indeed, covariantly constant spinors on $S^n$ were constructed explicitly in \cite{Lu:1998nu}
for a sphere parameterisation obtained by iteration of $ds^2_n = d\theta_n^2 + \sin^2\theta_nds^2_{n-1}$. The result can be written in terms of the $n$ angles $\theta^i=(\theta,\theta^{\8{\imath}})$ and the
antisymmetrised products of pairs of the constant d=10  Clifford matrices
$\Gamma_{a} =(\Gamma_{\theta},\Gamma_{\8{\imath}})$. For $n=5$, defining $\Gamma_{\hat 5}\equiv
\Gamma_{\theta}$, these equal
\begin{equation}
\chi\,= e^{\frac{\theta}{2}\,\Gamma_{r\theta}}\,\,
\prod_{\8{\jmath}=1}^4\,\,
e^{-\frac{\theta^{\8{\jmath}}}{2}\,\Gamma_{\8{\jmath}{\8{\jmath}+1}}}\,\,
\epsilon_0
\label{baryonkilling}
\end{equation}
where $\epsilon_0$ satisfies (\ref{eq:d3proj}). Even though there are additional Killing spinors in the near-horizon limit,
the associated extra supersymmetries will be broken by the baryonic
D5-brane probe configuration I am about to construct, so these can be ignored.

Plugging the ansatz in the kappa matrix (\ref{eq:kappabaryon}), the supersymmetry preserving condition (\ref{spc}) reduces, after some algebra, to
\begin{eqnarray}
& U\,\sqrt{\det\left[r^2\bar g_{ij}  + \partial_ir\partial_j r
- F_{0i}F_{0j}\right]}\, \epsilon = \nonumber \\
& \big[ Ur^5  \sqrt{{\rm det}\,\5g}\, \Gamma_{0}\gamma_*\tau_1 -
Ur^3\sqrt{{\rm det}\,\5g}\, F_{0j}\6_i r
\hat\gamma^{ij}\gamma_*\Gamma_{r}(i\tau_2)  \nonumber\\ &
\qquad
+\, Ur^4\sqrt{{\rm det} \bar g}\, \hat\gamma^i \gamma_*
\left(F_{0i} (i\sigma_2) + \6_i
r \Gamma_{0r}\tau_1\right)\big]\, \epsilon
\label{kapmatrix}
\end{eqnarray}
where $\hat\gamma^i= \5g^{ij}\hat\gamma_j$ and $\gamma_*=\Gamma_{45678}$.

Given the physical interpretation of the sought solitons, one imposes two supersymmetry projections on the constant Killing spinors $\epsilon_0$ :
\begin{eqnarray}
\Gamma_{0}\gamma_*\tau_1\epsilon_0 &=&\epsilon_0\, ,\label{susyprobe} \\
\Gamma_{0r}\, \tau_3\epsilon_0 &=& \epsilon_0 \,.\label{bion}
\end{eqnarray}
These are expected from the {\sl local} preservation of 1/2 supersymmetry by the
D5-brane and the IIB string in the radial direction, respectively. These projections imply
\begin{eqnarray}
\Gamma_{0r}\,\tau_3\,\epsilon\,&=&
\left[\,{\rm cos}\,\theta\,-\,
{\rm sin}\,\theta\,\Gamma_{r\theta}\,\right]\,\epsilon \nonumber\\
\Gamma_{0}\, \gamma_*\,\tau_1\,\epsilon\,&=&
\left[\,{\rm cos}\,\theta\,-\,
{\rm sin}\,\theta\,\Gamma_{r\theta}\,\right]\,\epsilon \nonumber \\
\Gamma_{i}\,\,\gamma_*\,i\tau_2\,\epsilon &=&
\,-\,\Gamma_{ri}\,\epsilon \nonumber\\
\Gamma_{i}\,\,\gamma_*\,
\Gamma_{0r}\,\tau_1\,\epsilon &=&
\Gamma_{ri}\,
e^{-\theta\Gamma_{r\theta}}\,\,\epsilon\nonumber\\
\gamma_*\,\,\Gamma_{r}\,\,i\tau_2\,\epsilon &=&
\,-\,\epsilon\,.
\label{baryoncond2}
\end{eqnarray}
Using these relations, one can rewrite the right hand side of (\ref{kapmatrix}) as
\begin{eqnarray}
& \Delta_5\,\bigg[ \left(r\sin\theta\right)'
 + \Gamma_{r\theta}\left((r\cos\theta)'-F_{0\theta}\right)
+ \Gamma_{r}\hat\gamma^{\8{\imath}}\left(\6_{\8{\imath}}r \cos\theta -
F_{0\8{\imath}}\right) \nonumber\\
& \, + \hat\gamma^{\8{\imath}\8{\jmath}}\,\frac{1}{r}
\left(\6_{\8{\imath}}r F_{0\8{\jmath}} -\6_{\8{\jmath}}r F_{0\8{\imath}}\right)
+ \hat\gamma^{\8{\imath}}\Gamma_{\theta}\,\frac{1}{r}\left(
\6_{\8{\imath}}rF_{0\theta}-r' F_{0\8{\imath}} + r\6_{\8{\imath}}r \sin\theta\right)\bigg]\, ,
\label{baryonpart}
\end{eqnarray}
where $\Delta_5=U\,r^4\,\sqrt{{\rm det}\,\bar g}$. The coefficients of $\Gamma_{r\theta}$ and
$\Gamma_{r}\hat\gamma^{\8{\imath}}$ in (\ref{baryonpart}) vanish when
\begin{equation}
F_{0i}\,=\partial_{i}\left(r\,\cos\theta\right).
\label{bpsbound}
\end{equation}
Furthermore, the ones of $\hat\gamma^{\8{\imath}\8{\jmath}}$ and $\hat\gamma^{\8{\imath}}
\Gamma_{\theta}$ also do. I will eventually interpret (\ref{bpsbound}) as the BPS equation for a world volume BIon. 
One concludes that (\ref{kapmatrix}) is satisfied as a consequence of (\ref{bpsbound}) provided that
\begin{equation}
U\,\sqrt{\det\left[r^2\bar g_{ij}  + \partial_ir\partial_j r -
F_{0i}F_{0j}\right]} = \Delta_5  (r\sin\theta)' .
\label{fin}
\end{equation}
It can be checked that this is indeed the case whenever (\ref{bpsbound}) holds. 

\paragraph{Hamiltonian analysis :} Solving the hamiltonian constraint ${\cal H}=0$ in \ref{hamu} allows to write the hamiltonian density for static configurations as \cite{Gomis:1999xs}
\begin{equation}\label{hsq}
{\cal H}^2 = U^{-\frac{1}{2}}\left[ \tilde E^i\tilde E^j g_{ij}
+ \det g \right]
\end{equation}
where $\tilde E^i$ is a covariantised electric field density related to $F_{0i}$ by
\begin{equation}\label{relef}
({\rm det g}) \, F_{0i} = \sqrt{-{\rm det}\,(\CG+F)}\,  \tilde E^j g_{ij}.
\end{equation}
For the ansatz (\ref{eq:baryonansatz}), this reduces to
\begin{equation}\label{introlast1}
\tilde E^i\,=\,U^{{1/4}}\,\, \frac{\sqrt{{\rm det} g}}{\sqrt{1\,-\,U^{{1/2}}\,g^{mn}\,F_{0m}\,F_{0n}}}\,\,
g^{ij}\,F_{0j}.
\end{equation}
It was shown in \cite{Gomis:1999xs} that one can rewrite the energy density (\ref{hsq}) as
\begin{equation}\label{ED2}
{\cal H}^2 = {\cal Z}_5^2 + \left[\Delta_5\left(r\cos\theta \right)'
-\7E^i\6_i \left(r\sin\theta\right)\right]^2 +
|\Delta_5 \5g^{\8{\imath}\8{\jmath}}\6_{\8{\jmath}}r -r\,\7E^{\8{\imath}} |^2
\end{equation}
where $||^2$ indicates contraction with $g_{\8{\imath}\8{\jmath}}$, and
\begin{equation}\label{topo}
{\cal Z}_5= \Delta_5 \left(r\sin\theta\right)' + \7E^i\6_i \left(r\cos\theta
\right).
\end{equation}
To achieve this, the 5-sphere metric was written as
\begin{equation}
ds^2 = d\theta^2 + \sin^2\theta\, d\Omega_4^2
\label{smetric}
\end{equation}
where $d\Omega_4^2$ is the $\SO(5)$ invariant metric on the 4-sphere,
which one takes to have coordinates $\theta^{\8{\imath}}$. In this way, all primes above refer to derivatives with respect to $\theta$ and $\bar g^{\8{\imath}\8{\jmath}}$ are the $\8{\imath}\8{\jmath}$ components of
the inverse S${}^5$ metric $\5g^{ij}$.

Using the Gauss' law constraint
\begin{equation}
\partial_i\tilde E^i = -4\,R^4 \sqrt{{\rm det} \bar g}\,,
\label{gausslaw}
\end{equation}
which has a non-trivial source term due to the \RR\ 5-form flux background, one can show that ${\cal Z}_5=\6_i {\cal Z}_5^i$ where $\vec{{\cal Z}_5}$ has components
\begin{eqnarray}
{\cal Z}_5^\theta & = & \7E^\theta\,r\cos\theta +\sqrt{\det\5g}\,\sin\theta
\left(a\,{r^5\over 5} + r\,R^4\right) \nonumber \\
{\cal Z}_5^{\8{\imath}} & = & \,\7E^{\8{\imath}}\,r\cos\theta \, .
\label{div1}
\end{eqnarray}
{}From (\ref{ED2}), and the divergence nature of ${\cal Z}_5$, one deduces the bound
\begin{equation}
{\cal H} \geq  |{\cal Z}_5|.
\label{Hbound}
\end{equation}
The latter is saturated when
\begin{eqnarray}
\7E^{\8{\imath}} & = & \Delta_5 {\5g^{\8{\imath}\8{\jmath}}\6_{\8{\jmath}}r \over r}
\label{BPSe1} \\
\7E^\theta & = & {\Delta_5 \over \left(r\sin\theta\right)'}
\left(\left(r\cos\theta\right)' - {\5g^{\8{\imath}\8{\jmath}}\6_{\8{\imath}}r\6_{\8{\jmath}}
\left(r\sin\theta\right)\over r}\right) \, .
\label{BPSe2}
\end{eqnarray}
Combining (\ref{BPSe1}) and (\ref{BPSe2}) with the Gauss law (\ref{gausslaw})
yields the equation
\begin{equation}\label{pde}
\partial_{\hat{\imath}}\left(\Delta_5\bar g^{\hat i\hat j}
{\partial_{\hat{\jmath}}r\over r}\right) + \partial_\theta \left[{\Delta_5 \over
(r\sin\theta)'}\left((r\cos\theta)' - \bar g^{\hat i\hat j}\partial_{\hat{\imath}}
r\partial_{\hat{\jmath}}r{\sin\theta\over r}\right)\right] = -4R^4\sqrt{{\rm det}\,\bar g}.
\end{equation}
{\it Any} solution to this equation gives rise to a 1/4 supersymmetric baryonic brane.

For a discussion of the first-order equations (\ref{BPSe1}) and ({\ref{BPSe2}) for a=1, see \cite{Callan:1999zf,Camino:1999xx}. Here, I will focus on the near horizon geometry corresponding to a=0. The hamiltonian density bound (\ref{Hbound}) allows us to establish an analogous one for the total energy $E$
\begin{equation}\label{inequality}
E \ge \int d^5 \sigma |{\cal Z}_5\big| \ge
\bigg|\int d^5 \sigma {\cal Z}_5\,\bigg|.
\end{equation}
While the first inequality is saturated under the
same conditions as above, the second requires ${\cal Z}_5$ to not changing sign within the integration region.
For this configuration to describe a baryonic brane, one must identify this
region with a 5-sphere having some number of singular points removed.
Assuming the second inequality {\sl is} saturated when the first one is, the total energy equals
\begin{equation}\label{kbound}
E = \lim_{\delta\rightarrow 0} \sum_k \int_{B_k} d\vec{S}\cdot \vec{{\cal Z}}
\end{equation}
where $B_k$ is a 4-ball of radius $\delta$ having the k'th singular point as its
centre. This expression suggests to interpret the k'th term in the sum as the energy of the IIB
string(s) attached to the k'th singular point. No explicit solutions to (\ref{pde}) with these boundary conditions are known though.

Consider $\SO(5)$ invariant configurations (for a discussion on less symmetric configurations see \cite{Gomis:1999xs}). In this case $\tilde E^{\8{\imath}}=0$,
\begin{equation}
\tilde E^\theta = \sqrt{\det g^{(4)}}\,\tilde E(\theta),
\end{equation}
and $r=r(\theta)$. The BPS condition (\ref{BPSe2}) reduces to \cite{Imamura:1998gk,Callan:1998iq,Craps:1999nc}
\begin{equation}
\frac{r'}{r} = \frac{\Delta\sin\theta + \tilde E\,\cos\theta}{\Delta\cos\theta - \tilde E\,\sin\theta}\,,
\label{oldbound}
\end{equation}
where $\Delta=\,R^4\,\sin^4\theta$, while the Gauss' law (\ref{gausslaw}) equals
\begin{equation}\label{lawgauss}
\tilde E' = -4\,R^4\,\sin^4\theta.
\label{q1}
\end{equation}
Its solution was first found in \cite{Callan:1998iq}
\begin{equation}\label{expressE}
\tilde E = \frac{1}{2}R^4\left[\,\,3\left(\nu\pi-\theta\right)+3 \sin\theta
\cos\theta+ 2\sin^3\theta \cos \theta\,\,\right]
\end{equation}
where $\nu$ is an integration constant restricted to lie in the interval $[0,1]$.

Given this explicit solution, let me analyse whether the second inequality in (\ref{inequality})
is saturated when the first one is, as I assumed before. Notice
\begin{equation}
{\cal Z}_5 = \sqrt{\det g^{(4)}}\, {\cal Z}(\theta)\quad {\rm with} \quad
{\cal Z}(\theta)= r\, \frac{\left(\Delta\cos\theta -\tilde E\sin\theta\right)^2 +
\left(\Delta\sin\theta + \tilde E\cos\theta\right)^2}{
\left(\Delta\cos\theta - \tilde E\sin\theta\right)},
\end{equation}
where I used (\ref{oldbound}). The sign of ${\cal Z}$ is determined by the sign of the denominator. Thus, it will not change
if it has no singularities within the region $\theta\in [0,\pi]$ (except, possibly, at the endpoints
$\theta=0,\pi$). Since
\begin{equation}\label{denom}
\Delta\cos\theta - \tilde E\sin\theta = \,\frac{3}{2} \,R^4\sin\theta\,\eta(\theta) \quad {\rm with} \quad \eta(\theta)\equiv \theta-\nu\pi -\sin\theta \cos\theta,
\end{equation}
one concludes that the denominator for ${\cal Z}$ vanishes at the endpoints $\theta=0,\pi$ but is otherwise positive provided $\eta(\theta)$ is. This condition is only satisfied for $\nu=0$, in which case
 (\ref{expressE}) becomes
\begin{equation}\label{expressEE}
\tilde{E} = \frac{1}{2}R^4\left[3 \left(\sin\theta\cos\theta -\theta\right) +
2\sin^3\theta \cos \theta\,\,\right].
\end{equation}
Integrating the differential equation (\ref{oldbound}) for $r(\theta)$ after substituting (\ref{expressEE}), one finds \cite{Callan:1998iq}
\begin{equation}
r= r_0\left({6\over 5}\right)^{1\over3} ({\rm cosec}\,\theta)(\theta
-\sin\theta\cos\theta)^{1\over3}\, ,
\end{equation}
where $r_0$ is the value of $r$ at $\theta=0$. It was shown in
\cite{Callan:1998iq} that this configuration corresponds to N fundamental strings
attached to the D5-brane at the point $\theta=\pi$, where $r(\theta)$ diverges. 

Solutions to (\ref{oldbound})  for $\nu\neq0$ were also obtained in  \cite{Callan:1998iq}. The range of the angular variable $\theta$ for which these solutions make physical sense is smaller than $[0,\pi]$ because the D5-brane does not completely wrap the 5-sphere. Consequently, the D5 probe captures only part of the five form flux. This suggests to interpret these spike configurations as corresponding to a number of strings less than $N$. In fact, it was argued in \cite{Brandhuber:1998xy,Imamura:1998hf} that baryonic multiquark states with $k<N$ quarks in ${\cal N}$=4 d=4 SYM correspond to $k$ strings connecting the D5-brane to $r=\infty$ while the remaining $N-k$ strings connect it to $r=0$. Since the $\nu\neq 0$ D5-brane solutions do reach $r=0$, it is tempting to speculate whether they correspond to these baryonic multi-quark states.

\paragraph{Related work :} There exists similar work in the literature. Besides the study of non-$\SO(5)$ invariant baryonic branes in AdS${}_5\times$S${}^5$, \cite{Gomis:1999xs} also carried the analysis for baryonic branes in M-theory.
Similar BPS bounds were found for D4-branes in D4-brane backgrounds or more generically, for D-branes in a D-brane background \cite{Callan:1999zf,Camino:1999xx} and D3-branes in (p,q)5-branes \cite{Simon:1999wy,Llatas:1999zr}. Baryon vertex configurations have also been studied in AdS${}_5\times$T${}^{1,1}$ \cite{Arean:2004mm}, AdS${}_5\times$Y${}^{p,q}$ \cite{Canoura:2005uz} and were extended to include the presence of magnetic flux \cite{Janssen:2006sc}. For a more general analysis of supersymmetric D-brane probes either in AdS or its pp-wave limit, see \cite{Skenderis:2002vf}.

\subsection{Giant gravitons \& superstars}
\label{eq:giants}

It was mentioned in subsection \ref{sec:supertube} that angular momentum can stabilise an expanded brane carrying the same quantum numbers as a lower dimensional brane. I will now review an example of such phenomena, involving supersymmetric expanding branes in AdS, the so called {\it giant gravitons} \cite{McGreevy:2000cw}. In this case, a rotating pointlike graviton in AdS expands into a rotating brane due to the \RR\ flux supporting the AdS supergravity solution \cite{Myers:1999ps}. Its angular momentum prevents the collapse of the expanding brane and it can actually make it supersymmetric \cite{Grisaru:2000zn,Hashimoto:2000zp}. 

Consider type IIB string theory in AdS${}_5\times$S${}^5$. It is well known that this theory has BPS graviton excitations rotating on the sphere at the speed of light. In the dual ${\cal N}=4$ d=4 SYM theory, these states correspond to single trace operators belonging to the chiral ring \cite{Andrianopoli:1998ut,Corley:2001zk,Berenstein:2004kk}. When their momentum becomes of order $N$, it is energetically favourable for these gravitons to expand into rotating spherical D3-branes, i.e. {\it giant gravitons}. The $N$ scaling is easy to argue for :  the conformal dimension must be proportional to the D3-brane tension times the volume of the wrapped cycle, which is controlled by the AdS radius of curvature $L_4$, thus giving
\begin{equation}
  \Delta  \propto T_{{\rm D3}} L_4^4 = N\,.
\end{equation}
Similar considerations apply in different AdS${}_{p+1}$ realisations of this phenomena \cite{Grisaru:2000zn,Mandal:2007ug}.
The field theory interpretation of these states was given in \cite{Balasubramanian:2001nh} in terms of subdeterminant operators.

Let us construct these configurations in AdS${}_5\times$S${}^5$. The bosonic background has a constant dilaton and non-trivial metric and \RR\ 4-form potential given by
\begin{eqnarray}
ds_{{\rm 10}}^2 &=& - \left(1+\frac{r^2}{L_4^2}\right)\,dt^2 + \frac{dr^2}{1+\frac{r^2}{L_4^2}} + r^2\,d\tilde\Omega_3^2 + L_4^2\,\left(d\theta^2 + \cos^2\theta\,d\phi^2 + \sin^2\theta\,d\Omega_3^2\right), \nonumber \\
C_4 &=& L_4^4\,\sin^4\theta\,d\phi\wedge\omega_3,
\end{eqnarray}
where $\omega_3$ stands for the volume form of the 3-sphere in S${}^5$ and it is understood $dC_4$ is made self-dual to satisfy the type IIB equations of motion\footnote{I do not write this term explicitly here because it will not couple to our D3-brane probes.}. Giant gravitons consist of D3-branes wrapping such 3-spheres and rotating in the $\phi$ direction to carry R-charge from the dual CFT perspective. Thus, one considers the bosonic ansatz
\begin{eqnarray}
  \sigma^0 &=& t, \quad \sigma^i = \omega^i, \nonumber \\
  \theta &=& \theta_0, \quad \phi = \phi (\tau), \quad r=0.
\end{eqnarray}
The D3-brane lagrangian density evaluated on this ansatz and integrating over the 3-sphere world volume is \cite{Grisaru:2000zn}
\begin{equation}
{\cal L}=\frac{N}{ L_4}\left[-\sin^3\theta\,\sqrt{1-L_4^2 \cos^2 \theta 
\,\dot{\phi}^2}+ L_4\sin^4 \theta \,\dot{\phi}\right]\,.
\label{lag2}
\end{equation}
Since $k=\partial_\phi$ is a Killing vector, the conjugate momentum $P_\phi$ is conserved
\begin{equation}
P_\phi=N\left[\frac{L_4\sin^3\theta\cos^2\theta\,\dot{\phi}}{\sqrt{1-L_4^2\cos^2\theta
\,\dot{\phi}^2}}+\sin^4\theta\right]\equiv N\,p,
\label{angmom}
\end{equation}
where the constant $p$ was defined. Computing the hamiltonian density
\begin{equation}
{\cal E} =P_\phi\dot{\phi}-{\cal L} = \frac{N}{L_4}\,\sqrt{p^2+\tan^2\theta\,(p-\sin^2\theta)^2},
\label{ham2}
\end{equation}
allows us to identify the stable configurations by extremising (\ref{ham2}). Focusing on finite size configurations, one finds 
\begin{equation}
\sin\theta_0= \sqrt{p} \quad \Longrightarrow \quad \dot{\phi}=\frac{1}{L_4} \quad \Longrightarrow \quad 
{\cal E} = \frac{P_\phi}{L_4}\,.
\label{hamx}
\end{equation}
Notice the latter equality saturates the BPS bound, $\Delta \equiv {\cal E}L_4 = P_\phi$, as expected from supersymmetry considerations.

To check whether the above configuration indeed preserves some supersymmetry, one must check whether there exists a subset of target space Killing spinors solving the kappa symmetry preserving condition (\ref{spc}). The 32 Killing spinors for the 
maximally supersymmetric AdS${}_5\times$S${}^5$ background were computed in \cite{Lu:1998nu,Grisaru:2000zn}. They are of the form $\epsilon = M\,\epsilon_\infty$ where $M$ is a non-trivial Clifford valued matrix depending on the bulk point and $\epsilon_\infty$ is an arbitrary constant spinor. It was shown in \cite{Grisaru:2000zn} that (\ref{spc}) reduces to
\begin{equation}
(\Gamma_{t\phi}-1)\epsilon_\infty=0.
\end{equation}
Thus, giant gravitons preserve half of the spacetime supersymmetry. Furthermore, they preserve the {\it same} supercharges as a pointlike graviton rotating in the $\phi$ direction.

\paragraph{General supersymmetric giant graviton construction : } There exist more general giant gravitons charged under the full  $\U(1)^3$ Cartan subalgebra of the full R-symmetry group $\SO(6)$. The general construction of such supersymmetric probes was done in \cite{Mikhailov:2000ya}. The main idea is to embed the bulk 5-sphere into an auxiliary embedding $\bC^3$ space with complex coordinates $z_i$ $i=1,2,3$ and AdS${}_5$ into $\bC^{1,2}$. In the probe calculation, the $Z_i$ become dynamical scalar fields subject to the defining quadric constraint $\sum_i |Z_i|^2 = 1$. To prove these configurations are supersymmetric one can use the well known isomorphism between geometric Killing spinors on both the 5-sphere and AdS${}_5$ and parallel spinors in $\bC^3$ and $\bC^{1,2}$, respectively. This is briefly reviewed in appendix \ref{sec:cone}. The conclusion of such analysis is that {\it any holomorphic} function $F(Z_1,Z_2,Z_3)$ gives rise to a supersymmetric giant graviton configuration  \cite{Mikhailov:2000ya} defined
\begin{eqnarray}
  |Z_1|^2 +  |Z_2|^2 + |Z_3|^2  &=& 1, \nonumber \\
  F(e^{-it/L_4}Z_1,e^{-it/L_4}Z_2,e^{-it/L_4}Z_3)&=& 0,
\end{eqnarray}
as the intersection of the 5-sphere with a holomorphic hypersurface properly evolved in world volume time. The latter involves rotations in each of the $\bC$ planes in $\bC^3$ at the speed of light (in $1/L_4$ units), which is a consequence of supersymmetry and a generalisation of the condition explicitly found in (\ref{hamx}).

\paragraph{Geometric quantisation and BPS counting :} The above construction is classical and applies to backgrounds of the form AdS${}_5\times {\cal M}_5$. In \cite{Beasley:2002xv}, the classical moduli space of holomorphic functions mentioned above was originally quantised and some of its BPS spectrum matched against the spectrum of chiral operators in ${\cal N}=4$ d=4 SYM. Later, in \cite{Biswas:2006tj,Mandal:2006tk}, the full partition function was derived and seen to agree with that of N noninteracting bosons in a 3d harmonic potential. Similar work and results were obtained for the moduli space of dual giant gravitons\footnote{Dual giant gravitons are spherical rotating D3-branes in which the 3-sphere wrapped by the brane is in AdS${}_5$. See \cite{Grisaru:2000zn} for a proper construction of these configurations and some of its properties.} when ${\cal M}_5$ is an Einstein-Sasaki manifold \cite{Martelli:2006vh}. The BPS partition functions derived from these geometric quantisation schemes agree with purely gauge theory considerations \cite{Berenstein:2005aa,Kinney:2005ej} and with the more algebraic approach to counting chiral operators followed in the plethystics program \cite{Benvenuti:2006qr,Feng:2007ur}.

\paragraph{Related work :} There exists an extensive amount of work constructing world volume configurations describing giant gravitons in different backgrounds to the ones mentioned above. This includes non-supersymmetric giant gravitons with NSNS fields \cite{Camino:2001ti}, M-theory giants with 3-form potnetial field \cite{Camino:2001jg}, giants in deformed backgrounds \cite{Pirrone:2006iq} or electric/magnetic field deformed giants in Melvin geometries \cite{Huang:2006kua}. For discussions on supersymmetric D3, fractional D5 and D7-brane probes in AdS${}_5\times$L${}^{abc}$, see \cite{Canoura:2006es}. There is also interesting work on bound states of giant gravitons \cite{Prokushkin:2004pv} and on the effective field theory description of many of such giants (a non-abelian world volume description) with the inclusion of higher moment couplings responsible for their physical properties \cite{Janssen:2002cf,Janssen:2003ri}.

\subsubsection{Giant gravitons as black hole constituents}

Individual giant gravitons carry conformal dimension of order $N$ and according to the discussion above, they exhaust the spectrum of chiral operators in the dual CFT, whereas R-charged AdS black holes carry mass of order $N^2$. The idea that supersymmetric R-charged AdS black holes could be interpreted as {\it distributions} of giant gravitons was first discussed in \cite{Myers:2001aq}, where these bulk configurations were coined as {\it superstars}. The main idea behind this identification comes from two observations 
\begin{itemize}
\item[1.] The existence of naked singularities in these black holes located where giant gravitons sit in AdS suggests the singularity is due to the presence of an {\it external source}.
\item[2.] Giant gravitons do {\it not} carry D3-brane charge, but they do locally couple to the \RR\ 5-form field strength giving rise to some D3-brane {\it dipole} charge. This means \cite{Myers:2001aq} that a small (five-dimensional) surface enclosing a portion of the giant graviton sphere will carry a net five-form flux proportional to the number of D3-branes enclosed. If this is correct, one should be able to determine the local density of giant gravitons at the singularity by analysing the net \RR\ 5-form flux obtained by considering a surface that is the boundary of a six-dimensional ball which only intersects the three-sphere of the giant graviton once at a point, very close to the singularity.
\end{itemize}

To check this interpretation, let us review these supersymmetric R-charged AdS${}_5$ black holes. These are solutions to ${\cal N}=2$ d=5 gauged supergravity with $\U(1)^3$ gauge symmetry \cite{Behrndt:1998ns,Behrndt:1998jd} properly embedded into type IIB \cite{Cvetic:1999xp}. Their metric is
\begin{eqnarray}
ds_{{\rm 10}}^2&=&\sqrt{\Delta}\left[-(H_1H_2H_3)^{-1}f dt^2+(f^{-1}dr^2+r^2 d\Omega_3^2)\right]\nonumber\\
& & + \frac{1}{\sqrt{\Delta}}\sum_{i=1}^3 H_i\left(L^2 d\mu_i^2+ \mu_i^2[L_4 d\phi_i+(H_i^{-1}-1)dt]^2\right),
\label{upten}
\end{eqnarray}
with the different scalar functions defined as
\begin{eqnarray}
  f&=&1 + \frac{r^2}{L_4^2}H_1 H_2 H_3 \qquad {\rm with} \qquad  H_i=1+\frac{q_i}{r^2}\,,  \nonumber \\
\Delta &=&H_1H_2H_3\sum_{i=1}^3 \frac{\mu_i^2}{H_i},\qquad {\rm with} \qquad \sum_{i=1}^3 \mu_i^2 = 1\,.
\label{upten2}
\end{eqnarray}
All these metrics have a naked singularity at the center of AdS that extends into the 5-sphere. Depending on the number of charges turned on, the rate at which curvature invariants diverges changes with the 5-sphere direction. Besides a constant dilaton, these BPS configurations also have a non-trivial \RR\ self-dual 5-form field strength $R_{5}=dC_{4}+*dC_{4}$ with
\begin{equation}
C_{4}=-\frac{r^4}{L_4}\Delta\, dt\wedge \omega_3
-L_4\sum_{i=1}^3 q_i \mu_i^2(L_4\,d\phi_i-dt)\wedge \omega_3\,,
\label{cfield}
\end{equation}
with $\omega_3$ being volume 3-form of the unit 3-sphere.

To test the microscopic interpretation for the superstar solutions, consider the single R-charged configuration with $q_2=q_3=0$.
This should correspond to a collection of giant gravitons rotating along $\phi_1$ with a certain distribution of sizes (specified by $\mu_1=\cos\theta_1$). To measure the density of giant gravitons sitting near a certain $\theta_1$, one must integrate $R_5$
over the appropriate surface. Describing the 3-sphere in AdS${}_5$ by
\begin{equation}
  d\Omega_3^2 = d\alpha_1^2 + \sin^2\alpha_1(d\alpha_2^2 + \sin^2\alpha_2d\alpha_3^2)\,,
\end{equation}
one can enclose a point on the brane at $\theta_1$ with a small five-sphere in the $\{r, \theta_1, \phi_1\,\alpha_i\}$ directions.
The relevant five-form component is
\begin{equation}
(R_5)_{\theta_1\phi_1\alpha_1\alpha_2\alpha_3}=
2 q_1 L_4^2 \sin\theta_1 \cos\theta_1\sin^2\alpha_1\sin\alpha_2\,,
\end{equation}
and by integrating the latter over the smeared direction $\phi_1$ and the 3-sphere, one infers the density of giants at a point $\theta_1$ \cite{Myers:2001aq}
\begin{equation}
\frac{d n_1}{d\theta_1}=\frac{N}{4\pi^3 L_4^4}\int
(R_5)_{\theta_1\phi_1\alpha_1\alpha_2\alpha_3}d\phi_1 d^3\alpha
=N\frac{q_1}{L_4^2}\sin 2\theta_1\,.
\label{dn1dth1}
\end{equation}
If this is correct, the total number of giant gravitons carried by the superstar is 
\begin{equation}
n_1=\int_0^{\pi/2}d\theta_1\,\frac{d n_1}{d\theta_1}=N \frac{q_1}{L_4^2}.
\end{equation}
The matching is achieved by comparing the microscopic momentum carried by a single giant at the location $\theta_1$, $P_{{\rm micro}}=N\sin^2\theta_1$, with the total mass of the superstar
\begin{equation}
M=\frac{N^2}{2}\frac{q_1}{L_4^3}\,.
\end{equation}
Indeed, by supersymmetry, the latter should equal the total momentum of the distribution
\begin{equation}
M = \frac{P_{\phi_1}}{L_4}=\int_0^{\pi/2}d\theta_1\,\frac{d n_1}{d\theta_1} \frac{P_{{\rm micro}}}{L_4}=\frac{N^2}{2}\frac{q_1}{L_4^3}\,,
\label{bakk}
\end{equation}
which establishes the physical correspondence. There exist extensions of these considerations when more than a single R-charge is turned on, i.e. when $q_2, q_3\neq 0$. See \cite{Myers:2001aq} for the specific details, though the conclusion remains the same.

\paragraph{1/2 BPS superstar \& smooth configurations :} Just as supertubes have smooth supergravity descriptions \cite{Emparan:2001ux} with U-dual interpretations in terms of chiral states in dual CFTs \cite{Lunin:2002iz} when some of the dimensions are compact, one may wonder whether a similar picture is available for chiral operators in ${\cal N}=4$ $ d=4$ SYM corresponding to collections of giant gravitons. For 1/2 BPS states, the supergravity analysis was done in \cite{Lin:2004nb}. The classical moduli space of smooth configurations was determined : it is characterised in terms of a single scalar function satisfying a Laplace equation. When the latter satisfies certain boundary conditions on its boundary, the entire supergravity solution is smooth. Interestingly, such boundary could be interpreted as the {\it phase space} of a single fermion in a 1d harmonic oscillator potential, whereas the boundary conditions correspond to exciting coherent states on it.  This matches the gauge theory description in terms of the eigenvalues of the adjoint matrices describing the gauge invariant operators in this 1/2 BPS sector of the full theory \cite{Corley:2001zk,Berenstein:2004kk}. Moreover, geometric quantisation applied on the subspace of these 1/2 BPS supergravity configurations also agreed with the picture of N free fermions in a 1d harmonic oscillator potential \cite{Grant:2005qc,Maoz:2005nk}. The singular superstar was interpreted as a {\it coarse-grained} description of the {\it typical} quantum state in that sector \cite{Balasubramanian:2005mg}, providing a bridge between quantum mechanics and classical geometry through the coarse-grained of quantum mechanical information. In some philosophically vague sense, these supergravity considerations provide some heuristic realisation of Wheeler's ideas \cite{Wheeler:1955zz,Wheeler:1957mu,Balasubramanian:2005kk}. Some partial progress was also achieved for similar M-theory configurations \cite{Lin:2004nb}. In this case, the quantum moduli space of BPS gauge theory configurations was identified in \cite{SheikhJabbari:2009kr} and some steps to identify the dictionary between these and the supergravity geometries were described in \cite{Donos:2010va}. Notice this set-up is also in agreement with the general framework illustrated in figure \ref{fig8}.

\paragraph{Less supersymmetric superstars : } Given the robustness of the results concerning the partition functions of 1/4 and 1/8 chiral BPS operators in ${\cal N}=4$ SYM and their description in terms of BPS giant graviton excitations, it is natural to study whether there exist smooth supergravity configurations preserving this amount of supersymmetry and the appropriate bosonic isometries to be interpreted as these chiral states. The classical moduli space of these configurations was given in \cite{Chen:2007du}, extending previous work \cite{Donos:2006iy,Donos:2006ms}. The equations describing these moduli spaces are far more complicated than its 1/2 BPS sector cousin,
\begin{itemize}
\item 1/4 BPS configurations depend on a 4d K\"ahler manifold with K\"ahler potential satisfying a non-linear Monge-Amp\`ere equation \cite{Chen:2007du}
\item 1/8 BPS configuration depend on a 6d manifold whose scalar curvature satisfies a non-linear equation in the scalar curvature itself and the square of the Ricci tensor \cite{Kim:2005ez}
\end{itemize}
Some set of necessary conditions for the smoothness of these configurations was discussed in \cite{Chen:2007du}. A more thorough analysis for the 1/4 BPS configurations was performed in \cite{Lunin:2008tf}, where it was argued that a set of extra consistency conditions were required, the latter constraining the location of the sources responsible for the solutions.
Interestingly, these constraints were found to be in perfect agreement with the result of a probe analysis. This reemphasises the usefulness of probe techniques when analysing supergravity matters in certain BPS contexts.

\subsection{Deconstructing black holes}

Both supertubes and giant gravitons are examples of supersymmetric states realised as classical solitons in brane effective actions and interpreted as the microscopic constituents of small black holes. The bulk entropy is matched after geometric quantization of their respective classical moduli spaces. This framework, which is summarised in figure \ref{fig8}, suggests the idea of {\it deconstructing} the black hole into zero-entropy, minimally charged bits, reinterpreting the initial black hole entropy as the ground state degeneracy of the quantum mechanics on the moduli space of such deconstructions (bits). 

\epubtkImage{constituent.png}{%
\begin{figure}[h]
  \centerline{\includegraphics[width=150mm]{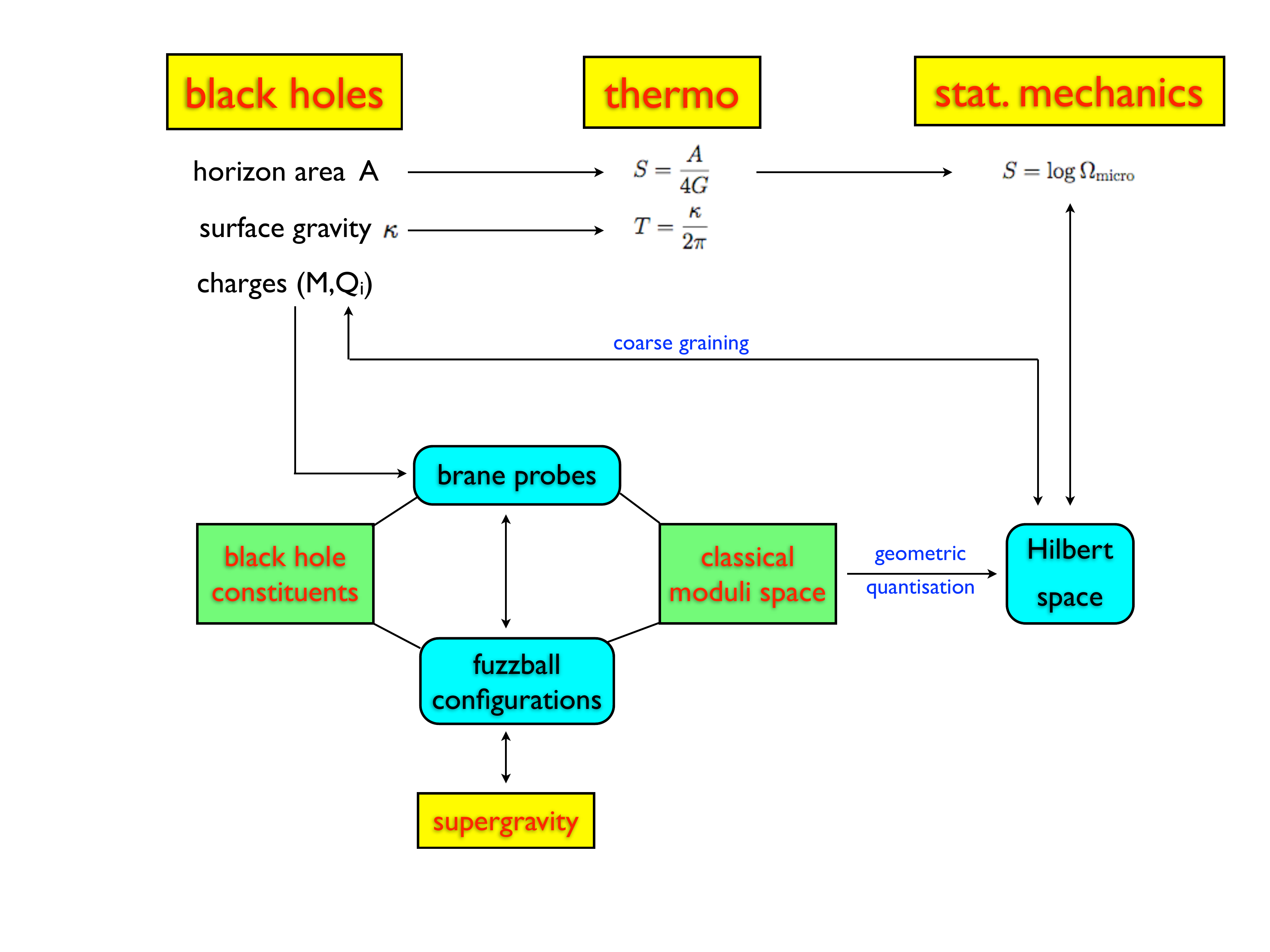}}
   \caption{Relation between the quantisation of the classical moduli space of certain supersymmetric probe configurations, their supergravity realisations and their possible interpretation as black hole constituents.}
  \label{fig8}
\end{figure}}

In this subsection, I briefly mention some work in this direction concerning large supersymmetric AdS${}_5\times$S${}^5$ black holes, deconstructions of supersymmetric asymptotically flat black holes in terms of constituent excitations living at the horizon of these black holes and constituent models for extremal static non-BPS black holes.

\paragraph{Large supersymmetric AdS${}_5$ black holes :} Large supersymmetric AdS${}_5\times$S${}^5$ black holes require the addition of angular momentum in AdS${}_5$, besides the presence of R-charges, to achieve a regular macroscopic horizon while preserving a generic 1/16 of the vacuum supersymmetries. The first examples were reported in
\cite{Gutowski:2004ez}. Subsequent work involving more general (non-)BPS black holes can be found in \cite{Gutowski:2004yv,Chong:2005da,Kunduri:2006ek}.

Given the success in identifying the degrees of freedom for R-charged black holes, it is natural to analyse whether the inclusion of angular momentum in AdS${}_5$ can be accomplished by more general (dual) giant graviton configurations carrying the same charges as the black hole. This task was initiated in \cite{Kim:2005mw}. Even though their work was concerned with configurations preserving 1/8 of the supersymmetry, the importance of a non-trivial Poynting vector on the D3-brane world volume to generate angular momentum was already pointed out, extending the mechanism used already for supertubes. In \cite{Kim:2006he}, the first extension of these results to 1/16 world volume configurations was considered. 
The equations satisfied for the most general 1/16 dual giant D3-brane probe in AdS${}_5\times$S${}^5$ were described in 
\cite{Ashok:2008fa} whereas explicit supersymmetric electromagnetic waves on (dual) giants were constructed in \cite{Ashok:2010jv}. Similar interesting work describing giant gravitons in the pp-wave background  with non-trivial electric fields was reported in \cite{AliAkbari:2007jr}.

All these configurations have interest on their own, given their supersymmetry and the conserved charges they carry, but further evidence is required to interpret them as bulk black hole constituents. This task was undertaken in \cite{Sinha:2007ni}. Instead of working in the vacuum, these authors studied the spectrum of classical supersymmetric (dual) giant gravitons in the near horizon geometries of these black holes in \cite{Sinha:2006sh}, following similar reasonings for 
asymptotically flat black holes \cite{Denef:2007yt}. The partial quantisation of this classical moduli space \cite{Sinha:2007ni} is potentially consistent with the identification of dual giants as the constituents of these black holes, but this remains an open question. In the same spirit, \cite{Ashok:2008fa} quantised the moduli space of the so called wobbling dual giants, 1/8 BPS configurations with two angular momentum in AdS${}_5$ and one in S${}^5$ and agreement was found with the gauge theory index calculations carried in \cite{Kinney:2005ej}. 

There have also been more purely field theoretical approaches to this problem. In \cite{Grant:2008sk}, cohomological methods were used to count operators preserving 1/16 of the supersymmetries in ${\cal N}=4$ d=4 SYM, whereas in
 \cite{Berkooz:2006wc}, explicit operators were written down, based on Fermi surface filling fermions models and working in the limit of large angular momentum in AdS${}_5$. These attempted to identify the pure states responsible for the entropy of the black hole and their counting agreed, up to order one coefficients, with the Hawking-Bekenstein classical entropy.

\paragraph{Large asymptotically flat BPS black holes :} There exists a large literature on the construction of supersymmetric configurations with the same asymptotics and charges as a given large BPS black hole, but having the latter carried by different {\it constituent} charges located at different "centers"\footnote{These configurations appeared in \cite{Bena:2005va,Berglund:2005vb}, extending earlier seminal work \cite{Denef:2000nb,Bates:2003vx}.}. The center locations are non-trivially determined by solving a set of constraint equations, the so called {\it bubble} equations. The latter is believed to ensure the global smoothness and lack of horizon of the configuration. These constraints do reflect the intrinsic bound state nature of these configurations. The identification of a subset of 1/2 BPS centers as the fundamental {\it constituents} for large black holes was further developed in \cite{Balasubramanian:2006gi}.

One of the new features in these deconstructions is that the charges carried by the different constituents do not have to match the charges carried by the hole, i.e. a constituent can carry D6-brane charge even if the hole does not, provided there exists a second center with anti-D6-brane charge, cancelling the latter.

This idea of deconstructing a given black hole in terms of maximally entropic configurations of constituent objects\footnote{What is meant by maximally entropic here is that given a large black hole, there may be more than one possible deconstruction of the total charge in terms of constituents with different charge composition. By maximally entropic I mean the choice of charge deconstruction whose moduli space of configurations carries the largest contribution to the entropy of the system.} was tested for the standard D0-D4 black hole in \cite{Denef:2007yt}. The hole was deconstructed
 in terms of $D6$ and ${\overline D6}$ branes with world volume fluxes turned on, inducing further D4-D2-D0 charges, and a large set of D0-branes. Working in a regime of charges where the distance between centers scales to zero, i.e. the so called {\it scaling} solution, all D0-branes become equidistant to the D6-branes, forming some sort of accretion disk and the geometry deep inside this ring of D0-branes becomes that of global AdS${}_3\times$S${}^2$, when lifting the configuration to M-theory. Using the microscopic picture developed in \cite{Gaiotto:2004ij}, where it was argued that the entropy of this black hole came from the degeneracy of states due to non-abelian D0-branes that expand into D2-branes due to the Myers' effect \cite{Myers:1999ps}, the authors in \cite{Denef:2007yt} manage to extend the near horizon wrapping M2-branes found in \cite{Simons:2004nm} to M2-branes wrapping supersymmetric cycles of the full geometry. It was then argued that the same counting done \cite{Gaiotto:2004ij}, based on the degeneracy of the lowest Landau level quantum mechanics problem emerging from the effective magnetic field on the transverse Calabi-Yau due to the coupling of the D2-D0 bound states to the background \RR\ 4-form field strength, would apply in this case.

The same kind of construction and logic was applied to black rings \cite{Emparan:2001wn,Elvang:2004ds} in \cite{Gimon:2007mha}. Further work on stable brane configurations in the near horizon on brane backgrounds can be found in \cite{Camino:2001at}.

\paragraph{Extremal non-BPS deconstructions :} These ideas are also applicable to non-supersymmetric systems, though one expects to have less control there. For the subset of {\it static} extremal non-BPS black holes in the STU model \cite{Cremmer:1984hj,STU,Behrndt:1996hu}, these methods turned out to be successful. The most general static black hole solution, including non-trivial moduli at infinity, was found in \cite{Gimon:2007mh,LopesCardoso:2007ky}. It was pointed out in \cite{Gimon:2007mh} that the mass of these black holes equals the sum of four mutually local 1/2 BPS constituents for {\it any} value of the background moduli fields and in {\it any} U-duality frame. Using probe calculations, it was shown that such constituents do not feel any force in the presence of these black holes \cite{Gimon:2009gk}. This suggested that extra quanta could be added to the system and located anywhere. This is consistent with the multi-center extremal non-BPS solutions found in \cite{Gaiotto:2007ag} : their centers are completely arbitrary but the charge vectors carried by each center are constrained to be the ones of the constituents identified in \cite{Gimon:2009gk} (or their linear combinations). This model identifies the same constituents as the ones used to account for the entropy of neutral black holes in \cite{Emparan:2006it} and extends it to the presence of fluxes. No further dynamical understanding of the open string degrees of freedom is available in terms of non-supersymmetric quiver gauge theories.

As soon as angular momentum is added to the system, while keeping extremality, the location of the deconstructed constituents gets constrained according to non-linear {\it bubble} equations that ensure the global smoothness of the full supergravity solution \cite{Bena:2009ev,Bena:2009en}. These are fairly recent developments and one expects further progress to be achieved in this direction in the future. For example, very recently, an analysis of stable, metastable and non-stable supertubes in smooth geometries being candidates for the microstates of black holes and black rings was presented in \cite{Bena:2011fc}. This includes configurations that would also be valid for non-extremal black holes.

\section{Some AdS/CFT related applications}
\label{sec:adscft}

This section is devoted to more dynamical applications of brane effective actions. More specifically, I will describe some well established reinterpretations of certain brane probe calculations in the context of the AdS/CFT correspondence \cite{Maldacena:1997re,Gubser:1998bc,Witten:1998qj,Aharony:1999ti}. I will mainly focus on two aspects :
\begin{itemize}
\item The use of {\it classical solitons} solving the brane (string) equations of motion in particular backgrounds and with specific boundary conditions, to holographically compute either the expectation value of certain gauge invariant operators or the spectrum in sectors of certain strongly coupled gauge theories.
\item The use of D-brane effective actions to describe the dynamics of a small number of degrees of freedom responsible either for {\it deforming} the original dual CFT to theories with less or no supersymmetry, or for capturing the interaction of massless modes among themselves and with other sectors of the system conveniently replaced by a supergravity background.
\end{itemize}

Covariance of brane effective actions allows to couple them to {\it any} on-shell supergravity background. In particular,
one can probe either AdS${}_5\times$S${}^5$, or black holes with these asymptotics, with branes, and according to the
AdS/CFT correspondence, one will be studying properties of the strongly coupled holographic theory in the vacuum or at finite temperature and chemical potentials, respectively. This set-up is illustrated in figure \ref{fig2}. The same interpretation will hold for non-relativistic versions of these backgrounds. Alternatively, and depending on the boundary conditions imposed on these probes, they can deform the theory towards less symmetric and more realistic physical systems. 

\epubtkImage{application1.png}{%
\begin{figure}[h]
  \centerline{\includegraphics[width=150mm]{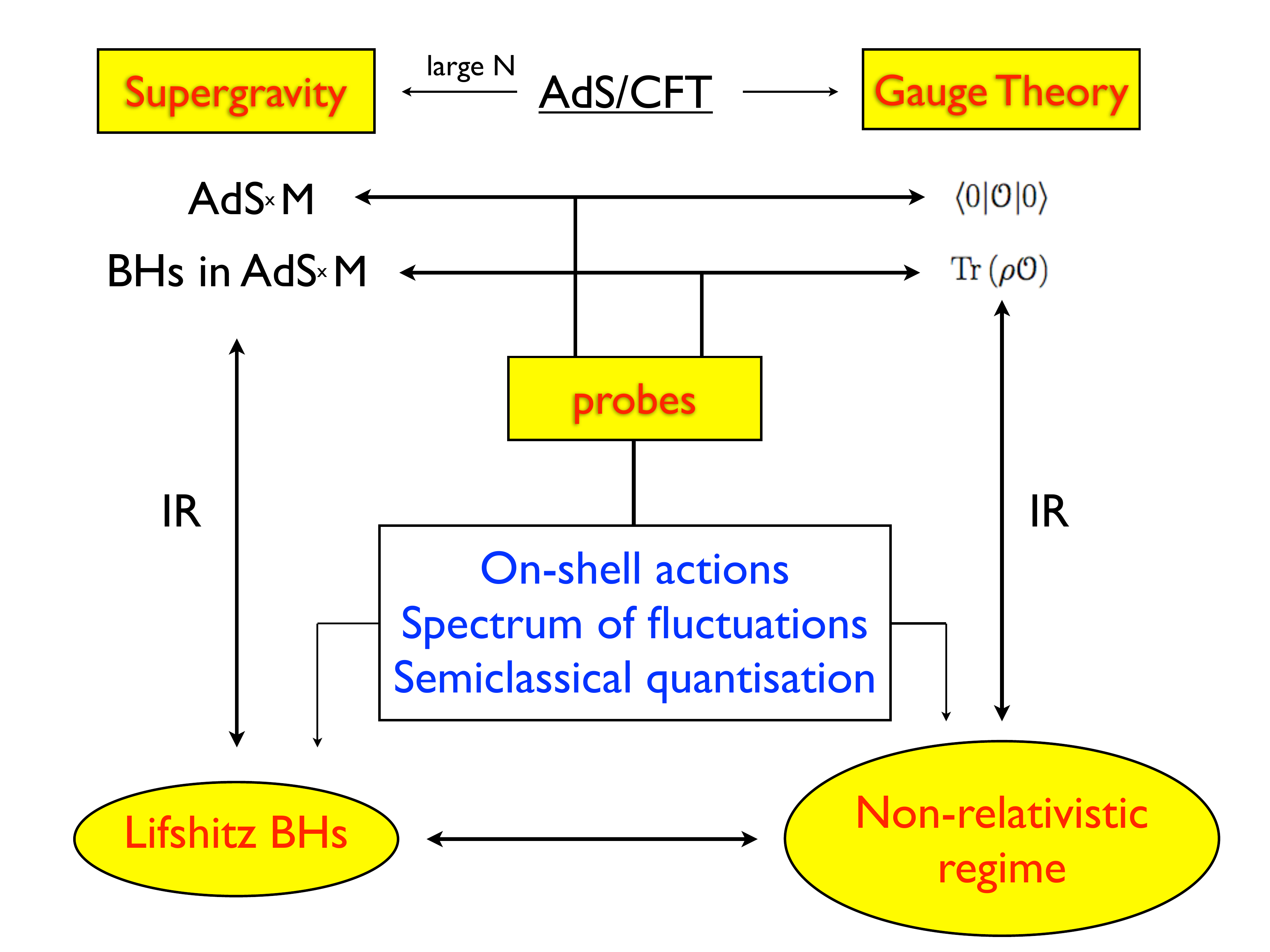}}
   \caption{General framework in which probe calculations in appropriate backgrounds with suitable boundary conditions can be reinterpreted as strongly coupled observables and spectrum in non-abelian gauge theories using the AdS/CFT correspondence.}
  \label{fig2}
\end{figure}}

In the following, I will review the calculation of Wilson loop expectation values, the use of worldsheet string solitons to study the spectrum of states with large charges in ${\cal N}=4$ SYM and the use of D-brane probes to either add flavour to the AdS/CFT correspondence or describe the dynamics of massless excitations in non-relativistic (thermal) set-ups which could be of relevance for strongly coupled condensed matter physics.

\subsection{Wilson loops} 
\label{sec:wloops}

As a first example of the use of classical solutions to brane effective actions to compute the expectation values of gauge invariant operators at strong coupling, I will review the prescription put forward in \cite{Maldacena:1998im,Rey:1998ik} for Wilson loop operators in ${\cal N}=4$ SYM. 

Wilson loop operators \cite{Wilson:1974sk} in $\SU(N)$ Yang-Mills theories are non-local gauge invariant operators
\begin{equation}
\label{wilsonloop}
W({\cal C} ) =\frac{1}{N} {\rm Tr P} e^{i  \oint_{\cal C} A}\,,
\end{equation}
depending on a closed loop in spacetime ${\cal C}$ and where the trace is over the fundamental representation of the gauge group. This operator allows to extract the energy $E(L)$ of a quark-antiquak pair separated a distance $L$. Indeed,  consider a rectangular closed loop in which the pair evolves in euclidean time $T$. In the limit $T\to\infty$, the expectation value of this rectangular Wilson loop equals
\begin{equation}
\langle W({\cal C} ) \rangle  = A(L)  e^{ - T E(L) }\,.
\end{equation}

To understand the prescription in \cite{Maldacena:1998im,Rey:1998ik}, one must first introduce massive quarks in the theory. This is achieved by breaking the original gauge symmetry of the original ${\cal N}=4$ SYM according to
\begin{equation}
 \U(N+1) \to  \U(N) \times \U(1)
\end{equation}
The massive W-bosons generated by this process have a mass proportional to the norm of the Higgs field expectation value responsible for the symmetry breaking $(|\vec \Phi|)$ and transform in the fundamental representation of the $\U(N)$ gauge symmetry, as required. Furthermore at energy scales much lower than $|\vec \Phi|$, the $\U(N)$ theory decouples from the $\U(1)$ theory.

In this set-up, the massive W-boson interacts with the $\U(N)$ gauge fields, including the scalar adjoint fields $X^I$ \cite{Maldacena:1998im}, leading to the insertion of the operator
\begin{equation}
\label{wilour}
W({\cal C }) = \frac{1}{N} {\rm Tr P} e^{ i \oint_{\cal C} ds
[ A_\mu(\sigma) \dot \sigma^\mu + \theta^I(s)
X^I(\sigma)\sqrt{\dot \sigma^2 }]}\,.
\end{equation}
The contour ${\cal C}$ is parameterised by $\sigma^\mu(s)$ whereas the vector $\vec \theta(s)$  maps each point on the loop to a point on the five-sphere. 

The proposal made in \cite{Maldacena:1998im,Rey:1998ik} to compute the expectation value of (\ref{wilour}) was
\begin{equation}
\label{wiloursugra}
 \langle W({\cal C }) \rangle  \sim   e^{ -S_{\rm string}}\,.
\end{equation}
This holds in the large $g_sN$ limit and $S_{\rm string}$ stands for the proper area of a fundamental string describing the loop ${\cal C}$ at the boundary of AdS${}_5$ and lying along $\theta^I(s)$ on S$^{}5$. Notice that a quantum mechanical calculation at strong coupling reduces to determining a minimal worldsheet surface in AdS${}_5$, i.e. solving the worldsheet equations of motion with appropriate boundary conditions, and then solving for the worldsheet energy as a function of the separation $L$ between the quark-antiquark. After subtracting the regularised mass of the W-boson one obtains the quark-antiquark potential energy
\begin{equation}
E(L) =-  \frac{4 \pi^2 ( 2 g^2_{{\rm YM}} N)^{1/2}}{\Gamma(\frac{1}{4})^4 L}\,, 
\end{equation}
which differs from the linear perturbative dependence on $g^2_{{\rm YM}} N$.

If one considers multiply wrapped Wilson loops, the many coincident strings will suffer from self-interactions. This may suggest that a more appropriate description of the system is in terms of a D3-brane with non-trivial world volume electric flux accounting for the fundamental strings. This is the approach followed in \cite{Drukker:2005kx}, where it was shown that for linear and circular loops the D3-brane action agreed with the string worldsheet result at weak coupling, but captures all the higher genus corrections at leading order in $\alpha'$.

\subsection{Quark energy loss in a thermal medium} 

Having learnt how to describe a massive quark in ${\cal N}=4$ SYM in terms of an string, this opens up the possibility of describing its energy loss as it propagates through a thermal medium. One can think of this process 
\begin{itemize}
\item[1.] either from the bulk perspective, where the thermal medium gets replaced by a black hole and energy flows down the string towards its horizon,
\item[2.]  or from the gauge theory perspective, where energy and momentum emanate from the quark and eventually thermalise.
\end{itemize}
In this section, I will take the bulk point of view originally discussed in \cite{Herzog:2006gh,Gubser:2006bz}, with a related fluctuation analysis in \cite{CasalderreySolana:2006rq}. The goal is to highlight the power of the techniques developed in previous sections rather than being self-contained. For a more thorough discussion, the reader should check the review on this particular topic \cite{Gubser:2009sn}. 

The thermal medium is holographically described in terms of the AdS${}_5$-Schwarzschild black hole 
\begin{equation}
  ds^2 = g_{mn}dx^mdx^n = \frac{L^2}{z^2} \left( -h(z) dt^2 + d\vec{x}^2 + \frac{dz^2}{h(z)} \right) \,, 
\label{E:LineElement}
\end{equation}
where $h(z)= 1 - \frac{z^4}{z_H^4}$ determines the horizon size $z_H$ and the black hole temperature $T = \frac{1}{\pi z_H}$. The latter coincides with the gauge theory temperature \cite{Witten:1998qj}. Notice $z=0$ is the location of the conformal boundary and $L$ is the radius of AdS${}_5$.

If one is interested in describing the dragging effect suffered by the quark due to the interactions with the thermal medium, one considers a non-static quark whose trajectory in the boundary satisfies $X^1(t) = vt$, assuming motion takes place only in the $x^1$ direction. One can parameterise the bulk trajectory as
\begin{equation}
  X^1(t,z) = vt + \xi(z)  \label{E:TrailingAnsatz}
 \end{equation}
where $\xi(z)$ satisfies $\xi \to 0$ as $z \to 0$. To determine $\xi(z)$, one must solve the classical equations of motion of the bosonic worldsheet action (\ref{eq:gsusystring}) in the background (\ref{E:LineElement}). These reduce to a set of conserved equations of the form
\begin{equation}
  \nabla_\mu \pi^\mu{}_m= 0\,, \qquad \qquad {\rm where} \qquad \qquad \pi^\mu{}_m \equiv -\frac{1}{2\pi\alpha'} \CG^{\mu\nu}g_{mn} \partial_\nu X^n 
\label{E:WSconservation}
\end{equation}
is the worldsheet momentum current conjugate to the position $X^m$. Plugging the ansatz (\ref{E:TrailingAnsatz}) into (\ref{E:WSconservation}), one finds
 \begin{equation}
  \frac{d\xi}{dz} = \frac{\pi_\xi}{h} \sqrt{\frac{h-v^2}{\frac{L^4}{z^4} h - \pi_\xi^2}} \,,
    \label{E:XiShape}
 \end{equation}
where $\pi_\xi$ is an integration constant. A priori, there are several allowed possibilities compatible with the reality of
the trailing function $\xi(z)$. These were analysed in \cite{Gubser:2009sn} where it was concluded the relevant physical solution is given by
\begin{equation}
\pi_\xi = - \frac{L^2}{z_*^2} \sqrt{h(z_*)}
    = - \frac{v}{\sqrt{1-v^2}} \frac{L^2}{z_H^2} \qquad \Longrightarrow \qquad
 \xi = -\frac{z_H v}{4i}
    \left( \log \frac{1 - iy}{1 + iy} + 
      i \log \frac{1 + y}{1 - y} \right) \,,
        \label{E:TrailingString}
\end{equation}
where $y$ is a rescaled depth variable $y = z/z_H$.

To compute the rate at which quark momentum is being transferred to the bath, one can simply integrate the conserved current $p^\mu{}_m$ over a line-segment and given the stready-state nature of the trailing string configuration, one infers 
\cite{Gubser:2009sn}
 \begin{equation}
  \frac{dp_m}{dt} = -\sqrt{-g} \, p^z{}_m.
 \end{equation}
This allows us to define the drag force as
\begin{equation}
  F_{\rm drag} = \frac{dp_1}{dt} = -\frac{L^2 }{2\pi z_H^2 \alpha'} \frac{v}{\sqrt{1-v^2}}=-\frac{\pi\sqrt\lambda}{2} T^2 \frac{v}{\sqrt{1-v^2}} \quad {\rm with} \quad \lambda = g_{\rm YM}^2 N = \frac{L^4}{\alpha'^2}.
    \label{DragForce}
\end{equation} 
For a much more detailed discussion on the physics of this system see \cite{Gubser:2009sn,CasalderreySolana:2011us}. The latter also includes a discussion of the same physical effect for a finite, but large quark mass, and the possible implications of these results and techniques for QCD. 

More recently, it was argued in \cite{Fiol:2011zg} that one can compute the energy loss by radiation of an infinitely massive half-BPS charged particle to all orders in $1/N$ using a similar construction to the one mentioned at the end of subsection \ref{sec:wloops}. This involved the use of classical D5-brane and D3-brane world volume reaching the AdS${}_5$ boundary to describe particles transforming in the antisymmetric and symmetric representations of the gauge group, respectively.

\subsection{Semiclassical correspondence} 

It is an extended idea in theoretical physics that states in quantum mechanics carrying large charges can be well approximated by a classical or semiclassical description. This idea gets realised in the AdS/CFT correspondence too. 
Consider the worldsheet sigma model description of a fundamental string in AdS${}_5\times$S${}^5$. One expects its 
{\it perturbative} oscillations to be properly described by supergravity, whereas {\it solitons} with large conformal dimension
\begin{equation}
  \Delta \sim \frac{1}{\sqrt{\lambda}}\,, \qquad \qquad \lambda = g^2_{{\rm YM}} N=g_s N
\end{equation}
and the spectrum of their semiclassical excitations may approximate the spectrum of highly excited string states in ${\cal N}=4$ SYM. This is the approach followed in \cite{Gubser:2002tv} where it was originally applied to rotating folded strings carrying large bare spin charge.

To get an heuristic idea on the analytic power behind this technique, let me reproduce the spectrum of large R-charge operators obtained in \cite{Berenstein:2002jq} using a worldsheet quantisation in the pp-wave background by considering the bosonic part of the worldsheet action describing the AdS$_{5}\times S_{5}$ sigma model \cite{Gubser:2002tv} 
\begin{equation} \label{saction}
   S=\frac{1}{2\alpha}\int d^2\sigma \, \sqrt{g} ((\nabla_\alpha n)^{2}+
   (\nabla_\alpha K)^{2})+...
  \end{equation}
where $n$ is a unit vector describing S${}^5$, $K$ is a hyperbolic unit vector describing AdS${}_5$,  the sigma model coupling $\alpha$ is $\alpha =\frac{1}{\sqrt{\lambda }}$\footnote{This overall coupling constant is derived from the string tension $1/2\pi\alpha^\prime$ and the overall $L^2\sim g_sN\,\alpha^\prime$ scale from the AdS$_{5}\times S_{5}$ background geometry.}   and I have ignored all fermionic and \RR\ couplings.

Consider a solution to the classical equations of motion describing a collapsed rotating closed string at the equator 
\begin{equation} \label{triv}
\theta =0\,,\qquad \psi =\omega \tau \,,
\end{equation}
where $\theta $ and $\psi $ are the polar and azimuthal angles on
S${}^2$ in S${}^5$. Its classical worldsheet energy equals
\begin{equation}
E =\frac{1}{2\alpha }\omega ^{2} = \frac{\alpha}{2}J^2 \qquad {\rm where} \qquad J = \frac{\omega}{\alpha}\,.
\end{equation}

Next, consider the harmonic fluctuations around this classical soliton. Focusing on the quadratic $\theta$ oscillations, 
\begin{equation}
\alpha L=\frac{1}{2}
\left [ (\nabla \theta )^{2}+\omega ^{2}\cos ^{2}\theta \right ]\simeq
\frac{1}{2} \left [
\left( \nabla \theta \right) ^{2}-\omega ^{2}\theta ^{2}+\omega^2 \right ]\,.
\end{equation}
one recognises the standard harmonic oscillator. Using its spectrum, one derives the corrections to the classical energy \begin{equation}
\delta =\frac{\alpha}{2 }J^{2}+\sum_{n}N_{n}\sqrt{n^{2}+\alpha ^{2}J^{2}}\,,
\end{equation}
where $N_{n}$ is the excitation number of the n-th such oscillator. There is a similar contribution from 
the AdS part of the action, obtained by the change $\alpha $ to $-\alpha$. Both contributions must satisfy the on-shell
condition
\begin{equation}
\delta ({\rm S}{}^5)+\delta ({\rm AdS}{}_5)\approx 0
\end{equation}
This is how one reproduces the spectrum derived in \cite{Berenstein:2002jq}
\begin{equation} \label{ppdim}
\Delta = J +\sum_{n=-\infty}^\infty N_n \sqrt {1 + \frac{\lambda n^2}{J^2}}\,.
\end{equation}

The method outlined above is far more general and it can be applied to study other operators. For example, one can study the relation between conformal dimension and AdS${}_5$ spin, as done in \cite{Gubser:2002tv}, by analysing the behaviour of solitonic closed strings rotating in AdS. Using global AdS${}_5$
\begin{equation}
  ds_{{\rm 5}} = L^2\left[-\cosh^2\rho\, dt^2 + d\rho^2 + \sinh^2\rho\left(d\theta^2 + \sin^2\theta d\phi^2 + \cos^2\theta d\psi^2\right)\right]\,,
\end{equation}
as the background where the bosonic string propagates and working in the gauge $\tau = t$, allows to identify
the worldsheet energy with the conformal dimension in the dual CFT. Consider a closed string at the equator of the 3-sphere while rotating in the azimuthal angle
\begin{equation}  
\label{phit}
   \phi = \omega t \,.
\end{equation}
For configurations $\rho=\rho(\sigma)$, the Nambu-Goto bosonic action reduces to
\begin{equation} 
S_{{\rm string}} = - 4 \frac{L^2}{2\pi \alpha'}\int dt \int_0^{\rho_0} d\rho \, \sqrt{\cosh^2\rho - (\dot \phi)^2\sinh^2 \rho}\,,
\end{equation}
where $\rho_0$ stands for the maximum radial coordinate and the factor of 4 arises because of the four string segments stretching from 0 to $\rho_0$ determined by the condition
 \begin{equation}\coth^2 \rho_0 = \omega^2\, .
\end{equation}
The energy ${\rm E}$ and spin ${\rm S}$ of the string are conserved charges given by
\begin{eqnarray} 
  E &=&  4 \frac{L^2}{2\pi \alpha'} \int_0^{\rho_0} d\rho \, \frac{\cosh^2\rho}{\sqrt{\cosh^2\rho - \omega^2\sinh^2 \rho}}, \label{EnergyS} \\
  S  &=&  4 \frac{L^2}{2\pi \alpha'} \int_0^{\rho_0} d\rho \, \frac{\omega \sinh^2\rho}{\sqrt{\cosh^2\rho - \omega^2\sinh^2 \rho}}. \label{SpinS}
\end{eqnarray}
Notice the dependence of ${\rm E}/\sqrt\lambda$ on $S/\sqrt \lambda$ is in parametric form since $L^4=\lambda \alpha'^2$.
One can obtain approximate expressions in the limits where the string is much
shorter or longer than the radius of curvature $L$ of AdS${}_5$.

\paragraph{Short strings :}  For large $\omega$, the maximal string stretching is $\rho_0\approx 1/\omega$. Thus, strings are shorter than the radius of curvature $L$. Calculations reduce to strings in flat space for which the parametric dependence is \cite{Gubser:2002tv}
 \begin{equation} 
 E= \frac{L^2}{\alpha' \omega}\, ,\qquad  S= \frac{L^2}{2\alpha' \omega^2}\ , \qquad \Longrightarrow \qquad
 E^2 = L^2 \frac{2 S}{\alpha'}\,.
\label{stand}
\end{equation}
Using the AdS/CFT correspondence, the conformal dimension equals the energy, i.e. $\Delta = E$. Furthermore, $S\ll \sqrt\lambda$ for large $\omega$. Thus,
\begin{equation}
\Delta^2 \approx m^2 L^2\ , \qquad {\rm where} \qquad m^2 = \frac{2 (S-2)}{\alpha'}
\end{equation}
for the leading closed string Regge trajectory, which reproduces the AdS/CFT result.
 
\paragraph{Long strings :}  The opposite regime takes place when $\omega$ is close to one (from above)
\begin{equation}
\omega = 1+ 2 \eta \,, \quad \eta \ll 1\qquad \Longrightarrow \qquad \rho_0 \to \frac{1}{2} \log \frac{1}{\eta}\,, \,\, S\gg \sqrt \lambda\,, 
\end{equation}
so that the string is sensitive to the AdS boundary metric. The string energy and spin become
\begin{eqnarray}
 E &=& \frac{L^2}{2\pi\alpha'} \left(\frac{1}{\eta} + \log \frac{1}{\eta} +\ldots \right)\ , \\
 S &=& \frac{L^2}{2\pi\alpha'} \left(\frac{1}{\eta} - \log  \frac{1}{\eta} +\ldots \right)\ ,
\end{eqnarray}
so that its difference approaches
\begin{equation} 
  E- S = \frac{\sqrt{\lambda}}{\pi} \log \frac{S}{\sqrt \lambda}+ \ldots
\end{equation}
This logarithmic asymptotics is qualitatively similar to the one appearing in perturbative gauge theories. For a more thorough discussion on this point, see \cite{Gubser:2002tv}.

Applying semiclassical quantisation methods to these classical solitons \cite{Frolov:2002av}, it was realised that one can interpolate the results for $E-S$ to the weakly couple regime. It should be stressed that these techniques allow to explore the AdS/CFT correspondence in non-supersymmetric sectors \cite{Frolov:2003xy}, appealing to the correspondence principle associated to large charges. It is also worth mentioning that due to the seminal work on the integrability of planar ${\cal N}=4$ SYM at one loop \cite{Minahan:2002ve,Beisert:2003yb}, much work has been devoted to using these semiclassical techniques in relation to integrability properties \cite{Arutyunov:2003uj}. Interested readers are encouraged to check the review \cite{Beisert:2004ry} on integrability and references therein.

\subsection{Probes as deformations \& gapless excitations in complex systems}

The dynamical regime in which brane effective actions hold is particularly suitable to describe physical systems made of several interacting subsystems in which one of them has a much smaller number of degrees of freedom. Assume the larger subsystems allow an approximate description in terms of a supergravity background. Then, focusing on the dynamics of this smaller subsector while keeping the dynamics of the larger subsystems frozen, corresponds to probing 
the supergravity background with the effective action describing the smaller subsystem. This conceptual framework is illustrated in figure \ref{fig3}.

\epubtkImage{application2.png}{%
\begin{figure}[h]
  \centerline{\includegraphics[width=150mm]{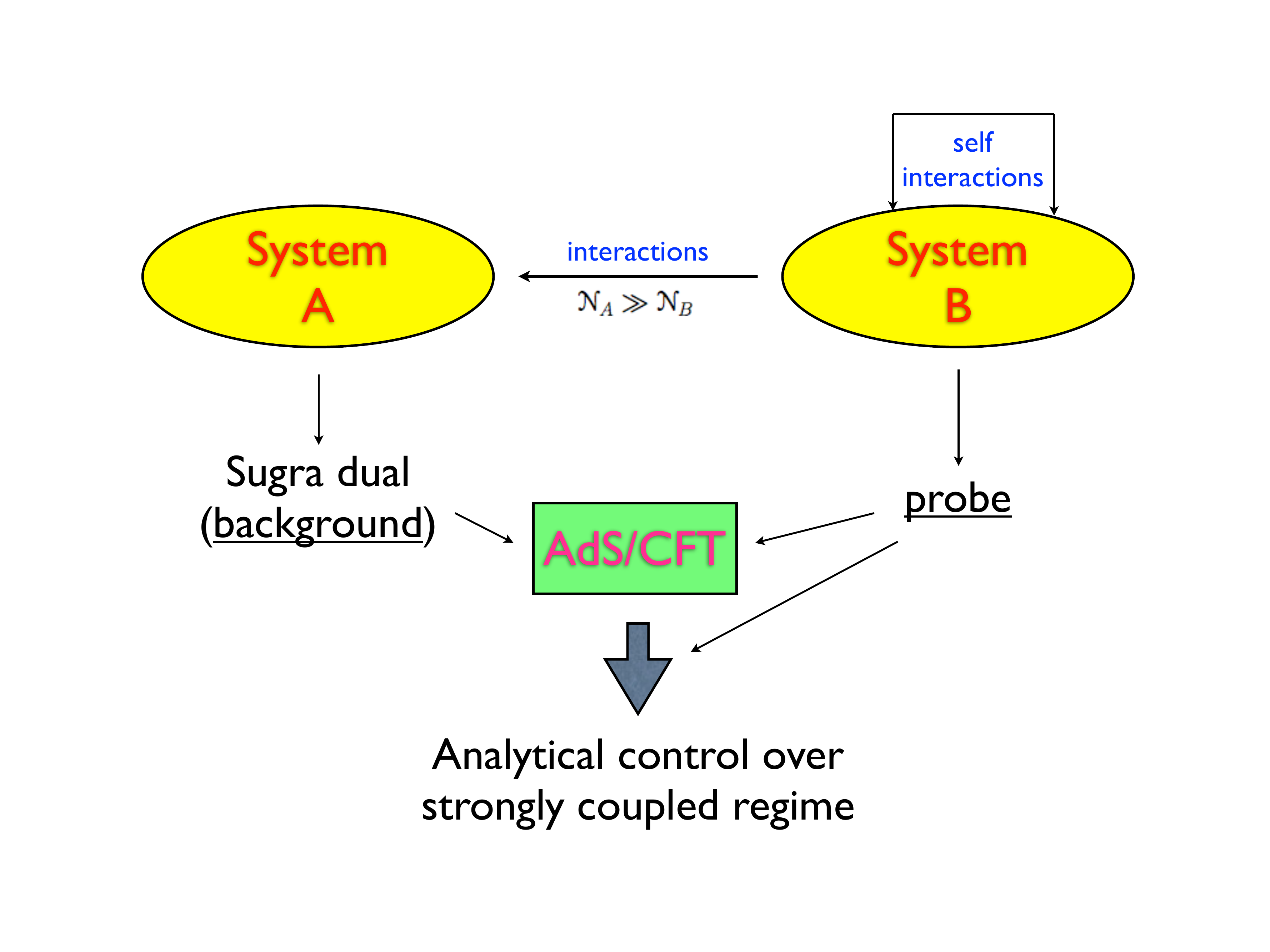}}
   \caption{Conceptual framework in which the probe approximation captures the dynamics of small subsystems interacting with larger ones that have reliable gravity duals.}
  \label{fig3}
\end{figure}}

This set-up occurs when the brane degrees of freedom are responsible for either breaking the symmetries of the larger system or describing an interesting isolated set of massless degrees of freedom whose interactions among themselves and with the background one is interested in studying. In the following, I very briefly describe how the first approach was used to introduce flavour in the AdS/CFT correspondence, and how the second one can be used to study physics reminiscent of certain phenomena in condensed matter systems.

\paragraph{Probing deformations of the AdS/CFT :} 

Deforming the original AdS/CFT allows to come up with set-ups with less or no supersymmetry. Whenever there is an small number of degrees of freedom responsible for the dynamics (typically D-branes), one may approximate the latter by the effective actions described in this review. This provides a reliable and analytical tool gravity describing the strongly coupled behaviour of the {\it deformed} gauge theory.

As an example, consider the addition of flavour in the standard AdS/CFT. It was argued in \cite{Karch:2002sh} that this could be achieved by adding $k$ D7-branes to a background of $N$ D3-branes. The D7-branes give rise to $k$ fundamental hypermultiplets arising from the lightest modes of the 3-7 and 7-3 strings, in the brane array
\begin{equation}
\begin{array}{rcccccccccl}
\mbox{D3:}\,\,\, & 1 & 2 & 3 & \_ & \_ & \_ & \_ & \_ & \_ & \, \\
\mbox{D7:}\,\,\, & 1 & 2 & 3 & 4 & 5 & 6 & 7 & \_ & \_ & \, . 
\end{array}
\end{equation}
The mass of these {\it dynamical} quarks is given by $m_q = L/2\pi \alpha'$, where $L$ is the distance between the 
D3- and the D7-branes in the 89-plane. If $g_s N \gg 1$ the D3-branes may be replaced (in the appropriate decoupling limit)  by an AdS${}_5\times$S${}^5$ geometry, as in the standard AdS/CFT argument, whereas if, in addition, $N \gg k$ then the back-reaction of the D7-branes on this geometry may be neglected. Thus, one is left, in the gravity description, with $k$ D7-brane probes in AdS${}_5\times$S${}^5$. In the particular case of $k=1$, one can use the effective action described before. This specific set-up was used in \cite{Kruczenski:2003be} to study the linearised fluctuation equations for the different excitations on the D7-probe describing different scalar and vector excitations to get analytical expressions for the spectrum of mesons in ${\cal N}=2$ SYM, at strong coupling.

This logic can be extended to non-supersymmetric scenarios\footnote{For an analysis of supersymmetric D5-branes in a supergravity background dual to ${\cal N}=1$ SYM, see \cite{Nunez:2003cf}.}. For example, using the string theory realisation of four-dimensional QCD with $N_c$ colours and $N_f \ll N_c$ flavours discussed in \cite{Witten:1998zw}. The latter involves $N_f$ D6-brane probes in the supergravity background dual to $N_c$ D4-branes compactified on
a circle with supersymmetry-breaking boundary conditions and in the limit in which all the resulting Kaluza--Klein modes decouple. For $N_f=1$ and for massless quarks, spontaneous chiral symmetry breaking by a quark condensate was exhibited in \cite{Kruczenski:2003uq} by working on the D6-brane effective action in the near horizon geometry of the $N_c$ D4-branes.

Similar considerations apply at finite temperature by using appropriate black hole backgrounds \cite{Witten:1998zw} in the relevant probe action calculations. This allows to study phase transitions associated with the thermodynamic properties of the probe degrees of freedom as a function of the probe location. This can be done in different theories, with flavour \cite{Mateos:2006nu}, and for different ensembles \cite{Kobayashi:2006sb,Mateos:2007vc}.

The amount of literature in this topic is enormous. I refer the reader to the reviews on the use of gauge-gravity duality to understand hot QCD and heavy ion collisions \cite{CasalderreySolana:2011us} and meson spectroscopy \cite{Erdmenger:2007cm}, and references therein. These explain the tools developed to apply the AdS/CFT correspondence in these set-ups.

\paragraph{Condensed matter \& strange metallic behaviour :} There has been a lot of work in using the AdS/CFT framework towards condensed matter applications. The reader is encouraged to read some of the excellent reviews on the subject \cite{Hartnoll:2009sz,Herzog:2009xv,McGreevy:2009xe,Hartnoll:2009qx,Hartnoll:2011fn}, and references therein. My goal in these paragraphs is to emphasise the use of IR probe branes to extract dynamical information about certain observables in quantum field theories in a state of finite charge density at low temperatures.

Before describing the string theory set-ups, it is worth attempting to explain why any AdS/CFT application may be able to capture any relevant physics for condensed matter systems. Consider the standard Fermi liquid theory, describing, among others, the conduction of electrons in regular metals. This theory is an example of an IR free fixed point, independent of the UV electron interactions, describing the lowest energy fermionic excitations taking place at the Fermi surface $k=k_f$. Despite its success, there is experimental evidence for the existence of different "states of matter" which are not described by this effective field theory. This could be explained by additional gapless bosonic excitations, perhaps arising as collective modes of the UV electrons. For them to be massless, the system must either be tuned to a quantum critical point or there must exist a kinematical constraint leading to a critical phase. 

One interesting possibility involving this mechanism consists on the emergence of gauge fields ("photons") at the onset of such critical phases. For example, 2+1 Maxwell theory in the presence of a Fermi surface (chemical potential $\mu$)
\begin{equation}
  {\cal L} = -\frac{1}{4}F^2 + \bar\psi \Gamma\cdot\left((i\partial + A) + \Gamma^0\,\mu\right) \psi\,,
\end{equation}
is supposed to describe at energies below $\mu$, the interactions between gapless bosons (photons) with the fermionic excitations of the Fermi surfaces. The one-loop correction to the classical photon propagator at low energy $\omega$ and momenta $k$ is
\begin{equation}
  D(\omega,\,k)^{-1} = \gamma\frac{\omega}{|k|} + |k|^2\,,
\end{equation}
Due to the presence of the chemical potential, this result manifestly breaks Lorentz invariance, but there exists
a non-trivial IR scaling symmetry (Lifshitz scale invariance)
\begin{equation}
  t\to \lambda^3\,t \,,\qquad \qquad |x|\to \lambda |x|\,,
\end{equation}
with dynamical exponent $z=3$, replacing the UV scaling $\{t,\,|x|\}\to \lambda \{t,\,|x|\}$. Since these systems are believed to be strongly interacting, it is an extremely challenging theoretical task to provide a proper explanation for them. It is this strongly coupled character and the knowledge of the relevant symmetries that suggest to search for similar behaviour in "holographic dual" descriptions.

The general set-up, based on the discussions appearing, among others, in \cite{Karch:2007pd,Myers:2008me,Hartnoll:2009ns}, is as follows. One considers a small set of charged degrees of freedom, provided by the probe "flavour" brane, interacting among themselves and with a larger set of neutral quantum critical degrees of freedom having Lifshitz scale invariance with dynamical critical exponent $z$. As in previous applications, the latter is replaced by a gravitational holographic dual with Lifshitz asymptotics \cite{Kachru:2008yh}
\begin{equation}
\label{eq:IRmetric}
ds^2_{{\rm IR}} = L^2 \left( - \frac{dt^2}{v^{2z}} + \frac{dv^2}{v^2} + \frac{dx^2 + dy^2}{v^2} \right) \,,
\end{equation}
where $v$ will play the role of the holographic radial direction. Turning on non-trivial temperature corresponds to considering black holes having the above asymptotics \cite{Danielsson:2009gi,Mann:2009yx,Bertoldi:2009vn,Balasubramanian:2009rx}
\begin{equation}
\label{eq:IRmetricT}
ds^2_{{\rm IR}} = L^2 \left( - \frac{f(v) dt^2}{v^{2z}} + \frac{dv^2}{f(v) v^2} + \frac{dx^2 + dy^2}{v^2} \right) \,,
\end{equation}
where the function $f(v)$ depends on the specific solution and characterises the thermal nature of the system.

In practice, one embeds the probe "flavour" brane in the spacetime holographic dual, which may include some non-trivial cycle wrapping in internal dimensions when embedded in string theory, and turns on some non-trivial electric $(\Phi(v))$ and magnetic fluxes $(B)$ on the brane
\begin{equation}
V = \Phi(v) dt + B x dy \,.
\label{eq:A1}
\end{equation}
At low energies and in a quantum critical system, the only available scales are external, i.e. given by temperature $T$, electric and magnetic fields $\{E,\,B\}$ and the density of charge carriers $J^t$. Solving the classical equations of motion for the world volume gauge field, allows to integrate $\Phi(v)$, whose constant behaviour at infinity, i.e. at $v\to 0$ in the above coordinate system, defines the chemical potential $\mu$ of the system.  Working in an ensemble of fixed charge carrier density $J^t$, which is determined by computing the variation of the action with respect to $\delta V_t^{(0)}=\delta \mu$, the free energy density $f$ is given by
\begin{equation}
f \equiv \frac{F}{{\rm Vol}_2} = \frac{TS_{\rm Dp}}{{\rm Vol}_2} +\mu J^t\,,
\label{eq:bfreeenergy}
\end{equation}
where ${\rm Vol}_2$ stands for the volume of the non-compact 2-space spanned by $\{x,\,y\}$ and $S_{\rm Dp}$ is the on-shell Dp-brane action. As in any thermodynamic system, observables such as specific heat or magnetic susceptibility can be computed from (\ref{eq:bfreeenergy}) by taking appropriate partial derivatives. Additionally, transport observables, such 
such as DC, AC or DC Hall conductivities can also be computed and studied as a function of the background, probe embedding and the different constants controlling the world volume gauge field (\ref{eq:A1}). 

More than the specific physics, which is nicely described in  \cite{Karch:2007pd,Myers:2008me,Hartnoll:2009ns}, what is important to stress, once more, is that using the appropriate backgrounds, exciting the relevant degrees of freedom and considering the adequate boundary conditions make the methods described in this review an extremely powerful tool to learn about physics in regimes of parameters that would otherwise be very difficult to handle, both analytically and conceptually.

\section{Multiple branes}
\label{sec:nonabelian}

The physics of multiple overlapping branes provides a connection between {\it brane physics} and {\it non-abelian} supersymmetric field theories. Thus, it has played a crucial role in the geometrisation of the latter and the interplay between string and field theory dualities.

An heuristic argument suggesting that the abelian description may break down comes from the analysis of BIons. All half-BPS probe branes described in this review feel {\it no force} when probing the background describing $N-1$ parallel branes of the same nature \cite{Tseytlin:1996hi}. This means they can sit at any distance $\ell$. Consider a Dp-brane in the background of $N-1$ parallel Dp-branes. As soon as the probe approaches the location of the Dp-branes sourcing the geometry, the properly regularised mass of the open string (BIon) stretching between the probe D-brane and the background D-branes will tend to zero \cite{Gauntlett:1999xz}. This suggests the potential emergence of extra massless modes in the spectrum of these open strings. If so, this would signal a breakdown in the effective action, since these extra modes were not included in the former. U-duality guarantees that similar considerations apply to other brane set-ups not having a microscopic theory where to test this phenomena.

In this section, I will briefly discuss the supersymmetric effective actions describing $N$ coincident Dp-branes and M2-branes in a Minkowski background. These correspond to non-abelian superYang-Mills (SYM) theories in different dimensions and certain d=3 superconformal field theories with non-dynamical gauge fields having Chern-Simons actions, respectively.

\subsection{D-branes}
\label{sec:mdbrane}

The perturbative description of D-branes in terms of opens strings \cite{Polchinski:1995mt} allows to answer the question regarding the enhancement of massless modes raised above in a firmer basis, at least at weak coupling. Consider the spectrum of open strings in the presence of two parallel Dp-branes separated by a physical distance $\ell$. As the latter approaches zero, i.e. it becomes smaller than the string scale, there is indeed an enhancement in the number of massless modes. Its origin is in the sector of open strings stretching between D-branes, which is precisely the one captured by the BIon argument. This enhancement  is consistent with an enhancement in the gauge symmetry from $\U(1)\times \U(1)$, corresponding to the two separated D-branes, to $\U(2)$, corresponding to the overlapping D-branes. The spectrum of massless excitations is then described by a {\it non-abelian} vector supermultiplet in the adjoint representation. To understand how this comes about, consider the set of massless scalar excitations. These are described by $(X^i)_{rs}$,  where $i$ labels the transverse directions to the brane, as in the abelian discussion, and the subindeces $r,s$ label the D-branes where the open strings are attached. This is illustrated in figure \ref{fig3b}. Since the latter are oriented, there exist $N^2-N$ such excitations, which arrange themselves into a matrix $X^i = X^i{}_aT^a$, with $T^a$ being generators of $\U(2)$ in the adjoint representation. The conclusion is valid for any number $N$ of D-branes of world volume dimension $p+1$ \cite{Witten:1995im}. 

\epubtkImage{non-abelian.png}{%
\begin{figure}[h]
  \centerline{\includegraphics[width=150mm]{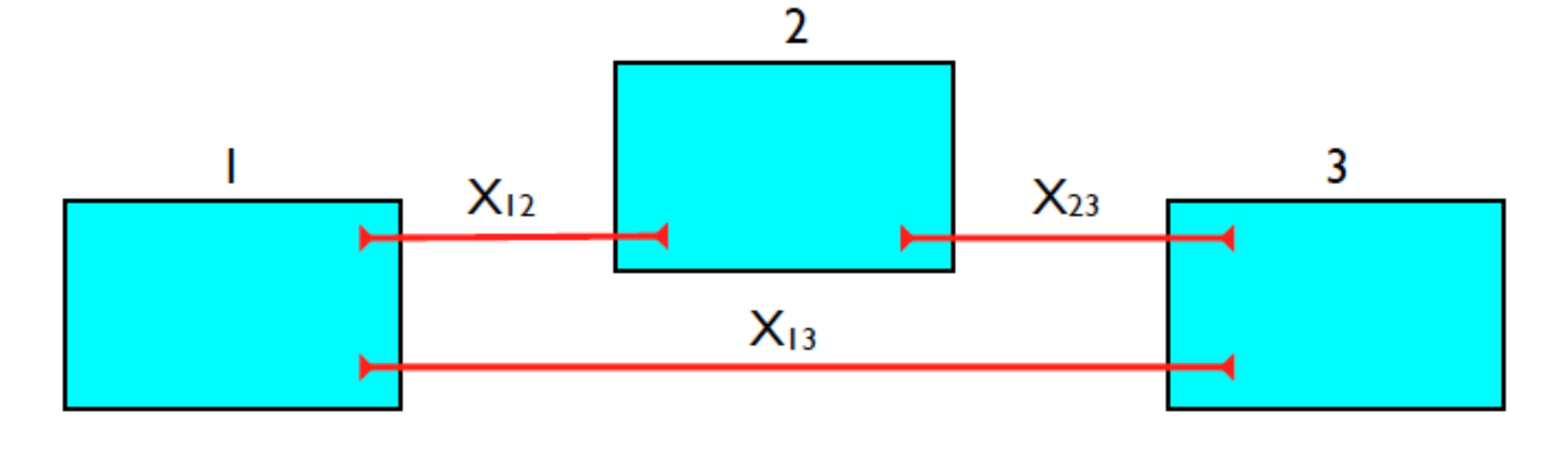}}
   \caption{Open strings stretched between multiple branes and their matrix representation.}
  \label{fig3b}
\end{figure}}

\paragraph{SuperYang-Mills action :} The previous discussion identifies the appropriate degrees of freedom to describe the low energy dynamics of multiple D-branes in Minkowski at weak coupling as non-abelian vector supermultiplets. Thus, multiple brane effective actions must correspond to supersymmetric non-abelian gauge field theories in p+1 dimensions. 
At lowest order in a derivative expansion, these are precisely SuperYang-Mills (SYM) theories. For simplicity of notation, let me focus on d=10 $\U(N)$ SYM with classical action
\begin{equation}
S =\int d^{10}\sigma \; \left( -\frac{1}{4} {\rm Tr}\; F_{\mu \nu}F^{\mu \nu}
+\frac{i}{2}  {\rm Tr}\; \bar{\psi} \Gamma^\mu D_{\mu} \psi \right)
\label{eq:SYM}
\end{equation}
where the field strength
\begin{equation}
F_{\mu \nu} = \partial_\mu A_\nu -\partial_\nu A_\mu -i g_{YM}[A_\mu, A_\nu]
\end{equation}
is the curvature of a $U(N)$ hermitian gauge field $A_\mu$ and $\psi$ is a 16-component Majorana-Weyl spinor of $\SO(1,9)$.
Both fields, $A_\mu$ and $\psi$, are in the adjoint representation of $U(N)$. The covariant derivative $D_\mu$ of $\psi$ is given by
\begin{equation}
D_\mu \psi = \partial_\mu \psi -i g_{\rm YM}[A_\mu, \psi]
\end{equation}
where $g_{{\rm YM}}$ is the Yang-Mills coupling constant. This action is also usually written in terms of rescaled fields, by absorbing a factor of $g_{\rm YM}$ in both $A_\mu$ and $\psi$, to pull an overall coupling constant dependence in front of the full action
\begin{equation}
S =\frac{1}{4g_{\rm YM}^2} \int d^{10}\sigma \; \left( - {\rm Tr}\;
F_{\mu \nu}F^{\mu \nu} +2i {\rm Tr}\; \bar{\psi} \Gamma^\mu
D_{\mu} \psi \right),
\label{eq:rgym}
\end{equation}
where $D_\mu\psi = \partial_\mu \psi -i[A_\mu\,,\psi]$.

The action (\ref{eq:SYM}) is invariant under the supersymmetry transformation
\begin{eqnarray}
\delta A_\mu & = & \frac{i}{2} \bar{\epsilon} \Gamma_\mu \psi \\
\delta \psi & = & -\frac{1}{4}  F_{\mu \nu} \Gamma^{\mu \nu} \epsilon
\label{eq:Yang-Mills-SUSY}
\nonumber
\end{eqnarray}
where $\epsilon$ is a constant Majorana-Weyl spinor in $\SO(1,9)$, giving rise to 16 independent supercharges. Classically, this is a well defined theory; quantum mechanically, it is anomalous. From the string theory perspective, as explained in section \ref{sec:validity}, this is just an effective field theory, valid at low energies $E\sqrt{\alpha^\prime}\ll 1$ and weak coupling $g_s\ll 1$.

\paragraph{Dimensional reduction :} The low energy effective action for multiple parallel Dp-branes in Minkowski is SYM in p+1 dimensions. This theory can be obtained by dimensional reduction of the ten-dimensional  super Yang-Mills theory introduced above. Thus, one proceeds as described in subsection \ref{sec:ccheck} : assume all fields are independent of coordinates $\sigma^{p + 1}, \ldots, \sigma^9$.  After dimensional reduction, the ten dimensional gauge field $A_\mu$ decomposes into a $(p +
1)$-dimensional gauge field $A_\alpha$ and $9-p$ adjoint scalar fields $X^I=2\pi\alpha'\Phi^I$\footnote{$\Phi^I$ is the natural adjoint scalar field after dimensional reduction. The rescaling by $2\pi\alpha^\prime$ is to match the natural scalar fields appearing in the abelian description provided by the DBI action. A similar rescaling occurs for the fermions omitted below, $\Psi=2\pi\alpha^\prime \psi$.}, describing the transverse fluctuations of the D-branes. The reduced action takes the form
\begin{equation}
S = \frac{1}{4g_{\rm YM}^2}   \int d^{p + 1} \sigma \;
{\rm Tr}\;(-F_{\alpha \beta} F^{\alpha \beta}-2(D_\alpha \Phi^I)^2
+[\Phi^I, \Phi^J]^2 +{\rm fermions}).
\label{eq:reduced}
\end{equation}
The p+1 dimensional YM coupling $g_{\rm YM}^2$ can be fixed by matching the expansion of the square root in the gauge fixed abelian D-brane action in a Minkowski background (\ref{superdbi}}) and comparing it with Maxwell's theory in the field normalisation used in (\ref{eq:rgym})
\begin{equation}
g_{{\rm YM}}^2 = \frac{1}{4 \pi^2 \alpha'^2 T_{\rm Dp}} 
= \frac{g_s}{\sqrt{\alpha'}}  (2 \pi \sqrt{\alpha'})^{p-2}.
\label{eq:ym-coupling}
\end{equation}
Notice also the appearance of a purely non-abelian interaction term in (\ref{eq:reduced}), the commutator $[\Phi^I, \Phi^J]^2$ that acts as a potential term. Indeed, its contribution is negative definite since $[\Phi^I,\Phi^J]^\dagger = [\Phi^J,\Phi^I]=-[\Phi^I,\Phi^J]$.

The classical vacuum corresponds to static configurations minimising the potential. This occurs when both the curvature $F_{\alpha\beta}$ and the fermions vanish, and for a set of commuting $\Phi^I$ matrices, at each point of the p+1 world volume.
In this situation, the fields $\Phi^I$ can be simultaneously diagonalised, so that one has
\begin{equation}
\Phi^I = \left(\begin{array}{cccc}
x^I_1 & 0 &0 &  \ddots\\
0 & x^I_2 &\ddots & 0\\
0 & \ddots & \ddots & 0\\
 \ddots & 0 & 0 & x^I_N
\end{array}\right)
\end{equation}
The $N$ diagonal elements of the matrix $\Phi^I$ are interpreted as the positions of $N$ distinct D-branes in
the $I$-th transverse direction \cite{Witten:1995im}.  Consider a vacuum describing $N-1$ overlapping Dp-branes and a single parallel D-brane separated in a transverse direction $\Phi$. This is equivalent to breaking the symmetry group to $\U(N-1)\times \U(1)$ by choosing a diagonal matrix for $\Phi$ with $x_0$ eigenvalue in the first $N-1$ diagonal entries and $x_N\neq x_0$ in the last diagonal entry. The off-diagonal components $\delta\Phi$ will acquire a mass, through the Higgs mechanism. This can be computed by expanding the classical action around the given vacuum. One obtains this mass is proportional
to the distance $|x_0-x_N|$ between the two sets of branes
\begin{equation}
  M^2 = \frac{(x_0- x_N)^2}{2\pi\alpha^\prime}, 
\end{equation}
according to the geometrical interpretation given to the eigenvalues characterising the vacuum. In the light of the open string interpretation, these off-diagonal components do precisely correspond to the open strings stretching between the different D-branes. The latter allow an alternative description in terms of the BIon configurations described earlier, by replacing the $N-1$ Dp-branes by its supergravity approximation, though the latter is only suitable at large distances compared to the string scale.

It can then be argued that the moduli space of classical vacua for $(p + 1)$-dimensional SYM is 
\begin{equation}
\frac{(\bR^{9-p})^N}{S_N}. 
\end{equation}
Each factor of $\bR$ stands for the position of the $N$ D-branes in
the $(9-p)$-dimensional transverse space, whereas the symmetry group $S_N$ is
the residual Weyl symmetry of the gauge group. The latter exchanges D-branes, indicating
they should be treated as indistinguishable objects.

A remarkable feature of this D-brane description is that a classical geometrical interpretation of D-brane configurations is only available when the matrices $\Phi^I$ are simultaneously diagonalisable. This provides a rather natural venue for non-commutative geometry to appear in D-brane physics at short distances, as first pointed out in \cite{Witten:1995im}.

The exploration of further kinematical and dynamical properties of these actions is beyond the scope of this review. There are excellent reviews on the subject, such as \cite{Polchinski:1996na,Taylor:1997dy,Johnson:2000ch}, where the connection to Matrix Theory \cite{Banks:1996vh} is also uncovered. If the reader is interested in understanding how T-duality acts on non-abelian D-brane effective actions, see \cite{Taylor:1996ik,Ganor:1996zk}. It is also particularly illuminating, especially for readers not used to the AdS/CFT philosophy, to appreciate that by integrating out N-1 overlapping D-branes at one loop, one is left with an abelian theory describing the remaining (single) D-brane. The effective dynamics so derived can be reinterpreted as describing a single D-brane in the background generated by the integrated N-1 D-branes, which is AdS${}_5\times$S${}^5$ \cite{Maldacena:1997kk}\footnote{There is a lot of work in this direction. For a review on the emergence of geometry and gravity in matrix models, in particular in the context of the IKKT matrix conjecture \cite{Ishibashi:1996xs}, see \cite{Steinacker:2010rh}. For more recent discussions, see \cite{Blaschke:2011qu}.}. This is illustrated in figure \ref{fig3bb}.

\epubtkImage{int-out.png}{%
\begin{figure}[h]
  \centerline{\includegraphics[width=150mm]{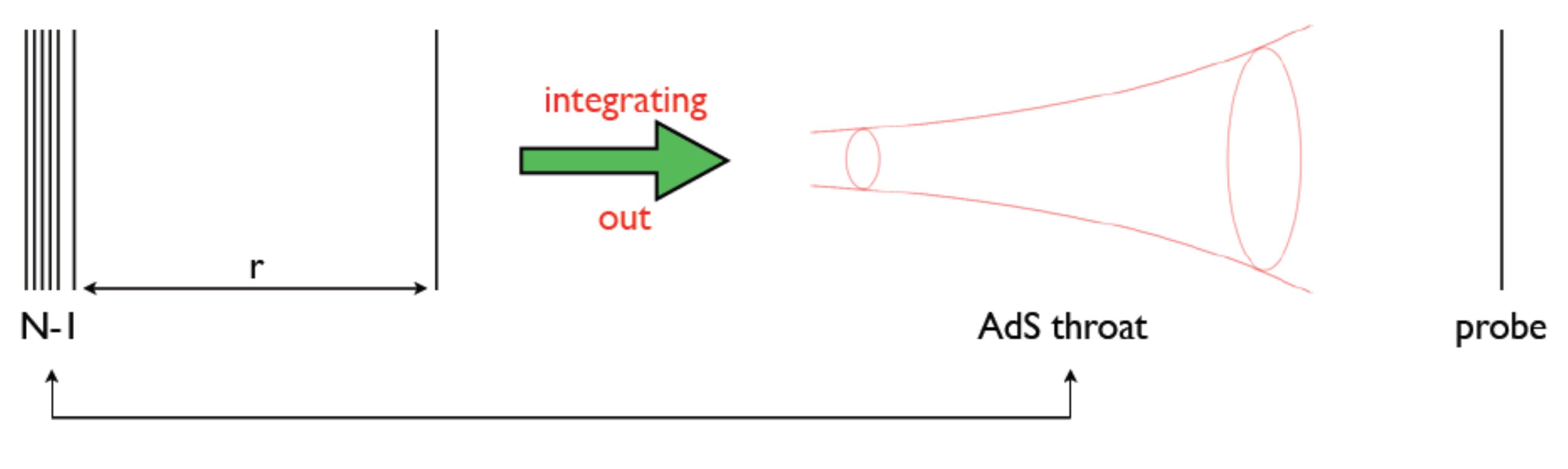}}
   \caption{Integrating out the degrees of freedom at one loop corresponding to N-1 of the D-branes gives rise to an effective action interpretable as an abelian gauge theory in an AdS throat.}
  \label{fig3bb}
\end{figure}}

Given the kinematical perspective offered in this review and the relevance of the higher order $\alpha^\prime$ corrections included in the abelian DBI action, I want to discuss two natural stringy extensions of the SYM description
\begin{itemize}
\item[1.] Keeping the background fixed, i.e. Minkowski, it is natural to consider the inclusion of higher order corrections in the effective action, matching the perturbative scattering amplitudes computed in the CFT description of open strings theory, and
\item[2.] Allowing to vary the background or equivalently, coupling the non-abelian degrees of freedom to curved background geometries. This is towards the direction of achieving a hypothetical covariant formulation of these actions, a natural question to ask given its relevance for the existence of the kappa invariant formulation of abelian D-branes.
\end{itemize}
In the following, I shall comment on the progress and the important technical and conceptual difficulties regarding the extensions of these non-abelian effective actions.

\paragraph{Higher order corrections :} In the abelian theory, it is well known that the DBI action captures {\it all} the higher order corrections in $\alpha^\prime$ to the open string effective action in the absence of field strength derivative terms\footnote{Using T-duality arguments this would also include acceleration and higher derivative corrections in the scalar sector $X^m$ describing the excitations of the D-brane along the transverse dimensions.} \cite{Fradkin:1985qd}. It was further pointed that such derivative corrections were compatible with a DBI expansion by requiring conformal invariance for the bosonic string in \cite{Abouelsaood:1986gd} and for the superstring in \cite{Bergshoeff:1987at}. 

In the non-abelian theory, such distinction is ambiguous due to the identity 
\begin{equation}
[D_\mu,\,D_\nu] F_{\rho\sigma} = [F_{\mu\nu},\,F_{\rho\sigma}],
\end{equation}
relating commutators with covariant derivatives. It was proposed by Tseytlin \cite{Tseytlin:1986ti} that the non-abelian extension of SYM including higher order $\alpha'$ corrections was given in terms of the {\it symmetrised} prescription. The latter consists on treating all $F_{\mu\nu}$ matrices as commuting. Equivalently, the action is completely symmetric in all monomial factors of F of the form ${\rm tr}(F\dots F)$. This reproduces the $F^2$ and $\alpha'^2 F^4$ terms of the full non-abelian action, but extends it to higher orders
\begin{equation}
  {\cal L}_{{\rm DBI}} \propto {\rm Str} \sqrt{\eta_{\mu\nu} + 2\pi\alpha' F_{\mu\nu}}.
\end{equation}
The notation ${\rm Str}$ defines this notion of symmetrised trace for each of the monomials appearing in the expansion of its arguments. For an excellent review describing the history of these calculations, motivating this prescription and summarising the most relevant properties of this action, see \cite{Tseytlin:1999dj}.

It is important to stress that, a priori, worldsheet calculations involving an arbitrary number of boundary disk insertions could determine this non-abelian effective action. Since this is technically hard, one can perform other consistency checks. For example, one can compare the D-brane BPS spectrum on tori in the presence of non-trivial magnetic fluxes. This is T-dual to intersecting D-branes, whose spectrum can be independently computed and compared with the fluctuation analysis of the proposed symmetrised non-abelian prescription. It was found in \cite{Hashimoto:1997gm,Denef:2000rj,Sevrin:2001ha} that the proposed prescription was breaking down at order $(\alpha')^4F^6$. Further checks at order $\alpha'^3$ and $\alpha'^4$ were carried over in \cite{Bilal:2001hb,Koerber:2001hk,Koerber:2001uu,Koerber:2002zb}. The proposal in \cite{Koerber:2001hk} was confirmed by a first principle five-gluon scattering amplitude at tree level in \cite{Medina:2002nk}. The conclusion is that the symmetrised prescription only works up to $F^4$
\begin{eqnarray}
 {\cal L} &=&  {\rm Str} \big[\frac{1}{4} F^2_{\mu\nu}  -  \frac{1}{8} (2\pi \alpha')^2
 \big(F^4 - \frac{1}{4} (F_{\mu\nu}^2)^2\big) + O(\alpha'^4) \big]  \\
  &=& {\rm tr} \big[
\frac{1}{4} F^2_{\mu\nu} -  \frac{1}{12}(2\pi\alpha')^2 \big(F_{\mu\nu}F_{\rho\nu}F_{\mu\lambda}F_{\rho\lambda} +
\frac{1}{2} F_{\mu\nu}F_{\rho\nu}F_{\rho\lambda} F_{\mu\lambda} \nonumber \\
& & - \  \frac{1}{4} F_{\mu\nu}F_{\mu\nu}F_{\rho\lambda}F_{\rho\lambda}
-\frac{1}{8} F_{\mu\nu}F_{\rho\lambda}F_{\mu\nu}F_{\rho\lambda} \big)
  + O(\alpha'^4) \big] \,.
\end{eqnarray}
These couplings were first found in its ${\rm Str}$ form in \cite{Gross:1986iv} and in its ${\rm tr}$ form in \cite{Tseytlin:1986ti}.
For further checks on Tseytlin's proposal using the existence of bound states and BPS equations, see the analysis in \cite{Brecher:1998su,Brecher:1998tv}.

\paragraph{Coupling to arbitrary curved backgrounds :} The above corrections attempted to include higher order corrections describing the physics of multiple D-branes in Minkowski. More generally, one is interested in coupling D-branes to arbitrary closed string backgrounds. In such situation, one would like to achieve a {\it covariant} formulation. This is non-trivial because as soon as the degrees of freedom become non-abelian, they lose their geometrical interpretation.
In the abelian case, $X^I$ described the brane location. In the non-abelian case, at most, only their eigenvalues $x_i^I$ may keep their interpretation as the location of the ith brane in the Ith direction. Given the importance and complexity of the problem, it is important to list a set of properties that one would like such formulation to satisfy. These are the so called D-geometry axioms \cite{Douglas:1997zw}. For the case of D0-branes, these are
\begin{itemize}
\item[1.] It must contain a {\it unique} trace since this is an effective action derived from string theory disk diagrams involving many graviton insertions in their interior and scalar/vector vertex operators on their boundaries. Since the disk boundary is unique, the trace must be unique.
\item[2.] It must reduce to N-copies of the particle action when the matrices $X^I$ are diagonal.
\item[3.] It must yield masses proportional to the geodesic distance for off-diagonal fluctuations.
\end{itemize}
Having in mind that we required spacetime gauge symmetries to be symmetries of the abelian brane effective actions, it would be natural to include in the above list invariance under target space diffeomorphisms. This was analysed for the effective action kinetic terms in \cite{DeBoer:2001uk}. Instead of discussing this here, I will discuss two non-trivial checks that any such formulation must satisfy
\begin{itemize}
\item[a)] to match the Matrix theory linear couplings to closed string backgrounds, and
\item[b)] to be T-duality covariant, extending the notion I discussed in section \ref{sec:tcov} for single D-branes.
\end{itemize}
The first was studied in \cite{Taylor:1999gq,Taylor:1999pr} and the second in \cite{Myers:1999ps}. Since the results derived from the latter turned out to be consistent with the former, I will focus on the implementation of T-duality covariance for non-abelian D-branes below.

As discussed in subsection \ref{sec:ccheck}, T-duality is implemented by a dimensional reduction. This was already applied for SYM in (\ref{eq:reduced}). Using the same notation introduced there and denoting the world volume direction along which one reduces by $\rho$, one learns that $F_{\mu\rho}\ \to\ D_\mu\Phi^p$, where $\Phi^p$ is the T-dual adjoint matrix scalar. Furthermore, covariant derivatives of transverse scalar fields $\Phi^I$ become
\begin{equation}
D_\rho \Phi^I=\partial_\rho\Phi^I+i[A_\rho,\Phi^I]=i[A_\rho,\Phi^I].
\label{nonsimple}
\end{equation}
Notice this contribution is purely non-abelian and it can typically contribute non-trivially to the potential terms in the effective action. To properly include these non-trivial effects, Myers \cite{Myers:1999ps} studied the consequences of requiring T-duality covariance taking as a starting point a properly covariantised version of the multiple D9-brane effective action, having assumed the symmetrised trace prescription described above. Studying T-duality along 9-p directions and imposing T-duality covariance of the resulting action, will generate all necessary T-duality compatible commutators, which would have been missed otherwise. This determines the DBI part of the effective action to be \cite{Myers:1999ps}
\begin{equation}
S_{{\rm DBI}}=-T_{\rm Dp} \int d^{p+1}\sigma\,{\rm STr} \left(e^{-\phi}\sqrt{-{\rm det}\left(
P\left[E_{\mu\nu}+E_{\mu I}(Q^{-1}-\delta)^{IJ}E_{J\nu}\right]+
\lambda\,F_{\mu\nu}\right)\,{\rm det}(Q^I{}_J)}
\right)\ ,
\label{finalbi}
\end{equation}
with 
\begin{equation}
E_{\mu\nu}=g_{\mu\nu}+B_{\mu\nu}\,,
\qquad \qquad
Q^I{}_J\equiv\delta^I{}_J+i\lambda\,[\Phi^I,\Phi^K]\,E_{KJ}, \qquad {\rm and} \qquad \lambda =2\pi\alpha^\prime.
\end{equation}
Here $\mu,\nu$ indices stand for world volume directions, whereas $I,J$ indices for transverse directions.
To deal with similar commutators arising from the WZ term, one considers \cite{Myers:1999ps} 
\begin{equation}
S_{{\rm WZ}}=T_{\rm Dp}\int {\rm STr}\left(P\left[e^{i\lambda\,{\rm i}_\Phi {\rm i}_\Phi} (
\sum C^{(n)}\,e^B)\right] e^{\lambda\,F}\right)\ .
\label{finalcs}
\end{equation}
where the interior product ${\rm i}_\Phi$ is responsible for their appearance, for example as in,
\begin{equation}
{\rm i}_\Phi {\rm i}_\Phi C_{2} = \Phi^J\Phi^I\,C^{2}_{IJ}=\frac{1}{2}C^{2}_{IJ}
\,[\Phi^J,\Phi^I]\,.
\label{nonabint}
\end{equation}
Notice one regards $\Phi^I$ as a vector field in the transverse space. In both actions (\ref{finalbi}) and (\ref{finalcs}), $P$ stands for pullback and it only applies to transverse brane directions since all longitudinal ones are non-physical. Its presence is confirmed by scattering amplitudes calculations \cite{Klebanov:1997kc,Gubser:1997yh,Garousi:1998fg}. Some remarks are in order
\begin{itemize}
\item[1.] There exists some non-trivial dependence on the scalars $\Phi^I$ through the arbitrary bosonic closed backgrounds appearing in the action. The latter is defined according to
\begin{eqnarray}
g_{\mu\nu}&=&\exp\left[\lambda\Phi^i\,{\partial_{x^i}}\right]g^0_{\mu\nu}
(\sigma^a,x^i)|_{x^i=0}
\label{slick}\\
&=&\sum_{n=0}^\infty \frac{\lambda^n}{n!}\,\Phi^{i_1}\cdots\Phi^{i_n}\,
(\partial_{x^{i_1}}\cdots\partial_{x^{i_n}})g^0_{\mu\nu}
(\sigma^a,x^i)|_{x^i=0}\, .
\nonumber
\end{eqnarray}
Analogous definitions apply to other background fields.
\item[2.] There exists a unique trace, because this is an open string effective action that can be derived from worldsheet disk amplitudes. The latter has a unique boundary. Thus, there must be a unique gauge trace \cite{Douglas:1997zw,Douglas:1997sm}. Above, the symmetrised prescription was assumed, not only because one is following Tseytlin and this was his prescription, but also because there are steps in the derivation of T-duality covariance that assumed this property and the scalar field $\Phi^I$ dependence on the background fields (\ref{slick}) is symmetric, by definition.
\item[3.] The WZ term (\ref{finalcs}) allows multiple Dp-branes to couple to \RR\ potentials with a form degree greater than the dimension of the world-volume. This is a purely non-abelian effect whose consequences will be discussed below.
\item[4.] There are different sources for the scalar potential : ${\rm det}\,Q^I{}_J$, its inverse in the first determinant of the DBI and contributions coming from commutators coupling to background field components in the expansion (\ref{slick}).
\end{itemize}

It was shown in detail in \cite{Myers:1999ps}, that the bosonic couplings described above were consistent with all the linear couplings of closed string background fields with Matrix Theory degrees of freedom, i.e. multiple D0-branes. These couplings were originally computed in \cite{Taylor:1999gq} and then extended to Dp-branes in \cite{Taylor:1999pr} using T-duality once more. We will not review this check here in detail, but as an illustration of the above formalism, present the WZ term for multiple D0-branes that is required to do such matching
\begin{eqnarray}
S_{{\rm WZ}}&=&\mu_0\int {\rm Tr}\left(P\left[C_{1}+
i\lambda\,{\rm i}_\Phi {\rm i}_\Phi\left(C_{3}+C_{1}\wedge B\right) \right.\right.
\label{cszero}\\
&&\qquad -\frac{\lambda^2}{2}({\rm i}_\Phi {\rm i}_\Phi)^2\left(C_{5}
+C_{3}\wedge B+\frac{1}{2}C_{1}\wedge B \wedge B\right)
\nonumber\\
&&\qquad -i\frac{\lambda^3}{6}({\rm i}_\Phi {\rm i}_\Phi)^3\left(
C_{7}+C_{5}\wedge B +\frac{1}{2}C_{3}\wedge B\wedge B +\frac{1}{6}C_{1}\wedge B \wedge B \wedge B\right)
\nonumber\\
&&\left.\left.+\frac{\lambda^4}{24}({\rm i}_\Phi {\rm i}_\Phi)^4\left(
C_{9}+\left(C_{7}+\frac{1}{2}C_{5}\wedge B
+\frac{1}{6}C_{3}\wedge B \wedge B + \frac{1}{24}C_{1}\wedge B \wedge B \wedge B\right) \wedge B\right)\right]\right)
\nonumber\\
&=&\mu_0\int dt\, {\rm Tr}\left(C_t^{1}+\lambda\,C_I^{1}D_t\Phi^I
+i\frac{\lambda}{2}(C_{tJK}^{3}\,[\Phi^K,\Phi^J]+\lambda\,C_{IJK}^{3}
\,D_t\Phi^I\,[\Phi^K,\Phi^J]) +\ldots \right)
\nonumber
\end{eqnarray}
Two points are worth emphasising about this matching :
\begin{itemize}
\item[1.] There is {\it no} ambiguity of trace in the {\it linear} Matrix theory calculations. Myers' suggestion is to extend this prescription to non-linear couplings.
\item[2.] Some transverse M5-brane charge couplings are unknown in Matrix theory, but these are absent in the lagrangian above. This is a prediction of this formulation.
\end{itemize}

One of the most interesting physical applications of the couplings derived above is the realisation of the {\it dielectric effect} in electromagnetism in string theory. As already mentioned above, the non-abelian nature of the degrees of freedom turns on new commutator couplings with closed string fields that can modify the scalar potential. If so, instead of the standard SYM vacua, one may find new potential minima with ${\rm Tr}\,\Phi^I=0$ but ${\rm Tr}(\Phi^I)^2\neq 0$. As a toy illustrative example of this phenomenum, consider N D0-branes propagating in Minkowski but in a constant background
\RR\ four-form field strength
\begin{equation}
R^{4}_{tIJK}=\left\lbrace\matrix{-2f \varepsilon_{IJK}&{\rm for}\ I,J,K\in
\lbrace 1,2,3\rbrace\cr
0&{\rm otherwise}\cr}
\right.
\label{backg}
\end{equation}
Due to gauge invariance, one expects a coupling of the form
\begin{equation}
\frac{i}{3}\lambda^2\mu_0\int dt\,{\rm Tr}\left(\Phi^I\Phi^J\Phi^K\right)
R^{4}_{tIJK}(t)\,.
\end{equation}
Up to total derivatives, this can indeed be derived from the cubic terms in the WZ action above. This coupling modifies the scalar potential to
\begin{equation}
V(\Phi)=-\frac{\lambda^2T_0}{4} {\rm Tr}([\Phi^I,\Phi^J]^2)
-\frac{i}{3}\lambda^2\mu_0 {\rm Tr}\left(\Phi^I\Phi^J\Phi^K\right)
R^{4}_{tIJK}(t)\,,
\label{potential}
\end{equation}
whose extremisation condition becomes
\begin{equation}
0=[[\Phi^I,\Phi^J],\Phi^K]+{i}\,f\varepsilon_{IJK}[\Phi^J,\Phi^K]\, .
\label{eqmot}
\end{equation}
The latter allows $\SU(2)$ solutions
\begin{equation}
\Phi^I=\frac{f}{2}\,\alpha^I \qquad {\rm with} \qquad [\alpha^I,\alpha^J]=2i\,\varepsilon_{IJK}\,\alpha^K\,,
\end{equation}
having lower energy than standard commuting matrices
\begin{equation}
V_{\rm N}=-\frac{\pi^2 \ell_s^3 f^4}{6g_s}N(N^2-1)\,.
\end{equation}

It is reassuring to compare the description above with the one available using the abelian formalism describing a single brane explained in section \ref{sec:bbrane}. I shall refer to the latter as dual brane description. For the particular example discussed above, since the D0-branes blow up into spheres due to the electric \RR\ coupling, one can look for on-shell configurations on the abelian D2-brane effective action in the {\it same} background corresponding to the expanded spherical D0-branes in the non-abelian description. These configurations exist, reproduce the energy $V_N$ up to $1/N^2$ corrections and carry {\it no} D2-brane charge \cite{Myers:1999ps}. Having reached this point, I am at a position to justify the expansion of pointlike gravitons into spherical D3-branes, giant gravitons, in the presence of the \RR\ flux supporting AdS${}_5\times$S${}^5$ described in section \ref{eq:giants}. The non-abelian description would involve non-trivial commutators in the WZ term giving rise to a fuzzy sphere extremal solution to the scalar potential. The abelian description reviewed in section \ref{eq:giants} corresponds to the dual D3-brane description in which by keeping the same background, one searches for on-shell spherical rotating D3-branes carrying the same charges as a pointlike graviton but
{\it no} D3-brane charge. For a more thorough discussion on the comparison between non-abelian solitons and their "dual" abelian descriptions, see \cite{Constable:1999ac,Constable:2001ag,Constable:2001kv,Myers:2003bw}.

\paragraph{Kappa symmetry and superembeddings : }  The covariant results discussed above did not include fermions. Whenever these were included in the abelian case, a further gauge symmetry was required, kappa symmetry, to keep covariance, manifest supersymmetry and describe the appropriate on-shell degrees of freedom. One suspects something similar may occur in the non-abelian case to reduce the number of fermionic degrees of freedom in a manifestly supersymmetric non-abelian formulation. It is important to stress that at this point world volume diffeomorphisms and kappa symmetry will no longer appear together. In all the discussions in this section, world volume diffeomorphisms are assumed to be fixed, in the sense that the only scalar adjoint matrices already correspond to the transverse directions to the brane.

Given the projective nature of kappa symmetry transformations, it may be natural to assume that there should be as many kappa symmetries as fermions. In \cite{Bergshoeff:2000ik}, a perturbative approach to determining such transformation
\begin{equation}
  \delta_\kappa \bar\theta^A = \bar\kappa^B(\sigma)\left(\mathbb{1}\delta^{BA} + \Gamma^{BA}(\sigma)\right)\,, \qquad \qquad A,B = 1,2,\dots N^2
\end{equation}
was analysed for multiple D-branes in SuperPoincar\'e. The idea was to expand the WZ term in covariant derivatives of the fermions and the gauge field strength $F$, involving some a priori arbitrary tensors. One then computes its kappa symmetry variation and attempts to identify the DBI term in the action at the same order by satisfying the requirement that the total action variation equals
\begin{equation}
  \delta_\kappa {\cal L} = -\delta_\kappa\bar\theta\left(1-\Gamma\right){\cal T}\,,
\end{equation}
order by order. In a sense, one is following the same strategy as in \cite{Aganagic:1996pe}, determining the different unknown tensors order by order. Unfortunately, it was later concluded in \cite{Bergshoeff:2001dc} that such approach could not work.

There exists some body of work constructing classical supersymmetric and kappa invariant actions involving non-abelian gauge fields representing the degrees of freedom of multiple D-branes. This started with actions describing branes of lower co-dimension propagating in lower dimensional spacetimes \cite{Sorokin:2001av,Sorokin:2002je,Drummond:2002kg}. It was later extended to multiple D0-branes in an arbitrary number of dimensions, including type IIA, in \cite{Panda:2003dj}. Here, both world volume diffeomorphisms and kappa symmetry were assumed to be abelian. It was checked that when the background is SuperPoincar\'e, the proposed action agreed with Matrix Theory \cite{Banks:1996vh}. Using the superembedding formalism \cite{Sorokin:1999jx}, actions were proposed reproducing the same features in \cite{Bandos:2009yp,Bandos:2009xy,Bandos:2009gk,Bandos:2010hc,Bandos:2010af}, some of them involving superparticle propagating in arbitrary eleven dimensional backgrounds. Finally, there exists a slightly different approach in which besides using the superembedding formalism, the world sheet Chan-Patton factors describing multiple D-branes are replaced by boundary fermions. The actions constructed in this way in \cite{Howe:2007eb}, based on earlier work \cite{Howe:2006rv}, have similar structure to the ones described in the abelian case, their proof of kappa symmetry invariance is analogous and they reproduce Matrix Theory when the background is SuperPoincar\'e and most of the features highlighted above for the bosonic couplings described by Myers.

\paragraph{Relation to non-commutative geometry :} There are at least two reasons why one may expect non-commutative geometry to be related to the description of multiple D-brane actions :
\begin{itemize}
\item[1.] D-brane transverse coordinates being replaced by matrices,
\item[2.] the existent non-commutative geometry description of D-branes in the presence of a $B$-field in space-time (or a magnetic field strength on the brane) \cite{Douglas:1997fm,Connes:1997cr,Seiberg:1999vs}.
\end{itemize}
The general idea behind non-commutative
geometry is to replace the space of functions by a non-commutative algebra. In the D-brane context, a natural candidate to consider would be the algebra
\begin{equation}
{\cal A} = C^{\infty}(M) \otimes M_N(C)\,.
\end{equation}
As customary in non-commutative geometry, the latter does not yet carry any metric information. Following Connes \cite{Connes:1994yd}, the construction of a Riemannian structure requires a spectral triple $({\cal A},{\cal H},D)$
which in addition to ${\cal A}$ also contains a Hilbert space ${\cal H}$ and a self-adjoint operator $D$ obeying certain
properties. It would be interesting to find triples $({\cal A},{\cal H},D)$ that describe in a natural
way metrics relevant for multiple D-branes, incorporating the notion of covariance. 

Regarding D-branes in the presence of a B-field, the main observation is that the structure of an abelian non-commutative
gauge theory is similar to that of a non-abelian commutative gauge theory. In both cases, fields no longer commute,
and the field strengths are non-linear. Moreover, non-commutative
gauge theories can be constructed starting from a non-abelian
commutative theory by expanding around suitable backgrounds and
taking $N\to \infty$ \cite{Seiberg:2000zk}. This connection suggests it may be
possible to relate the gravity coupling of non-commutative gauge theories
to the coupling of non-abelian D-brane actions to curved backgrounds (gravity). 
This was indeed the approach taken in \cite{Das:2000ur} where the stress-tensor of
non-commutative gauge theories was derived in this way. In \cite{Cornalba:2000ua}, constraints on the kinematical properties of non-abelian D-brane actions due to this connection were studied.

\subsection{M2-branes}
\label{sec:mm2}

In this section, I would like to briefly mention the main results involving the amount of progress recently achieved in the description of N parallel M2-branes, referring to the relevant literature when appropriate. This will be done taking the different available perspectives on the subject : a purely kinematic approach, based on supersymmetry and leading to {\it 3-algebras}, a purely field theory approach leading to three dimensional CFTs involving Chern-Simons terms, a brane construction approach, in which one infers the low energy effective description in terms of an intersection of branes and the connection between all these different approaches.

The main conclusion is that the effective theory describing N M2-branes is a $d=3$, $\U(N) \times \U(N)$ gauge
theory with four complex scalar fields $C_I$ ($I=1,2,3,4$) in the $\bf(N,\bar N)$ representation, their complex conjugate fields in the $\bf(\bar N , N)$ representation and their fermionic partners \cite{Aharony:2008ug}. The theory includes non dynamical gauge fields with a Chern-Simons action with levels $k$ and $-k$ for the two gauge groups. This gauge theory is weakly coupled in the large $k$ limit ($k \gg N$) and strongly coupled in the opposite regime $(k\ll N)$, for which a weakly coupled gravitational description will be available if $N\gg 1$.

\paragraph{Supersymmetry approach :} Inspection of the d=3 SYM supersymmetry transformations and the geometrical intuition coming from M2-branes suggest to look for a supersymmetric field theory with field content involving eight scalar fields $X^I=X^I_a T^a$\footnote{Eight is the number of transverse dimensions to the world volume of the M2-branes.} and their fermionic partners 
$\Psi = \Psi_a T^a$, and being invariant under a set of supersymmetry transformations whose most general form is 
\begin{eqnarray}\label{susygauged}
\nonumber \delta X^I_d &=& i\bar\epsilon\Gamma^I\Psi_d,\\
\nonumber \delta \Psi_d &=& \partial_\mu X^I_d\Gamma^\mu \Gamma^I\epsilon
-\frac{1}{6} X^I_aX^J_bX^K_c f^{abc}{}_{d}\Gamma^{IJK}\epsilon
+\frac{1}{2}X^J_aX^J_bX^I_c g^{abc}{}_d \Gamma^I\epsilon\,.
\end{eqnarray}
This was the original approach followed in \cite{Bagger:2006sk}, based on a {\it real} vector space with basis $T^a$, $a=1,\dots N$, endowed with a triple product
\begin{equation}
 [T^a,T^b,T^c] = f^{abc}{}_{d}\,T^d,
\end{equation} 
where the set of $f^{abc}{}_d$ are real, fully antisymmetric in $a,b,c$ and satisfy the fundamental identity
\begin{equation}\label{otherFI}
f^{[abc}{}_{e}f^{d]ef}{}_{g}=0\, .
\end{equation}
Closure of the supersymmetry algebra requires (\ref{otherFI}), but also shows the appearance of an extra gauge symmetry \cite{Bagger:2006sk}. To deal properly with the latter, one must introduce an additional (non-dynamical) gauge field $\tilde A_{\mu}{}^c{}_d$ requiring to consider a more general set of supersymmetry transformations \cite{Bagger:2007jr,Gustavsson:2007vu}
\begin{eqnarray}\label{eq:susygauged}
\nonumber \delta X^I_d &=& i\bar\epsilon\Gamma^I\Psi_d\\
\nonumber \delta \Psi_d &=& D_\mu X^I_d\Gamma^\mu \Gamma^I\epsilon
-\frac{1}{6} X^I_aX^J_bX^K_c f^{abc}{}_{d}\Gamma^{IJK}\epsilon
+\frac{1}{2}X^J_aX^J_bX^I_c g^{abc}{}_d \Gamma^I\epsilon \\
\delta\tilde A_{\mu}{}^c{}_d &=& i\bar\epsilon
\Gamma_\mu\Gamma_IX^I_a\Psi_b h^{abc}{}_{d}\,.
\end{eqnarray}
Here $D_\mu$ is a covariant derivative, whereas $g^{abc}{}_d$ and $h^{abc}{}_d$ define triple products on the algebra. 

Closure of the supersymmetry algebra determines a set of equations of motion that can be derived which from a lagrangian.
It was soon realised that under the assumptions of a real vector space, essentially the only 3-algebra is the one defined by
$f^{abcd}=f^{abc}{}_{e}h^{ed}$, with $h^{ab}={\rm Tr}\left(T^a,T^b\right)$ defining an inner product, and satisfying $f^{abcd}\propto \varepsilon^{abcd}$ \cite{2007arXiv0712.1398N,Papadopoulos:2008sk,Gauntlett:2008uf}. Interestingly, it was pointed out in \cite{VanRaamsdonk:2008ft} that such supersymmetric field theory could be rewritten as a Chern-Simons theory. The latter provided a link between a purely kinematic approach, based on supersymmetry considerations, and purely field theoric results that had independently been developed.

\paragraph{Field theory considerations :} Conformal field theories have many applications. In the particular context of Chern-Simons matter theories in d=3, they can describe interesting IR fixed points in condensed matter systems. Here, I am interested
in their supersymmetric versions to explore the AdS${}_4$/CFT${}_3$ conjecture.

Let me start this overview with ${\cal N}=2$ theories. ${\cal N}=2$ Chern-Simons theories coupled to matter\footnote{For a complete list of references, see \cite{Aharony:2008ug}.} include a vector multiplet $A$, the dimensional reduction of the four dimensional ${\cal N}=1$ vector multiplet, in the adjoint representation of the gauge group $G$, and chiral multiplets $\Phi_i$ in representations $R_i$ of the latter. Integrating out the D-term equation and the gaugino, one is left with the action
\begin{eqnarray}
S^{{\cal N}=2} = \int & & \frac{k}{4\pi}
{\rm Tr}(A \wedge dA + \frac{2}{3} A^3) + D_{\mu} \bar\phi_i D^{\mu}
\phi_i + i \bar\psi_i \gamma^{\mu} D_{\mu} \psi_i \nonumber \\
 & &- \frac{16\pi^2}{k^2}(\bar\phi_i T^a_{R_i} \phi_i) (\bar\phi_j T^b_{R_j} \phi_j) (\bar \phi_k
T^a_{R_k} T^b_{R_k} \phi_k) - \frac{4\pi}{k} (\bar\phi_i T^a_{R_i} \phi_i) (\bar\psi_j T^a_{R_j} \psi_j)  \label{eq:n2} \\
 & & -\frac{8\pi}{k} (\bar\psi_i T^a_{R_i} \phi_i) (\bar\phi_j T^a_{R_j} \psi_j),\nonumber
\end{eqnarray}
where $\phi_i$ and $\psi_i$ are the bosonic and fermionic components of the chiral superfield $\Phi_i$ and the gauge field $A$ is non-dynamical.

There are ${\cal N}=3$ generalisations, but since their construction is more easily argued for starting with the field content of an ${\cal N}=4$ theory, let me review the latter first. The field content of the ${\cal N}=4$ theories adds an auxiliary (non-dynamical) chiral multiplet $\varphi$ in the adjoint representation of $G$ and pairs chiral multiplets $\Phi_i,\,\tilde{\Phi}_i$ into a set of hypermultiplets by requiring them to transform in conjugate representations, as the notation suggests. The theory does {\it not} contain Chern-Simons terms, but a superpotential $W = \tilde{\Phi}_i \varphi \Phi_i$ for each pair. ${\cal N}=3$ theories are constructed by the addition of Chern-Simons terms, as in (\ref{eq:n2}), and the extra superpotential $W = -\frac{k}{8\pi} {\rm Tr}(\varphi^2)$. Integrating out $\varphi$ leads to a superpotential
\begin{equation}
W = \frac{4\pi}{k} (\tilde{\Phi}_i T^a_{R_i} \Phi_i) (\tilde{\Phi}_j T^a_{R_j} \Phi_j)\,.
\end{equation}
The resulting ${\cal N}=3$ theory has the same action as (\ref{eq:n2}) with the addition of the above superpotential.

In \cite{Aharony:2008ug}, an ${\cal N}=6$ theory based on the gauge group $\U(N)\times\U(N)$ was constructed. Its field content includes two hypermultiplets in the bifundamental and the Chern-Simons levels of the two gauge groups were chosen to be equal but opposite in sign. Denoting the bifundamental chiral superfields by $A_1, A_2$ and their anti-bifundamental by $B_1, B_2$, the superpotential then equals
\begin{equation}
W = \frac{k}{8\pi} {\rm Tr}(\varphi_{(2)}^2 - \varphi_{(1)}^2) + {\rm Tr}(B_i \varphi_{(1)} A_i)
+ {\rm Tr}(A_i \varphi_{(2)} B_i)\,.
\end{equation}
After integrating out the auxiliary fields $\varphi_{(i)}$,
\begin{equation}
W = \frac{2\pi}{k} {\rm Tr}(A_i B_i A_j B_j - B_i A_i B_j A_j) =
\frac{4\pi}{k} {\rm Tr}(A_1 B_1 A_2 B_2 - A_1 B_2 A_2 B_1).
\end{equation}
As discussed in \cite{Aharony:2008ug}, the four bosonic fields $C_I \equiv (A_1, A_2, B_1^*, B_2^*)$ transform in the ${\bf 4}$ of $\SU(4)$, matching the generic $\SO({\cal N})$ R-symmetry in $d=3$ super-CFTs. For a more thorough discussion of global symmetries and gauge invariant observables, see \cite{Aharony:2008ug}.

It was argued in \cite{Aharony:2008ug} that the ${\cal N}=6$ theory constructed above was dual to N M2-branes on $\bC^4/\bZ_k$ for $k\geq 3$. Below, I briefly review the brane construction in which their argument is based. This will provide a nice example of the notion of geometrisation (or engineering) of supersymmetric field theories provided by brane configurations.

\paragraph{Brane construction :} Following the seminal work \cite{Hanany:1996ie}, one can associate low energy effective field theories to the dynamics of brane configurations stretching between branes. Consider a set of N D3-branes wrapping the $x^6$ direction and ending on different NS5-branes according to the array 
\begin{equation}
\begin{array}{ccccccccccl}
{\rm NS}5: &1&2&3 & 4&5 &\_&\_&\_&\_ & \nn \\
{\rm NS}5: &1&2&3 &4 &\_&\_&\_&\_&\_ & \nn \\  
{\rm D}3: &1 & 2 &\_&\_&\_&6 &\_&\_&\_& \,.  
\end{array}
\label{arraym2}
\end{equation}
This gives rise to an ${\cal N}=4$ $\U(N)\times \U(N)$ gauge theory in d=1+2 dimensions, along the $\{x^1,\,x^2\}$ directions, whose field content includes a vector multiplet in the adjoint representation and 2 complex bifundamental hypermultiplets, describing the transverse excitations to both D3-branes and NS5-branes \cite{Hanany:1996ie}. 

Adding $k$ D5-branes as illustrated in the array below
\begin{equation}
\begin{array}{ccccccccccl}
{\rm NS}5: &1&2&3 & 4&5 &\_&\_&\_&\_ & \nn \\
{\rm NS}5: &1&2&3 &4 &5 &\_&\_&\_&\_ & \nn \\  
{\rm D}5: &1&2&3 &4 &\_&\_&\_&\_&9 & \nn \\  
{\rm D}3: &1 & 2 &\_&\_&\_&6 &\_&\_&\_& \,,  
\end{array}
\label{eq:arraym2}
\end{equation}
breaks supersymmetry to ${\cal N}=2$ and adds $k$ massless chiral multiplets in the ${\bf N}$ and ${\bf \bar{N}}$ representation of each of the $\U(N)$ gauge group factors. Field theorically, this ${\cal N}=2$ construction allows a set of mass deformations that can be mapped to different geometrical notions \cite{Hanany:1996ie,Bergman:1999na,Aharony:2008ug}:
\begin{itemize}
\item[1.] Moving the D5-branes along the 78-directions generates a complex mass parameter.
\item[2.] Moving the D5-branes along the 5-direction generates a real mass, of positive sign for the fields in the fundamental representation and of negative sign for the ones in the anti-fundamental.
\item[3.] Breaking the k D5-branes and NS5-branes along the 01234 directions and merging them into an intermediate (1,k) 5-brane bound state generates a real mass of the same sign for both, ${\bf N}$ and ${\bf \bar{N}}$, representations. This mechanism is a {\it web deformation} \cite{Bergman:1999na}. The merging is characterised by the angle $\theta$ relative to the original NS5-brane subtended by the bound state in the 59-plane. The final brane configuration is made of a single NS5-brane in the 012345 directions and a (1,k) 5-brane in the $01234[5,9]_\theta$, where $[5,9]_\theta$ stands for the $x^5\cos\theta + x^4\sin\theta$ direction. $\theta$ is fixed by supersymmetry \cite{Aharony:1997ju}.
\end{itemize}
After the web deformation and at low energies, one is left with an ${\cal N}=2$ $U(N)_k\times U(N)_{-k}$ Yang-Mills-Chern-Simons theory with four massless bi-fundamental matter multiplets (and their complex conjugates), and two massless adjoint
matter multiplets corresponding to the motion of the D3-branes in the directions 34 common to the two 5-branes. 

The enhancement to an ${\cal N}=3$ theory described in the purely field theoretical context is realised in the brane construction by rotating the (1,k) 5-brane in the 37 and 48-planes by the same amount as in the original deformation. Thus, one ends with
a single NS5-brane in the 012345 and a (1,k) 5-brane along $012[3,7]_\theta[4,8]_\theta[5,9]_\theta$.This particular mass deformation ensures all massive adjoint fields acquire the {\it same} mass, enhancing the symmetry to ${\cal N}=3$. Equivalently, there must exist an $\SO(3)_{\rm R}$ R-symmetry corresponding to the possibility of having the same $\SO(3)$ rotations in the 345 and 789 subspaces. Thus, the d=3 supersymmetric field theory must be ${\cal N}=3$.

The connection to ${\cal N}=6$ is obtained by flowing the ${\cal N}=3$ theory to the IR \cite{Aharony:2008ug}. Indeed, by integrating out all the massive fields, we recover the field content and interactions described in the field theoretical ${\cal N}=6$ construction. The enhancement to ${\cal N}=8$ for $k=1,2$ was properly discussed in \cite{Gustavsson:2009pm}.

It was realised in \cite{Aharony:2008ug} that under T-duality in the $x^6$ direction and uplifting the configuration to M-theory, the brane construction gets mapped to N M2-branes probing some configuration of KK-monopoles. These have a supergravity description in terms of hyper-K\"ahler geometries \cite{Gauntlett:1997pk}. Flowing to the IR in the dual gravitational picture is equivalent to probing the near horizon of these geometries, which includes the expected AdS${}_4$ factor times a quotient of the 7-sphere. 

The Chern-Simons theory has a $1/k$ coupling constant. Thus, large $k$ has a weakly coupled description.
At large $N$, it is natutal to consider the 't Hooft limit : $\lambda=N/k$ fixed. The gauge theory is weakly coupled for $k \gg  N$ and strongly coupled for $k \ll N$. In the latter situation, the supergravity description becomes reliable and weakly coupled for $N\gg 1$ 
 \cite{Aharony:2008ug}.

\paragraph{Matching field theory, branes and 3-algebra constructions :} The brane derivation of the supersymmetric field theory relevant to describe multiple M2-branes raised the natural question for what the connection was, if any, with the 3-algebra formulation that stimulated all these investigations. The answer was found in \cite{Bagger:2008se}. The main idea was to consider a 3-algebra based on a {\it complex} vector space endowed with a triple product
\begin{equation}
[T^a,T^b;{\overline T}^{\bar c}] = f^{ab{\bar c}}{}_{d}\,T^d,
\end{equation}
and an inner product
\begin{equation}
  h^{{\bar a}b} = {\rm Tr} \left({\overline T}^a T^b\right).
\end{equation}
The change in the notation points out antisymmetry only occurs in the first two indices. Furthermore, the constants
$f^{ab{\bar c}}{}_d$ satisfy the following
fundamental identity,
\begin{equation}\label{FI}
f^{ef{\bar g}}{}_bf^{cb{\bar a}}{}_d +f^{fe{\bar a}}{}_bf^{cb{\bar g}}{}_d+
f^{*{\bar g}{\bar a} f}{}_{\bar b} f^{ce{\bar b}}{}_d+f^{*{\bar a}{\bar g} e}{}_{\bar b} f^{cf{\bar b}}{}_d=0\,.
\end{equation}
It was proved in \cite{Bagger:2008se} that this set-up manages to close the algebra on the different fields giving rise to some set of equations of motion. In particular, the ${\cal N}=6$ conformal field theories described in \cite{Aharony:2008ug} could be rederived for the particular choices
\begin{equation}
  f^{ab\overline{c}\overline{d}}=-f^{ba\overline{c}\overline{d}}\,, \qquad {\rm and} \qquad f^{ab\overline{c}\overline{d}}=f^{*\overline{c}\overline{d}ab}.
\end{equation}
Thus, the 3-algebra approach based on complex vector spaces is also suitable to describe these string theory models. Furthermore, it provides with a mathematical formalism capable of describing more general set-ups.

\section{Related topics}
\label{sec:open}

There are several topics not included in previous sections that are also relevant to the subjects uncovered in this review.
The purpose of this last section is to mention some of them, mentioning their main ideas and/or approaches, and more importantly, referring the reader to some of the relevant references where they are properly developed and explained.

\paragraph{Superembedding approach :} The GS formulation consists in treating the bulk spacetime as a supermanifold while keeping the bosonic nature of the world volume. The {\it superembedding} formalism is a more symmetric formulation, in which both bulk and world volume are described as supermanifolds. As soon as the world volume formulation is extended into superspace, it incorporates extra degrees of freedom which are non-physical. There exists a geometrically natural interpretation for the set of constraints, first discussed in \cite{Sorokin:1989zi}, imposed to remove them.  Given a target space supervielbein $E^M(Z)=(E^a,\,E^\alpha)$ and world volume superconnection $e^A(\sigma,\eta)=(e^a,\,e^\alpha)$, where $\eta$ stands for the new world volume fermionic coordinates, then the pullback of the bosonic component can be expanded as
\begin{equation}
  E^a(Z(\sigma,\,\eta)) = e^bE_b^a + e^\alpha E_\alpha^a.
\end{equation}
The constraint consists in demanding
\begin{equation}
  E_\alpha^a(Z(\sigma,\eta))=0.
\label{eq:supcons1}
\end{equation}
This means that at any world volume point, the tangent space in the Grassmann directions forms a subspace of the Grassmann tangent space in the bulk.

There are many results in this subject, nicely reviewed in \cite{Sorokin:1999jx}. It is worth mentioning that some equations of motion for supersymmetric objects in different number of dimensions were actually first derived in this formalism rather than in the GS one, including \cite{Galperin:1992bw} for the d=10 superparticle, \cite{Bandos:1995zw} for the superstring and supermembrane, \cite{Howe:1996mx} for superbranes and \cite{Howe:1996yn} for the M5-brane\footnote{
The equivalence of the equations of motion obtained in the PST--formalism and the ones developed in the superembedding formalism was proved in \cite{Bandos:1997gm}.}. It is particularly relevant to stress the work done in formulating the M5-brane equations of motion covariantly \cite{Howe:1997fb,Howe:1997vn} and their use to identify supersymmetric world volume solitons \cite{Howe:1997ue,Howe:1997et}, and in pointing out the relation between superembeddings and non-linear realisations of supersymmetry \cite{Adawi:1997sq}.

\paragraph{MKK-monopoles and other exotic brane actions :} This review was focused on the dynamics of D-branes and M-branes. It is well known that string \& M theory have other extended objects, such as KK-monopoles or NS5-branes. There is a nice discussion regarding the identification of the degrees of freedom living on these branes in \cite{Hull:1997kt}. Subsequently, effective actions were written down to describe the dynamics of its bosonic sectors in \cite{Bergshoeff:1997gy,Bergshoeff:1998ef,Eyras:1998hn,Eyras:1998kx}. In particular, it was realised that gauged sigma models are able to encapsulate the right properties for KK monopoles. The results obtained in these references are consistent with the action of T-duality and S-duality.
It would of course be very interesting to include fermions in these actions and achieve kappa symmetry invariance.

\paragraph{Blackfolds :} The blackfold approach is suitable to describe the effective world volume dynamics of branes, still in the probe approximation, having a thermal population of excitations. In some sense, it describes the dynamics of these objects on length scales larger than the brane thickness. This formalism was originally developed in \cite{Emparan:2009cs,Emparan:2009at} and extended and embedded in string theory in \cite{Emparan:2011hg}. It was applied to the study of {\it hot} BIons in \cite{Grignani:2010xm,Grignani:2011mr}, emphasising the physical features not captured by the standard Dirac-Born-Infeld action, and to blackfolds in AdS \cite{Armas:2010hz}.

\paragraph{Non-relativistic kappa invariant actions :} All the branes described in this review are relativistic. It is natural to study their non-relativistic limits, both for its own sake, but also as an attempt to identifying new sectors of string theory that may be solvable. The latter is the direction originally pursued in \cite{Gomis:2000bd,Danielsson:2000gi} by considering closed strings in Minkowski. This was extended to closed strings in AdS${}_5\times$S${}^5$ in \cite{Gomis:2005pg}. At the level of brane effective actions in Minkoswki, non-relativistic diffeomorphism and kappa symmetry invariant versions of them were obtained in \cite{Gomis:2004pw} for D0-branes, fundamental strings and M2-branes, and later extended to general Dp-branes in \cite{Gomis:2005bj}. The consistency of these non-relativistic actions under the action of duality transformations was checked in \cite{Kamimura:2005rz}. This work was extended to non-relativistic effective D-brane actions in AdS${}_5\times$S${}^5$ in \cite{Brugues:2006yd,Sakaguchi:2006pg}.

\paragraph{Multiple M5-branes : } It is a very interesting problem to find the non-abelian formulation of the (2,0) tensor multiplet describing the dynamics of N M5-branes. Following similar ideas to the ones used in the construction of the multiple M2-brane action using 3-algebras, some non-abelian representation of the (2,0) tensor supermultiplet was found in \cite{Lambert:2010wm}. Their formulation includes a non-abelian analogue of the auxiliary scalar field appearing in the PST formulation of the abelian M5-brane. Closure of the superalgebra provides a set of equations of motion and constraints. Expanding the theory around a particular vacuum, gives rise to d=5 SYM along with an abelian (2,0) d=6 supermultiplet. This connection to d=5 SYM was further studied in \cite{Lambert:2010iw}. Some further work along this direction can be found in \cite{Honma:2011br}. Some of the BPS equations derived from this analysis were argued to be naturally reinterpreted in loop space \cite{Papageorgakis:2011xg}. There has been a different approach to the problem involving non-commutative versions of 3-algebras \cite{Gustavsson:2010nc}, but it seems fair to claim that this remains a very important open problem for the field.



\section{Acknowledgements}
\label{sec:acknow}
Some of the material uncovered in this review is based on the PhD {\it World Volume Approach to String Theory} defended by the author under the supervision of Prof. J.~Gomis at the University of Barcelona in May 2000. JS would like to thank his PhD advisor J.~Gomis for introducing him into this subject, to P.~K.~Townsend for sharing his extensive knowledge on many topics reviewed here, and to F.~Brandt, K.~Kamimura, O.~Lunin, D.~Mateos, A.~Ramallo and J.M.~Figueroa-O'Farrill for discussions and collaboration on part of the material reported in this work. The work of J.S. was partially supported by the Engineering and Physical Sciences Research Council [grant number EP/G007985/1].

\newpage

\begin{appendix}

\section{Target superspace formulation \& constraints}
\label{sec:appcons}

In this appendix, I very briefly mention the superspace formulation for ${\cal N}=2$ type IIA \cite{Carr:1986tk} and IIB 
 \cite{Howe:1983sra} and ${\cal N}=1$ d=11  \cite{Cremmer:1980ru,Brink:1980az} supergravity theories. The first goal is to set the relevant notation for the superfield components describing the physical massless fields coupling to the brane effective action degrees of freedom described in the main text. These are the physical fields appearing in the standard component formulation of these theories, i.e. eleven dimensional supergravity \cite{Cremmer:1978km}, its dimensional reduction \cite{Giani:1984wc} and type IIB \cite{Schwarz:1983qr,Howe:1983sra}. The second goal is to present
the set of constraints satisfied by these superfields ensuring both formulations are on-shell equivalent. The latter are
crucial to prove the kappa symmetry invariance of brane effective actions in curved backgrounds discussed in subsection \ref{sec:susycurve}.

\subsection{${\cal N}=2$ type IIA/B superspace}
\label{sec:iiab}

In components, ${\cal N}=2$ type IIA/B supergravities describe the dynamics of the gravity supermultiplet. The latter contains
\begin{itemize}
\item Type IIA : its bosonic sector contains metric $g_{mn}$, dilaton $\phi$, NS-NS 2-form $B_2$, RR potentials $C_r$ $r=1,3,5$, whereas its fermionic counterparts includes the dilatino $\lambda$ and the gravitino $\Psi_m$.
\item Type IIB : its bosonic sector contains metric $g_{mn}$, dilaton $\phi$, NS-NS 2-form $B_2$, RR potentials $C_r$ $r=0,2,4$, whereas its fermionic counterparts includes the dilatino $\lambda$ and the gravitino $\Psi_m$.
\end{itemize}
Both theories differ in the chiralities of their fermionic sectors and the dimensionality of their RR gauge potentials. Furthermore, the field strength of the RR 4-form potential in type IIB is {\it self-dual}.

To make the local supersymmetry of this component formalism manifest, one proceeds as in global supersymmetry by introducing the notion of superspace and superfields. The theory is defined on a {\it supermanifold} with local coordinates
$Z^M$ involving both bosonic $x^m$ and fermionic $\theta$ ones. The latter have chirality properties depending on the theory they are attached to. The physical content of the theory is described by superfields, tensors in superspace, defined as a polynomial expansion in the fermionic coordinates
\begin{equation}
  \Phi (x,\,\theta) = \phi(x) + \theta^\alpha\phi_\alpha(x) + \dots 
\end{equation}
whose components include the physical fields listed above. For an extensive and pedagogical introduction to the superfield and superspace formulation in supergravity, see \cite{Wess:1992cp}. 

A general feature of this formalism is that it achieves manifest invariance under supersymmetry at the expense of introducing an enormous amount of extra {\it unphysical} degrees of freedom, i.e. many of the different components of the superfields under consideration. If one wishes to establish an equivalence between these superspace formulations and the standard component ones, one must impose a set of constraints in the former to consistently, without breaking the manifest supersymmetry, reproduce the on-shell equations of motion from the latter. This relation appears schematically in figure \ref{fig7}. 

The superspace formulation of the ${\cal N}=2$ type IIA/B  supergravity multiplets is as follows:
\begin{itemize}
\item[1.] Given the presence of fermions, it is natural to work in local tangent frames. Thus, instead of using the metric variables $g_{mn}(x)$, one works in terms of bosonic vielbeins $E_m^a(x)$. These are then extended to a supervielbein 
$E_M^A(x,\theta)$, where $M=\{m,\,\alpha\}$ stands for the superspace curved indices, whereas $A=\{a,\,{\underline \alpha}\}$ describes both flat bosonic and fermionic tangent space indices. $E_M^A(x,\theta)$ already includes the gravitino $\Psi_m$ as a higher dimension component in its fermionic $\theta$ expansion. 
\item[2.] One extends all remaining bosonic fields to superfields with the same tensor structure, i.e. $B_2 = \frac{1}{2}B_{mn}(x)dx^m\wedge dx^n$ is extended to $B_2=\frac{1}{2}B_{AC}(x,\theta)E^A\wedge E^B$, where $E^{A}=dZ^{M}{E_{M}}^{A}$, and similarly for all other fields, including the dilaton.
\end{itemize}

The following discussion follows closely section 3 in \cite{Cederwall:1996ri}. As in Riemannian geometry, we can describe the geometry of a curved background in terms of a torsion and curvature two forms, but now in superspace:
\begin{eqnarray}
T^{A} &= & DE^{A}\equiv  dE^{A}+E^{B}\wedge{\omega_{B}}^{A}, \\
{R_{A}}^{B} &=& d{\omega_{A}}^{B}+{\omega_{A}}^{C}\wedge{\omega_{C}}^{B}.
\end{eqnarray}
The covariant derivative $D$ is defined in terms of a lorentzian connection one-form ${\omega_{A}}^{B}$, but in type IIB, it includes an additional $\U(1)$ connection defined on the coset space $\SU(1,1)/\U(1)$ where the set of type IIB scalars live
\cite{Howe:1983sra}. These superspace torsion and curvature forms satisfy the Bianchi identities
\begin{eqnarray}
DT^{A}&=&E^{B}\wedge{R_{B}}^{A}, \\
D{R_{A}}^{B}&=& 0. 
\label{TheFirstAndSecondBI}
\end{eqnarray}

The first of the constraints I was alluding to before is the, so called,
{\it Lorentzian assumption}. It amounts to the conditions
\begin{equation}
{\omega_{a}}^{{\underline \beta}}=0={\omega_{{\underline \alpha}}}^{b} \qquad \Rightarrow \qquad {R_{a}}^{{\underline \beta}}=0={R_{{\underline \alpha}}}^{b}\,.
\label{TheLorentzianAssumption}
\end{equation}
This guarantees the absence of non-trivial crossed terms between the bosonic and fermionic components of the connection and curvature in superspace. Conceptually, this is similar to the condition described in (\ref{eq:supcons1}) in the superembedding formalism \cite{Sorokin:1999jx}.

Some of the additional constraints involve the components of the super-field strengths of the different super-gauge potentials making up the superspace formulation for type IIA/B introduced above. Denote by $H_3$, the NS-NS super-three-form, by $R_n$, the RR super-$n$-forms, and define them as
\begin{equation}
H_{3} =dB_{2}, \qquad \qquad R= e^{B_{2}}\!\wedge d(e^{-B_{2}}\!\wedge C)\equiv\bigoplus_{n=1}^{10}
R_{n},
\label{eq:fsum}
\end{equation}
where I introduced the formal sum over all RR gauge potentials by $C\equiv\bigoplus_{n=0}^{9}C_{n}$ and proceeded analogously for their field strengths\footnote{The reader should keep in mind that the RR field strengths $R_{n}$ with $n\geq5$ are non-physical, in the sense that they are Hodge duality related to the physical propagating degrees of freedom  contained in the RR field strengths $R_{10-n}$ \cite{Douglas:1995bn,Duff:1992hu}.}. These obey the Bianchi identities
\begin{eqnarray}
dH_{3}&=&0, \\
dR-R\wedge H&=&0,
\label{BianchiIdentities}
\end{eqnarray}
and are invariant under a set of gauge transformations leaving the supergravity lagrangian invariant
\begin{eqnarray}
\delta B_{2}&=& d\lambda_1,\\
\delta C &=& e^{B_{2}}\,\wedge d\mu.
\label{TargetSpaceGaugeTransf}
\end{eqnarray}
Since the Bianchi identity (\ref{BianchiIdentities}) allows to set either the even or odd RR forms to zero, this reproduces the well known statement that on-shell type \II A \cite{Carr:1986tk} and \II B \cite{Howe:1983sra}
supergravities contain even and odd RR field strengths, respectively. To match the full on-shell supergravity formulation in standard components one must impose the following further set of constraints:
\begin{eqnarray}
& & \phantom{\hbox{\II A:\quad}}{T_{\underline\alpha\underline\beta}}^{c}
	=2i\Gamma_{\underline\alpha\underline\beta}^{c}\,, \qquad {T_{a\underline\beta}}^{c}=0\,,\\
& & \hbox{\II A:\quad}{T_{\underline\alpha\underline\beta}}^{\underline\gamma}=\frac{3}{2}{\delta_{(\underline\alpha}}^\Gamma\Lambda_{\underline\beta)}
	+2{(\Gamma_{\sharp} )_{(\underline\alpha}}^{\underline\gamma}(\Gamma_{\sharp} \Lambda)_{\underline\beta)}
	-\frac{1}{2}{(\Gamma_a)}_{\underline\alpha\underline\beta}(\Gamma^a\Lambda)^{\underline\gamma} \\
& & \quad\qquad\qquad +(\Gamma_a\Gamma_{\sharp} )_{\underline\alpha\underline\beta}(\Gamma^a\Gamma_{\sharp} \Lambda)^{\underline\gamma}	
	+\frac{1}{4}{(\Gamma_{ab})_{(\underline\alpha}}^{\underline\gamma}(\Gamma^{ab}\Lambda)_{\underline\beta)}\,,\\
& & \hbox{\II B:\quad}{T_{\underline\alpha\underline\beta}}^{\underline\gamma}=-{(J)_{(\underline\alpha}}^{\underline\gamma}(J\Lambda)_{\underline\beta)}
		+{(K)_{(\underline\alpha}}^{\underline\gamma}(K\Lambda)_{\underline\beta)} \\
& & \quad\qquad\qquad +\frac{1}{2}(\Gamma_aJ)_{\underline\alpha\underline\beta}(\Gamma^aJ\Lambda)^{\underline\gamma}
		-\frac{1}{2}(\Gamma_aK)_{\underline\alpha\underline\beta}(\Gamma^aK\Lambda)^{\underline\gamma}\,,\\
& & \nonumber \\
& & \hbox{\phantom{\II A:\quad}} H_{\underline\alpha\underline\beta\underline\gamma}=0\,,\\
& & \hbox{\II A:\quad}H_{a\underline\beta\underline\gamma}=-2i e^{\frac{\phi}{2}}(\Gamma_{\sharp} \Gamma_{a})_{\underline\beta\underline\gamma}\,,\\
& & \hbox{\phantom{\II A:\quad}}H_{ab\underline\gamma}
	=e^{\frac{\phi}{2}}(\Gamma_{ab} \Gamma_{\sharp} \Lambda)_{\underline\gamma}\,,\\
& & \hbox{\II B:\quad}H_{a\underline\beta\underline\gamma}=-2i e^{\frac{\phi}{2}}( K\Gamma_{a})_{\underline\beta\underline\gamma}\,,\\
& & \hbox{\phantom{\II A:\quad}}H_{ab\underline\gamma}
	=e^{\frac{\phi}{2}}(\Gamma_{ab}  K\Lambda)_{\underline\gamma}\,,\\
& & \nonumber \\
& & \hbox{\phantom{\II A:\quad}} R_{(n)\underline\alpha\underline\beta\underline\gamma A_{1}...A_{n-3}}=0\,,\\
& & \hbox{\II A:\quad}R_{(n)a_{1}...a_{n-2}\underline\alpha\underline\beta}=2i\,e^{\frac{n-5}{4}\phi}
(\Gamma_{a_{1}...a_{n-2}}(\Gamma_{\sharp} )^{\frac{n}{2}})_{\underline\alpha\underline\beta}\,,\\
& & \hbox{\phantom{\II A:\quad}}R_{(n)a_{1}...a_{n-1}\underline\alpha}
	=-\frac{n-5}{2}e^{\frac{n-5}{4}\phi}
	(\Gamma_{a_{1}...a_{n-1}}(-\Gamma_{\sharp} )^{\frac{n}{2}}\Lambda)_{\underline\alpha}\,,\\
& & \hbox{\II B:\quad}R_{(n)a_{1}...a_{n-2}\underline\alpha\underline\beta}=2i\,e^{\frac{n-5}{4}\phi}
	(\Gamma_{a_{1}...a_{n-2}}K^{\frac{n-1}{2}}I)_{\underline\alpha\underline\beta}\,,\\
& &  \hbox{\phantom{\II A:\quad}}R_{(n)a_{1}...a_{n-1}\underline\alpha}
	=-\frac{n-5}{2}e^{\frac{n-5}{4}\phi}
	(\Gamma_{a_{1}...a_{n-1}}K^{\frac{n-1}{2}}I\Lambda)_{\underline\alpha}\,.\\
& & \nonumber \\
& & \hbox{\phantom{\II A:\quad}}\Lambda_{\underline\alpha}=\frac{1}{2}\partial_{\underline\alpha}\phi\,.
\label{TheConstraints}
\end{eqnarray}
Here $\Gamma_\sharp=\Gamma_0\Gamma_1\dots \Gamma_9$ stands for the ten dimensional analogue of the $\gamma_5$ matrix in d=4, i.e. the chirality matrix, whereas $K$ and $J$ are $\SO(2)$ matrices appearing in the real formulation of type \II B supergravity \cite{Cederwall:1996ri}. In the last line, $\phi$ stands for the superfield containing the bulk dilaton, whereas $\Lambda_{\underline\alpha}$ has the appropriate dilatino as its leading component.

Even though the dual potential $B_6$ to the NS-NS 2-form $B_2$ does not explicitly appear in the kappa invariant D-brane effective actions reviewed in section \ref{sec:bbrane}, its field strength $H_7$ is relevant to understand
the solution to the Bianchi identities in type \II B, as explained in detail in \cite{Cederwall:1996ri}. For completeness, I include its definition below
\begin{eqnarray}
\hbox{\II A:\quad}&H_{7}= & dB_{6}-\frac{1}{2} C_{1}\wedge R_{6}
	+\frac{1}{2} C_{3}\wedge R_{4}-\frac{1}{2} C_{5}\wedge R_{2},\\
\hbox{\II B:\quad}&H_{7}= & dB_{6}+\frac{1}{2} C_{0}\wedge R_{7}
	-\frac{1}{2} C_{2}\wedge R_{5}+\frac{1}{2} C_{4}\wedge R_{3}
	-\frac{1}{2} C_{6}\wedge R_{1}.
\end{eqnarray}
By construction, these obey the constraints
\begin{eqnarray}
\hbox{\II A:\quad}&H_{a_{1}...a_{5}\underline\alpha\underline\beta}
	=& 2ie^{-\frac{\phi}{2}}(\Gamma_{a_{1}...a_{5}})_{\underline\alpha\underline\beta},\\
&H_{a_{1}...a_{6}\underline\alpha}= &-e^{-\frac{\phi}{2}}(\Gamma_{a_{1}...a_{6}}\Lambda)_{\underline\alpha},\\
\hbox{\II B:\quad}&H_{a_{1}...a_{5}\underline\alpha\underline\beta}
	=& 2ie^{-\frac{\phi}{2}}(\Gamma_{a_{1}...a_{5}}K)_{\underline\alpha\underline\beta},\\
&H_{a_{1}...a_{6}\underline\alpha}= &-e^{-\frac{\phi}{2}}(\Gamma_{a_{1}...a_{6}}K\Lambda)_{\underline\alpha},
\end{eqnarray}
and the Bianchi identities
\begin{eqnarray}
\hbox{\II A:\quad} dH_{7}+R_{2}\wedge R_{6}-\frac{1}{2} R_{4} \wedge R_{4}&=&0,\\
\hbox{\II B:\quad} dH_{7}+R_{1}\wedge R_{7}-R_{3}\wedge R_{5}&=&0.
\end{eqnarray}

\subsection{${\cal N}=1$ d=11 supergravity conventions}
\label{sec:d11app}

There is a similar discussion for ${\cal N}=1$ d=11 supergravity \cite{Cremmer:1978km} whose gravity supermultiplet involves metric $g_{mn}(x)$, a three gauge field potential $A_3(x)$, or its Hodge dual $A_6(x)$, and a gravitino $\Psi_m(x)$. When embedding this structure in ${\cal N}=1$ d=11 superspace  \cite{Cremmer:1980ru,Brink:1980az}, one uses local coordinates $Z^M = (x^m,\,\theta)$ where now $\theta$ stands for an eleven dimensional Majorana spinor having 32 real components. As before, the superfield encoding information about both the metric and gravitino is the supervielbein $E^A=dZ^{M}{E_{M}}^{A}$, the superfield extension of the bosonic vielbein ${E_{m}}^{a}$. The notation is as before, with the understanding that the current bosonic indices, both curved $(m)$ and tangent space $(a)$, run from 0 to 10.
Furthermore, $A_3(x)$ is extended into a superfield 3-form $A_3(x,\theta)$ with superspace components $A_{BCD}(x,\theta)$.

As in type \II A and B, it is natural to introduce the field strengths of these superfield potentials
\begin{eqnarray}
R_{4} &=& dA_{3}\, , \nonumber \\
R_{7} &=& dA_{6} +\frac{1}{2} A_{3}\wedge R_{4}\,,
\label{onea}
\end{eqnarray}
which are gauge invariant under the abelian gauge potential transformations
\begin{eqnarray}
\delta A_{3} &=& d\Lambda_{2}, \nonumber \\
\delta A_{6} &=& d\Lambda_{5} - \frac{1}{2} \Lambda_{2}\wedge R_{4}.
\label{aonea}
\end{eqnarray}

These superfields satisfy the set of constraints
\begin{eqnarray}
T^{\underline a} &=&-iE^{\underline \alpha}\wedge E^{\underline
\beta}\Gamma_{\underline{\alpha}\underline{\beta}}^{\underline a}+
E^{\underline b}\wedge E^{\underline \beta}\,
T_{\underline{b\beta}}^{\underline a}
+\frac{1}{2} E^{\underline b}\wedge E^{\underline c}
T_{\underline{b}\underline{c}}^{\underline a}\,, \\
R_{4}&=& \frac{1}{2}E^{\underline b}\wedge E^{\underline a}\wedge
E^{\underline\alpha}\wedge E^{\underline \beta} (\Gamma_{\underline{ab}})_{\underline{\alpha\beta}} 
+\frac{1}{{4!}}E^{\underline a}\wedge E^{\underline b}\wedge
E^{\underline c}\wedge E^{\underline d}R_{\underline{d}\underline{c}
\underline{b}\underline{a}}, \\
R_{7}&=& \frac{1}{{5!}}E^{\underline{a_1}}\wedge...\wedge
E^{\underline{a_5}}\wedge E^{\underline\alpha}\wedge E^{\underline \beta}
(\Gamma_{\underline{a_1}...\underline{a_5}})_{\underline{\alpha}
\underline{\beta}} + \frac{1}{7!}E^{{\underline{a}}_1} \dots E^{{\underline{a}}_7} R_{{\underline{a}}_7\dots{\underline{a}}_1}.
\end{eqnarray}
These allow to establish an equivalence between this superspace formulation and the on-shell supergravity component one.
They are also crucial to prove the kappa symmetry invariance of both M2 and M5-brane actions.

\section{Cone construction \& supersymmetry}
\label{sec:cone}

It is well known that S${}^d$ and AdS${}_d$ can be described as surfaces embedded in $\bR^{d+1}$ and $\bR^{2,d-1}$. What is less known, especially in the physics literature, is that geometric Killing spinors on the latter are induced from parallel spinors on the former. This was proved by B\"ar \cite{springerlink:10.1007/BF02102106} in the riemannian case and by Kath \cite{Kath} in the pseudo-riemannian case. In this appendix, I briefly review this result, based on some unpublished notes \cite{felipe-jose}.

Consider a riemannian spin manifold $(M,g)$ having geometric Killing spinors $\psi$ satisfying the differential equation
\begin{equation}
  \nabla_m \psi = -\frac{\epsilon}{2R}\Gamma_m\psi .
\end{equation}
$R$ is related to the curvature of the manifold and $\epsilon$ is a sign, to be spelled out below. From a physics point of view, the right hand side of this equation is the remnant of the gravitino supersymmetry transformation in the presence of non-trivial fluxes proportional to the volume form of the manifold $(M,g)$. Mathematically, it is a rather natural extension of the notion of covariantly constant Killing spinors. The statement that the manifold $(M,g)$ allows an embedding in a higher dimensional riemannian space $\widetilde M$ corresponds, metrically, to considering the metric of a cone $\widetilde g$ in $\widetilde M$ with base space $M$. Thus,
\begin{equation}
  \widetilde M = \bR^+ \times M \qquad {\rm and}\qquad \widetilde g =
  dr^2 + \left(\frac{r}{R}\right)^2 g~,
\end{equation}
where $R>0$ is the radius of curvature of $(M,g)$.  There exists a similar construction in the pseudo-riemannian case in which the cone is now along a timelike direction. In the following, I will distinguish two different cases, though part of the analysis will be done simultaneously:
\begin{itemize}
\item $(M^{d},g)$ riemannian with riemannian cone $(\widetilde
  M^{d+1},\widetilde g)$, and 
\item $(M^{1,d-1},g)$ lorentzian with pseudo-riemannian cone $(\widetilde M^{2,d-1},\widetilde g)$.
\end{itemize}

To establish an explicit map between Killing spinors in both manifolds, one needs to relate their spin connections. To do so, consider a local coframe $\theta^i$ for $(M,g)$ and $\widetilde\theta^a$ for $(\widetilde M,\widetilde g)$, defined as
\begin{equation}
  \widetilde\theta^r = dr \qquad {\rm and}\qquad \widetilde\theta^i =
  \frac{r}{R} \theta^i.
\end{equation}
The connection coefficients $\omega^i{}_j$ and $\widetilde\omega^a{}_b$ satisfy the corresponding structure equations
\begin{equation}
  d\theta^i + \omega^i{}_j\wedge \theta^j = 0 \qquad {\rm and}\qquad
  d\widetilde\theta^a + \widetilde\omega^a{}_b\wedge
  \widetilde\theta^b = 0~.
\end{equation}
Given the relation between coframes, the connections are related as
\begin{equation}
  \widetilde\omega^i{}_j = \omega^i{}_j \qquad {\rm and}\qquad
  \widetilde\omega^i{}_r = \frac1r \widetilde\theta^i = \frac1R
  \theta^i~.
\end{equation}
Let $\widetilde\nabla$ denote the spin connection on $(\widetilde M,
\widetilde g)$:
\begin{equation}
  \widetilde \nabla = d + \frac{1}{4} \widetilde\omega^{ab}
  \widetilde\Gamma_{ab}~,
\end{equation}
where $\widetilde\gamma_a$ are the gamma-matrices for the relevant
Clifford algebra. Plugging in the expression for the connection coefficients for the
cone, one finds
\begin{equation}
  \widetilde \nabla = d + \frac{1}{4} \omega^{ij}
  \widetilde\Gamma_{ij} + \frac{1}{2R} \theta^i
  \widetilde\Gamma_{ir}.
\end{equation}
To continue we have to discuss the embedding of Clifford algebras in
order to recognise the above connection intrinsically on $(M,g)$.
This requires distinguishing two cases, according to the signature of
$(M,g)$.

\subsection{$(M,g)$ riemannian}
\label{sec:riemcone}

When $(M,g)$ is riemannian, there exists a natural embedding of the Clifford algebra $\Cl(d,0)$ into
$\Cl(d+1,0)^{{\rm even}}$: 
\begin{equation}
  \Cl(d,0) \hookrightarrow \Cl(d+1,0)^{{\rm even}} \qquad {\rm where}\qquad
  \Gamma_i \mapsto \varepsilon \widetilde\Gamma_i \widetilde\Gamma_r.
\end{equation}
This embedding depends on a sign $\varepsilon$ (for ``embedding'') and has the property that $\Gamma_{ij} \mapsto
\widetilde\Gamma_{ij}$. Thus, it embeds $\fspin(d,0)$ into $\fspin(d+1,0)$.  When $d$ is odd, the dimension of the minimal spinor representation in $\widetilde M$ is double the one in $M$. In this case, the Clifford valued volume form $\nu$ in both manifolds is mapped as follows
\begin{equation}
  \nu_{d,0} \mapsto \varepsilon \nu_{d+1,0}.
\end{equation}
Thus, spinors in $M$ will be mapped to spinors of a definite chirality in $\widetilde M$.

Plugging this embedding into the expression for $\widetilde\nabla$, one
sees that a $\widetilde\nabla$-parallel spinor $\widetilde\psi$
in the cone, restricts to $(M,g)$ to a geometric Killing
spinor $\psi = \widetilde\psi|_{r=R}$ obeying
\begin{equation}
  \nabla_X \psi = - \frac{\varepsilon}{2R} X \cdot \psi.
\end{equation}
This is the defining equation for a geometric Killing 
Furthermore,
\begin{itemize}
\item if $d$ is even: there exists a one--to--one correspondence between parallel spinors $\widetilde\psi$ in $\widetilde M^{d+1}$
  and geometric Killing spinors $\psi$ in $M^{d}$; and
\item if $d$ is odd: there exists a one--to--one correspondence between parallel spinors $\widetilde\psi$ in $M^{d+1}$ of definite chirality\footnote{The chirality can be $\pm 1$ or $\pm i$, depending on the reality of the volume form eigenspace.} eigenvalues and geometric Killing spinors $\psi$ in $M^{d}$. 
\end{itemize}

\subsection{$(M,g)$ of signature $(1,d-1)$}
\label{sec:lorcone}

When $(M,g)$ is lorentzian, there also exists a natural embedding of the Clifford algebra $\Cl(1,d-1)$ into
$\Cl(2,d-1)^{{\rm even}}$ depending on a sign $\varepsilon$:
\begin{equation}
  \Cl(1,d-1) \hookrightarrow \Cl(2,d-1)^{{\rm even}} \qquad {{\rm where}}\qquad
  \Gamma_i \mapsto \varepsilon \widetilde\Gamma_i \widetilde\Gamma_r~.
\end{equation}
This embedding induces an embedding $\fspin(d,0) \hookrightarrow
\fspin(d+1,0)$, $\Gamma_{ij} \mapsto \widetilde\Gamma_{ij}$.  Moreover
if $d$ is odd, one has
\begin{equation}
  \nu_{1,d-1} \mapsto \varepsilon \nu_{2,d-1}.
\end{equation}
Plugging this into the expression for $\widetilde\nabla$, we
see that a $\widetilde\nabla$-parallel spinor $\widetilde\psi$
in the cone, restricts to $(M,g)$ to a geometric Killing
spinor $\psi = \widetilde\psi|_{r=R}$ obeying
\begin{equation}
  \nabla_X \psi = - \frac{\varepsilon}{2R} X \cdot \psi~.
\end{equation}
Furthermore,
\begin{itemize}
\item if $d$ is even: there exists a one--to--one correspondence between parallel spinors $\widetilde\psi$ in $\widetilde M^{2,d-1}$
  and geometric Killing spinors $\psi$ in $M^{1,d-1}$; and
\item if $d$ is odd: there exists a one--to--one correspondence between parallel spinors $\widetilde\psi$ in $\widetilde M^{2,d-1}$
  with definite chirality\footnote{Same comments as above.} and geometric Killing spinors $\psi$ in $M^{1,d-1}$.
\end{itemize}

\end{appendix}



\bibliography{simon-biblio}

\end{document}